\RequirePackage{fix-cm}
\documentclass[smallextended]{svjour3}       
\smartqed  
\usepackage[ruled,vlined]{algorithm2e}
\usepackage{bm}
\usepackage{bbm}
\usepackage{graphicx}
\usepackage[numbers,sort&compress]{natbib}
\usepackage{stackengine}
\usepackage{subfigure}
\usepackage{amssymb}
\usepackage{amsmath}
\usepackage{booktabs}
\usepackage{multirow}
\usepackage{xcolor}
\usepackage{tikz}
\usetikzlibrary{shapes.misc}
\usetikzlibrary{positioning,arrows,arrows.meta}
\tikzset{cross/.style={cross out, draw, 
         minimum size=2*(#1-\pgflinewidth), 
         inner sep=0pt, outer sep=0pt},
         cross/.default={0.2cm}}
\usepackage[margin=1.2in]{geometry}
\definecolor{PaleBlue}{HTML}{92DFF3}
\definecolor{PaleYellow}{HTML}{FFFBBD}
\definecolor{MacFolderBlue}{RGB}{94,174,213}

\newcommand{\half}{\frac{1}{2}}
\newcommand{\dt}{\Delta t}
\newcommand{\dx}{\Delta x}
\newcommand{\dy}{\Delta y}

\newcommand{\grad}{\nabla}
\newcommand{\divergence}{\nabla \cdot}

\newcommand{\Fv}{\mathbf{F}}
\newcommand{\Gv}{\mathbf{G}}
\newcommand{\Sv}{\mathbf{S}}
\newcommand{\Wv}{\mathbf{W}}
\newcommand{\Uv}{\mathbf{U}}
\newcommand{\Vv}{\mathbf{V}}
\newcommand{\Rv}{\mathbf{R}}
\newcommand{\Av}{\mathbf{A}}
\newcommand{\Bv}{\mathbf{B}}
\newcommand{\Cv}{\mathbf{C}}

\newcommand{\Fxv}{\mathbf{F}^x}
\newcommand{\Fyv}{\mathbf{F}^y}

\newcommand{\Gxv}{\mathbf{G}^x}
\newcommand{\Gyv}{\mathbf{G}^y}
\newcommand{\Gxhv}{\hat{\mathbf{G}}^x}
\newcommand{\Gyhv}{\hat{\mathbf{G}}^y}
\newcommand{\Gxtv}{\tilde{\mathbf{G}}^x}
\newcommand{\Gytv}{\tilde{\mathbf{G}}^y}

\newcommand{\Shv}{\hat{\mathbf{S}}}

\newcommand{\Qv}{\mathbf{Q}}
\newcommand{\Phiv}{\mathbf{\Phi}}
\newcommand{\Wcv}{\bm{\mathcal{W}}}

\newcommand{\vel}{\bm{u}}

\newcommand{\RN}[1]{%
  \textup{\uppercase\expandafter{\romannumeral#1}}%
}

\newcommand{\sign}{\mathrm{sign}}

\newtheorem{mylemma}{Lemma}

\begin{document}

\title{A positivity-preserving Eulerian two-phase approach with thermal relaxation for compressible flows with a liquid and gases}
\titlerunning{A positivity-preserving Eulerian two-phase approach for a liquid and gases}


\author{Man Long Wong         \and
        Jordan B. Angel       \and \\
        Cetin C. Kiris}

\authorrunning{M.L. Wong, J.B. Angel, and C.C. Kiris} 

\institute{Man Long Wong \at
              Science and Technology Corporation, Moffett Field, CA 94035, United States \\
              \email{manlong.wong@nasa.gov}           
           \and
           Jordan B. Angel \at
              NASA Ames Research Center, Moffett Field, CA 94035, United States
           \and
           Cetin C. Kiris \at
              NASA Ames Research Center, Moffett Field, CA 94035, United States
}

\date{Received: date / Accepted: date}

\maketitle

\begin{abstract}
A positivity-preserving fractional algorithm is presented for solving the four-equation homogeneous relaxation model (HRM) with an arbitrary number of ideal gases and a liquid governed by the stiffened gas equation of state. The fractional algorithm consists of a time step of the hyperbolic five-equation model by Allaire et al. and an algebraic numerical thermal relaxation step at an infinite relaxation rate. Interpolation and flux limiters are proposed for the use of high-order Cartesian finite difference or finite volume schemes in a general form such that the positivity of the partial densities and squared sound speed, as well as the boundedness of the volume fractions and mass fractions, are preserved with the algorithm. A conservative solution update for the four-equation HRM is also guaranteed by the algorithm which is advantageous for certain applications such as those with phase transition. The accuracy and robustness of the algorithm with a high-order explicit finite difference weighted compact nonlinear scheme (WCNS) using the incremental-stencil weighted essentially non-oscillatory (WENO) interpolation, are demonstrated with various numerical tests.
\keywords{diffuse interface method \and multi-phase \and fractional algorithm \and relaxation system \and shock-capturing \and weighted essentially non-oscillatory (WENO)}
\end{abstract}

\section{Introduction}

In this work, a two-phase flow model with the assumptions of mechanical and thermal equilibria is considered, i.e. all species are assumed to have the same velocity, pressure, and temperature.
This is a homogeneous relaxation model (HRM)~\cite{flaatten2011relaxation, lund2012hierarchy} that can be reduced from the seven-equation model by~\citet{baer1986two} in the limit of zero mechanical and thermal relaxation times.
The model is sometimes simply called four-equation model~\cite{allaire2002five,saurel2016general,chiapolino2017simple} as the model with two species only consists of four equations for one-dimensional (1D) flows. 
This model is popular in the literature on numerical methods for multi-material or multi-phase flows, such as~\cite{saurel1999simple, abgrall2001computations, kunz2000preconditioned, venkateswaran2002computation, lund2012two,le2014towards,saurel2016general,chiapolino2017simple}, where many of these works involve simulations of flows with phase transition.
The four-equation HRM is also commonly used for studying compressible interfacial instabilities or turbulence, such as Rayleigh--Taylor and Richtmyer--Meshkov instabilities/turbulence \cite{mellado2005large,olson2007rayleigh,reckinger2016comprehensive,hill2006large,thornber2010influence,wong2019high,wong2022analysis} where buoyancy, viscous, or subgrid-scale effects can also be modeled easily by including additional terms.
When this model is discretized in an Eulerian framework without any viscous effects, a small amount of artificial mixing is allowed at the interfaces between fluids through numerical dissipation, thus this numerical flow model belongs to the family of diffuse-interface methods~\cite{saurel2018diffuse}.
Low-order (second order or lower) numerical methods are commonly adopted to spatially discretize the four-equation model due to their robustness despite large dissipation errors at the interfaces. While high-order interface-capturing methods can be used theoretically for less smeared diffuse interfaces due to their higher resolution and more localized dissipation properties, numerical failures can occur easily for extreme problems such as cavitation and flashing flows that involve shocks or vacuum regions, even when shock-capturing techniques are employed. The numerical difficulties can be attributed to the mixture sound speed that is close to the sound speed formulation by Wood~\cite{wood1941textbook,wallis1969texbook},
which has non-monotonic variation with volume fraction for common two-phase flows, such as water-air flows.

We propose a robust numerical approach to obtain the solutions to the four-equation HRM using the five-equation model by~\citet{allaire2002five} with the numerical thermal relaxation at infinite relaxation rate.
The five-equation model has a continuity equation for each species, a mixture momentum equation, a mixture total energy equation, and a volume fraction advection equation for each species, except the last one. The species are only assumed to have the same velocity and pressure, i.e. are in mechanical equilibrium, in this five-equation model.
While the two models considered are termed four-equation HRM and five-equation model for simplicity in this work, the former and the latter in fact have $N + D + 1$ and $2N + D$ equations respectively for a problem with $D$ dimensions and $N$ species.
It is proved in this work that the five-equation model is reduced to the four-equation model in the limit of infinite thermal relaxation rate.
In order to obtain the solutions to the four-equation model, a fractional algorithm is proposed, 
where the five-equation model by Allaire et al. and stiff thermal relaxation are solved alternatively.
Our proposed method is different than the previous works mentioned earlier where the four-equation HRM is solved directly.
This approach is similar to the papers on relaxation methods~\cite{saurel2009simple,pelanti2014mixture,pelanti2019numerical} based on six-equation models, where a hyperbolic step of the models is advanced, followed by a sequence of mechanical, thermal or/and chemical relaxation steps during a full time stepping. Particularly, it is shown that the six-equation models can be reduced to another five-equation model by Kapila et al.~\cite{kapila2001two,murrone2005five} if only the stiff mechanical relaxation is carried out.
While both belong to five-equation models, the mixture sound speed of the five-equation model by Allaire et al. has monotonic characteristic for liquid-gas flows but
that by Kapila et al. is given by Wood's formulation. As explained in~\cite{saurel2009simple,pelanti2014mixture}, the non-monotonic behaviour of Wood's mixture sound speed leads to two sonic points in the numerical diffuse interface. This may affect the propagation of acoustics waves across the numerical interface and pose challenges of the construction of robust numerical methods. 
As the four-equation model also has the non-monotonic Wood-like mixture sound speed,
the fractional algorithm formed by hyperbolic solve of the five-equation model and stiff thermal relaxation is numerically more advantageous than solving the four-equation model directly.

In the case of a two-phase flow that is composed of many ideal gases and a liquid governed by the stiffened gas equation of state, we can construct a positivity-preserving fractional algorithm. The positivity-preserving method can preserve the positivity of partial densities and squared mixture sound speed, as well as the boundedness of volume fractions after each full time step.
A high-order positivity-preserving method for the five-equation model by Allaire et al. is presented in our earlier paper for ideal gas-liquid (stiffened gas) flows~\cite{wong2021positivity}, under a mild assumption that the ratio of specific heats of the stiffened gas is larger than that of the ideal gas. While this assumption is generally true for gas-liquid applications, the algorithm is also limited to the combination of one liquid with one gas. In this work, we have extended the method for flows that are composed of a stiffened gas and an arbitrary number of ideal gases. There is also no assumption on the relative magnitude between the ratios of specific heats of different species. This diffuse interface method is practical for many engineering applications such as the simulation of water-based acoustic suppression systems under extreme rocket launch environments that involve liquid water with many gases, such as air, rocket exhaust gases, water vapor, etc.
Through a numerical experiment, we can verify that the Wood-like mixture sound speed of the four-equation model can be obtained with the fractional approach.
The Wood-like mixture sound speed is a much better approximation to the Wood's sound speed for bubbly flows~\cite{wilson2008audible,wilson2005phase} than the monotonic sound speed of the five-equation model by Allaire et al. Therefore, the thermally relaxed four-equation model is believed to be more suitable for capturing sound propagation in dispersed multi-phase flows or under-resolved mixtures of liquids and gases.
Moreover, the thermal relaxation step is computationally efficient as it only requires an algebraic solve without stencil operations.

Unlike the quasi-conservative methods~\cite{karni1994multicomponent, abgrall1996prevent, housman2009time1, housman2009time2, johnsen2006implementation, nonomura2012numerical} that sacrifice the conservation of conservative variables for robustness, the fractional algorithm is a fully conservative method for the four-equation HRM while preserving positivity. The fully conservative property is advantageous to the extension of the method for more complicated multi-phase flows, such as cavitating, boiling, and evaporating flows with phase transition. The phase transition effects can be robustly modeled by simply adding a thermo-chemical relaxation step~\cite{pelanti2014mixture,chiapolino2017simple} after the thermal relaxation, since the state after thermal relaxation is always guaranteed to have bounded mass fractions and volume fractions.

The paper is organized as follows. A review of the five-equation model by~\citet{allaire2002five} is first provided in section~\ref{sec:five_eqn_model}. In section~\ref{sec:five_eqn_model_relaxation_model}, it is proved that the five-equation model with the infinitely fast thermal relaxation is reduced to the four-equation HRM. Then, a first order fractional algorithm to solve the five-equation model with the thermal relaxation and 
the extension of the positivity-preserving fractional algorithm to higher order spatial discretization with Runge--Kutta time integration are discussed in section~\ref{sec:fractional_algorithm}. In section~\ref{sec:test_problems}, numerical tests are shown to highlight the high accuracy and robustness of the proposed finite difference scheme. Finally, concluding remarks are given in section~\ref{sec:conclustion}.


\section{The five-equation model by Allaire et al.} \label{sec:five_eqn_model}

The single-velocity and single-pressure five-equation model proposed by~\citet{allaire2002five} for compressible multi-species/multi-phase flows is first introduced. If there are $N$ species, the model has the following form:
\begin{equation}
\begin{split}
    \partial_t \left( \alpha_k \rho_k \right)  + \nabla \cdot \left( \alpha_k \rho_k \bm{u} \right) & =0  \quad \forall k \in \left[1, N \right] , \\
    \partial_t \left( \rho \bm{u} \right) + \nabla \cdot \left( \rho \bm{u} \otimes \bm{u} \right) + \nabla p &= 0, \\
    \partial_t E + \nabla \cdot \left[ \left( E + p \right) \bm{u} \right] &= 0, \\
    \partial_t \alpha_k + \bm{u} \cdot \nabla \alpha_k &= 0  \quad \forall k \in \left[1, N-1 \right] ,
\end{split}
\label{eq:system_5eq}
\end{equation}
where $\vel$ and $p$ are the velocity vector and the pressure of the mixture respectively.
$\vel = u$ for one-dimensional (1D), $\vel = ( u \ v )^{T}$ for two-dimensional (2D), and $\vel = ( u \ v \ w)^{T}$ for three-dimensional (3D) cases.
$\alpha_k$ and
$\rho_k$ are respectively the volume fraction and species density of species $k$,
where $k=1,2,\cdots,N$. The partial densities of species $k$ are $\alpha_k \rho_k = \rho Y_k$  where $\rho$ is the mixture density
and $Y_k$ is mass fraction of species $k$.
$E = \rho ( e + \lvert \vel \rvert^2 / 2 )$ 
is the mixture total energy per unit volume, where $e$ is the mixture
specific internal energy. 
The constraint on volume fractions $\alpha_k$ and the definitions of the mixture density $\rho$ and the mixture specific internal energy $e$ are given by:
\begin{align}
    1      &= \sum_{k=1}^{N} \alpha_k, \\
    \rho   &= \sum_{k=1}^{N} \alpha_k \rho_k = \sum_{k=1}^{N} \rho Y_k , \\
    \rho e &= \sum_{k=1}^{N }\alpha_k \rho_k e_k ,
\end{align}
where $e_k$ is the species specific internal energy of species $k$. This model given by the system of equations~\eqref{eq:system_5eq} is named the five-equation model as there are five equations for 1D two-fluid flows. Although it is called five-equation model in this work, there are in fact $2N + D$ equations for a problem with $D$ dimensions and $N$ species.

The system of equations is closed with the equation of state of each species and the species pressure equilibrium assumption. In this work, each species is assumed to obey the modified form~\cite{le2004elaborating, saurel2008modelling} of stiffened gas equation of state~\cite{harlow1971fluid, menikoff1989riemann} in the following general expression:
\begin{equation}
  \frac{p_k}{\gamma_k - 1} + \frac{\gamma_k p_k^\infty}{\gamma_k - 1} = \rho_k (e_k - q_k) , \label{eq:species_stiffened_EOS}
\end{equation}
where $\gamma_k$, $p_k^\infty$, and $q_k$ are the fitting parameters for each of the fluids. The parameter $\gamma_k$ is the ratio of specific heats which is assumed greater than one and $p_k^\infty$ is a non-negative fluid property.
The species is an ideal gas if $p_k^\infty = 0$. In addition, the species temperature $T_k$ is given by
\begin{equation}
    T_k = \frac{p_k + p_k^\infty}{\left( \gamma_k - 1 \right)  \rho_k c_{v,k}} ,
\end{equation}
where $c_{v,k}$ is the species specific heat at constant volume of species $k$. $c_{v,k}$ is related to $c_{p,k}$, which is the species specific heat at constant pressure, through  the ratio of specific heats $\gamma_k$,
\begin{equation}
    \gamma_k = \frac{c_{p,k}}{c_{v,k}} .
\end{equation}
By assuming a single pressure among all species, i.e.
$p_1 = p_2 = \cdots = p_N = p$, we can obtain a mixture equation of state by multiplying each species equation of state by $\alpha_k$ and summing over all species. This gives us
\begin{equation}
  \frac{p}{\overline \gamma-1} + \frac{\overline \gamma ~\overline p^\infty}{\overline \gamma-1} = \rho (e - \bar{q}),
\end{equation}
where $\overline{\gamma}$, $\overline{p^\infty}$, and $\bar{q}$ are mixture properties that are defined by the following relations:
\begin{align}
  \frac{1}{\overline \gamma-1} &= \sum_{k=1}^{N} \frac{\alpha_k}{\gamma_k-1}, \\
  \frac{\overline \gamma ~\overline p^\infty}{\overline \gamma-1} &= \sum_{k=1}^{N} \frac{\alpha_k\gamma_k p_k^\infty}{\gamma_k-1}, \\
  \overline{q} &= \sum_{k=1}^{N} Y_k q_k . \label{eq:q_bar}
\end{align}
This model has a mixture speed of sound $c$ given by
\begin{align} 
  \rho c^2 &= \overline \gamma\left(p+\overline p^\infty\right) \\
           &= \overline \gamma(\overline \gamma -1)\left[\rho \left( e - \overline{q} \right) - \overline p^\infty\right].
\end{align}

The strictly admissible set of states, $G$, for this many-species model is defined by extending that in our previous paper~\cite{wong2021positivity}:
\begin{equation}
  G = \left\{\left.\Wv = 
  \begin{pmatrix}
    \alpha_1 \rho_1 \\
    \alpha_2 \rho_2 \\
    \vdots \\
    \alpha_{N} \rho_{N} \\
    \rho \bm{u} \\
    E      \\
    \alpha_1 \\
    \alpha_2 \\
    \vdots \\
    \alpha_{N-1} \\
  \end{pmatrix}
 \ \right| \  \alpha_k \geq 0,
 \ \alpha_k \rho_k \geq 0,\ \forall k \in \left[1, N \right],\ \rho c^2 > 0 \right\},
 \label{eq:G_set}
\end{equation}
where $\Wv$ is the conservative variable vector given by
\begin{equation}
\Wv = \left( \alpha_1 \rho_1 \ \alpha_2 \rho_2 \ \cdots \alpha_N \rho_N \ \rho \bm{u} \ E \ \alpha_1 \ \alpha_2 \cdots \alpha_{N-1} \right)^T\footnote{Strictly speaking, $\alpha_k$ are not conservative variables.}.
\end{equation}
This means that a solution state is only considered physically admissible if all volume fractions, all partial densities. This also implies that all mass fractions $Y_k$ and volume fractions $\alpha_k$ are bounded between zero and one.
The squared sound speed $c^2$ is also required to be positive such that the system of equations remains hyperbolic with real wave speeds.

The positivity-preserving procedures on the five-equation model in the previous work~\cite{wong2021positivity} require that the set $G$ be convex. This can be guaranteed if $\rho c^2$ or $\left[ \rho \left( e - \overline{q} \right) - \overline p^\infty \right]$ is a concave function of the conservative variables $\Wv$. In the case of flows consisting of two species, this condition is satisfied (even with the more general form of stiffened gas equation of state in this work) if one of the species is an ideal gas and the ratio of specific heats of the liquid (stiffened gas) is larger than that of the ideal gas\footnote{The condition is always satisfied for flows with only ideal gases.}. However, this is not generally true for many-species flows, even if only one of the species is stiffened gas and others are ideal gases.

As will be discussed in a later section, positivity-preserving numerical methods can still be constructed if the five-equation model is solved with an algebraic thermal relaxation step for flows with one stiffened gas and an arbitrary number of ideal gases. This requires the use of a less restrictive admissible set $\mathcal{G}$ that is convex for any compositions of species described by the modified form of stiffened gas equation of state. $\mathcal{G}$ is defined as:
\begin{equation}
  \mathcal{G} = \left\{\left.\Wv = 
  \begin{pmatrix}
    \alpha_1 \rho_1 \\
    \alpha_2 \rho_2 \\
    \vdots \\
    \alpha_{N} \rho_{N} \\
    \rho \bm{u} \\
    E      \\
    \alpha_1 \\
    \alpha_2 \\
    \vdots \\
    \alpha_{N-1} \\
  \end{pmatrix}
 \ \right| \ \alpha_k \geq 0,
 \ \alpha_k \rho_k \geq 0,\ \forall k \in \left[1, N \right],\ \rho(e - \overline{q}) > 0 \right\},
 \label{eq:mathcal_G_set}
\end{equation}
It should be noted that $\rho(e - \overline{q})$ can be written in terms of the conservative variables as
\begin{equation}
    \rho(e - \overline{q}) = \left( E - \frac{1}{2} \frac{\lvert \rho \bm{u} \rvert ^2}{\rho} - \sum_{k=1}^{N} \alpha_k \rho_k q_k \right) .
\end{equation}
It is obvious that $\alpha_k$ and $\alpha_k \rho_k$ are both concave functions of the conserved variables $\Wv$. In order to prove that $\mathcal{G}$ is a convex set, 
it is also needed to show that $\rho \left( e - \overline{q} \right)$ is a concave function of $\Wv$.
\begin{mylemma}
  The function $\rho \left( e - \overline{q} \right)$ is a concave function of the conserved variables $\Wv$.
  \label{lem:concave}
\end{mylemma}
\begin{proof}
  The non-zero eigenvalues of the Hessian matrix of the function are:
  \begin{equation}
    \left\{-\frac{1}{\sum_{k=1}^{N} \alpha_k \rho_k}, \ -\frac{N \lvert \rho \bm{u} \rvert ^2 + \left( \sum_{k=1}^{N} \alpha_k \rho_k \right)^2}
          {\left(\sum_{k=1}^{N} \alpha_k \rho_k \right)^3} \right\} .
  \end{equation}
  The first non-zero eigenvalue does not exist for the 1D case. Since all eigenvalues are either negative or zero, the lemma is proved.
\end{proof}

By using Jensen's inequality and the fact that all the left hand sides of the inequalities in equation~\eqref{eq:mathcal_G_set} are concave functions of $\Wv$, the set $\mathcal{G}$ is proved to be convex.


\section{Proposed relaxation model: five-equation model by Allaire et al. with infinitely fast thermal relaxation} \label{sec:five_eqn_model_relaxation_model}

The following analytical form of the augmented five-equation model by~\citet{allaire2002five} with the thermal relaxation term is proposed as the governing equations:
\begin{equation}
\begin{split}
    \partial_t \left( \alpha_k \rho_k \right) + \nabla \cdot \left( \alpha_k \rho_k \bm{u} \right) & =0 \quad \forall k \in \left[1, N \right] , \\
    \partial_t \left( \rho \bm{u} \right) + \nabla \cdot \left( \rho \bm{u} \otimes \bm{u} \right) + \nabla p &= 0 , \\
    \partial_t E + \nabla \cdot \left[ \left( E + p \right) \bm{u} \right] &= 0 , \\
    \partial_t \alpha_k + \bm{u} \cdot \nabla \alpha_k &= \sum_{i = 1, i \neq k}^{N} H_{i,k} \left( T_i - T_k \right) \quad \forall k \in \left[1, N-1 \right] .
\end{split}
\label{eq:relaxation_system_5eq}
\end{equation}
Artificial relaxation source terms $\sum_{i = 1, i \neq k}^{N} H_{i,k} \left( T_i - T_k \right)$ are added to the right hand sides of the volume fraction equations compared to the original five-equation model,
where $T_k$ is the temperature of species $k$ and the relaxation coefficients $H_{i,j}$ have the following properties:
\begin{equation}
    H_{i,j} \geq 0, \quad H_{i,j} = H_{j,i} \quad \forall i,j .
\end{equation}

In this work, we aim at obtaining the solutions to equation~\eqref{eq:relaxation_system_5eq} with infinite relaxation rate or zero relaxation time, i.e. $H_{ij} \to \infty$. With this condition, the relaxation system can be proved to reduce to the four-equation homogeneous relaxation model, or HRM:
\begin{equation}
\begin{split}
    \partial_t \left( \alpha_k \rho_k \right) + \nabla \cdot \left( \alpha_k \rho_k \bm{u} \right) & =0 \quad \forall k \in \left[1, N \right] , \\
    \partial_t \left( \rho \bm{u} \right) + \nabla \cdot \left( \rho \bm{u} \otimes \bm{u} \right) + \nabla p &= 0 , \\
    \partial_t E + \nabla \cdot \left[ \left( E + p \right) \bm{u} \right] &= 0 ,
\end{split}
\label{eq:relaxed_system_4eq}
\end{equation}
where all species are at pressure and thermal equilibria, i.e. $T_1 = T_2 = \cdots = T_N = T$ and $p_1 = p_2 = \cdots = p_N = p$.

To prove that the relaxed system of equation~\eqref{eq:relaxation_system_5eq} is the four-equation HRM given by equation~\eqref{eq:relaxed_system_4eq}, we follow the method by~\citet{flaatten2010wave}. For simplicity, only the 1D version of equation~\eqref{eq:relaxation_system_5eq} is considered but the proof can be easily generalized for multi-dimensional space.
A 1D hyperbolic system in a general relaxation form similar to~\cite{chen1994hyperbolic,flaatten2010wave} is given by:
\begin{equation}
    \partial_t \Wv + \partial_x \left[ \Fv \left( \Wv \right) \right] + \Cv \partial_x \Phiv = \frac{1}{\varepsilon} \Rv \left( \Wv \right) ,
    \label{eq:relaxation_system}
\end{equation}
where $\Wv$ is the solution vector of conservative variables with size $M$ and $\varepsilon$ is the relaxation time. The system is given a $m \times M$ constant coefficient matrix $\Qv$ with rank $m < M$ such that
\begin{equation}
    \Qv \Rv \left( \Wv \right) = \mathbf{0} \quad \forall \Wv.
\end{equation}
It is further assumed that $\Qv \Cv d \Phiv$ is a perfect differential:
\begin{equation}
    \Qv \Cv d \Phiv = d \Gv \left( \Wv \right) .
\end{equation}
Therefore, we obtain a conservation law for the reduced variable $\Wcv = \Qv \Wv$ by multiplying both sides of equation~\eqref{eq:relaxation_system} by $\Qv$:
\begin{equation}
    \partial_t \Wcv + \partial_x \left[ \Qv \Fv \left( \Wv \right) + \Gv \left( \Wv \right) \right] = \mathbf{0}.
    \label{eq:relaxed_system_1}
\end{equation}
In addition, it is assumed that each $\Wcv$ uniquely determines a local equilibrium value of the relaxation system given by equation~\eqref{eq:relaxation_system} through function $\bm{\mathcal{E}}$, i.e. $\Wv = \bm{\mathcal{E}} \left( \Wcv \right)$ and $\Rv \left( \bm{\mathcal{E}} \left( \Wcv \right) \right) = \mathbf{0}$ is satisified as well as
\begin{equation}
    \Qv \bm{\mathcal{E}} \left( \Wcv \right) = \Wcv \quad \forall \Wcv .
    \label{eq:Q_requirement}
\end{equation}
Finally, equation~\eqref{eq:relaxed_system_1} can be closed as a reduced system by imposing the local equilibrium condition for $\Wv$, i.e.
\begin{align}
    \Wv &= \bm{\mathcal{E}} \left( \Wcv \right) , \\
    \partial_t \Wcv + \partial_x \bm{\mathcal{F}} \left( \Wcv \right) &= \mathbf{0} ,
    \label{eq:relaxed_system_2}
\end{align}
where $\bm{\mathcal{F}}$ is the reduced flux given by
\begin{equation}
    \bm{\mathcal{F}} \left( \Wcv \right) = \Qv \Fv \left( \bm{\mathcal{E}} \left( \Wcv \right) \right) + \Gv \left( \bm{\mathcal{E}} \left( \Wcv \right) \right) .
\end{equation}

For the 1D version of the proposed five-equation relaxation model, it can be easily seen that:
\begin{equation}
    \Wv = 
  \begin{pmatrix}
    \alpha_1 \rho_1 \\
    \alpha_2 \rho_2 \\
    \vdots \\
    \alpha_{N} \rho_{N} \\
    \rho \bm{u} \\
    E      \\
    \alpha_1 \\
    \alpha_2 \\
    \vdots \\
    \alpha_{N-1} \\
  \end{pmatrix}, \quad
  \Fv \left( \Wv \right) =
  \begin{pmatrix}
    \alpha_1 \rho_1 u \\
    \alpha_2 \rho_2 u \\
    \vdots \\
    \alpha_{N} \rho_{N} u \\
    \rho u^2 + p \\
    (E + p) u  \\
    0 \\
    0 \\
    \vdots \\
    0 \\
  \end{pmatrix}, \quad
  \Phiv =
  \begin{pmatrix}
    \alpha_1 \\
    \alpha_2 \\
    \vdots \\
    \alpha_{N-1} \\
  \end{pmatrix} .
\end{equation}
The $(2N+1) \times (N-1)$ matrix $\Cv$ and the relaxation source, $\Rv$, are given by
\begin{equation}
    \Cv =
    u
    \begin{bmatrix}
      0      & 0      & \cdots & 0 \\
      0      & 0      & \cdots & 0 \\
      \vdots & \vdots & \ddots & \vdots \\
      0      & 0      & \cdots & 0 \\
      1      & 0      & \cdots & 0 \\
      0      & 1      & \cdots & 0 \\
      \vdots & \vdots & \ddots & \vdots\\
      0      & 0      & \cdots & 1 \\
    \end{bmatrix},  \quad
    \Rv \left( \Wv \right) = 
    \begin{pmatrix}
        0 \\
        0 \\
        \vdots \\
        \sum_{i = 1, i \neq 1}^{N} h_{i,1} \left( T_i - T_1 \right) \\
        \sum_{i = 1, i \neq 2}^{N} h_{i,2} \left( T_i - T_2 \right) \\
        \vdots \\
        \sum_{i = 1, i \neq N-1}^{N} h_{i,N-1} \left( T_i - T_{N-1} \right) \\
    \end{pmatrix} ,
\end{equation}
where
\begin{equation}
    \frac{h_{i,j}}{\varepsilon} = H_{i,j}.
\end{equation}
When the coefficients $H_{i,j}$ tend to infinity or relaxation time $\varepsilon$ tends to zero, we achieve instantaneous thermal equilibrium for the reduced system, i.e.:
\begin{equation}
    T_i = T_j = T \quad \forall i, j .
\end{equation}
Lastly, the $(N+2) \times (2N + 1)$ matrix $\Qv$ for the relaxation system is given by
\begin{equation}
    \Qv =
    \begin{bmatrix}
        1      & 0      & \cdots & 0      & 0      & \cdots & 0 \\
        0      & 1      & \cdots & 0      & 0      & \cdots & 0 \\
        \vdots & \vdots & \ddots & \vdots & \vdots & \ddots & \vdots \\
        0      & 0      & \cdots & 1      & 0      & \cdots & 0 \\
    \end{bmatrix} .
\end{equation}
The assumption given by equation~\eqref{eq:Q_requirement} is satisfied. By assuming that there are more than two species, i.e. $N \geq 2$, the required assumption on the rank of $\Qv$ is also met since
\begin{equation}
    \mathrm{rank} \left( \Qv \right) = N + 2 < M = 2N + 1 .
\end{equation}
Furthermore, we have:
\begin{equation}
    \Wcv = 
    \begin{pmatrix}
        \alpha_1 \rho_1 \\
        \alpha_2 \rho_2 \\
        \vdots \\
        \alpha_{N} \rho_{N} \\
        \rho u \\
        E      \\
    \end{pmatrix} , \quad
    \bm{\mathcal{F}} =
    \begin{pmatrix}
        \alpha_1 \rho_1 u \\
        \alpha_2 \rho_2 u \\
        \vdots \\
        \alpha_{N} \rho_{N} u \\
        \rho u^2 + p \\
        (E + p) u   \\
    \end{pmatrix} .
\end{equation}
Therefore, it is proved that the four-equation HRM is the reduced system of the five-equation relaxation model with infinite thermal relaxation rates, i.e. $H_{i,j} \to \infty$.

The four-equation HRM is closed with a mixture equation of state with the isothermal and isobaric equilibrium assumptions. Given an equation of state for each species, the mixture pressure and temperature are computed with:
\begin{align}
    p = p \left( \alpha_1 \rho_1, \alpha_2 \rho_2, \dots , \alpha_N \rho_N, e \right) &=
        p_1 \left( \rho_1, e_1 \right) =
        p_2 \left( \rho_2, e_2 \right) = 
        \cdots = 
        p_N \left( \rho_N, e_N \right) , \\
    T = T \left( \alpha_1 \rho_1, \alpha_2 \rho_2, \dots , \alpha_N \rho_N, e \right) &=
        T_1 \left( \rho_1, e_1 \right) =
        T_2 \left( \rho_2, e_2 \right) =
        \cdots =
        T_N \left( \rho_N, e_N \right) .
\end{align}
The sound speed of the four-equation HRM, $c_{\mathrm{4}\text-\mathrm{eqn}}$, is given by:
\begin{equation}
    {c_{\mathrm{4}\text-\mathrm{eqn}}}^2 = \sum_{k=1}^{N} Y_k \Psi_k + \Gamma \frac{p}{\rho} , \label{eq:4_eqn_sos}
\end{equation}
where the partial derivatives of the mixture pressure, $\Psi_k$ and $\Gamma$ (Gr\"{u}neisen parameter), are given by:
\begin{equation}
    \Psi_k = \frac{\partial p}{\partial \left( \alpha_k \rho_k \right)}, \quad
    \Gamma = \frac{1}{\rho} \frac{\partial p}{\partial e} .
\end{equation}
For a mixture with one stiffened gas and an arbitrary number of ideal gases, the explicit forms of mixture pressure and temperature given by the equation of state and the partial derivatives for the mixture sound speed are given in a later section for the four-equation HRM.


\section{Fractional algorithm for the four-equation homogeneous relaxation model (HRM)} \label{sec:fractional_algorithm}

In the previous section, it is proved that the five-equation model relaxes to the four-equation HRM when the relaxation source terms have infinite rates. In order to obtain the numerical solutions to the four-equation HRM robustly, a fractional approach that is composed of a hyperbolic step of the five-equation model and an infinitely fast numerical thermal relaxation step is proposed. With the use of first order Harten--Lax--van Leer contact (HLLC) scheme and forward Euler time stepping, this approach can be shown to be positivity-preserving for a stiffened gas with an arbitrary number of ideal gases. In other words, the solutions that are initially in set $G$ are advanced to a state in set $G$ after a full step with the fractional algorithm.
It will also be shown that the positivity-preserving algorithm can also be extended for any Cartesian high-order finite difference or finite volume schemes in a general form and strong stability preserving Runge--Kutta (SSP-RK) time stepping methods~\cite{shu1988total,gottlieb2009high,gottlieb2001strong} when the proposed
positivity-preserving limiters are used.

\subsection{Hyperbolic step}

During the hyperbolic step, the solutions of the five-equation model by Allaire et al. without the relaxation source terms are computed numerically. The underlying partial differential equations (PDEs) are given by equation~\eqref{eq:system_5eq}.
For simplicity, the 2D governing equations with domain $[x_a, x_b]\times[y_a, y_b]$ is considered in this section.
Furthermore, the domain is discretized uniformly into a Cartesian grid with $N_x \times N_y$ grid points. Hence, the domain is covered by cells $I_{i,j}=\left[x_{i-1/2},\ x_{i+1/2}\right]\times\left[y_{j-1/2},\ y_{j+1/2}\right]$ for $1 \leq i \leq N_x$, $1 \leq j \leq N_y$, where the grid midpoints are given by:
\begin{equation}
  x_{i+\half} = x_a + i \Delta x, \quad
  y_{j+\half} = y_a + j \Delta y
\end{equation}
\noindent and
\begin{equation}
    \Delta x = \frac{x_b - x_a}{N_x}, \quad \Delta y = \frac{y_b - y_a}{N_y} .
\end{equation}

Numerical discretizations in different directions can be treated independently, with the use of exact or approximate Riemann solver for quasi-1D Riemann problems in the corresponding directions. For a quasi-1D Riemann problem in the $x$ direction, we can drop the index in the $y$ direction and the first order approximate solutions at grid $i$ from step $n$ to step $n+1$ with time step size $\dt$ are given by the cell-averaged value:
\begin{equation}
\begin{split}
    \Wv_i^{n+1} &= \frac{1}{\dx} \int_{x_{i-\half}}^{x_{i}} R\left( \frac{x - x_{i-\half}}{\dt}, \Wv_{i-1}^{n}, \Wv_i^{n} \right) dx \\
    &\quad + \frac{1}{\dx} \int_{x_{i}}^{x_{i+\half}}R\left(\frac{x - x_{i+\half}}{\dt}, \Wv_{i}^{n}, \Wv_{i+1}^{n} \right) dx ,
\end{split} \label{eq:soln_convex_averaging}
\end{equation}
where $R \left( x/t, \ \Wv_L, \ \Wv_R \right)$ are the self-similar solutions given by the Riemann solver with left and right states, $\Wv_L$ and $\Wv_R$. The approximate solutions are only valid if the approximate waves generated at the two midpoints $x_{i \pm 1/2}$ 
do not interact under a suitable Courant-–Friedrichs-–Lewy (CFL) condition.
In our previous paper~\cite{wong2021positivity}, the positivity-preservation of the HLLC Riemann solver for the five-equation model with a liquid (stiffened gas) and an ideal gas was mathematically proved. In this work,
the solver is extended for flows with one liquid and many ideal gases. It will also be shown that the extended HLLC Riemann solver is quasi-positivity-preserving 
such that
solutions that are originally in set $G$ are advanced to states in set $\mathcal{G}$ after the hyperbolic stepping.

When a Riemann solver is used for a conservative hyperbolic system, the approximate conservative flux at each cell midpoint can be obtained using the divergence theorem. As the system of equations of the five-equation model is quasi-conservative, it is more convenient to define a quasi-conservative flux vector at each midpoint for the numerical discretization. The quasi-conservative flux vector is conservative for all equations except the advection equations of volume fractions.
In order to derive the quasi-conservative numerical flux, the system given by equation~\eqref{eq:system_5eq} is first rewritten in the following flux-source form:
\begin{equation}
    \partial_t \Wv + \partial_x \Gxv \left( \Wv \right) + \partial_y \Gyv \left( \Wv \right) = \Sv \left( \Wv, \grad \Wv \right), \label{eq:W_G_eqn}
\end{equation}
where
\begin{equation}
\begin{split}
    \Gxv \left( \Wv \right) =
        \begin{pmatrix}
            \alpha_1 \rho_1 u       \\
            \alpha_2 \rho_2 u       \\
            \vdots                  \\
            \alpha_N \rho_N u       \\
            \rho u^2 + p            \\
            \rho v u                \\
            \left( E + p \right) u  \\
            \alpha_1 u              \\
            \alpha_2 u              \\
            \vdots                  \\
            \alpha_{N-1} u          \\
        \end{pmatrix}, \quad
    \Gyv \left( \Wv \right) =
        \begin{pmatrix}
            \alpha_1 \rho_1 v       \\
            \alpha_2 \rho_2 v       \\
            \vdots                  \\
            \alpha_N \rho_N v       \\
            \rho u v                \\
            \rho v^2 + p            \\
            \left( E + p \right) v  \\
            \alpha_1 v              \\
            \alpha_2 v              \\
            \vdots                  \\
            \alpha_{N-1} v          \\
        \end{pmatrix}, \quad
    \Sv \left( \Wv, \grad \Wv \right) =
        \begin{pmatrix}
            0 \\
            0 \\
            \vdots \\
            0 \\
            0 \\
            0 \\
            0 \\
            \alpha_1 \divergence \vel \\
            \alpha_2 \divergence \vel \\
            \vdots \\
            \alpha_{N-1} \divergence \vel
        \end{pmatrix} .
\end{split}
\end{equation}
The conservative flux vectors $\Fxv$ and $\Fyv$ are defined as
\begin{equation}
\begin{split}
    \Fxv \left( \Wv \right) =
        \begin{pmatrix}
            \alpha_1 \rho_1 u       \\
            \alpha_2 \rho_2 u       \\
            \vdots                  \\
            \alpha_N \rho_N u       \\
            \rho u^2 + p            \\
            \rho v u                \\
            \left( E + p \right) u  \\
            0                       \\
            0                       \\
            \vdots                  \\
            0                       \\
        \end{pmatrix}, \quad
    \Fyv \left( \Wv \right) =
        \begin{pmatrix}
            \alpha_1 \rho_1 v       \\
            \alpha_2 \rho_2 v       \\
            \vdots                  \\
            \alpha_N \rho_N v       \\
            \rho u v                \\
            \rho v^2 + p            \\
            \left( E + p \right) v  \\
            0                       \\
            0                       \\
            \vdots                  \\
            0                       \\
        \end{pmatrix}. \label{eq:F_x_F_y_def}
\end{split}
\end{equation}
It can be easily seen that flux vectors $\Fxv$ ($\Fyv$) and $\Gxv$ ($\Gyv$) are related by:
\begin{align}
    \Gxv &= \Fxv + \left( 0\ 0\ \cdots \ 0\ 0\ 0\ 0\ f^x_{\alpha_1}\ f^x_{\alpha_2}\ \cdots\ f^x_{\alpha_{N-1}} \right)^T, \\
    \Gyv &= \Fyv + \left( 0\ 0\ \cdots \ 0\ 0\ 0\ 0\ f^y_{\alpha_1}\ f^y_{\alpha_2}\ \cdots\ f^y_{\alpha_{N-1}} \right)^T,
\end{align}
where $f^x_{\alpha_k} = \alpha_k u$ and $f^y_{\alpha_k} = \alpha_k v$.

The fully discretized form of equation~\eqref{eq:W_G_eqn} with first order accurate forward Euler time integration is given by:
\begin{equation}
    \frac{\Wv^{n+1}_{i,j} - \Wv^{n}_{i,j}}{\dt}
    + \frac{\Gxhv_{i+\half,j}
            - \Gxhv_{i-\half,j}}{\dx} 
      + \frac{\Gyhv_{i,j+\half}
            - \Gyhv_{i,j-\half}}{\dy} =
      \Shv_{i,j} , \label{eq:fully_discretized_W_G_eqn}
\end{equation}
where
\begin{equation}
    \Shv_{i,j} =
      \begin{pmatrix}
              0 \\
              0 \\
              \vdots \\
              0 \\
              0 \\
              0 \\
              0 \\
              \alpha_{1,i,j}^n \left(
        \frac{\hat{u}_{i+\half,j}
            - \hat{u}_{i-\half,j}}{\dx} 
      + \frac{\hat{v}_{i,j+\half}
            - \hat{v}_{i,j-\half}}{\dy}
            \right) \\
              \alpha_{2,i,j}^n \left(
        \frac{\hat{u}_{i+\half,j}
            - \hat{u}_{i-\half,j}}{\dx} 
      + \frac{\hat{v}_{i,j+\half}
            - \hat{v}_{i,j-\half}}{\dy}
            \right) \\
            \vdots \\
              \alpha_{N-1,i,j}^n \left(
        \frac{\hat{u}_{i+\half,j}
            - \hat{u}_{i-\half,j}}{\dx} 
      + \frac{\hat{v}_{i,j+\half}
            - \hat{v}_{i,j-\half}}{\dy}
            \right)
          \end{pmatrix} . \label{eq:discretized_S}
\end{equation}
$\hat{\mathbf{G}}^{x}_{i\pm 1/2,j}$ ($\hat{\mathbf{G}}^{y}_{i,j\pm 1/2}$) is obtained through the approximation of $\hat{\mathbf{F}}^{x}_{i\pm 1/2,j}$ ($\hat{\mathbf{F}}^{y}_{i,j\pm 1/2}$) and $f^{x}_{\alpha,i\pm 1/2,j}$ ($f^{y}_{\alpha,i,j\pm 1/2}$) since
\begin{align}
\begin{split}
  \Gxhv_{i-\half,j} &=
      \Fv^{x}_{i-\half,j}
      + ( 0 \ 0 \ \cdots \ 0 \ 0 \ 0 \ 0 \ 
      f^{x}_{\alpha_1,i-\half,j} \ 
      f^{x}_{\alpha_2,i-\half,j} \ 
      \cdots \ 
      f^{x}_{\alpha_{N-1},i-\half,j} \ 
      )^T ,
\end{split}
  \\
\begin{split}
  \Gxhv_{i+\half,j} &=
      \Fv^{x}_{i+\half,j}
      + ( 0 \ 0 \ \cdots \ 0 \ 0 \ 0 \ 0 \ 
      f^{x}_{\alpha_1,i+\half,j} \ 
      f^{x}_{\alpha_2,i+\half,j} \ 
      \cdots \ 
      f^{x}_{\alpha_{N-1},i+\half,j} \ 
      )^T .
\end{split}
\end{align}

\subsubsection{First order HLLC Riemann solver}

The first order approximation of $\hat{\mathbf{F}}^{x}_{i\pm 1/2,j}$, $\hat{\mathbf{F}}^{y}_{i,j\pm 1/2}$, $f^{x}_{\alpha,i\pm 1/2,j}$ and $f^{y}_{\alpha,i,j\pm 1/2}$ with the HLLC Riemann solver is discussed in this sub-section by extending the previous paper~\cite{wong2021positivity}.

The HLLC approximate solutions at a midpoint for a quasi-1D Riemann problem in the $x$ direction are given by:
\begin{equation}
  \Wv^{\mathrm{HLLC}} 
  = \left\{ \begin{array}{ll} 
      \Wv_L,     &  \text{if }  s_L > 0,\\ 
      \Wv_{*,L}, &  \text{if }  s_L \leq 0 < s_*, \\
      \Wv_{*,R}, &  \text{if }  s_* \leq 0 \leq s_R,\\
      \Wv_R,     &  \text{if }  s_R < 0 , 
             \end{array}\right  .
      \label{eq:HLLC_states}
\end{equation}
where $L$ and $R$ are the left and right states respectively at the midpoint. With $K = L$ or $R$, the star state for the five-equation model for a 2D problem is given by:
\begin{equation}
  \Wv_{*,K}  = 
  \begin{pmatrix}
    \chi_{*,K} \left(\alpha_1 \rho_1\right)_K \\
    \chi_{*,K} \left(\alpha_2 \rho_2\right)_K \\
    \vdots                                    \\
    \chi_{*,K} \left(\alpha_N \rho_N\right)_K \\
    \chi_{*,K} \rho_K s_*                     \\
    \chi_{*,K} \rho_K v_K                     \\
    \chi_{*,K}\left[ E_K + (s_*-u_K)\left(\rho_K s_* + \frac{p_K}{s_K-u_K}\right) \right] \\
    \alpha_{1,K}                              \\
    \alpha_{2,K}                              \\
    \vdots                                    \\
    \alpha_{N-1,K}                            \\
  \end{pmatrix} .
\end{equation}
\noindent $\chi_{*K}$ is defined as:
\begin{equation}
    \chi_{*K} = \frac{s_K - u_K}{s_K - s_*}, 
\end{equation}
and the wave speeds are given by~\citet{einfeldt1991godunov,batten1997choice}:
\begin{align}
    s_{L} \left( \Wv_L, \Wv_R \right) &= \min{\left( \bar{u} - \bar{c}, u_L - c_L \right)} , \\
    s_{R} \left( \Wv_L, \Wv_R \right) &= \max{\left( \bar{u} + \bar{c}, u_R + c_R \right)} , \\
    s_{*} \left( \Wv_L, \Wv_R \right) &= \frac{p_R - p_L + \rho_L u_L \left( s_L - u_L \right) - \rho_R u_R \left( s_R - u_R \right)}{\rho_L \left( s_L - u_L \right) - \rho_R \left( s_R - u_R \right) } , \label{eq:s_star_definition}
\end{align}
where $\bar{u}$ and $\bar{c}$ can be approximated with the arithmetic averages from the left and right states.

The first order accurate approximations of $\hat{\mathbf{G}}^{x}_{i\pm 1/2,j}$ with the first order HLLC solutions are given by:
{
\small
\begin{align}
\begin{split}
  \Gxhv_{i-\half,j} &=
  \Gv^{x,\mathrm{HLLC}}_{i-\half,j}
  \left( \Wv_{i-1,j}, \Wv_{i,j} \right) \\
      &=
      \Fv^{x,\mathrm{HLLC}}_{i-\half,j} \left(\Wv_{i-1,j}, \Wv_{i,j}\right) \\
      &\quad + ( 0 \ 0 \ \cdots \ 0 \ 0 \ 0 \ 0 \ \\
      &\quad \quad f^{x, \mathrm{HLLC}}_{\alpha_1,i-\half,j} \left( \Wv_{i-1,j}, \Wv_{i,j} \right) \ 
      f^{x, \mathrm{HLLC}}_{\alpha_2,i-\half,j} \left( \Wv_{i-1,j}, \Wv_{i,j} \right) \ 
      \cdots \ 
      f^{x, \mathrm{HLLC}}_{\alpha_{N-1},i-\half,j} \left( \Wv_{i-1,j}, \Wv_{i,j} \right) )^T ,
\end{split}
  \\
\begin{split}
  \Gxhv_{i+\half,j} &=
  \Gv^{x,\mathrm{HLLC}}_{i+\half,j}
  \left( \Wv_{i,j}, \Wv_{i+1,j} \right) \\
      &=
      \Fv^{x,\mathrm{HLLC}}_{i+\half,j} \left(\Wv_{i,j}, \Wv_{i+1,j}\right) \\
      &\quad + ( 0 \ 0 \ \cdots \ 0 \ 0 \ 0 \ 0 \ \\
      &\quad \quad f^{x, \mathrm{HLLC}}_{\alpha_1,i+\half,j} \left( \Wv_{i,j}, \Wv_{i+1,j} \right) )^T \ 
      f^{x, \mathrm{HLLC}}_{\alpha_2,i+\half,j} \left( \Wv_{i,j}, \Wv_{i+1,j} \right) )^T \ 
      \cdots \ 
      f^{x, \mathrm{HLLC}}_{\alpha_{N-1},i+\half,j} \left( \Wv_{i,j}, \Wv_{i+1,j} \right) )^T ,
\end{split}
\end{align}
}%
where
\begin{align}
    \Fv^{x,\mathrm{HLLC}}_{i-\half,j} \left(\Wv_{i-1,j}, \Wv_{i,j}\right)  &= \Fv^{x,\mathrm{HLLC}} \left(R^{\mathrm{HLLC}}\left(0, \Wv_{i-1,j}, \Wv_{i,j}\right)\right) , \\
    \Fv^{x,\mathrm{HLLC}}_{i+\half,j} \left(\Wv_{i,j}, \Wv_{i+1,j}\right) &= \Fv^{x,\mathrm{HLLC}} \left(R^{\mathrm{HLLC}}\left(0, \Wv_{i,j}, \Wv_{i+1,j}\right)\right) .
\end{align}
The conservative HLLC flux vectors $\Fv^{x,\mathrm{HLLC}}_{i\pm 1/2,j}$ can be obtained with the approximate HLLC solutions using the divergence theorem~\cite{batten1997choice}:
\begin{equation}
  \begin{split}
    & \Fv^{x,\mathrm{HLLC}}\left(R^{\mathrm{HLLC}}\left(0, \Wv_L, \Wv_R\right)\right) = \\
    &\quad \frac{1+\sign(s_*)}{2}\left(\Fv_L + s_-\left(\Wv_{*,L}-\Wv_L\right)\right) 
     + \frac{1-\sign(s_*)}{2}\left(\Fv_R + s_+\left(\Wv_{*,R}-\Wv_R\right)\right) , 
  \end{split}
\end{equation}
where
\begin{equation}
    s_{-} = \min{\left( 0, s_L \right)}, \quad s_{+} = \max{\left( 0, s_R \right)} .
\end{equation}
Note that the last $N-1$ components of $\Fv^{x,\mathrm{HLLC}}$ for the advection equations are zero. The discretization of the advection equations are contributed by $\hat{u}_{i\pm 1/2,j}$ and $f^{x, \mathrm{HLLC}}_{\alpha_k,i\pm 1/2,j}$ which are given by the first order accurate approximations as:
\begin{align}
    \hat{u}_{i-\half,j} &= u_{*,i-\half,j} =
    u^{\mathrm{HLLC}}_{*} \left( \Wv_{i-1,j}, \Wv_{i,j} \right) = s_{*} \left( \Wv_{i-1,j}, \Wv_{i,j} \right) , \\
    \hat{u}_{i+\half,j} &= u_{*,i+\half,j} =
    u^{\mathrm{HLLC}}_{*} \left( \Wv_{i,j}, \Wv_{i+1,j} \right) = s_{*} \left( \Wv_{i,j}, \Wv_{i+1,j} \right) ,
\end{align}
and
\begin{align}
\begin{split}
    f^{x,\mathrm{HLLC}}_{\alpha_k,i-\half,j} \left( \Wv_{i-1,j}, \Wv_{i,j} \right) &=
        \frac{1+\sign(s_{*,i-\half,j})}{2}(\alpha_{k,i-1,j} s_{*,i-\half,j}) \\
        &\quad + \frac{1-\sign(s_{*,i-\half,j})}{2}(\alpha_{k,i,j} s_{*,i-\half,j}) ,
\end{split}
    \\
\begin{split}
    f^{x,\mathrm{HLLC}}_{\alpha_k,i+\half,j} \left( \Wv_{i,j}, \Wv_{i+1,j} \right) &=
        \frac{1+\sign(s_{*,i+\half,j})}{2}(\alpha_{k,i,j} s_{*,i+\half,j}) \\
        &\quad + \frac{1-\sign(s_{*,i+\half,j})}{2}(\alpha_{k,i+1,j} s_{*,i+\half,j}) .
\end{split}
\end{align}
The expressions given above form the first order accurate solutions of volume fractions given by:
\begin{equation}
    {\alpha_k}_i^{n+1} = {\alpha_k}_i^n - \frac{u_{+,i-\half} \dt}{\dx}\left({\alpha_k}^n_i - {\alpha_k}^n_{i-1}\right) - \frac{u_{-,i+\half} \dt}{\dx}\left({\alpha_k}^n_{i+1} - {\alpha_k}^n_i\right) \quad \forall k \in \left[1, N-1 \right] ,
\end{equation}
where $u_{+,i-1/2} = \max\{0,s_{*,i-1/2}\}$ and $u_{-,i+1/2} = \min\{0,s_{*,i+1/2}\}$. Note that this time advancement of the volume fractions satisfies the finite volume approximation given by equation~\eqref{eq:soln_convex_averaging}.

Finally, the quasi-conservative flux vector $\hat{\mathbf{G}}^{x,\pm}_{i\mp 1/2,j}$ (similarly for $\hat{\mathbf{G}}^{y,\pm}_{i,j\mp 1/2}$) is introduced:
\begin{equation}
  \hat{\mathbf{G}}^{x,\pm}_{i\mp\half,j} =
    \Gxhv_{i\mp\half,j}
    - \hat{u}_{i\mp\half,j}  \ (0\ 0\ \cdots \ 0\ 0\ 0\ 0\ \alpha_{1,i,j} \ \alpha_{2,i,j} \ \cdots \ \alpha_{N-1,i,j} )^T , \label{eq:G_pm}
\end{equation}
where equation~\eqref{eq:fully_discretized_W_G_eqn} can be simplified to:
\begin{equation}
    \frac{\Wv^{n+1}_{i,j} - \Wv^{n}_{i,j}}{\dt}
    + \frac{\hat{\mathbf{G}}^{x,-}_{i+\half,j}
          - \hat{\mathbf{G}}^{x,+}_{i-\half,j}}{\dx} 
    + \frac{\hat{\mathbf{G}}^{y,-}_{i,j+\half}
          - \hat{\mathbf{G}}^{y,+}_{i,j-\half}}{\dy} =
    \mathbf{0} . \label{eq:fully_discretized_W_G_pm_eqn}
\end{equation}
While $\hat{\mathbf{G}}^{x,\pm}_{i\mp 1/2,j}$ are called quasi-conservative flux vectors, the components of $\hat{\mathbf{G}}^{x,\pm}_{i\mp 1/2,j}$, or $\hat{\mathbf{G}}^{x,\mathrm{HLLC},\pm}_{i\mp 1/2,j}$ more precisely in this sub-section, for all conservative equations, except the advection equations, are conservative numerical fluxes for the corresponding equations and are equivalent to the corresponding components of $\Fv^{x,\mathrm{HLLC}}_{i\mp 1/2,j}$.

\subsubsection{Proof of quasi-positivity-preservation for first order HLLC solver}

For a quasi-1D Riemann problem, it is shown in equation~\eqref{eq:soln_convex_averaging} that the solution update is the convex averaging of the approximate solutions to the problem. 
As a result, with an initial state in $G$, the HLLC Riemann solver gives quasi-positivity-preserving solution in set $\mathcal{G}$ if all possible states in equation~\eqref{eq:HLLC_states} can be proved to be in set $\mathcal{G}$, using Jensen's inequality for integral equations. Only the left star state is considered here but the right star state can be proved to be quasi-positivity-preserving by symmetry. 

From the definition of $s_*$ given by equation~\eqref{eq:s_star_definition}, it can be easily determined that $s_L < s_*$~\cite{batten1997choice}. Moreover, since  $s_L = \min{\left( \bar{u} - \bar{c}, u_L - c_L \right)}$, we have $s_L < u_L$. This means all partial densities in the star state are positive:
\begin{align}
  \left( \alpha_k \rho_k \right)_{*,L} &= \frac{s_L-u_L}{s_L-s_*} \left( \alpha_k \rho_k \right)_L \geq 0 \quad \forall k \in \left[1, N \right] .
\end{align}
The mixture density is also positive as $\rho_{*,L} = \sum_{k=1}^{N} \left( \alpha_k \rho_k \right)_{*,L}$ and hence:
\begin{equation}
  \rho_{*,L} = \frac{s_L-u_L}{s_L-s_*}\rho_L \geq 0 .
\end{equation}
The positivity of mixture and partial densities implies that all mass fractions are bounded between zero and one. As for the volume fraction, since $\alpha_{k,*,L} = \alpha_{k,L}$,
\begin{equation}
    0 \leq \alpha_{1,*,L} \leq 1 \quad \forall k \in \left[1, N \right].
\end{equation}
The last requirement to complete the proof of quasi-positivity-preserving for the HLLC Riemann solver is to show $[\rho (e - \overline{q})]_{*,L} > 0$. From the definition of $[\rho (e - \overline{q})]_{*,L}$, we have:
\begin{align}
  [\rho (e - \overline{q})]_{*,L} &=
    E_{*,L} - \half\frac{\lvert \rho \bm{u}_{*,L} \rvert^2}{\rho_{*,L}} - (\rho \overline{q})_{*,L} =
    E_{*,L} - \half\frac{\left(\rho u \right)_{*,L}^2 + \left( \rho v \right)_{*,L}^2}{\rho_{*,L}} - (\rho \overline{q})_{*,L} \\
\begin{split}
  [\rho (e - \overline{q})]_{*,L} &=
    \frac{s_L-u_L}{s_L-s_*} \left( E_L - \rho_L \overline{q}_L \right)  + \frac{s_L-u_L}{s_L-s_*}(s_*-u_L)\left(\rho_L s_* + \frac{p_L}{s_L-u_L}\right) \\
  &\quad - \half \frac{s_L-u_L}{s_L-s_*}\rho_L \left( s_*^2 + v_L^2 \right) .
\end{split}
\end{align}
Therefore, we require the following inequality:
{
\small
\begin{align}
&\frac{s_L-u_L}{s_L-s_*} \left( E_L - \rho_L \overline{q}_L \right)  + \frac{s_L-u_L}{s_L-s_*}(s_*-u_L)\left(\rho_L s_* + \frac{p_L}{s_L-u_L}\right)
- \half \frac{s_L - u_L}{s_L - s_*} \rho_L \left( s_*^2 + v_L^2 \right) > 0 \\
&\left( E_L - \rho_L \overline{q}_L \right)  + (s_*-u_L)\left(\rho_L s_* + \frac{p_L}{s_L-u_L}\right)
-\half \rho_L \left( s_*^2 + v_L^2 \right) > 0 \\
&(\rho e)_L + \half \rho_L \left( u_L^2 + v_L^2 \right) - \rho_L \overline{q}_L + (s_*-u_L)\left(\rho_L s_* + \frac{p_L}{s_L-u_L}\right)
-\half \rho_L \left( s_*^2 + v_L^2 \right) > 0 \\
&\half \rho_L(s_*-u_L)^2 - \frac{p_L}{u_L-s_L}(s_*-u_L) + \rho_L \left( e_L - \overline{q}_L \right) > 0.
\end{align}
}%
Let us define $\beta = s_* - u_L$ such that the inequality is a quadratic function of $\beta$. We can show that this
quadratic function has no real roots by ensuring the discriminant is negative. That is,
\begin{equation}
  \left(\frac{p_L}{u_L-s_L}\right)^2 - 2\rho_L\left[ \rho_L \left( e_L - \overline{q}_L \right) \right] < 0.
\end{equation}
This implies that we require the following to complete the proof:
\begin{equation}
  s_L < u_L - \frac{p_L}{\sqrt{2 \rho_L\left[ \rho_L \left( e_L - \overline{q}_L \right) \right]}} .
  \label{eq:s_L_ineqaulity_constraint}
\end{equation}
Note that
\begin{align*}
  \frac{p_L}{\sqrt{2 \rho_L\left[ \rho_L \left( e_L - \overline{q}_L \right) \right]}} &= 
  \sqrt{ \left( \frac{\overline \gamma_L-1}{2 \rho_L} \right) \left( \frac{p_L^2}{p_L + \overline \gamma_L \overline p^\infty_L} \right) }  \\
  &= \sqrt{ \left( \frac{\overline\gamma_L - 1}{2 \overline\gamma_L} \right) \left( \frac{p_L}{ p_L + \overline\gamma_L \overline p^\infty_L} \right) \left( \frac{p_L}{ p_L + \overline p^\infty_L } \right) \left( \frac{\overline \gamma_L(p_L + \overline p^\infty_L)}{\rho_L} \right) } \\
  &< \sqrt{\frac{\overline \gamma_L(p_L + \overline p^\infty_L)}{\rho_L}}  \\
  &= c_L.
\end{align*}
This means the right hand side of equation~\eqref{eq:s_L_ineqaulity_constraint} is larger than $u_L - c_L$ while the left hand side of the equation is equal to or smaller than $u_L - c_L$ since $s_L = \min{(\bar{u} - \bar{c}, u_L - c_L )}$.
Therefore,
the required constraint on $s_L$ given by equation~\eqref{eq:s_L_ineqaulity_constraint} is already satisfied and we have proved that the solutions given by the left star state have positive partial densities and $\rho (e - \overline{q})$. Besides, the volume fractions are bounded. Thus, the HLLC Riemann solver is quasi-positivity-preserving to give solutions in set $\mathcal{G}$ if the left and right initial states are in set ${G}$ and the approximate waves generated at the midpoints do not interact under a suitable CFL condition for the equation~\eqref{eq:soln_convex_averaging} to be valid. To satisfy the latter condition for a 1D problem, the time step size $\dt$ should be computed with the following equation:
\begin{equation}
  \dt = \frac{CFL}{\left( \left| u \right| + c \right)_{\mathrm{max}}} \cdot \dx ,
\end{equation}
where the CFL number, $CFL$, should be equal or smaller than 0.5.
Note that if the initial states are in $\mathcal{G}$, there may be numerical failure since $c_L$ or $c_R$ may be imaginary. We should also emphasize that the solutions given by the HLLC Riemann solver are not guaranteed to be in set $G$ even if the initial states are in set ${G}$.

To prove that the HLLC solver with the first order interpolation is quasi-positivity-preserving for a 2D problem,
equation~\eqref{eq:fully_discretized_W_G_pm_eqn} can be re-written as:
\begin{equation}
  \Wv_{i,j}^{n+1} =
    \sigma_x \Wv_{i,j}^{x} + \sigma_y \Wv_{i,j}^{y} ,
    \label{eq:convex_split_2D}
\end{equation}
where $\Wv_{i,j}^{x}$ and $\Wv_{i,j}^{y}$ are the contributions in the $x$ and $y$ directions to the overall solution and are defined as:
\begin{align}
  \Wv_{i,j}^{x} &= \Wv_{i,j}^{n} + \lambda_x \left(
      \hat{\mathbf{G}}^{x,+}_{i-\half,j}
      - \hat{\mathbf{G}}^{x,-}_{i+\half,j}
    \right) , \\
  \Wv_{i,j}^{y} &= \Wv_{i,j}^{n} + \lambda_y \left( 
      \hat{\mathbf{G}}^{y,+}_{i,j-\half}
      - \hat{\mathbf{G}}^{y,-}_{i,j+\half}
    \right) .
\end{align}
$\sigma_x$ and $\sigma_y$ are positive and are partitions of the contribution in the $x$ and $y$ directions respectively where $\sigma_x + \sigma_y = 1$. They are defined as~\cite{hu2013positivity,wong2021positivity}:
\begin{equation}
  \sigma_x = \frac{\tau_x}{\tau_x + \tau_y}, \quad
  \sigma_y = \frac{\tau_y}{\tau_x + \tau_y}, \quad
  \tau_x = \frac{\left( \left| u \right| + c \right)_{\mathrm{max}}}{\dx}, \quad
  \tau_y = \frac{\left( \left| v \right| + c \right)_{\mathrm{max}}}{\dy} .
\end{equation}
$\tau_x$ and $\tau_y$ defines the time step size $\Delta t$ with the CFL number, $CFL$:
\begin{equation}
  \dt = \frac{CFL}{\tau_x + \tau_y} .
\end{equation}
As a result, this implies that $\tau_x$ and $\tau_y$ are given as:
\begin{equation}
  \lambda_x = \frac{CFL}{\left( \left| u \right| + c \right)_{\mathrm{max}}} = \frac{\dt^x}{\dx} , \quad
  \lambda_y = \frac{CFL}{\left( \left| v \right| + c \right)_{\mathrm{max}}} = \frac{\dt^y}{\dy} .
\end{equation}
$\dt^x$ and $\dt^y$ are the equivalent 1D time step sizes in different directions.
Since $\Wv_{i,j}^{x}$ and $\Wv_{i,j}^{y}$ are obtained independently using quasi-1D Riemann problems, they can be further decomposed as
\begin{align}
  \Wv_{i,j}^{x} &= \half \Wv_{i,j}^{x,-} + \half \Wv_{i,j}^{x,+} , \\
  \Wv_{i,j}^{y} &= \half \Wv_{i,j}^{y,-} + \half \Wv_{i,j}^{y,+} ,
\end{align}
where
{
\small
\begin{alignat}{2}
    \half \Wv_{i,j}^{x,-} &= \frac{1}{\dx} \int_{x_{i-\half}}^{x_{i}}
      R \left( \frac{x - x_{i-\half}}{\dt}, \Wv_{i-1,j}^{n}, \Wv_{i,j}^{n} \right) dx &&= \frac{1}{2} \left[
      \Wv_{i,j}^{n} + 2 \lambda_x \left(
        \hat{\mathbf{G}}^{x,+}_{i-\half,j}
        - \Fxv_{i,j}
      \right) \right] , \\
    \half \Wv_{i,j}^{x,+} &= \frac{1}{\dx} \int_{x_{i}}^{x_{i+\half}}
      R\left(\frac{x - x_{i+\half}}{\dt}, \Wv_{i,j}^{n}, \Wv_{i+1,j}^{n} \right) dx &&= \frac{1}{2} \left[
      \Wv_{i,j}^{n} - 2 \lambda_x \left(
        \hat{\mathbf{G}}^{x,-}_{i+\half,j}
        - \Fxv_{i,j}
      \right) \right] .
\end{alignat}
}%
$\Wv_{i,j}^{x,-}$ and $\Wv_{i,j}^{x,+}$ are the time-advanced finite volume solutions in the half cells from $x_{i-1/2}$ to $x_i$ and from $x_i$ to $x_{i+1/2}$ respectively given by the two components in equation~\eqref{eq:soln_convex_averaging}. It is similar for $\Wv_{i,j}^{y,\pm}/2$.
$CFL \leq 0.5$ is required for equation~\eqref{eq:soln_convex_averaging} to be valid to give $\Wv_{i,j}^{x}$ (and $\Wv_{i,j}^{y}$). With $CFL \leq 0.5$, both $\Wv_{i,j}^{x}$ and $\Wv_{i,j}^{y}$ are in set $\mathcal{G}$ if the solutions in the previous step are in $G$. As equation~\eqref{eq:convex_split_2D} is also a convex averaging of $\Wv_{i,j}^{x}$ and $\Wv_{i,j}^{y}$, $\Wv_{i,j}^{n+1}$ after the time stepping is also in $\mathcal{G}$.

\subsection{Thermal relaxation step}

A thermal relaxation step with infinite rates follows after each hyperbolic step where the input to this step should already be in set $\mathcal{G}$.
Mathematically, this means the following ordinary differential equations (ODE's) are solved with $H_{i,j} \to \infty$:
\begin{equation}
\begin{split}
    D_t \left( \alpha_k \rho_k \right) &= 0 \quad \forall k \in \left[1, N \right] , \\
    D_t \left( \rho \bm{u} \right) &= 0 , \\
    D_t E &= 0 , \\
    D_t \alpha_k &= \sum_{i = 1, i \neq k}^{N} H_{i,k} \left( T_i - T_k \right) \quad \forall k \in \left[1, N-1 \right] .
\end{split}
\label{eq:relaxation_system_5eq_thermal_step}
\end{equation}
It can be easily seen that the mixture density, momentum, and total energy are conserved:
\begin{equation}
     \rho^{*} = \rho^{0}, \quad \rho^{*}\bm{u}^{*} = \rho^{0}\bm{u}^{0}, \quad E^{*} = E^{0} ,
\end{equation}
where the variables after the hyperbolic step but before the relaxation step are denoted by $\left(\cdot\right)^{0}$ and those after the relaxation step is denoted by $\left(\cdot\right)^{*}$. The partial density of each species is also conserved:
\begin{equation}
    \rho^{*} Y^{*}_k = \rho^{0} Y^{0}_k \ \forall k \in \left[1, N \right] .
\end{equation}
This implies that the mass fractions and the mixture specific internal energy are conserved:
\begin{align}
    Y^{*} &= Y^{0},\ \forall k \in \left[1, N \right] , \\
    e^{*} &= e^{0} .
\end{align}

\begin{figure}[!ht]
  \centering
  \begin{tikzpicture}[thick,scale=0.8, every node/.style={transform shape}]
  
  \useasboundingbox (0cm,0cm)  rectangle (10cm, 5cm);
  
  \draw[draw=none,fill=PaleBlue!60] (0.0cm,0.0cm) -- (0.0cm,4.0cm) -- (3.5cm,4.0cm) -- (0.5cm,0.0cm) -- cycle;
  \draw[draw=none,fill=PaleYellow!80] (0.5cm,0.0cm) -- (3.5cm,4.0cm) -- (4.0cm,4.0cm) -- (4.0cm,0.0cm) -- cycle;
  
  \draw[draw=none,fill=PaleBlue!60] (7.0cm,0.0cm) -- (7.0cm,4.0cm) -- (10.5cm,4.0cm) -- (7.5cm,0.0cm) -- cycle;
  \draw[draw=none,fill=PaleYellow!80] (7.5cm,0.0cm) -- (10.5cm,4.0cm) -- (11.0cm,4.0cm) -- (11.0cm,0.0cm) -- cycle;
  
  \draw[black] (0.0cm,0.0cm) rectangle ++(4cm,4cm);
  \draw[black] (7.0cm,0.0cm) rectangle ++(4cm,4cm);
  
  \draw[dashed,gray] (0.5cm,0.0cm) -- ( 3.5cm,4cm);
  \draw[dashed,gray] (7.5cm,0.0cm) -- (10.5cm,4cm);
  
  \node[align=center, text width=3cm, gray, font=\itshape] at (2.0cm,3.6cm) {liquid};
  \node[align=center, text width=3cm, gray, font=\itshape] at (2.0cm,0.30cm) {gases};
  
  \node[align=center, text width=3cm] at (2.0cm,2.4cm) {$p^0 = p_1^0 = p_2^0 = p_3^0$};
  \node[align=center, text width=3cm] at (2.0cm,1.8cm) {$T_1^0 \neq T_2^0 \neq T_3^0$};
  
  \node[text width=3cm] at (1.75cm,3.60cm) {$\rho_1^0$, $\alpha_1^0$};
  \node[text width=3cm] at (4.40cm,0.85cm) {$\rho_2^0$, $\alpha_2^0$};
  \node[text width=3cm] at (4.40cm,0.40cm) {$\rho_3^0$, $\alpha_3^0$};
  
  
  \node[align=center, text width=3cm, gray, font=\itshape] at (9.0cm,3.6cm) {liquid};
  \node[align=center, text width=3cm, gray, font=\itshape] at (9.0cm,0.30cm) {gases};
  
  \node[align=center, text width=3cm] at (9.0cm,2.4cm) {$p^* = p_1^* = p_2^* = p_3^*$};
  \node[align=center, text width=4cm] at (9.0cm,1.8cm) {$T^* = T_1^* = T_2^* = T_3^*$};
  
  \node[text width=3cm] at ( 8.75cm,3.60cm) {$\rho_1^*$, $\alpha_1^*$};
  \node[text width=3cm] at (11.40cm,0.85cm) {$\rho_2^*$, $\alpha_2^*$};
  \node[text width=3cm] at (11.40cm,0.40cm) {$\rho_3^*$, $\alpha_3^*$};
  
  
  \draw[-{Straight Barb[angle'=60,scale=3]}] (4.25cm,2.0cm) -- (6.75cm,2.0cm);
  \node[align=center, text width=3cm] at (5.5cm,2.8cm) {thermal \\ relaxation};
    
  \end{tikzpicture}
  \caption{Schematic diagram showing thermal relaxation of a mixture with a liquid and two gases. Note that the gases can be thought as bubbles in the liquid for under-resolved dispersed flows although there is only one interface drawn within the grid cell in this example.}
  \label{fig:schematic_thermal_relaxation}
  
\end{figure}
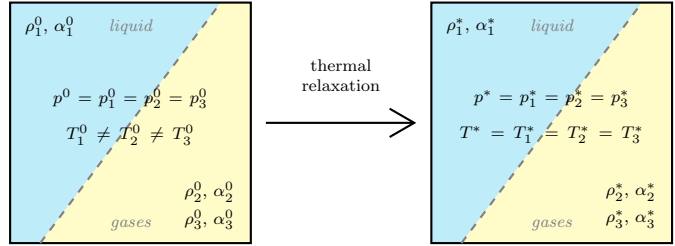

The state after the thermal relaxation is assumed to be in thermal and mechanical equilibria. That means that all species have same pressure, $p^{*}=p_1^{*}=p_2^{*}=\cdots=p_N^{*}$, and same temperature, $T^{*}=T_1^{*}=T_2^{*}=\cdots=T_N^{*}$. Also, it should be reminded that the constraints on volume fractions and mixture internal energy should still be obeyed:
\begin{align}
    1      &= \sum_{k=1}^{N} \alpha_k^{*} , \label{eq:thermal_relax_constraint_1} \\
    \rho^{0} e^{0} &= \sum_{k=1}^{N} \alpha_k^{*} \rho_k^{*} e_k^{*} . \label{eq:thermal_relax_constraint_2}
\end{align}
Using the fact that $\alpha_k^{*} = \rho^{0} Y_k^{0} / \rho_k^{*}$, the two constraints above can be reduced to:
\begin{align}
    \frac{1}{\rho^{0}} &= \sum_{k=1}^{N} \frac{Y_k^{0}}{\rho_k^{*}} , \\
    e^{0} &= \sum_{k=1}^{N} Y_k^{0} e_k^{*} .
\end{align} \label{eq:thermal_relax_constraints}
From the equation of state:
\begin{align}
    \rho_k^{*} \left(p_k^{*}, T_k^{*} \right) &= \frac{p_k^{*} + p_k^\infty}{\left( \gamma_k - 1 \right) c_{v,k} T_k^{*}} , \label{eq:rho_k_thermally_relaxed} \\
    e_k^{*} \left(p_k^{*}, T_k^{*} \right) &= \frac{p_k^{*} + \gamma_k p_k^\infty}{p_k^{*} + p_k^\infty} c_{v,k} T_k^{*} + q_k .
\end{align}
Substituting the two equations above into the constraint equations~\eqref{eq:thermal_relax_constraint_1} and \eqref{eq:thermal_relax_constraint_2} and using the assumptions of pressure and temperature equilibria~\cite{chiapolino2017simple}, we get:
\begin{align}
    \frac{1}{\rho^{0}} &= \left[ \frac{Y_1^{0} \left( \gamma_1 - 1 \right) c_{v,1}}{p^{*} + p_1^\infty} + \frac{1}{p^{*}} \sum_{k=2}^{N} Y_k^{0} \left( \gamma_k - 1 \right) c_{v,k} \right] T^{*} , \\
    e^{0} &= \left[ \frac{Y_1^{0} \left( p^{*} + \gamma_1 p_1^\infty \right) c_{v,1}}{p^{*} + p_1^\infty} + \sum_{k=2}^{N} Y_k^{0} c_{v,k} \right] T^{*} + \sum_{k=1}^{N} Y_k^{0} q_k .
\end{align}
Note that this is derived based on the assumption that all species other than the first species are ideal gases, i.e. $p_k^\infty = 0 \ \forall k \in \left[2, N \right]$.
The equations can be further simplified to
\begin{align}
    T^{*} &= \frac{1}{\rho^{0}} \left\{ \frac{Y_1^{0} \left( \gamma_1 - 1 \right) c_{v,1}}{p^{*} + p_1^\infty} + \frac{1}{p^{*}} \left[ \overline{c_p}^{0} - \overline{c_v}^{0} - Y_1^{0} \left( c_{p,1} - c_{v,1} \right) \right] \right\} ^{-1} , \label{eq:T_relax_rho} \\
    T^{*} &= \left( e^{0} - \overline{q}^{0} \right) \left[ \frac{Y_1^{0} \left( p^{*} + \gamma_1 p_1^\infty \right) c_{v,1}}{p^{*} + p_1^\infty} + \overline{c_v}^{0} - Y_1^{0} c_{v,1}\right]^{-1} \label{eq:T_relax_e} ,
\end{align}
where
\begin{equation}
    \overline{c_p} = \sum_{k=1}^{N} Y_k c_{p,k} , \quad
    \overline{c_v} = \sum_{k=1}^{N} Y_k c_{v,k} ,
\end{equation}
and $\overline{q}$ is given by equation~\eqref{eq:q_bar}.
Equating the two equations, we get a quadratic equation for $p^{*}$:
\begin{equation}
    A_p \left( \Wv^{0} \right) \cdot {p^{*}}^2 + B_p \left( \Wv^{0} \right) \cdot p^{*} + C_p \left( \Wv^{0} \right) = 0,
\end{equation}
where
\begin{align}
    A_p \left( \Wv \right) &= \overline{c_v} , \label{eq:Ap_Bp_Cp_1}\\
    B_p \left( \Wv \right) &= - \left[ \rho \left( e - \overline{q} \right) \left( \overline{c_p} - \overline{c_v} \right) - p_1^\infty \overline{c_v} - p_1^\infty Y_1 \left( c_{p,1} - c_{v,1} \right) \right] , \label{eq:Ap_Bp_Cp_2}\\
    C_p \left( \Wv \right) &= - \rho \left( e - \overline{q} \right) p_1^\infty \left[ \left( \overline{c_p} - \overline{c_v} \right) - Y_1 \left( c_{p,1} - c_{v,1} \right) \right] . \label{eq:Ap_Bp_Cp_3}
\end{align}
Since the states before the thermal relaxation are in set $\mathcal{G}$, all mass fractions are bounded between zero and one, $\rho^{0} > 0$, and $[\rho (e - \overline{q})]^{0} > 0$ (or $( e^{0} - \overline{q}^{0} ) > 0$), one of the roots is strictly negative and another one is strictly positive for the mixture pressure which is given by:
\begin{equation}
    p^{*} = \frac{-B_p \left( \Wv^{0} \right) + \sqrt{\left[ B_p\left( \Wv^{0} \right) \right]^2 - 
        4 A_p\left( \Wv^{0} \right) \cdot C_p\left( \Wv^{0} \right)}}{2A_p\left( \Wv^{0} \right)} . \label{eq:p_relax}
\end{equation}
The corresponding mixture temperature $T^{*}$ is also strictly positive and can be computed by either equations \eqref{eq:T_relax_rho} or \eqref{eq:T_relax_e}.
Since the mixture pressure and temperature of the state after the thermal relaxation are positive, the species densities $\rho_k^{*}$ given by equation~\eqref{eq:rho_k_thermally_relaxed} are all positive. This means $\alpha_k^* = \rho^{0} Y_k^{0} / \rho_k^{*} \geq 0$, $\overline{\gamma}^{*} > 1$, and $\overline {p^\infty}^{*} \geq 0$. The squared sound speed (or $(\rho c^2)^{*}$) of the five-equation model are positive since $(\rho c^2)^{*} = \overline \gamma^{*} (p^{*} + \overline{p^\infty}^{*})$. 
The state after the thermal relaxation step is therefore proved to be in set $G$. Figure~\ref{fig:schematic_thermal_relaxation} shows the schematic diagram of the thermal relaxation step of a mixture with a liquid and two gases at a grid cell. The state before the thermal relaxation can be viewed as a state after the hyperbolic step of the five-equation model where the species do not have the same temperature. After the thermal relaxation, $\rho_k$, $\alpha_k$, $p_k$, and $T_k$ are all changed for each species but $Y_k$ and $\alpha_k \rho_k$ are conserved, as well as the mixture density, momentum, and total energy.
Note that when the four-equation HRM is solved directly, equations \eqref{eq:p_relax} and \eqref{eq:T_relax_rho} (or \eqref{eq:T_relax_e}) also serve as the explicit forms of the equation of state for the mixture pressure and temperature given the conservative variable vector (excluding the volume fractions).

\begin{figure}[!ht]
\centering
\includegraphics[width=0.6\textwidth]{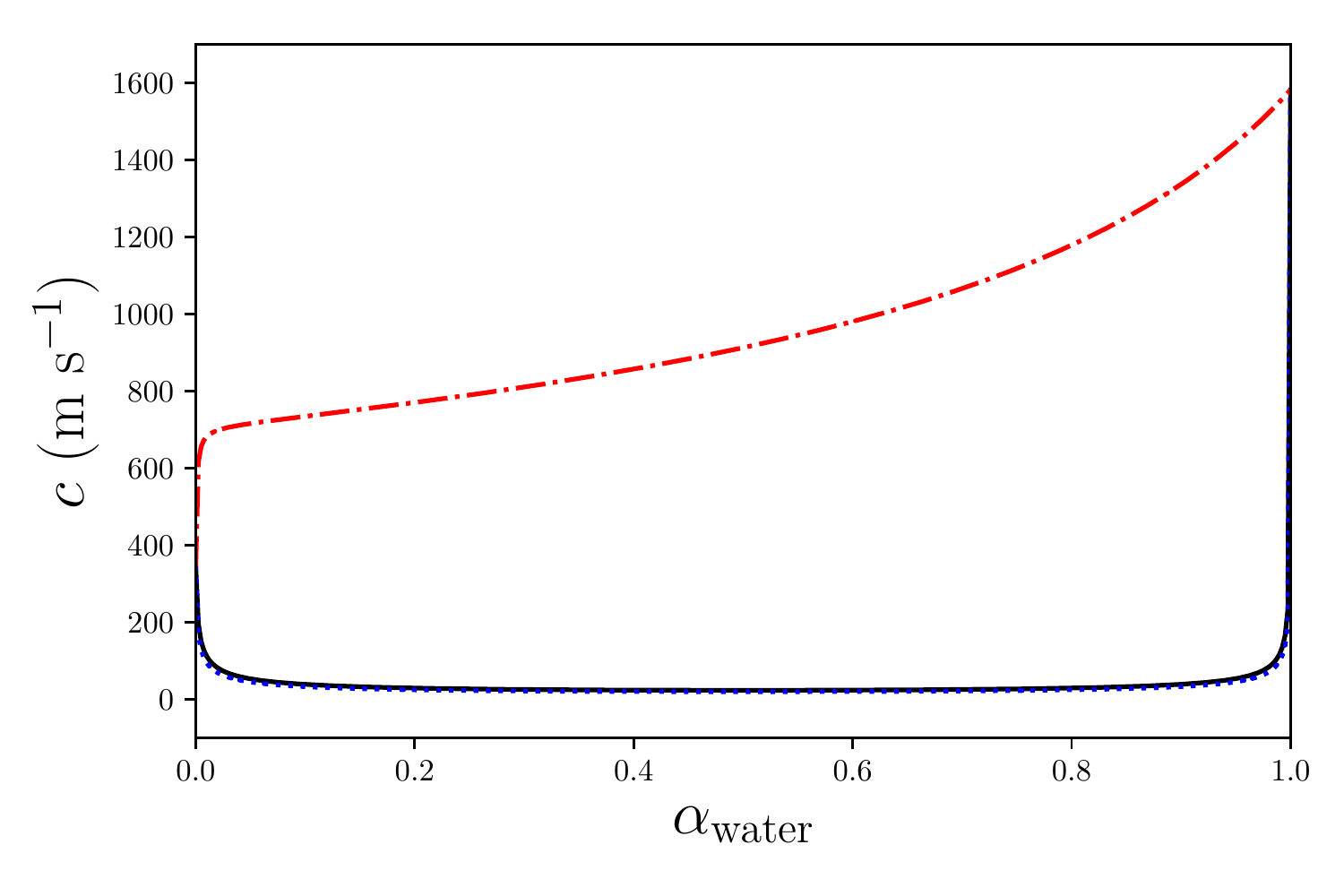}
\caption{Comparison of the analytical sound speeds of water-air mixture of different flow models at pressure $p = 101325\ \mathrm{Pa}$ and temperature $T = 298\ \mathrm{K}$. Black solid line: Wood's sound speed; red dash-dotted line: five-equation model by~\citet{allaire2002five}; blue dotted line: four-equation HRM. The properties of the liquid water and air are given by table~\ref{table:fluid_properties} in section~\ref{sec:test_problems}.} 
\label{fig:compare_sound_speed_flow_models}
\end{figure}

While the mixture sound speed of the four-equation HRM given by equation~\eqref{eq:4_eqn_sos} is not used for the fractional algorithm,
it is the correct sound speed for the underlying numerical flow model.
The partial derivatives for computing the sound speed of the four-equation HRM are given by:
\begin{align}
    \Psi_k &= - \frac{ \frac{\partial A_p}{\partial \left( \alpha_k \rho_k \right)} \left( \Wv \right) \cdot p^2 +
         \frac{\partial B_p}{\partial \left( \alpha_k \rho_k \right)} \left( \Wv \right) \cdot p + 
         \frac{\partial C_p}{\partial \left( \alpha_k \rho_k \right)} \left( \Wv \right) }
         {2 A_p \left( \Wv \right) \cdot p + B_p \left( \Wv \right)} ,\\
    \Gamma &= - \frac{\frac{\partial B_p}{\partial e} \left( \Wv \right) \cdot p + \frac{\partial C_p}{\partial e} \left( \Wv \right)}
        {2 A_p \left( \Wv \right) \cdot p + B_p \left( \Wv \right)} .
\end{align}
Note that the above partial derivatives are derived for the case when the first species is a stiffened gas governed by equation~\eqref{eq:species_stiffened_EOS} and other species are ideal gases, which is first assumed in equations~\eqref{eq:T_relax_rho} and \eqref{eq:T_relax_e}.
The four-equation HRM is also the reduced model of another popular five-equation model by~\citet{kapila2001two} with the thermal relaxation.
It is proved~\cite{flaatten2011relaxation,lund2012hierarchy} that the mixture sound speed of the relaxed four-equation model given by equation~\eqref{eq:4_eqn_sos} is smaller than or equal to that of Kapila's model, which is given by the speed of sound formulation by~\citet{wood1941textbook}.
Figure~\ref{fig:compare_sound_speed_flow_models} compares the analytical sound speeds of the water-air mixture, $c$,  against the volume fraction of liquid water, $\alpha_{\mathrm{water}} = \alpha_1$, from different models at pressure and temperature equilibria, where $p = 101325\ \mathrm{Pa}$ and $T = 298\ \mathrm{K}$. It can be seen that the sound speed of the reduced four-equation model is Wood-like, as it is similar to and is slightly smaller than that of the Wood's sound speed while the sound speed of the the five-equation model by Allaire et al. is monotonic and is much larger than the sound speeds of the other two models in the mixture region. Although the sound speed of the four-equation HRM is not exactly the Wood's sound speed, it is a much better approximation of the Wood's sound speed compared to that of the five-equation model by Allaire et al. for bubbly flows~\cite{wilson2008audible,wilson2005phase}. This is beneficial in the situation when the gas bubbles are not well resolved by the grid, or the multi-phase flows are dispersed.

\subsection{High-order finite difference scheme and positivity-preserving procedures for the hyperbolic step}

The hyperbolic stepping of the five-equation model with the first order HLLC scheme followed by the thermal relaxation step is positivity-preserving for the time advancement of flows with a stiffened gas and an arbitrary number of ideal gases. 
However, the first order HLLC scheme is numerically dissipative and has large dispersion error, thus not ideal as a diffuse interface method. In this section, a high-order discretization of the five-equation model is introduced using a discontinuity-capturing finite difference scheme. Positivity-preserving limiters for the high-order scheme that can guarantee that the time advancement is free from numerical failures are also discussed.

\subsubsection{Weighted compact nonlinear scheme}

Following our previous work~\cite{wong2021positivity}, a high-order accurate scheme of the family of weighted compact nonlinear schemes~\cite{deng2000developing, nonomura2007increasing, zhang2008development, nonomura2009effects, nonomura2012numerical,wong2017high, wong2019thesis, wong2021positivity} (WCNSs) in the explicit form is chosen for the discretization of the spatial derivatives.
A semi-discretized form of the equation~\eqref{eq:W_G_eqn} with finite differencing is first considered for the hyperbolic time stepping:
\begin{equation}
    \frac{\partial \Wv}{\partial t} \bigg|_{i,j}
      + \widehat{ \frac{\partial \Gxv}{\partial x} } \bigg|_{i,j}
      + \widehat{ \frac{\partial \Gyv}{\partial y} } \bigg|_{i,j} = \Shv_{i,j} , \label{eq:semi_W_G_eqn}
\end{equation}
where $\widehat{ \partial \Gxv / \partial x}  |_{i,j}$ and $\widehat{ \partial \Gyv / \partial y} |_{i,j}$ are high-order approximations of the flux derivatives and 
$\hat{\Sv}_{i,j}$ is the high-order approximation of the source term,
which is given by:
\begin{equation}
    \Shv_{i,j} = \begin{pmatrix}
      0 \\
      0 \\
      \vdots \\
      0 \\
      0 \\
      0 \\
      0 \\
      \alpha_{1,i,j}
       \left( \left. \widehat{ \frac{\partial u}{\partial x} } \right|_{i,j} +
       \left. \widehat{ \frac{\partial v}{\partial y} } \right|_{i,j}
      \right) \\
      \alpha_{2,i,j}
       \left( \left. \widehat{ \frac{\partial u}{\partial x} } \right|_{i,j} +
       \left. \widehat{ \frac{\partial v}{\partial y} } \right|_{i,j}
      \right) \\
      \vdots \\
      \alpha_{N-1,i,j}
       \left( \left. \widehat{ \frac{\partial u}{\partial x} } \right|_{i,j} +
       \left. \widehat{ \frac{\partial v}{\partial y} } \right|_{i,j}
      \right)
      \end{pmatrix} .
\end{equation}
$\widehat{ \partial u / \partial x} |_{i,j}$ and $\widehat{ \partial v / \partial y} |_{i,j}$ are high-order approximations of the velocity derivatives.

A high-order discretization of the fluxes is consistent and conservative for the conservative equations if:
\begin{equation}
    \widehat{ \frac{\partial \Gv^{x}}{\partial x} } \bigg|_{i,j} =
      \frac{\Gxhv_{i+\half,j}
            - \Gxhv_{i-\half,j}}{\dx} , \quad
    \widehat{ \frac{\partial \Gv^{y}}{\partial y} } \bigg|_{i,j} =
      \frac{\Gyhv_{i,j+\half}
            - \Gyhv_{i,j-\half}}{\dy}.
            \label{eq:flux_diff_form}
\end{equation}
Similarly, it is assumed that the finite differencing of the velocity can be rewritten in the following form as:
\begin{equation}
    \widehat{ \frac{\partial u}{\partial x} } \bigg|_{i,j} =
      \frac{\hat{u}_{i+\half,j} - \hat{u}_{i-\half,j}}{\dx}, \quad
    \widehat{ \frac{\partial v}{\partial y} } \bigg|_{i,j} =
      \frac{\hat{v}_{i,j+\half} - \hat{v}_{i,j-\half}}{\dy} .
\end{equation}
Thus, the equation~\eqref{eq:semi_W_G_eqn} can be rewritten as:
\begin{equation}
    \frac{\partial \Wv}{\partial t} \bigg|_{i,j}
      + \frac{\Gxhv_{i+\half,j}
            - \Gxhv_{i-\half,j}}{\dx} 
      + \frac{\Gyhv_{i,j+\half}
            - \Gyhv_{i,j-\half}}{\dy} =
      \Shv_{i,j} , \label{eq:semi_W_G_eqn_cons}
\end{equation}

The sixth order accurate explicit scheme from the hybrid cell-midpoint and cell-node compact scheme (HCS)~\cite{deng2011new} family is used for the approximation of the first order derivatives $\widehat{ \partial \Gxv / \partial x}  |_{i,j}$ and $\widehat{ \partial \Gyv / \partial y} |_{i,j}$. The sixth order explicit HCS formulation is given by:
\begin{equation}
\begin{split}
    \widehat{ \frac{\partial \Gxv}{\partial x} } \bigg|_{i,j}
    &= \frac{1}{\dx}  \left[
      \psi \left(\Gxtv_{i+\half,j} - \Gxtv_{i-\half,j} \right)
      - \frac{175 \psi - 192}{256} \left(\Gxv_{i+1,j} - \Gxv_{i-1,j} \right) \right. \\
      &\quad \left. + \frac{35 \psi - 48}{320} \left(\Gxv_{i+2,j} - \Gxv_{i-2,j} \right) - \frac{45 \psi - 64}{3840} \left(\Gxv_{i+3,j} - \Gxv_{i-3,j} \right) \right]. \label{eq:HCS6}
\end{split}
\end{equation}
$\psi$ is a user chosen parameter and $\psi = 256/175$ is adopted in this work such that the HCS formulation becomes eighth order accurate. The finite differencing is similar in the $y$ direction.

Following~\cite{wong2019thesis,subramaniam2019high}, the HCS given by equation~\eqref{eq:HCS6} has the reconstructed flux $\hat{\mathbf{G}}^{\mathrm{HCS},x}_{i+1/2,j}$ given by:
\begin{equation}
\begin{split}
    \hat{\mathbf{G}}^{\mathrm{HCS},x}_{i+\half,j}
    &= \psi \Gxtv_{i+\half,j} - \left( \frac{75 \psi}{128} - \frac{37}{60} \right) \left( \Gxv_{i,j} + \Gxv_{i+1,j} \right) \\
    &\quad + \left( \frac{25 \psi}{256} - \frac{2}{15} \right) \left( \Gxv_{i-1,j} + \Gxv_{i+2,j} \right)
    - \left( \frac{3 \psi}{256} - \frac{1}{60} \right) \left( \Gxv_{i-2,j} + \Gxv_{i+3,j} \right) \label{eq:reconstructed_G}.
\end{split}
\end{equation}

The numerical derivatives of the velocity components are also given by the same finite difference scheme for the flux derivatives:
\begin{equation}
\begin{split}
    \frac{\partial u}{\partial x} \bigg|_{i,j} \approx
    \widehat{ \frac{\partial u}{\partial x} } \bigg|_{i,j} &=
    \frac{1}{\dx}  \left[
      \psi \left(\tilde{u}_{i+\half,j} - \tilde{u}_{i-\half,j} \right)
      - \frac{175 \psi - 192}{256} \left(u_{i+1,j} - u_{i-1,j} \right) \right. \\
      &\quad \left. + \frac{35 \psi - 48}{320} \left(u_{i+2,j} - u_{i-2,j} \right) - \frac{45 \psi - 64}{3840} \left(u_{i+3,j} - u_{i-3,j} \right) \right].
\end{split}
\end{equation}
The reconstructed velocity component $\hat{u}^{\mathrm{HCS}}_{i+1/2,j}$ is given by:
\begin{equation}
\begin{split}
    \hat{u}^{\mathrm{HCS}}_{i+\half,j}
    &= \psi \tilde{u}_{i+\half,j} - \left( \frac{75 \psi}{128} - \frac{37}{60} \right) \left( u_{i,j} + u_{i+1,j} \right) + \left( \frac{25 \psi}{256} - \frac{2}{15} \right) \left( u_{i-1,j} + u_{i+2,j} \right) \\
    &\quad - \left( \frac{3 \psi}{256} - \frac{1}{60} \right) \left( u_{i-2,j} + u_{i+3,j} \right) \label{eq:reconstructed_vel} .
\end{split}
\end{equation}

In order to capture the shocks and the material interfaces, a robust interpolation with an accurate Riemann solver is needed to compute the midpoint fluxes, $\Gxtv_{i+1/2,j}$ and $\Gytv_{i,j+1/2}$, and the midpoint velocities, $\tilde{u}_{i+1/2,j}$ and $\tilde{v}_{i,j+1/2}$.
We have chosen the fifth order accurate incremental-stencil weighted essentially non-oscillatory (WENO) interpolation in our previous paper~\cite{wong2021positivity} due to its robustness for flows involving interactions of strong shocks and gas-liquid interfaces with high density ratios. The variables chosen to interpolate are the primitive variables with the pressure and velocity projected to the characteristic fields. 
This approach is shown to suppress pressure oscillations at material interfaces~\cite{johnsen2006implementation,nonomura2012numerical,wong2017high}. 
The left-biased and right-biased interpolated variables are then given as inputs to the HLLC Riemann solver for obtaining the midpoint fluxes and velocities.
The recipe to approximate the midpoint fluxes and velocities with the WCNS finite differencing approach is detailed in the previous paper~\cite{wong2021positivity}. Also, the incremental-stencil (IS) interpolation and the characteristic decomposition are briefly discussed in the \ref{appendix:weno_IS} and \ref{appendix:char_decomp} respectively.

After the high-order midpoint flux vector $\Gxhv_{i+1/2,j} = \hat{\mathbf{G}}^{\mathrm{HCS},x}_{i+1/2,j}$ and the velocity $\hat{u}_{i+1/2,j} = \hat{u}^{\mathrm{HCS}}_{i+1/2,j}$ are reconstructed, the high-order accurate version of quasi-conservative flux vector $\hat{\mathbf{G}}^{x,\pm}_{i+1/2,j}$ can be obtained by re-using the forms given by equation~\eqref{eq:G_pm}:
  \begin{alignat}{4}
  \hat{\mathbf{G}}^{-}_{i+\half,j} &=
      \Gxhv_{i+\half,j} -
      \hat{u}_{i+\half,j} \ (0\ 0\ \cdots\ 0\ 0\ 0\ 0\ \alpha_{1,i,j} &&\ \alpha_{2,i,j} \ &\cdots &\  \alpha_{N-1,i,j} )^T , \\
  \hat{\mathbf{G}}^{+}_{i+\half,j} &=
      \Gxhv_{i+\half,j} -
      \hat{u}_{i+\half,j} \ (0\ 0\ \cdots\ 0\ 0\ 0\ 0\ \alpha_{1,i+1,j} &&\ \alpha_{2,i+1,j} \ &\cdots &\  \alpha_{N-1,i+1,j} )^T .
  \end{alignat}
Finally, the semi-discretized form with the high-order quasi-conservative flux is written as:
\begin{equation}
    \frac{\partial \Wv}{\partial t} \bigg|_{i,j}
      + \frac{\hat{\mathbf{G}}^{x,-}_{i+\half,j}
          - \hat{\mathbf{G}}^{x,+}_{i-\half,j}}{\dx} 
      + \frac{\hat{\mathbf{G}}^{y,-}_{i,j+\half}
          - \hat{\mathbf{G}}^{y,+}_{i,j-\half}}{\dy} =
    \mathbf{0} . \label{eq:semi_W_G_pm_eqn}
\end{equation}

To further improve the accuracy and robustness near shocks, the high-order reconstructed flux and velocity, $\hat{\mathbf{G}}^{\mathrm{HCS},x}_{i+1/2,j}$ and $\hat{u}^{\mathrm{HCS}}_{i+1/2,j}$, can be blended with the second order reconstructed flux and velocity, i.e. $\Gxtv_{i+1/2,j}$ and $\tilde{u}_{i+1/2,j}$:
\begin{align}
    \Gxhv_{i+\half,j} &= \sigma^s_{i+\half,j}  \ \Gxtv_{i+\half,j} + (1 - \sigma^s_{i+\half,j}) \ \hat{\mathbf{G}}^{\mathrm{HCS},x}_{i+\half,j} ,  \label{eq:flux_blending} \\
    \hat{u}_{i+\half,j} &= \sigma^s_{i+\half,j} \ \tilde{u}_{i+\half,j} + (1 - \sigma^s_{i+\half,j}) \ \hat{u}^{\mathrm{HCS}}_{i+\half,j} . \label{eq:velocity_blending}
\end{align}
$\sigma^s_{i+1/2,j}$ is a shock sensor at the midpoint. The sensor has the following form based on the mixture density $\rho$ and the pressure $p$:
\begin{equation}
    \sigma^s_{i+\half,j} = \tanh { \left\{  C_{s} \left[ \sigma_{i+\half,j} \left( \rho \right) \sigma_{i+\half,j} \left( p \right) \right]^q  \right\} },
\end{equation}
where for a variable $u$, normalized fourth order second derivatives are used to compute $\sigma_{i+1/2,j}(u)$:
\begin{align}
    \sigma_{i+\half,j} \left( u \right) &= \half \left[ \sigma^{+}_{i+\half,j} \left( u \right) + 
    \sigma^{-}_{i+\half,j} \left( u \right) \right] , \\
    \sigma^{+}_{i+\half,j} \left( u \right) &= 
    \frac{\left| - u_{i-1,j} + 16 u_{i,j} - 30 u_{i+1,j} + 16 u_{i+2,j} - u_{i+3,j} \right|}
         {\left|   u_{i-1,j} + 16 u_{i,j} + 30 u_{i+1,j} + 16 u_{i+2,j} + u_{i+3,j} \right|} , \\
    \sigma^{-}_{i+\half,j} \left( u \right) &= 
    \frac{\left| - u_{i-2,j} + 16 u_{i-1,j} - 30 u_{i,j} + 16 u_{i+1,j} - u_{i+2,j} \right|}
         {\left|   u_{i-2,j} + 16 u_{i-1,j} + 30 u_{i,j} + 16 u_{i+1,j} + u_{i+2,j} \right|} .
\end{align}
The exponents $q=2$ and constant $C_s=1.0\mathrm{e}{12}$ are chosen in this work to localize the flux blending around the shocks such that
the spatial formal order of accuracy is minimally affected in the smooth regions. While it is not able to prove the spatial formal order of accuracy with the flux blending, the fifth order spatial convergence rate of the underlying scheme is shown with a smooth advecton problem in a later section.
Although the overall scheme is slightly modified from the finite difference scheme in our previous paper~\cite{wong2021positivity} with the flux blending,
this new version of finite difference scheme is still called WCNS-IS in this work.

\subsubsection{Positivity-preserving procedures}

The finite difference WCNS-IS has the shock- and interface-capturing capabilities for compressible multi-phase simulations. However, the scheme is not free from numerical failures in low pressure regions close to vacuum or in regions with interactions between strong shocks and high density gradient material interfaces, where the volume fractions can become out of bounds, or partial densities and squared speed of sound can turn negative.
Positivity-preserving techniques are critical to ensure the admissibility of the numerical solutions. In the next two subsections, positivity-preserving interpolation and quasi-positivity-preserving flux limiters are introduced which are the extension of those in our previous work~\cite{wong2021positivity} from two-species liquid-gas (and also gas-gas) flows to flows with a liquid (stiffened gas) and an arbitrary number of ideal gases. The former can limit the WENO interpolated values to prevent failures when the HLLC Riemann solver is used, while the latter ensures that the limited flux can advance the solutions from a state in set $G$ to a state in set $\mathcal{G}$.

\paragraph{Positivity-preserving interpolation limiter}

A general high-order interpolation such as the WENO interpolation is not positivity-preserving for partial densities and squared sound speed. Also, it is not bounded for volume fractions. This may cause numerical failure during the sound speed calculation in the HLLC Riemann solver for the midpoint fluxes and velocities, if either or both of the left-biased and right-biased interpolated midpoint values are not in the set $G$.
However, the first order interpolation with the left and right node values are in the admissible set $G$. Therefore, the high-order left (right)-biased WENO interpolation can be limited with the first order interpolation using the left (right) node values. For simplicity, this sub-section only discusses the limiting procedures for the left-biased WENO interpolation in the $x$ direction for the governing equations. Hence, the subscripts ``$L$" and ``$j$" are omitted.

The positivity-preserving limiting for interpolation involves several stages.
In the first stage,
a limited conservative variable vector with positive partial densities, $\tilde{\Wv}_{i+1/2}^{*}$, is obtained at each midpoint.
As described in algorithm~\ref{alg:interp_limit_densities}, $\tilde{\Wv}_{i+1/2}^{*}$ is first initialized as the WENO interpolated conservative variable vector. Each partial density in $\tilde{\Wv}_{i+1/2}^{*}$ is then limited independently through convex combination of itself with the corresponding partial density in the first order interpolated conservative variable vector using a user-defined small threshold $\epsilon_{\alpha_k \rho_k}$. 
After limiting each partial density, we apply a limiting procedure on the volume fractions by first initializing $\tilde{\Wv}_{i+1/2}^{**}$ as $\tilde{\Wv}_{i+1/2}^{*}$. However, unlike the partial densities, the volume fractions in $\tilde{\Wv}_{i+1/2}^{*}$ cannot be limited to be positive independently with those in the first order interpolated conservative variable vector using a small threshold $\epsilon_{\alpha_k}$. Otherwise, the limited volume fractions may violate the constraint that they should sum to one, or may fail to preserve their bounds. One possible strategy is to perform successive convex combinations of all volume fractions together with those in the first order conservative variable vector using the blending factors $\theta^{\alpha_k}_{i+1/2}$ for the volume fractions one by one. Nevertheless, since the volume fractions linearly depend on the conservative variable vector, a simpler algorithm that can obtain the mathematically same limited conservative variable vector is to perform a single convex combination with a blending factor $\theta_{i+1/2}$ that is
small enough to limit the volume fractions all at once.
The limiting algorithm for the volume fractions is detailed in algorithm~\ref{alg:interp_limit_vol_frac}.
We have chosen the tolerances for the limiting procedures as $\epsilon_{\alpha_k \rho_k} = \epsilon_{\alpha_k} = 1.0\mathrm{e}{-10}$.
After this stage, the limited conservative variable vector $\tilde{\Wv}_{i+1/2}^{**}$ should have all partial densities (including mixture density) positive and all volume fractions bounded between zero and one. This also means that all mass fractions are bounded between zero and one. The remaining quantity to limit is the squared speed of sound. Unfortunately, for the cases with one stiffened gas and an arbitrary number of ideal gases,
the squared speed of sound is not a concave function of the conservative variables and $G$ is not a convex set in general. Therefore, the limiting procedure using a convex combination of the $\tilde{\Wv}_{i+1/2}^{**}$ with the first order interpolated variable vector in our previous work~\cite{wong2021positivity} is not effective here. In case the limited conservative variable vector has negative squared sound speed, a hard switch can be used to set $\tilde{\Wv}_{i+1/2}^{**} = \Wv_i$. Here, the hard switch is turned on when:
\begin{align*}
    \alpha_k \rho_k \left( \tilde{\Wv}_{i+\half}^{**} \right) &< \min \left( \epsilon^{\mathrm{HS}}_{\alpha_k \rho_k}, \ \alpha_k \rho_k \left( \Wv_i \right) \right) , \\
    \alpha_k \left( \tilde{\Wv}_{i+\half}^{**} \right) &< \min \left( \epsilon^{\mathrm{HS}}_{\alpha_k}, \ \alpha_k \left( \Wv_i \right) \right) , \\
    \rho c^2 \left( \tilde{\Wv}_{i+\half}^{**} \right) &< \min \left( \epsilon^{\mathrm{HS}}_{\rho c^2 }, \ \rho c^2 \left( \Wv_i \right) \right) ,
\end{align*}
where $\epsilon^{\mathrm{HS}}_{\alpha_k \rho_k} = 1.0\mathrm{e}{-11}$,  $\epsilon^{\mathrm{HS}}_{\alpha_k} = 1.0\mathrm{e}{-11}$, and $\epsilon^{\mathrm{HS}}_{\rho c^2 } = 1.0\mathrm{e}{-9}$.
The possibility of having negative squared sound speed is low if all partial densities are positive and all volume fractions are bounded after the limiting. Therefore, the effect of hard switch on the spatial formal order of accuracy is minimal.
The hard switch is also turned on when partial densities or volume fractions are out of bound to prevent errors due to numerical round-off during the limiting.
Note that this interpolation limiting method can also be applied for other high-order interpolations or any high-order reconstructions in WENO finite volume schemes. 
The proof of the preservation of the formal order of accuracy of a high-order interpolation or reconstruction method in smooth regions by the limiting method can be found in our previous paper~\cite{wong2021positivity} when the hard switch is not used.

\begin{algorithm}[!ht]
\SetAlgoLined
\footnotesize
  \For{\textup{all midpoints}}{
    Initialize $\tilde{\Wv}_{i+\half}^{*} = \tilde{\Wv}_{i+\half}$\;
    \For{\textup{all species} $k = 1,2,\cdots,N$}{
      \uIf{$\alpha_k \rho_k \left( \tilde{\Wv}_{i+\half} \right) < \min \left(\epsilon_{\alpha_k \rho_k}, \ \alpha_k \rho_k \left( \Wv_i \right) \right) $}{
        Solve $\theta_{i+\half}$ from the formula:
        \begin{equation*}
          \left( 1 - \theta_{i+\half} \right) \alpha_k \rho_k \left( \Wv_i \right) + \theta_{i+\half} \alpha_k \rho_k \left( \tilde{\Wv}_{i+\half} \right) = \min \left(\epsilon_{\alpha_k \rho_k}, \ \alpha_k \rho_k \left( \Wv_i \right) \right) ; 
        \end{equation*}
      }
      \Else{
        $\theta_{i+\half} = 1$\;
      }
      Perform convex combination on the corresponding $\alpha_k \rho_k$:
      \begin{equation*}
        \alpha_k \rho_k \left( \tilde{\Wv}_{i+\half}^{*} \right) = \left( 1 - \theta_{i+\half} \right) \alpha_k \rho_k \left( \Wv_i \right) + \theta_{i+\half} \alpha_k \rho_k \left( \tilde{\Wv}_{i+\half} \right) ;
      \end{equation*}
    }
  }
 \caption{Left-biased interpolation limiting procedure for positive partial densities.}
 \label{alg:interp_limit_densities}
\end{algorithm}

\begin{algorithm}[!!ht]
\SetAlgoLined
\footnotesize
  \For{\textup{all midpoints}}{
    Initialize $\tilde{\Wv}_{i+\half}^{**} = \tilde{\Wv}_{i+\half}^{*}$, $\theta_{i+\half} = 1$\;
    \For{\textup{all species} $k = 1,2,\cdots,N$}{
      \uIf{$\alpha_k \left( \tilde{\Wv}_{i+\half}^{*} \right) < \min \left(\epsilon_{\alpha_k}, \ \alpha_k \left( \Wv_i \right) \right) $}{
        Solve $\theta^{\alpha_k}_{i+\half}$ from the formula:
        \begin{equation*}
          \left( 1 - \theta^{\alpha_k}_{i+\half} \right) \alpha_k \left( \Wv_i \right) + \theta^{\alpha_k}_{i+\half} \alpha_k \left( \tilde{\Wv}_{i+\half}^{*} \right) = \min \left(\epsilon_{\alpha_k}, \ \alpha_k \left( \Wv_i \right) \right);
        \end{equation*}
      }
      \Else{
        $\theta^{\alpha_k}_{i+\half} = 1$\;
      }
      
      Set $\theta_{i+\half} = \min \left( \theta_{i+\half} , \ \theta^{\alpha_k}_{i+\half} \right)$\;
    }
    Perform convex combination on all $\alpha_k$:
    
    \For{\textup{species} $k = 1,2,\cdots,N-1$}{
      \begin{equation*}
        \alpha_k \left( \tilde{\Wv}_{i+\half}^{**} \right) = \left( 1 - \theta_{i+\half} \right) \alpha_k \left( \Wv_i \right) + \theta_{i+\half} \alpha_k \left( \tilde{\Wv}_{i+\half}^{*} \right) ;
      \end{equation*}
    }
  }
 \caption{Left-biased interpolation limiting procedure for bounded volume fractions.}
 \label{alg:interp_limit_vol_frac}
\end{algorithm}

\paragraph{Quasi-positivity-preserving flux limiter}

The hyperbolic time stepping using the first order flux vector from the HLLC Riemann solver can guarantee that the time-advanced conservative variable vector is in $\mathcal{G}$, i.e. is positivity- and boundedness-preserving with regard to $\mathcal{G}$, or quasi-positivity-preserving, when the initial state is in $G$. However, the high-order flux reconstruction stage for $\hat{\mathbf{G}}^{\pm}_{i+1/2}$ using the HLLC Riemann solver and the HCS finite differencing is not quasi-positivity-preserving for the time stepping in general even when the left-biased and right-biased interpolated values passed into the HLLC Riemann solver are in $G$, and thus this may cause numerical failures during the thermal relaxation step. Therefore, a flux limiter is critical to ensure that the reconstructed flux vectors used for the hyperbolic time stepping generate a state in $\mathcal{G}$. Following the same splitting idea introduced in the section of HLLC Riemann solver, if first order forward Euler time stepping is used with the WCNS for full discretization, we can obtain the following equation:
\begin{equation}
  \Wv_{i,j}^{n+1} =
    \sigma_x \underbrace{ \left( \half \Wv_{i,j}^{x,-} + \half \Wv_{i,j}^{x,+} \right) }_{\Wv_{i,j}^{x}} + \sigma_y \underbrace{ \left( \half \Wv_{i,j}^{y,-} + \half \Wv_{i,j}^{y,+} \right) }_{\Wv_{i,j}^{y}} ,
\end{equation}
where
\begin{align}
  \Wv_{i,j}^{x,\mp} &= 
    \Wv_{i,j}^{n} \pm 2 \lambda_x \left(
      \hat{\mathbf{G}}^{x,\pm}_{i\mp\half,j}
      - \Fxv_{i,j}
    \right) , \\
  \Wv_{i,j}^{y,\mp} &= 
    \Wv_{i,j}^{n} \pm 2 \lambda_y \left(
      \hat{\mathbf{G}}^{y,\pm}_{i,j\mp\half}
      - \Fyv_{i,j}
    \right) .
\end{align}
Note that $\hat{\mathbf{G}}^{x,\pm}_{i\mp1/2,j}$ and $\hat{\mathbf{G}}^{y,\pm}_{i,j\mp1/2}$ here are the high-order reconstructed fluxes in contrast to the first order HLLC fluxes, $\hat{\mathbf{G}}^{x,\mathrm{HLLC},\pm}_{i\mp1/2,j}$ and $\hat{\mathbf{G}}^{y,\mathrm{HLLC},\pm}_{i,j\mp1/2}$, used in the section of first order HLLC scheme.
It should also be reminded that the definitions of $\Fxv_{i,j}$ and $\Fyv_{i,j}$ are given by equation~\eqref{eq:F_x_F_y_def}.
Since both $\sigma_x$ and $\sigma_y$ are positive and $\sigma_x + \sigma_y = 1$, $\Wv_{i,j}^{n+1}$ is a convex combination of $\Wv_{i,j}^{x,\pm}$ and $\Wv_{i,j}^{y,\pm}$. If all four conservative variable vectors are in the set $\mathcal{G}$, $\Wv_{i,j}^{n+1}$ is also in the the set $\mathcal{G}$. For simplicity, only quasi-positivity-preserving flux limiting in $x$ direction for $\Wv_{i,j}^{x,\pm}$ is discussed here and hence ``$x$" superscript and ``$j$" subscript are dropped in the following part.

If the first order flux vector from the approximate Riemann solver is quasi-positively preserving such that all intermediate states given by the approximate Riemann solutions are in the set $\mathcal{G}$, such as the HLLC Riemann solver presented in the earlier section, a quasi-positivity flux limiting procedure with two stages can be constructed such that the time-advanced conservative variable vector is also in the set $\mathcal{G}$. The algorithms are extended from those in our previous work~\cite{wong2021positivity}.
The first stage is to obtain the limited flux $\hat{\Gv}_{i+1/2}^{*,\pm}$ through a convex combination of the high-order solution $\Wv_i^{\pm}$ and the first order solution $\Wv_i^{\mathrm{HLLC}, \pm}$ such that all partial densities and volume fractions are positive in the limited solutions $\Wv_i^{*,\pm}$, where
\begin{equation}
  \Wv_{i}^{*,\mp} = 
    \Wv_{i}^{n} \pm 2 \lambda \left(
      \hat{\mathbf{G}}^{*,\pm}_{i\mp\half}
      - \Fv_{i}
    \right) .
\end{equation}
The limiting procedure of this stage is detailed in algorithm~\ref{alg:flux_limit_densities_vol_frac} using tolerances $\epsilon_{\alpha_k \rho_k}$ and $\epsilon_{\alpha_k}$. The algorithm makes use of the fact that both partial densities and volume fractions are linear functions of the conservative variable vector, thus the limiting can be achieved for the partial densities and volume fractions all at once by only using one pair of convex combinations with a blending factor $\theta_{i+1/2}$.
After this stage, both solutions $\Wv_i^{*, \pm}$ time-advanced with the limited fluxes have all partial densities (including mixture density) positive and all volume fractions bounded.
Note that in the last step of the algorithm, the intention is to hybridize $( \hat{\Gv}_{i + 1/2}^{\pm} - \Fv_i )$ with $( \hat{\Gv}_{i + 1/2}^{\mathrm{HLLC},\pm} - \Fv_i )$ 
to obtain $( \hat{\Gv}_{i + 1/2}^{*,\pm} - \Fv_i )$
but the $\Fv_i$ on both sides of the equation cancel each other. Also, the flux limiting process is conservative for all equations except the advection equations of volume fractions.

In the next stage, the positivity flux limiting for $\rho \tilde{e} = \rho(e - \overline{q})$ is carried out similarly. This step is described in algorithm~\ref{alg:flux_limit_energy} with tolerance $\epsilon_{\rho \tilde{e}}$ and utilizes the fact that the set $\mathcal{G}$ of the conservative variable vector is convex to obtain the final limited flux $\hat{\Gv}_{i+1/2}^{**,\pm}$. The quantity $\rho \tilde{e}$ of $\Wv_i^{**, \pm}$ is limited to be positive, while all partial densities and volume fractions are maintained to be positive, where
\begin{equation}
  \Wv_{i}^{**,\mp} = 
    \Wv_{i}^{n} \pm 2 \lambda \left(
      \hat{\mathbf{G}}^{**,\pm}_{i\mp\half}
      - \Fv_{i}
    \right) .
\end{equation}
The tolerances for the limiting procedures are chosen as $\epsilon_{\alpha_k \rho_k} = 1.0\mathrm{e}{-10}$, $\epsilon_{\alpha_k} = 1.0\mathrm{e}{-10}$, and $\epsilon_{\rho \tilde{e}} = 1.0\mathrm{e}{-8}$. While $\Wv_i^{**, \pm}$ should be in set $\mathcal{G}$ mathematically after all the flux limiting procedures, a hard switch is suggested after the flux limiting procedures to prevent errors due to numerical round-off. The hard switch sets $\hat{\Gv}_{i+1/2}^{**,\pm} = \hat{\Gv}_{i+1/2}^{\mathrm{HLLC,\pm}}$ if either $\Wv_{i}^{**,+}$ or $\Wv_{i+1}^{**,-}$ are at states that are not bounded by smaller tolerances:
\begin{align*}
    \alpha_k \rho_k \left( \Wv_{i}^{**,+} \right) &< \min \left( \epsilon^{\mathrm{HS}}_{\alpha_k \rho_k}, \ \alpha_k \rho_k \left( \Wv_i^{\mathrm{HLLC}, +}  \right) \right) , \\
    \alpha_k \left( \Wv_{i}^{**,+} \right) &< \min \left( \epsilon^{\mathrm{HS}}_{\alpha_k}, \ \alpha_k \left( \Wv_i^{\mathrm{HLLC}, +} \right) \right) , \\
    \rho \tilde{e} \left( \Wv_{i}^{**,+} \right) &< \min \left( \epsilon^{\mathrm{HS}}_{\rho \tilde{e} }, \ \rho \tilde{e} \left( \Wv_i^{\mathrm{HLLC}, +} \right) \right) , \\
    \alpha_k \rho_k \left( \Wv_{i+1}^{**,-} \right) &< \min \left( \epsilon^{\mathrm{HS}}_{\alpha_k \rho_k}, \ \alpha_k \rho_k \left( \Wv_{i+1}^{\mathrm{HLLC}, -} \right) \right) , \\
    \alpha_k \left( \Wv_{i+1}^{**,-} \right) &< \min \left( \epsilon^{\mathrm{HS}}_{\alpha_k}, \ \alpha_k \left( \Wv_{i+1}^{\mathrm{HLLC}, -} \right) \right) , \\
    \rho \tilde{e} \left( \Wv_{i+1}^{**,-} \right) &< \min \left( \epsilon^{\mathrm{HS}}_{\rho \tilde{e} }, \ \rho \tilde{e} \left( \Wv_{i+1}^{\mathrm{HLLC}, -} \right) \right) ,
\end{align*}
where $\epsilon^{\mathrm{HS}}_{\alpha_k \rho_k} = 1.0\mathrm{e}{-11}$,  $\epsilon^{\mathrm{HS}}_{\alpha_k} = 1.0\mathrm{e}{-11}$, and $\epsilon^{\mathrm{HS}}_{\rho \tilde{e} } = 1.0\mathrm{e}{-9}$.

\begin{algorithm}[!ht]
\SetAlgoLined
\footnotesize
  \For{\textup{all midpoints}}{
    Compute $\Wv_i^{+}$ and $\Wv_{i+1}^{-}$, and initialize $\theta_{i+\half} = 1$\;
    \For{\textup{all species} $k = 1,2,\cdots,N$}{
      \uIf{$\alpha_k \rho_k \left( \Wv_i^{+} \right) < \min \left( \epsilon_{\alpha_k \rho_k}, \ \alpha_k \rho_k \left( \Wv_i^{\mathrm{HLLC}, +} \right) \right)$}{
        Solve $\theta_{i+\half}^{\alpha_k \rho_k, +}$ from the formula:
        \begin{equation*}
          \left( 1 - \theta_{i+\half}^{\alpha_k \rho_k, +} \right) \alpha_k \rho_k \left( \Wv_i^{\mathrm{HLLC}, +} \right) + \theta_{i+\half}^{\alpha_k \rho_k, +} \alpha_k \rho_k \left( \Wv_i^{+} \right) = \min \left( \epsilon_{\alpha_k \rho_k}, \ \alpha_k \rho_k \left( \Wv_i^{\mathrm{HLLC}, +} \right) \right); 
        \end{equation*}
      }
      \Else{
        $\theta_{i+\half}^{\alpha_k \rho_k, +}=1$\;
      }
      
      \uIf{$\alpha_k \rho_k \left( \Wv_{i+1}^{-} \right) < \min \left( \epsilon_{\alpha_k \rho_k}, \ \alpha_k \rho_k \left( \Wv_{i+1}^{\mathrm{HLLC}, -} \right) \right)$}{
        Solve $\theta_{i+\half}^{\alpha_k \rho_k, -}$ from the formula:
        \begin{equation*}
          \left( 1 - \theta_{i+\half}^{\alpha_k \rho_k, -} \right) \alpha_k \rho_k \left( \Wv_{i+1}^{\mathrm{HLLC}, -} \right) + \theta_{i+\half}^{\alpha_k \rho_k, -} \alpha_k \rho_k \left( \Wv_{i+1}^{-} \right) = \min \left( \epsilon_{\alpha_k \rho_k}, \ \alpha_k \rho_k \left( \Wv_{i+1}^{\mathrm{HLLC}, -} \right) \right); 
        \end{equation*}
      }
      \Else{
        $\theta_{i+\half}^{\alpha_k \rho_k, -}=1$\;
      }
      
      Set $\theta_{i+\half}=\min\left( \theta_{i+\half}, \ \theta_{i+\half}^{\alpha_k \rho_k, +}, \ \theta_{i+\half}^{\alpha_k \rho_k, -} \right)$\;
    }
    \For{\textup{all species} $k = 1,2,\cdots,N$}{
      \uIf{$\alpha_k \left( \Wv_i^{+} \right) < \min \left( \epsilon_{\alpha_k}, \ \alpha_k \left( \Wv_i^{\mathrm{HLLC}, +} \right) \right)$}{
        Solve $\theta_{i+\half}^{\alpha_k,+}$ from the formula:
        \begin{equation*}
          \left( 1 - \theta_{i+\half}^{\alpha_k,+} \right) \alpha_k \left( \Wv_i^{\mathrm{HLLC}, +} \right) + \theta_{i+\half}^{\alpha_k,+} \alpha_k \left( \Wv_i^{+} \right) = \min \left( \epsilon_{\alpha_k}, \ \alpha_k \left( \Wv_i^{\mathrm{HLLC}, +} \right) \right); 
        \end{equation*}
      }
      \Else{
        $\theta_{i+\half}^{\alpha_k,+}=1$\;
      }
      
      \uIf{$\alpha_k \left( \Wv_{i+1}^{-} \right) < \min \left( \epsilon_{\alpha_k}, \ \alpha_k \left( \Wv_{i+1}^{\mathrm{HLLC}, -} \right) \right)$}{
        Solve $\theta_{i+\half}^{\alpha_k,-}$ from the formula:
        \begin{equation*}
          \left( 1 - \theta_{i+\half}^{\alpha_k,-} \right) \alpha_k \left( \Wv_{i+1}^{\mathrm{HLLC}, -} \right) + \theta_{i+\half}^{\alpha_k,-} \alpha_k \left( \Wv_{i+1}^{-} \right) = \min \left( \epsilon_{\alpha_k}, \ \alpha_k \left( \Wv_{i+1}^{\mathrm{HLLC}, -} \right) \right); 
        \end{equation*}
      }
      \Else{
        $\theta_{i+\half}^{-}=1$\;
      }
      
      Set $\theta_{i+\half}=\min\left( \theta_{i+\half}, \ \theta_{i+\half}^{\alpha_k, +}, \ \theta_{i+\half}^{\alpha_k, -} \right)$\;
    }
    Perform convex combination:
    \begin{align*}
      \hat{\Gv}_{i+\half}^{*,+} = \left( 1 - \theta_{i+\half} \right) \hat{\Gv}_{i+\half}^{\mathrm{HLLC,+}} + \theta_{i+\half} \hat{\Gv}_{i+\half}^{+}; \\
      \hat{\Gv}_{i+\half}^{*,-} = \left( 1 - \theta_{i+\half} \right) \hat{\Gv}_{i+\half}^{\mathrm{HLLC,-}} + \theta_{i+\half} \hat{\Gv}_{i+\half}^{-};
    \end{align*}
  }
 \caption{Flux limiting procedure for positive partial densities and bounded volume fractions.}
 \label{alg:flux_limit_densities_vol_frac}
\end{algorithm}

\begin{algorithm}[!ht]
\SetAlgoLined
\footnotesize

    \For{\textup{all midpoints}}{
      Compute $\Wv_i^{*, +}$ and $\Wv_{i+1}^{*, -}$\;
      \uIf{$\rho \tilde{e} \left( \Wv_i^{*, +} \right) < \min \left( \epsilon_{\rho \tilde{e} }, \ \rho \tilde{e} \left( \Wv_i^{\mathrm{HLLC}, +} \right) \right)$}{
        Solve $\theta_{i+\half}^{+}$ from the formula:
        \begin{equation*}
          \left( 1 - \theta_{i+\half}^{+} \right) \rho \tilde{e} \left( \Wv_i^{\mathrm{HLLC}, +} \right) + \theta_{i+\half}^{+} \rho \tilde{e} \left( \Wv_i^{*, +} \right) = \min \left( \epsilon_{\rho \tilde{e} }, \ \rho \tilde{e} \left( \Wv_i^{\mathrm{HLLC}, +} \right) \right); 
        \end{equation*}
      }
      \Else{
        $\theta_{i+\half}^{+}=1$\;
      }
      
      \uIf{$\rho \tilde{e} \left( \Wv_{i+1}^{*, -} \right) < \min \left( \epsilon_{\rho \tilde{e} }, \ \rho \tilde{e} \left( \Wv_{i+1}^{\mathrm{HLLC}, -} \right) \right)$}{
        Solve $\theta_{i+\half}^{-}$ from the formula:
        \begin{equation*}
          \left( 1 - \theta_{i+\half}^{-} \right) \rho \tilde{e} \left( \Wv_{i+1}^{\mathrm{HLLC}, -} \right) + \theta_{i+\half}^{-} \rho \tilde{e} \left( \Wv_{i+1}^{*, -} \right) = \min \left( \epsilon_{\rho \tilde{e} }, \ \rho \tilde{e} \left( \Wv_{i+1}^{\mathrm{HLLC}, -} \right) \right); 
        \end{equation*}
      }
      \Else{
        $\theta_{i+\half}^{-}=1$\;
      }
      Set $\theta_{i+\half}=\min\left( \theta_{i+\half}^{+}, \ \theta_{i+\half}^{-} \right)$\;
      Perform convex combination:
      \begin{align*}
        \hat{\Gv}_{i+\half}^{**,+} &= \left( 1 - \theta_{i+\half} \right) \hat{\Gv}_{i+\half}^{\mathrm{HLLC,+}} + \theta_{i+\half} \hat{\Gv}_{i+\half}^{*,+}; \\
        \hat{\Gv}_{i+\half}^{**,-} &= \left( 1 - \theta_{i+\half} \right) \hat{\Gv}_{i+\half}^{\mathrm{HLLC,-}} + \theta_{i+\half} \hat{\Gv}_{i+\half}^{*,-};
      \end{align*}
    }
 \caption{Flux limiting procedure for positive $\rho \tilde{e}$.}
 \label{alg:flux_limit_energy}
\end{algorithm}

The proof of the preservation of the formal order of accuracy of a high-order finite difference or finite volume method in the general form given by equation~\eqref{eq:semi_W_G_pm_eqn} with the reconstructed flux in smooth regions by the limiting method is detailed in our previous paper~\cite{wong2021positivity} when the hard switch is not used.
The overall WCNS-IS spatial discretization with the positivity-preserving interpolation and the quasi-positivity-preserving flux limiters is called PP-WCNS-IS method in the following sections.

\subsection{The proposed fractional algorithm with strong stability preserving Runge--Kutta time stepping}

The fractional algorithm is positivity- and boundedness-preserving with regard to $G$ for first order HLLC and fifth order PP-WCNS-IS (or any high-order finite difference and finite volume schemes in the suitable form given by equation~\eqref{eq:semi_W_G_pm_eqn} with the limiting) when the first order explicit Euler time stepping is used. The extension of the positivity- and boundedness-preserving fractional algorithm with the high-order SSP-RK time stepping methods~\cite{shu1988total,gottlieb2009high,gottlieb2001strong} is trivial since SSP-RK time stepping schemes are convex combinations of Euler forward steps. However, the numerical thermal relaxation needs to be applied at the end of each stage of the time stepping, instead of at the end of the full time step. Nevertheless, the upper limit of CFL number is still constrained by 0.5.

It should also be mentioned that the fractional algorithm gives conservative solutions to the four-equation HRM since (1) the components of the quasi-conservative flux vectors for the transport equations of partial densities, mixture momentum, and mixture total energy are conservative during the hyperbolic advancement of the five-equation model, and (2) the algebraic thermal relaxation step preserves the partial densities, mixture momentum, and mixture total energy.

\subsection{Advantages of the fractional algorithm}

The fractional algorithm is computationally more expensive than solving the four-equation HRM directly since the former approach has $N - 1$ additional equations for flows with $N$ species during the hyperbolic stepping. 
However, the fractional approach has various advantages that are summarized below:
\begin{itemize}
    \item The fractional algorithm with high-order shock-capturing schemes is robust for compressible multi-phase flows with shocks and strong expansion waves, especially physically admissible solutions can be guaranteed with the use of the proposed positivity-preserving limiters for flows composed of one stiffened gas and many ideal gases. This is beneficial as higher order schemes in general have lower dispersion and dissipation errors at the diffuse interfaces, compared to the generally more robust low order shock-capturing schemes.
    \item The speed of sound of the four-equation HRM, which is close to the Wood's sound speed can be obtained asymptotically with the algorithm. This is also verified in one of the tests in section~\ref{sec:test_problems}. The Wood-like sound speed is a better approximation of the sound speed for dispersed flows than the sound speed of five-equation model by Allaire et al.
    \item Many phase transition models~\cite{saurel2016general,chiapolino2017simple} are developed on the four-equation HRM with the thermal equilibrium assumption of the solutions. Therefore, the numerical thermal relaxation step can serve as a bridge between the numerical methods developed for the five-equation model by Allaire et al. and the phase transition models.
    \item Compared to the fractional algorithms using six-equation models~\cite{saurel2009simple,pelanti2014mixture} with both infinitely stiff pressure and thermal relaxation, our fractional algorithm using the five-equation model by Allaire et al. with the thermal relaxation is a computationally more efficient method to obtain solutions to the four-equation HRM for many-species flows since the former has $N-1$ more equations than the latter.
    \item The algorithm is a conservative method for the four-equation HRM and hence it is advantageous for the extension of the method to flows with multi-physics phenomena, such as combustion or phase transition where mass and energy conservation is important. 
\end{itemize}

One well-known disadvantage of the five-equation model by Allaire et al. is that the existence of a mathematical entropy cannot be proved with the pressure equilibrium assumption~\cite{allaire2002five} (unlike the five-equation reduced model by~\citet{kapila2001two} with the existence of a mathematical entropy proved in \cite{murrone2005five}). 
However, we should emphasize that the five-equation model only serves as a step-model in the fractional algorithm to solve the four-equation HRM and there is in fact the existence of the mathematical entropy for the four-equation HRM~\cite{saurel2016general}.
The speed of sound of the four-equation HRM, which is close to the Wood's sound speed, can also be obtained asymptotically with the algorithm, since the five-equation with infinite thermal relaxation rate reduces analytically to the four-equation HRM. This is also verified in one of the tests in the next section.

\clearpage


\section{Test problems} \label{sec:test_problems}

Various 1D and 2D test problems are chosen to test the robustness and accuracy of the positivity-preserving fractional algorithm with PP-WCNS-IS for liquid-gas(es) flows.
Since the main purpose of this work is to show the robustness of the proposed fractional algorithm with a high-order shock-capturing scheme, most of the test problems involve interactions between shock waves and material interfaces and may even be conducted under very extreme conditions.
For a more fundamental analysis of the accuracy, robustness, and positivity-preservation in particular, of the proposed method,
none of these test problems involve phase transition in this work.
Note that the algorithm with phase transition has been tested successfully with some benchmark problems and applications of space launch environments in our another work~\cite{lavaSciTech2023}.
Unless specified otherwise, the species numbers and the fluid properties of the fluids used in the test problems are given in table~\ref{table:fluid_properties}. These problems have either two or three species where the first species is assigned as stiffened gas, while others are ideal gases. Note that when there are three species in the problem, the third species is either sulphur hexafluoride or artificial detonation products.
In all of these problems, the species are required to be in the mechanical and thermal equilibrium conditions initially.
The three-stage third order SSP-RK scheme (SSP-RK3), which is also referred as the third order total variation diminishing Runge--Kutta scheme (TVD-RK3)~\cite{shu1988total}, is used for time stepping in all simulations
except for the temporal convergence test problem.
The CFL number is chosen to be 0.5 to satisfy the constraint for positivity-preservation, unless constant time step size is used\footnote{For the algorithm to be strictly positivity-preserving, the time step size for each full time advancement needs to satisfy the CFL constraints from all intermediate Runge--Kutta solutions within the full step. However, numerical failures are not experienced in any of the tests when the time step size is only computed using the solutions at the beginning of the first Runge--Kutta stage.}. When constant time step size is used, the corresponding CFL number is verified to be always less than but not much smaller than 0.5 until the end of the simulations,
except for the convergence test problems.
We have also verified that the solutions of the algorithm with PP-WCNS-IS stay admissible with positive partial densities, mixture temperature, mixture pressure, squared sound speeds for the five-equation model and four-equation HRM, and bounded volume fractions
over the course of each test simulation.

\begin{table}[!ht]
\small
  \begin{center}
  \begin{tabular}{@{}c | ccccc@{}}\toprule
    Fluid & \textnormal{Species number, $k$} & $\gamma_k$ & \addstackgap{$c_{p,k}\ (\mathrm{J\ kg^{-1} K^{-1}})$} & \addstackgap{$p^{\infty}_k\ (\mathrm{Pa})$} &
    $q_k\ (\mathrm{J\ kg^{-1}})$ \\ \midrule
    Liquid water                             & 1  & 3.0  & 4200.0 & $8.533\mathrm{e}{8}$ & $-0.1148\mathrm{e}{7}$ \\
    Air                                      & 2  & 1.4  & 1007.0 & 0                    & 0                    \\
    Sulphur hexafluoride ($\mathrm{SF_6}$) / & \multirow{2}{*}{3} & 1.1/  & 664.0/ & \multirow{2}{*}{0} & \multirow{2}{*}{0} \\
    Detonation products                      & & 1.25 & 1000.0 & & \\ \bottomrule
  \end{tabular}
  \caption{Properties of the fluids. The properties of the liquid water are obtained from~\cite{barberon2005finite}.}
  \label{table:fluid_properties}
  \end{center}
\end{table}


\subsection{Temporal convergence study}

To test the temporal order of accuracy of the fractional algorithm with the five-equation model by Allaire et al. and the numerical thermal relaxation, the advection of volume fraction disturbance in a 1D periodic domain $[-1, 1) \ \mathrm{m}$ is used similarly as the convergence test in our previous work~\cite{wong2021positivity}. The pressure and temperature fields are uniformly at $p=101325$ Pa and $T=298$ K respectively.
The exact solutions are given by table~\ref{table:exact_1D_convergence} and the initial conditions are given by the solutions at $t=0$.


\begin{table}[!ht]
\small
  \begin{center}
    \begin{tabular}{@{}ccccccc@{}}\toprule
    \addstackgap{\stackanchor{$\rho_1$}{$(\mathrm{kg\ m^{-3}})$}} &
    \stackanchor{$\rho_2$}{$(\mathrm{kg\ m^{-3}})$} &
    \stackanchor{$u$}{$(\mathrm{m\ s^{-1}})$} &
    \stackanchor{$p$}{$(\mathrm{Pa})$} &
    \stackanchor{$T$}{$(\mathrm{K})$} &
    $\alpha_1$ \\ \midrule
    \addstackgap{1022.7724412751677} & 1.1817862212832324 & 10 & 101325 & 298 & $0.5 + 0.25 \sin \left[ \pi (x - 10t) \right]$ \\\bottomrule
    \end{tabular}
  \end{center}
  \caption{Exact solutions of the 1D temporal convergence problem.}
  \label{table:exact_1D_convergence}
\end{table}

Since the exact solutions are known, the derivatives of the convective flux can also be computed exactly and they are derived as:
\begin{align}
    \partial_x \left( \alpha_1 \rho_1 u \right) &= \rho_{1,\mathrm{cst}} \, u_{\mathrm{cst}} \, \partial_x \alpha_1 , \\
    \partial_x \left( \alpha_2 \rho_2 u \right) &= \rho_{2,\mathrm{cst}} u_{\mathrm{cst}} \, \partial_x \alpha_2 , \\
    \partial_x \left( \rho u^2 + p \right) &= u_{\mathrm{cst}}^2 \, \partial_x \rho , \\
    \partial_x \left[ \left( E + p \right) u \right] &= u_{\mathrm{cst}} \, \partial_x \left( \rho e \right) + \frac{u_{\mathrm{cst}}^3}{2} \, \partial_x \rho, \\
    u \partial_x \alpha_1 &= u_{\mathrm{cst}} \, \partial_x \alpha_1 ,
\end{align}
where
\begin{align}
    \partial_x \alpha_1 &= 0.25 \pi \cos \left[ \pi (x - 10t) \right] , \quad
    \partial_x \alpha_2 = - \partial_x \alpha_1 , \\
    \partial_x \rho     &= \rho_{1,\mathrm{cst}} \, \partial_x \alpha_{1} + 
    \rho_{2,\mathrm{cst}} \, \partial_x \alpha_{2} , \\
    \partial_x \left( \rho e \right) &=
    \left( \rho_1 e_1 \right)_{\mathrm{cst}} \partial_x \alpha_1 +
    \left( \rho_2 e_2 \right)_{\mathrm{cst}} \partial_x \alpha_2 .
\end{align}
$\left( \rho_k e_k \right)_{\mathrm{cst}}$ of species $k$ is given by
\begin{equation}
    \left( \rho_k e_k \right)_{\mathrm{cst}} =
    \frac{p_{\mathrm{cst}}}{\gamma_k - 1} + \frac{\gamma_k p^{\infty}_k}{\gamma_k - 1} + \rho_{k,\mathrm{cst}} q_k .
\end{equation}
The constant values of the species density $\rho_{k,\mathrm{cst}}$, mixture velocity $u_{\mathrm{cst}}$, and mixture pressure $p_{\mathrm{cst}}$ are given by the exact solutions in table~\ref{table:exact_1D_convergence}.

Simulations using different time stepping schemes with the thermal relaxation step are conducted with number of grid cells $N_x = 64$ until $t = 0.1\ \mathrm{s}$ with time step refinements from 10 steps to 80 steps. The time stepping schemes chosen are the forward Euler, two-stage second order SSP-RK (SSP-RK2) scheme~\cite{gottlieb2001strong}, and SSP-RK3 scheme~\cite{shu1988total}.
The derivatives for the SSP-RK2 at the second stage are computed at $t^n + \dt$ and those for SSP-RK3 at the second and third stages are computed at $t^n + \dt$ and $t^n + \dt/2$ respectively, where $t^n$ is the start time of the time step.
Tables~\ref{table:1D_errors_and_rate_of_temporal_convergence_advection_euler}, \ref{table:1D_errors_and_rate_of_temporal_convergence_advection_TVDRK2}, and \ref{table:1D_errors_and_rate_of_temporal_convergence_advection_TVDRK3} show the $L_2$ and $L_{\infty}$ errors together with the corresponding measured rates of convergence of mixture density $\rho$ and volume fraction $\alpha_1$ at $t = 0.1\ \mathrm{s}$ using different time stepping schemes. The $L_2$ error of mixture density is computed as:
\begin{align}
    L_2\ \mathrm{error}&= \sqrt{ \sum_{i=0}^{N-1} \Delta x \left( {\rho}_{i} - {\rho}_{i}^{\mathrm{exact}} \right)^2 / \sum_{i=0}^{N-1} \Delta x },
\end{align}
where ${\rho}_{i}^{\mathrm{exact}}$ is the exact solution of mixture density at the corresponding grid point. The computation of $L_2$ error of volume fraction is similar.
It can be seen from the table that the forward Euler and SSP-RK2 schemes achieve the expected rates of convergence in time. It is interesting to see that one higher order of accuracy is obtained by SSP-RK3 in this problem. While it is not clear why the measured order of accuracy of SSP-RK3 is higher than the expected one, at least none of the time stepping schemes have order degradation in this test because of the thermal relaxation step.  

\begin{table}[!ht]
\centering
\small
\begin{tabular}{@{}c | cccc | cccc@{}}\toprule
              & \multicolumn{4}{c|}{Mixture density, $\rho$} & \multicolumn{4}{c}{Volume fraction, $\alpha_1$} \\
    \cmidrule(r){2-9}
    \addstackgap{\stackanchor{Number of}{time steps}} & \addstackgap{\stackanchor{$L_2$ error}{$(\mathrm{kg\ m^{-3}})$}} & $L_2$ order & \addstackgap{\stackanchor{$L_\infty$ error}{$(\mathrm{kg\ m^{-3}})$}} & $L_\infty$ order & $L_2$ error & $L_2$ order & $L_\infty$ error & $L_\infty$ order \\ \midrule
    10 & $5.681\mathrm{e}{+01}$ &      & $8.035\mathrm{e}{+01}$ &      & $5.561\mathrm{e}{-02}$ &      & $7.865\mathrm{e}{-02}$ &      \\
    20 & $2.838\mathrm{e}{+01}$ & 1.00 & $4.012\mathrm{e}{+01}$ & 1.00 & $2.778\mathrm{e}{-02}$ & 1.00 & $3.927\mathrm{e}{-02}$ & 1.00 \\
    40 & $1.418\mathrm{e}{+01}$ & 1.00 & $2.005\mathrm{e}{+01}$ & 1.00 & $1.389\mathrm{e}{-02}$ & 1.00 & $1.962\mathrm{e}{-02}$ & 1.00 \\
    80 & $7.092\mathrm{e}{+00}$ & 1.00 & $1.002\mathrm{e}{+01}$ & 1.00 & $6.942\mathrm{e}{-03}$ & 1.00 & $9.809\mathrm{e}{-03}$ & 1.00 \\
\end{tabular}
\caption{Errors and temporal orders of convergence of mixture density and volume fraction for the 1D advection problem using the fractional algorithm with the exact spatial derivatives and 
forward Euler time stepping scheme
at $t = 0.1\ \mathrm{s}$.}
\label{table:1D_errors_and_rate_of_temporal_convergence_advection_euler}
%
\centering
\small
\begin{tabular}{@{}c | cccc | cccc@{}}\toprule
              & \multicolumn{4}{c|}{Mixture density, $\rho$} & \multicolumn{4}{c}{Volume fraction, $\alpha_1$} \\
    \cmidrule(r){2-9}
    \addstackgap{\stackanchor{Number of}{time steps}} & \addstackgap{\stackanchor{$L_2$ error}{$(\mathrm{kg\ m^{-3}})$}} & $L_2$ order & \addstackgap{\stackanchor{$L_\infty$ error}{$(\mathrm{kg\ m^{-3}})$}} & $L_\infty$ order & $L_2$ error & $L_2$ order & $L_\infty$ error & $L_\infty$ order \\ \midrule
    10 & $2.976\mathrm{e}{+00}$ &      & $4.203\mathrm{e}{+00}$ &      & $2.913\mathrm{e}{-03}$ &      & $4.114\mathrm{e}{-03}$ &      \\
    20 & $7.430\mathrm{e}{-01}$ & 2.00 & $1.049\mathrm{e}{+00}$ & 2.00 & $7.273\mathrm{e}{-04}$ & 2.00 & $1.027\mathrm{e}{-03}$ & 2.00 \\
    40 & $1.857\mathrm{e}{-01}$ & 2.00 & $2.623\mathrm{e}{-01}$ & 2.00 & $1.818\mathrm{e}{-04}$ & 2.00 & $2.567\mathrm{e}{-04}$ & 2.00 \\
    80 & $4.642\mathrm{e}{-02}$ & 2.00 & $6.557\mathrm{e}{-02}$ & 2.00 & $4.544\mathrm{e}{-05}$ & 2.00 & $6.418\mathrm{e}{-05}$ & 2.00 \\
\end{tabular}
\caption{Errors and temporal orders of convergence of mixture density and volume fraction for the 1D advection problem using the fractional algorithm with the exact spatial derivatives and 
two-stage second order SSP-RK (SSP-RK2) time stepping scheme
at $t = 0.1\ \mathrm{s}$.}
\label{table:1D_errors_and_rate_of_temporal_convergence_advection_TVDRK2}
%
\centering
\small
\begin{tabular}{@{}c | cccc | cccc@{}}\toprule
              & \multicolumn{4}{c|}{Mixture density, $\rho$} & \multicolumn{4}{c}{Volume fraction, $\alpha_1$} \\
    \cmidrule(r){2-9}
    \addstackgap{\stackanchor{Number of}{time steps}} & \addstackgap{\stackanchor{$L_2$ error}{$(\mathrm{kg\ m^{-3}})$}} & $L_2$ order & \addstackgap{\stackanchor{$L_\infty$ error}{$(\mathrm{kg\ m^{-3}})$}} & $L_\infty$ order & $L_2$ error & $L_2$ order & $L_\infty$ error & $L_\infty$ order \\ \midrule
    10 & $1.225\mathrm{e}{-03}$ &      & $1.731\mathrm{e}{-03}$ &      & $1.199\mathrm{e}{-06}$ &      & $1.694\mathrm{e}{-06}$ &      \\
    20 & $7.641\mathrm{e}{-05}$ & 4.00 & $1.079\mathrm{e}{-04}$ & 4.00 & $7.479\mathrm{e}{-08}$ & 4.00 & $1.056\mathrm{e}{-07}$ & 4.00 \\
    40 & $4.773\mathrm{e}{-06}$ & 4.00 & $6.742\mathrm{e}{-06}$ & 4.00 & $4.672\mathrm{e}{-09}$ & 4.00 & $6.599\mathrm{e}{-09}$ & 4.00 \\
    80 & $2.983\mathrm{e}{-07}$ & 4.00 & $4.213\mathrm{e}{-07}$ & 4.00 & $2.920\mathrm{e}{-10}$ & 4.00 & $4.124\mathrm{e}{-10}$ & 4.00 \\
\end{tabular}
\caption{Errors and temporal orders of convergence of mixture density and volume fraction for the 1D advection problem using the fractional algorithm with the exact spatial derivatives and 
three-stage third order SSP-RK (SSP-RK3) time stepping scheme
at $t = 0.1\ \mathrm{s}$.}
\label{table:1D_errors_and_rate_of_temporal_convergence_advection_TVDRK3}
\end{table}

\subsection{Spatial convergence study}

In order to verify the spatial order of accuracy of the fractional algorithm using PP-WCNS-IS, the previous advection problem is extended to a 2D periodic domain $[-1, 1) \ \mathrm{m} \times[-1, 1) \ \mathrm{m}$. Following the previous convergence test, pressure and temperature fields are uniformly at $p=101325$ Pa and $T=298$ K.
The exact solutions are given by table~\ref{table:exact_2D_convergence} and the problem is initialized at $t=0$.


\begin{table}[!ht]
\small
  \begin{center}
    \begin{tabular}{@{}ccccccc@{}}\toprule
    \addstackgap{\stackanchor{$\rho_1$}{$(\mathrm{kg\ m^{-3}})$}} &
    \stackanchor{$\rho_2$}{$(\mathrm{kg\ m^{-3}})$} &
    \stackanchor{$u$}{$(\mathrm{m\ s^{-1}})$} &
    \stackanchor{$v$}{$(\mathrm{m\ s^{-1}})$} &
    \stackanchor{$p$}{$(\mathrm{Pa})$} &
    \stackanchor{$T$}{$(\mathrm{K})$} &
    $\alpha_1$ \\ \midrule
    \addstackgap{1022.7724412751677} & 1.1817862212832324 & 10 & 10 & 101325 & 298 & $0.5 + 0.25 \sin \left[ \pi (x + y - 20t) \right]$ \\\bottomrule
    \end{tabular}
  \end{center}
  \caption{Exact solutions of the 2D spatial convergence problem.}
  \label{table:exact_2D_convergence}
\end{table}

\noindent Simulations using PP-WCNS-IS with SSP-RK3 are conducted up to $t = 0.1\ \mathrm{ms}$ with mesh refinements from $N_x = N_y = 8$ to $N_x = N_y = 256$. All simulations are run with very small constant time steps in order to observe the spatial order of accuracy of PP-WCNS-IS. $\Delta t / \Delta x = 4.0\mathrm{e}{-5} \ \mathrm{s\ m^{-1}}$ is used.

Table~\ref{table:2D_errors_and_rate_of_convergence_advection} shows the measured rates of convergence of mixture density $\rho$ and volume fraction $\alpha_1$ from $L_2$ and $L_{\infty}$ errors at $t = 0.1\ \mathrm{ms}$. The $L_2$ error of mixture density is computed as:
\begin{align}
    L_2\ \mathrm{error}&= \sqrt{ \sum_{i=0}^{N-1} \sum_{j=0}^{N-1} \Delta x \Delta y \left( {\rho}_{i, j} - {\rho}_{i, j}^{\mathrm{exact}} \right)^2 / \sum_{i=0}^{N-1} \sum_{j=0}^{N-1} \Delta x \Delta y },
\end{align}
where ${\rho}_{i, j}^{\mathrm{exact}}$ is the exact solution of mixture density at the corresponding grid point. The computation of $L_2$ error of volume fraction is similar.
It can be seen from the table that the spatial discretization achieves the expected rates of convergence in this 2D test problem.

\begin{table}[!ht]
\small
\centering
\begin{tabular}{@{}c | cccc | cccc@{}}\toprule
              & \multicolumn{4}{c|}{Mixture density, $\rho$} & \multicolumn{4}{c}{Volume fraction, $\alpha_1$} \\
    \cmidrule(r){2-9}
    \addstackgap{\stackanchor{Number of}{grid points}} & \addstackgap{\stackanchor{$L_2$ error}{$(\mathrm{kg\ m^{-3}})$}} & $L_2$ order & \addstackgap{\stackanchor{$L_\infty$ error}{$(\mathrm{kg\ m^{-3}})$}} & $L_\infty$ order & $L_2$ error & $L_2$ order & $L_\infty$ error & $L_\infty$ order \\ \midrule
      $8^2$ & $3.203\mathrm{e}{-02}$ &      & $3.594\mathrm{e}{-02}$ &      & $3.136\mathrm{e}{-05}$ &      & $3.518\mathrm{e}{-05}$ &     \\
     $16^2$ & $1.550\mathrm{e}{-04}$ & 7.69 & $3.109\mathrm{e}{-04}$ & 6.85 & $1.517\mathrm{e}{-07}$ & 7.69 & $3.044\mathrm{e}{-07}$ & 6.85\\
     $32^2$ & $4.960\mathrm{e}{-06}$ & 4.97 & $7.385\mathrm{e}{-06}$ & 5.40 & $4.855\mathrm{e}{-09}$ & 4.97 & $7.229\mathrm{e}{-09}$ & 5.40\\
     $64^2$ & $1.715\mathrm{e}{-07}$ & 4.85 & $2.432\mathrm{e}{-07}$ & 4.92 & $1.679\mathrm{e}{-10}$ & 4.85 & $2.381\mathrm{e}{-10}$ & 4.92\\
    $128^2$ & $5.498\mathrm{e}{-09}$ & 4.96 & $7.781\mathrm{e}{-09}$ & 4.97 & $5.382\mathrm{e}{-12}$ & 4.96 & $7.616\mathrm{e}{-12}$ & 4.97\\
    $256^2$ & $1.751\mathrm{e}{-10}$ & 4.97 & $2.574\mathrm{e}{-10}$ & 4.92 & $1.713\mathrm{e}{-13}$ & 4.97 & $2.538\mathrm{e}{-13}$ & 4.91\\ \bottomrule
\end{tabular}
\caption{Errors and spatial orders of convergence of mixture density and volume fraction for the 2D advection problem using the fractional algorithm
with the PP-WCNS-IS at $t = 0.1\ \mathrm{ms}$.}
\label{table:2D_errors_and_rate_of_convergence_advection}
\end{table}

While the disturbances in this and the previous advection test cases are smooth, a problem that involves 1D advection of discontinuous material interfaces is also considered and discussed in \ref{appendix:1D_material_advection}.


\subsection{One-dimensional acoustic wave propagation}

As proved in the previous section, the five-equation model with the infinitely fast thermal relaxation can be analytically reduced to the four-equation HRM.
The purpose of this sub-section is to shown that the fractional algorithm composed of the time advancement of five-equation model by Allaire et al. 
using first order HLLC or PP-WCNS-IS methods
and the numerical thermal relaxation has the expected mixture sound speed of four-equation HRM.
To achieve that, a numerical experiment is set up to verify that the fractional algorithm can yield acoustic waves propagating at the mixture sound speed given by the four-equation model. This is conducted in a 1D domain, $x \in \left[-1, 1 \right] \ \mathrm{m}$, that is split into two sub-domains initially at $x_i = -0.5\ \mathrm{m}$. The left region is essentially composed of air with $\alpha_1 \approx 1.0\mathrm{e}{-8}$ and the right region is a uniform mixture with $\alpha_1 = \alpha_2 = 0.5$ at the beginning of the simulations.

The mixture density and pressure fields in the left region are initially perturbed:
\begin{align}
    \rho &= \rho_{\mathrm{ref}} + \rho^{\prime}, \\
    p    &=  p_{\mathrm{ref}}   + p^{\prime},
\end{align}
where the background pressure and temperature are $p_{\mathrm{ref}} = 101325\ \mathrm{Pa}$ and $T_{\mathrm{ref}} = 298\ \mathrm{K}$ respectively. The background mixture density $\rho_{\mathrm{ref}}$ is obtained with the background volume fraction $\alpha_{1,\mathrm{ref}} = 1.0\mathrm{e}{-8}$. The perturbed mixture density and pressure fields are given by:
\begin{align}
    \rho^{\prime} &= A \exp \left[ - \left( \frac{x - x_{\mathrm{center}}}{\sigma} \right)^2 \right] , \\
    p^{\prime} &= c_{\mathrm{ref}}^2 \rho^{\prime},
\end{align}
where $c_{\mathrm{ref}} = 346.46\ \mathrm{m\ s^{-1}}$ is the mixture sound speed given by the four-equation model using the background conditions and is basically the sound speed of pure air. $x_{\mathrm{center}} = -0.75\ \mathrm{m}$ is the location of the peak of the initial Gaussian pulse. After the mixture density and pressure fields are computed, the mixture temperature in this left region can then obtained. The species densities are computed with the perturbed mixture temperature and pressure. At the end, the mass fractions and volume fractions can be obtained after the species densities are evaluated. The initial conditions are given by table~\ref{table:IC_1D_acoustic_wave_propagation_problem}. Extrapolations are used at the domain boundaries for waves leaving the domain.

While analytically the five-equation model is reduced to the four-equation HRM with infinitely fast thermal relaxation, it is important to verify that five-equation model with numerical thermal relaxation can yield sound speed that asymptotically converges to the sound speed of the four-equation model. $A = 0.01$ and $\sigma = 0.05$ are chosen for the numerical experiment. Two acoustic waves travel through the computational domain in the left and right directions from the initial pulse. Part of the right travelling pulse is reflected at the interface at $x_i$ and part of it is transmitted into the mixture region. The sound speed of the mixture region is measured as the wave speed of the transmitted acoustic pulse from $t = 30\ \mathrm{ms}$ to $t = 50\ \mathrm{ms}$. The distance of the acoustic pulse travelled is computed with its centroid at the two times, where a Gaussian shape of the transmitted pulse is assumed.

\begin{figure}[!ht]
\centering
\includegraphics[width=0.6\textwidth]{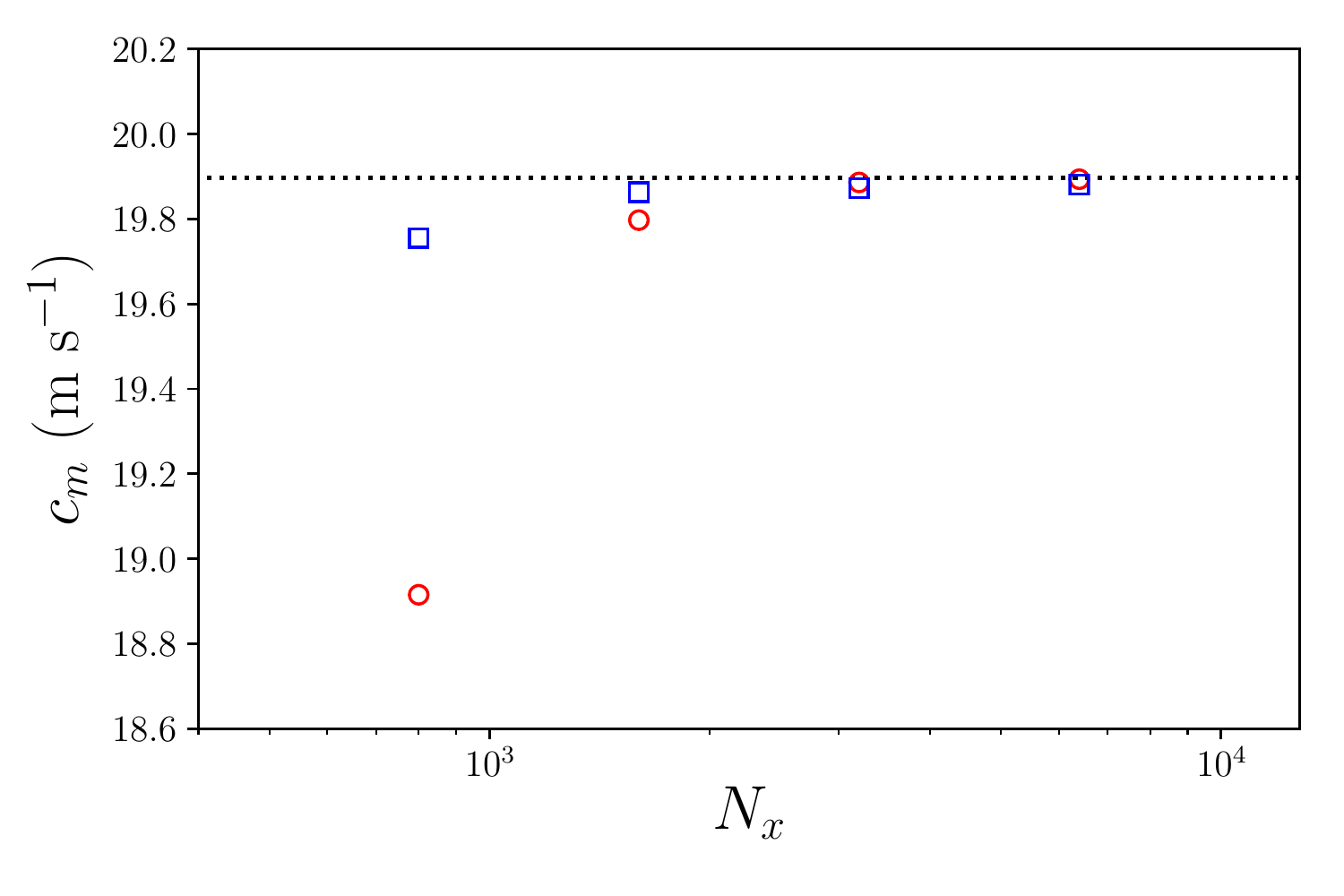}
\caption{Grid convergence of measured sound speed of the fractional algorithm using the five-equation model and the numerical thermal relaxation with different schemes for the acoustic wave propagation problem. Black dotted line: analytical sound speed of four-equation model; red circles: measured sound speed with the first order HLLC; blue squares: measured sound speed with the PP-WCNS-IS.}
\label{fig:compare_sound_speed_five_eqn_thermal_relax}
\end{figure}

\begin{figure}[!ht]
\centering
\includegraphics[width=0.6\textwidth]{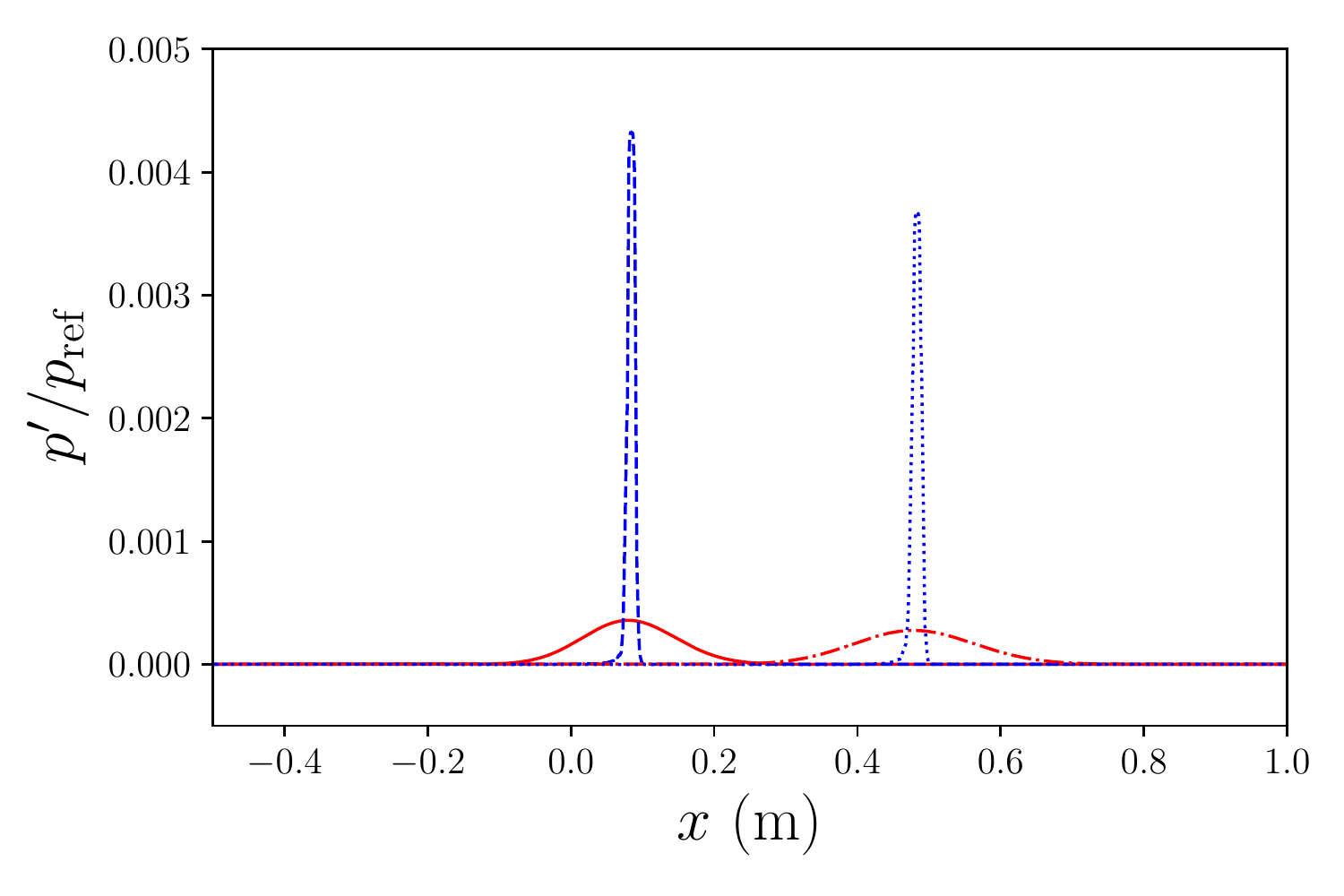}
\caption{Normalized pressure fluctuations computed with the fractional algorithm using the five-equation model and the numerical thermal relaxation with different schemes at two different times on a grid with 6400 points of the acoustic wave propagation problem. Red solid line: first order HLLC at $t = 30\ \mathrm{ms}$; blue dashed line PP-WCNS-IS at $t = 30\ \mathrm{ms}$; red dash-dotted line: first order HLLC at $t = 50\ \mathrm{ms}$; blue dotted line PP-WCNS-IS at $t = 50\ \mathrm{ms}$.}
\label{fig:compare_pressure_five_eqn_thermal_relax}
\end{figure}

Figure~\ref{fig:compare_sound_speed_five_eqn_thermal_relax} shows the grid sensitivity test of the measured sound speed $c_{m}$ of the fractional algorithm using the first order HLLC scheme and PP-WCNS-IS with the number of grid points varying from $N_x = 800$ to $N_x = 6400$. With the volume fraction values of the right sub-domain, the sound speeds given by the Wood's formulation, five-equation model by Allaire et al., and the four-equation HRM are respectively $23.540\ \mathrm{m\ s^{-1}}$, $913.05\ \mathrm{m\ s^{-1}}$, and $19.897\ \mathrm{m\ s^{-1}}$. From the plot, it can be seen that the measured sound speeds from both schemes converge to the analytical value of the four-equation HRM. The sound speeds given by PP-WCNS-IS converges at a faster rate due to smaller dispersion error compared to the first order HLLC scheme. Figure~\ref{fig:compare_pressure_five_eqn_thermal_relax} compares the normalized pressure fluctuations from the background pressure inside the mixture region at times $t = 30\ \mathrm{ms}$ and $t = 50\ \mathrm{ms}$ computed with the two schemes. It can be seen that the pressure pulses of the HLLC scheme have smaller peaks and larger spreads compared to those of the PP-WCNS-IS as the first order scheme has much larger dissipation error.

\begin{table}[!ht]
\small
  \begin{center}
    \begin{tabular}{@{}c | ccccc@{}}\toprule
    $(m)$ &
    \addstackgap{\stackanchor{$\alpha_1 \rho_1$}{$(\mathrm{kg\ m^{-3}})$}} &
    \stackanchor{$\alpha_2 \rho_2$}{$(\mathrm{kg\ m^{-3}})$} &
    \stackanchor{$u$}{$(\mathrm{m\ s^{-1}})$} &
    \stackanchor{$p$}{$(\mathrm{Pa})$} &
    $\alpha_1$ \\ \midrule
    \addstackgap{$x < -0.5$} & \addstackgap{\stackanchor{$1.3689201806025044\mathrm{e}{-5}$}{$+ \left( \alpha_1 \rho_1 \right)^{\prime}$}} & \addstackgap{\stackanchor{$1.1817862094653702$}{$ + \left( \alpha_2 \rho_2 \right)^{\prime}$}} & 0 & $101325 + p^{\prime}$ & $1.0\mathrm{e}{-8} + \alpha_1^{\prime}$ \\
    \addstackgap{$x \geq -0.5$} & $684.4600903012522$ & $0.5908931106416162$ & 0 & $101325$ & $0.5$ \\ \bottomrule
    \end{tabular}
  \end{center}
  \caption{Initial conditions of 1D acoustic wave propagation problem.}
  \label{table:IC_1D_acoustic_wave_propagation_problem}
\end{table}

\subsection{One-dimensional gas/liquid shock tube problem}

This gas/liquid shock tube problem is similar to the ones by~\citet{chen2008flow} and \citet{wang2018incremental}. Initially there is a discontinuity at $x = 0.8\ \mathrm{m}$. On each side of the discontinuity, the mixture is at pressure and temperature equilibria, where the pressure values on the left and right sides are respectively $1.0\mathrm{e}{9}\ \mathrm{Pa}$ and $1.0\mathrm{e}{5}\ \mathrm{Pa}$. The left and right portions of the shock tube are mainly composed by liquid water and air respectively, where the species density are $\rho_1 = 1000\ \mathrm{kg\ m^{-3}}$ and $\rho_2 = 20\ \mathrm{kg\ m^{-3}}$.
This gives temperature values of $661.89\ \mathrm{K}$ and $17.378\ \mathrm{K}$ on the left and right sides respectively.
The initial conditions are given by table~\ref{table:IC_1D_gas_liquid_shock_tube_problem}.
Extrapolations are applied at both boundaries. The spatial domain is $x \in \left[0, 1.5 \right] \ \mathrm{m}$ and the final time is at $t = 3\mathrm{e}{-4} \ \mathrm{s}$.
A shock wave, a material discontinuity, and a rarefaction wave are generated from the initial discontinuity, where the two discontinuities propagate to the right direction while the rarefaction wave travels to the left.
Three cases of simulations are conducted with different combinations of numerical algorithms and spatial schemes, which are respectively four-equation HRM using first order HLLC, and fractional algorithm with five-equation model and thermal relaxation using first order HLLC and PP-WCNS-IS. When the first order HLLC scheme is used, $\Delta t = 5.0\mathrm{e}{-8} \ \mathrm{s}$ on a uniform grid with 5000 grid points is chosen.
As for the PP-WCNS-IS method, two uniform grids with 500 and 1000 grid points are used with $\Delta t = 5.0\mathrm{e}{-7} \ \mathrm{s}$ and $\Delta t = 2.5\mathrm{e}{-7} \ \mathrm{s}$ respectively.

\begin{table}[!ht]
\small
  \begin{center}
    \begin{tabular}{@{}c | ccccc@{}}\toprule
    $(m)$ &
    \addstackgap{\stackanchor{$\alpha_1 \rho_1$}{$(\mathrm{kg\ m^{-3}})$}} &
    \stackanchor{$\alpha_2 \rho_2$}{$(\mathrm{kg\ m^{-3}})$} &
    \stackanchor{$u$}{$(\mathrm{m\ s^{-1}})$} &
    \stackanchor{$p$}{$(\mathrm{Pa})$} &
    $\alpha_1$ \\ \midrule
    \addstackgap{$x < 0.8$} & $9.9999998999999991\mathrm{e}{2}$ & $5.2511071660640894\mathrm{e}{-5}$ & 0 & $1.0\mathrm{e}{9}$ & $1 - 1.0\mathrm{e}{-8}$ \\
    \addstackgap{$x \geq 0.8$} & $1.7538240816326533\mathrm{e}{-4}$ & $1.9999999799999998\mathrm{e}{1}$ & 0 & $1.0\mathrm{e}{5}$ & $1.0\mathrm{e}{-8}$ \\ \bottomrule
    \end{tabular}
  \end{center}
  \caption{Initial conditions of the 1D gas/liquid shock tube problem.}
  \label{table:IC_1D_gas_liquid_shock_tube_problem}
\end{table}

\begin{figure}[!ht]
\centering
\subfigure[Partial density of liquid profile]{%
\includegraphics[width=0.45\textwidth]{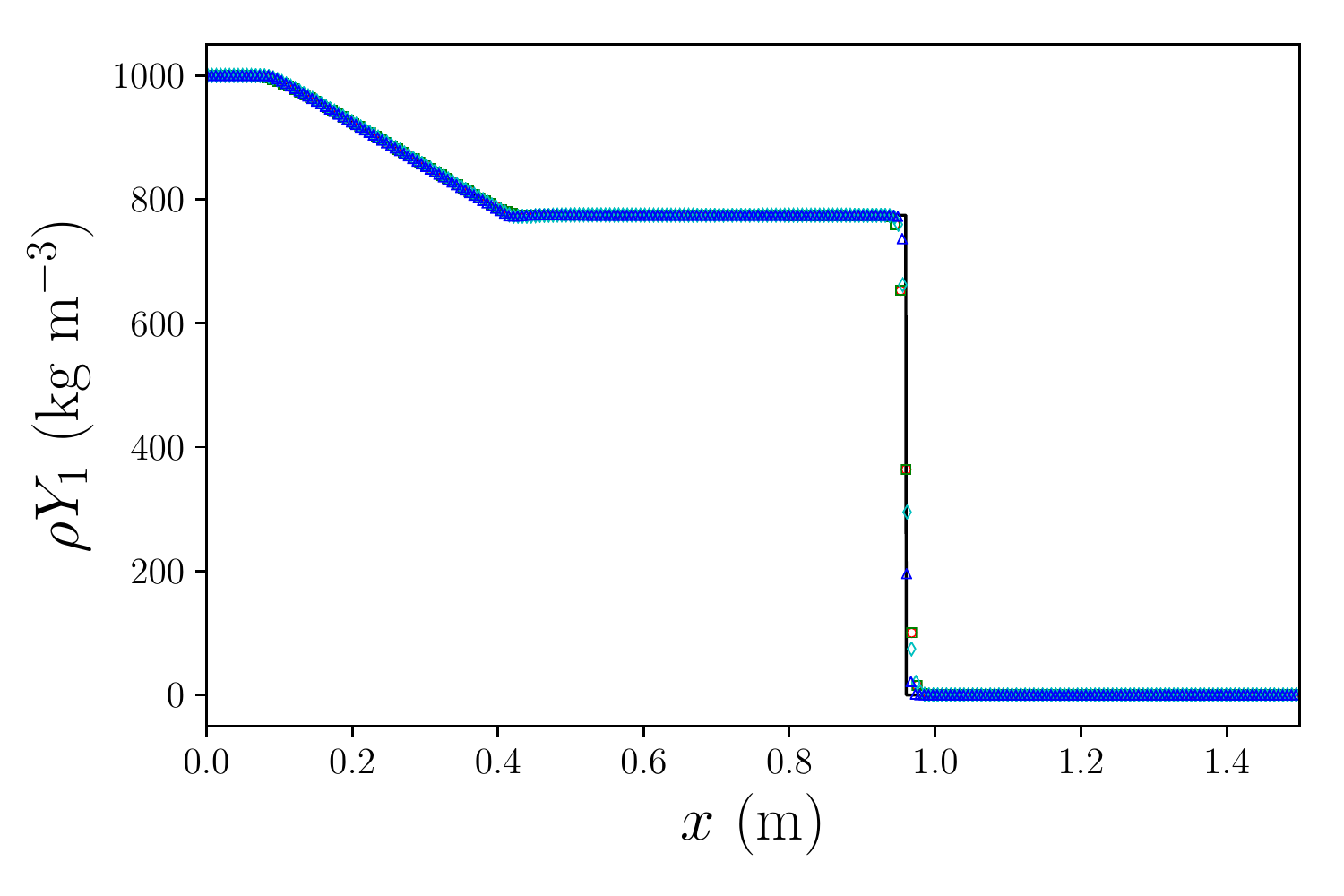}
\label{fig:compare_gas_liquid_rhoY1_global}}
\subfigure[Partial density of gas profile]{%
\includegraphics[width=0.45\textwidth]{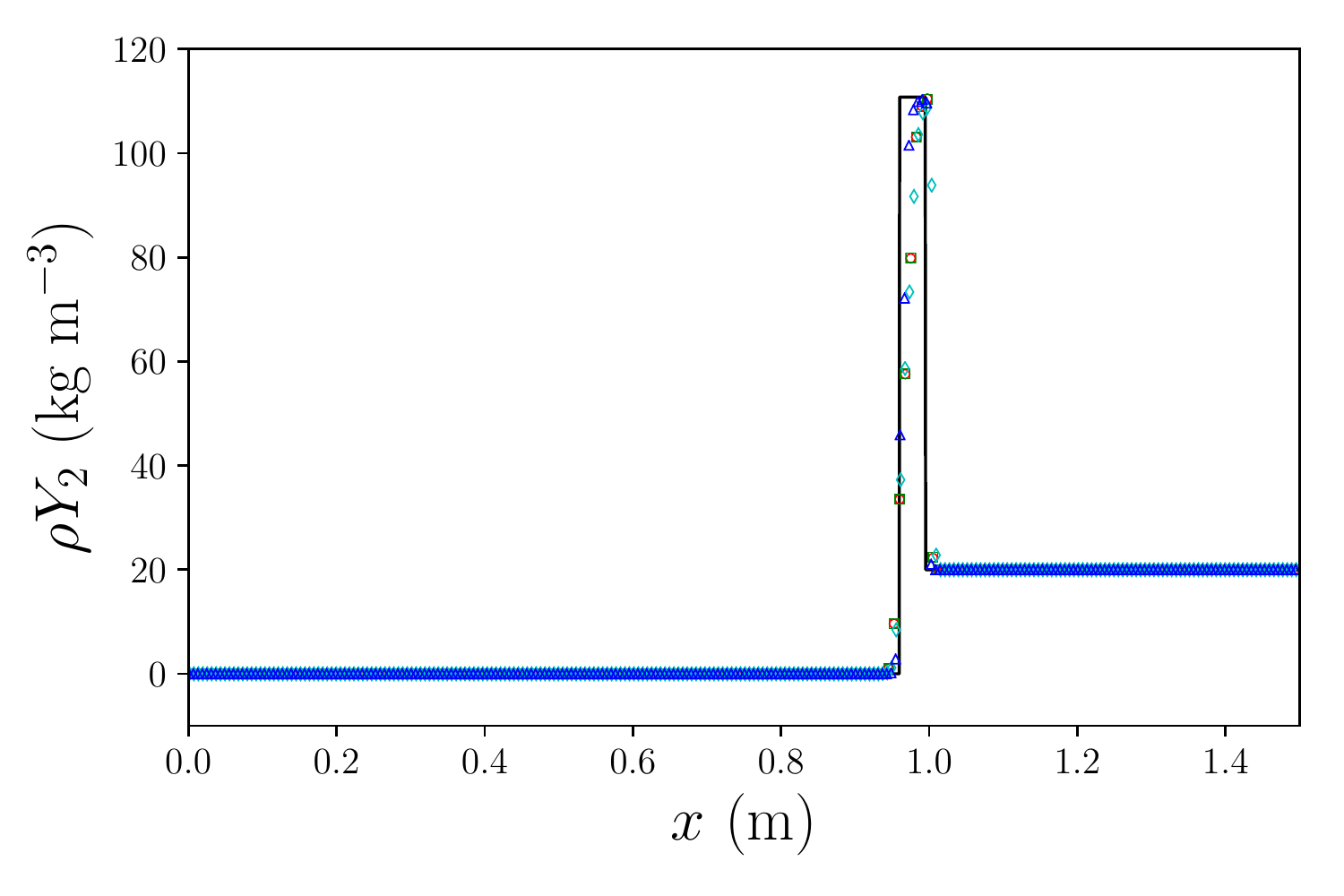}
\label{fig:compare_gas_liquid_rhoY2_global}}
\subfigure[Velocity profile]{%
\includegraphics[width=0.45\textwidth]{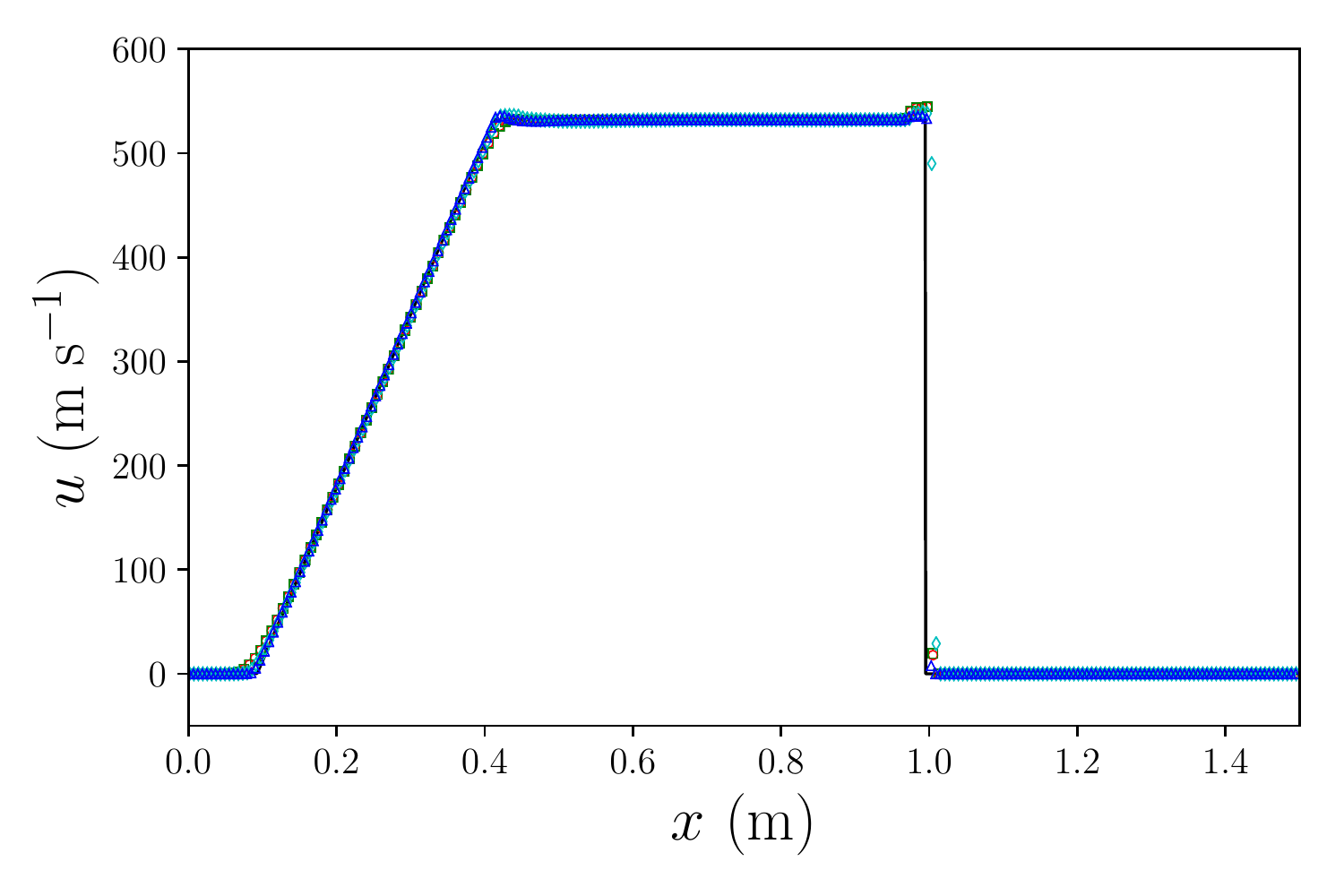}
\label{fig:compare_gas_liquid_u_global}}
\subfigure[Pressure profile]{%
\includegraphics[width=0.45\textwidth]{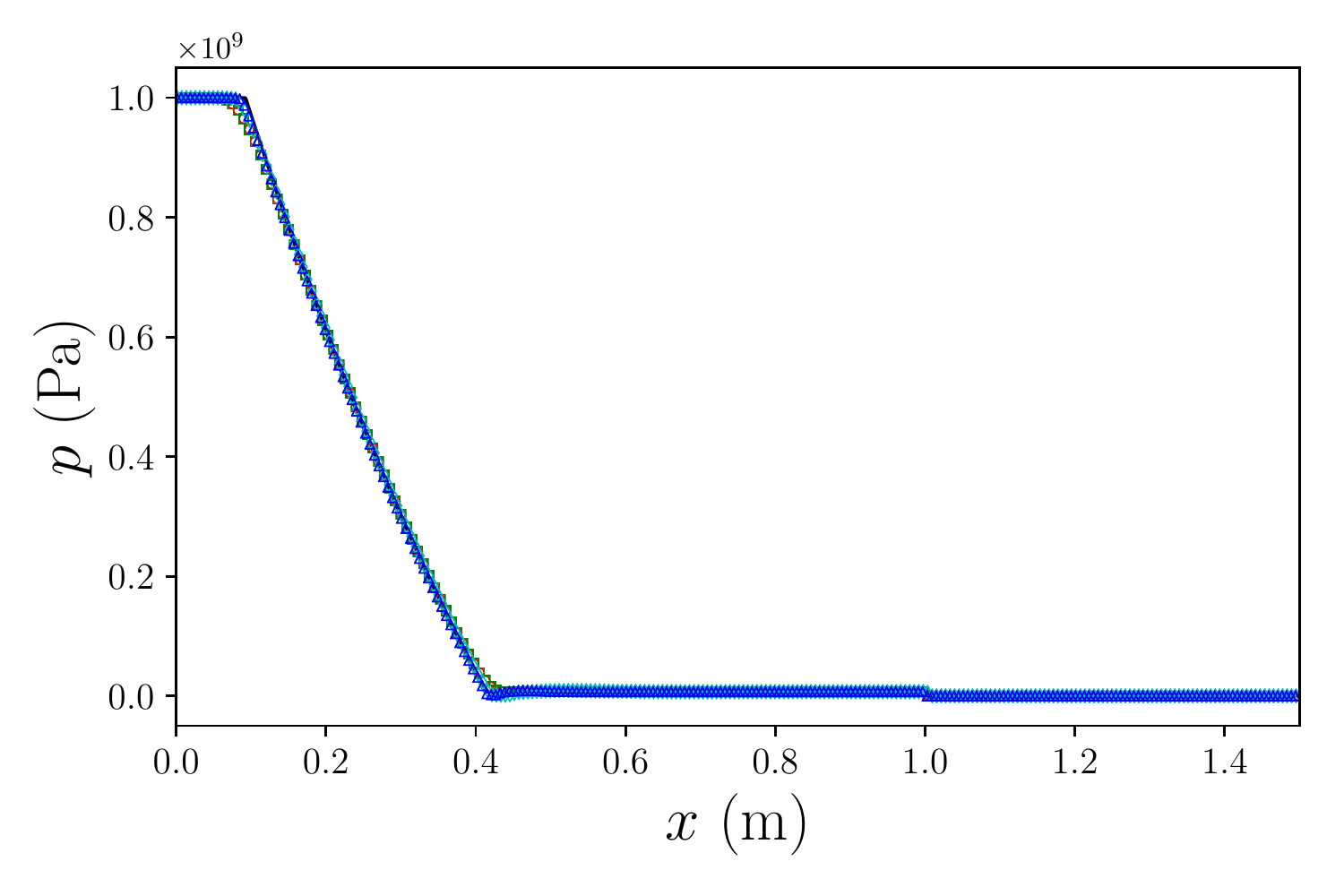}
\label{fig:compare_gas_liquid_p_global}}
\subfigure[Temperature profile]{%
\includegraphics[width=0.45\textwidth]{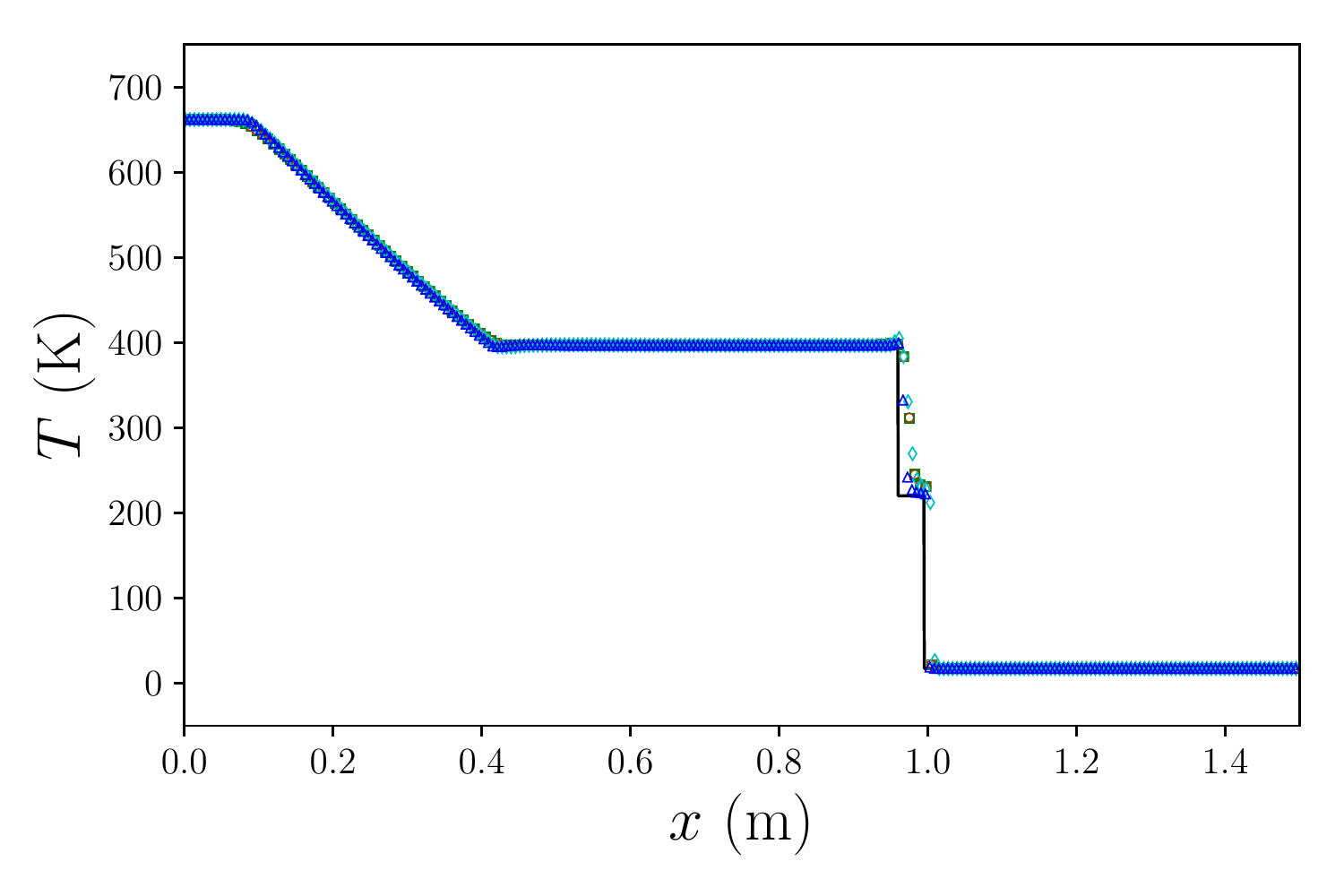}
\label{fig:compare_gas_liquid_T_global}}
\subfigure[Volume fraction of liquid profile]{%
\includegraphics[width=0.45\textwidth]{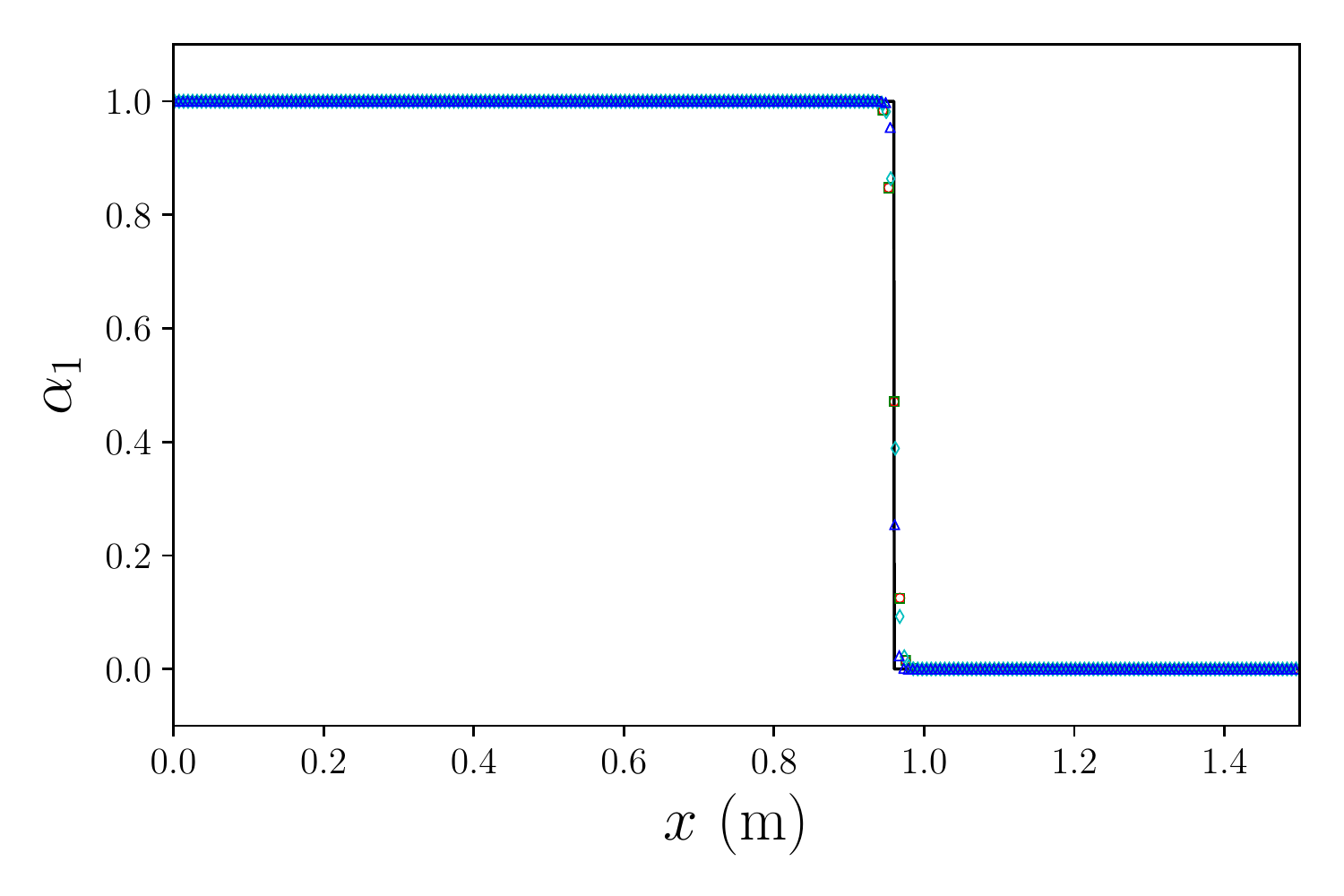}
\label{fig:compare_gas_liquid_vol_frac_global}}
\caption{Gas/liquid shock tube problem at $t = 3\mathrm{e}{-4} \ \mathrm{s}$ using different combinations of numerical algorithms and schemes. Black solid line: exact; red circles: four-equation HRM with the first order HLLC; green squares: five-equation model with the numerical thermal relaxation using the first order HLLC; cyan diamonds: five-equation model with the numerical thermal relaxation using the PP-WCNS-IS with 500 points; blue triangles: five-equation model with the numerical thermal relaxation using the PP-WCNS-IS with 1000 points. 1 out of 25 grid points is plotted for the first order HLLC.
1 out of 2 grid points and 1 out of 4 grid points are plotted for the PP-WCNS-IS with 500 and 1000 points respectively.}
\label{fig:compare_gas_liquid}
\end{figure}

\begin{figure}[!ht]
\centering
\subfigure[Local velocity profile]{%
\includegraphics[width=0.45\textwidth]{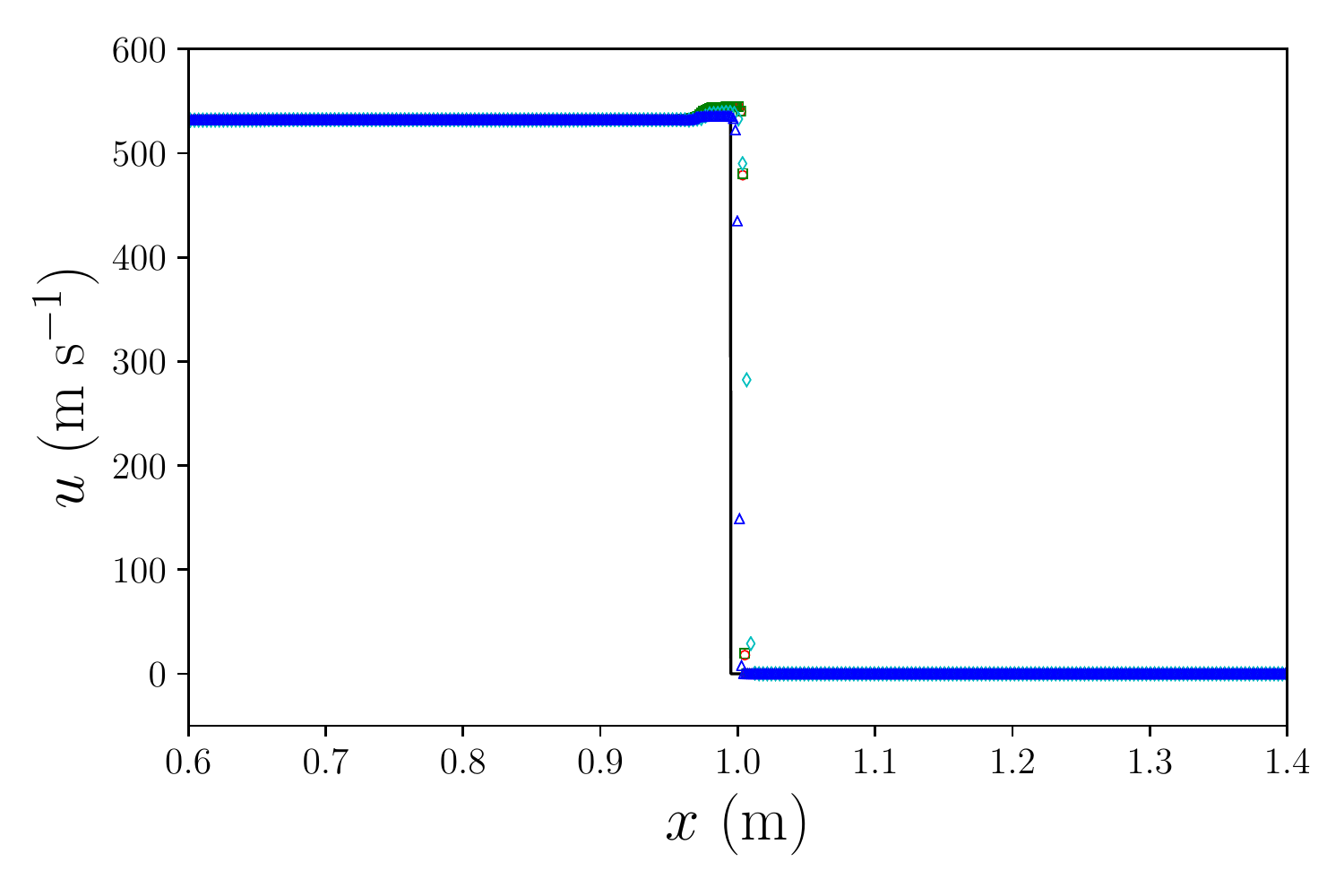}
\label{fig:compare_gas_liquid_u_local}}
\subfigure[Local pressure profile]{%
\includegraphics[width=0.45\textwidth]{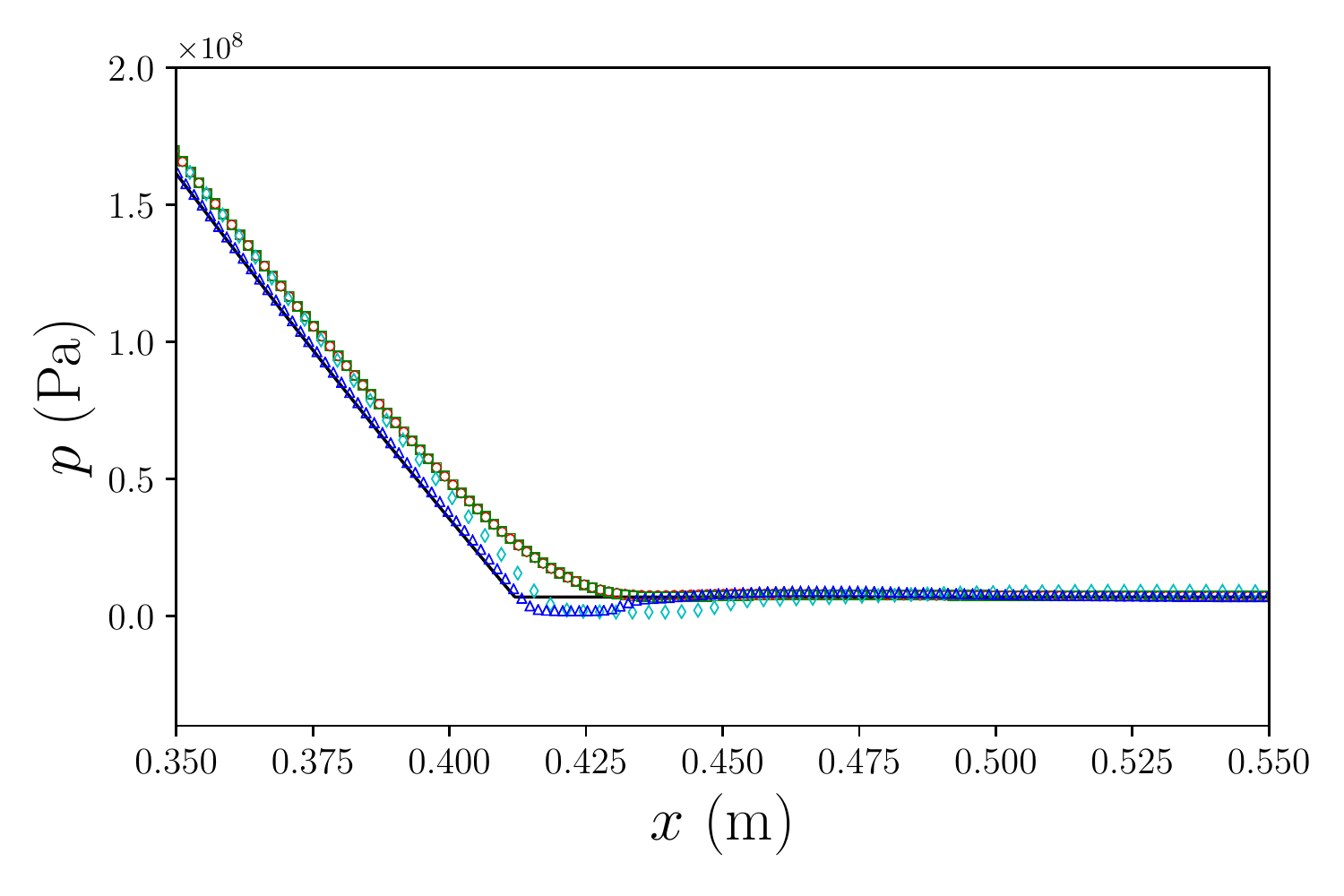}
\label{fig:compare_gas_liquid_p_local}}
\caption{Gas/liquid shock tube problem at $t = 3\mathrm{e}{-4} \ \mathrm{s}$ using different combinations of numerical algorithms and schemes. Black solid line: exact; red circles: four-equation HRM with the first order HLLC; green squares: five-equation model with the numerical thermal relaxation using the first order HLLC; cyan diamonds: five-equation model with the numerical thermal relaxation using the PP-WCNS-IS with 500 points; blue triangles: five-equation model with the numerical thermal relaxation using the PP-WCNS-IS with 1000 points.1 out of 5 grid points and all grid point are plotted for the first order HLLC and PP-WCNS-IS respectively.}
\label{fig:compare_gas_liquid_local}
\end{figure}

Figure~\ref{fig:compare_gas_liquid} compares the numerical solutions from the different combinations of numerical algorithms and schemes with the exact solutions. In figure~\ref{fig:compare_gas_liquid_rhoY2_global} which shows the solutions of the partial density of air, it can be seen that PP-WCNS-IS with the fractional algorithm can capture both the material interface (left density jump) and the shock (right density jump) reasonably well with different grid resolutions even though the two discontinuities are quite close to each other.
The first order HLLC schemes is too dissipative and requires one order of magnitude more grid cells to capture both features as well as the PP-WCNS-IS with 500 grid points no matter which numerical algorithms are used for the first order scheme.
The jumps in the temperature field caused by the material interface and shock are also captured reasonably well by all methods with the chosen grid resolutions.
Grid convergence towards to the exact solution is seen by comparing the PP-WCNS-IS solutions with the two different grid resolutions and this is most clear in the regions containing the discontinuities representing the material interface and shock respectively for the fields of gas partial density and temperature, which are shown in figures~\ref{fig:compare_gas_liquid_rhoY2_global} and \ref{fig:compare_gas_liquid_T_global} respectively.
In figure~\ref{fig:compare_gas_liquid_u_local}, it can be seen that the fractional algorithm with the PP-WCNS-IS can capture the velocity jump at the shock well and the solution computed with 500 grid points is already comparable with those from the first order methods using many more grid cells.
When the grid is refined, a smaller overshoot around the shock can be seen for the PP-WCNS-IS.
While small undershoots can be seen in both pressure field solutions obtained with PP-WCNS-IS around the starting location of the rarefaction wave in figure~\ref{fig:compare_gas_liquid_p_local} using different mesh resolutions, the overall PP-WCNS-IS solutions around that region are more accurate than those from the first order HLLC methods, where the numbers of grid cells are also much smaller for the former.

\subsection{One-dimensional extreme gas/liquid shock tube problem \label{sec:1D_extreme_gas_liquid_Sod}}

This is a more extreme version of the previous shock tube problem with much larger pressure and temperature ratios across the initial discontinuity.
The initial left and right pressure values are $1.0\mathrm{e}{12}\ \mathrm{Pa}$ and $1.0\mathrm{e}{5}\ \mathrm{Pa}$ respectively, while the initial left and right temperatures are respectively $357448\ \mathrm{K}$ and $17.378\ \mathrm{K}$. The initial conditions are given by table~\ref{table:IC_1D_extreme_gas_liquid_shock_tube_problem}. Similar to the previous problem, simulations of same combinations of spatial schemes and numerical algorithms are conducted. When the first order HLLC scheme is used, $\Delta t = 8.0\mathrm{e}{-10} \ \mathrm{s}$ on a uniform grid with 10000 grid points is chosen. As for the PP-WCNS-IS, $\Delta t = 8.0\mathrm{e}{-9} \ \mathrm{s}$ on a uniform grid with 1000 grid points is used. 
This problem is numerically challenging and failures are experienced when the positivity-preserving limiters are turned off for the PP-WCNS-IS method.

\begin{figure}[!ht]
\centering
\subfigure[Partial density of liquid profile]{%
\includegraphics[width=0.45\textwidth]{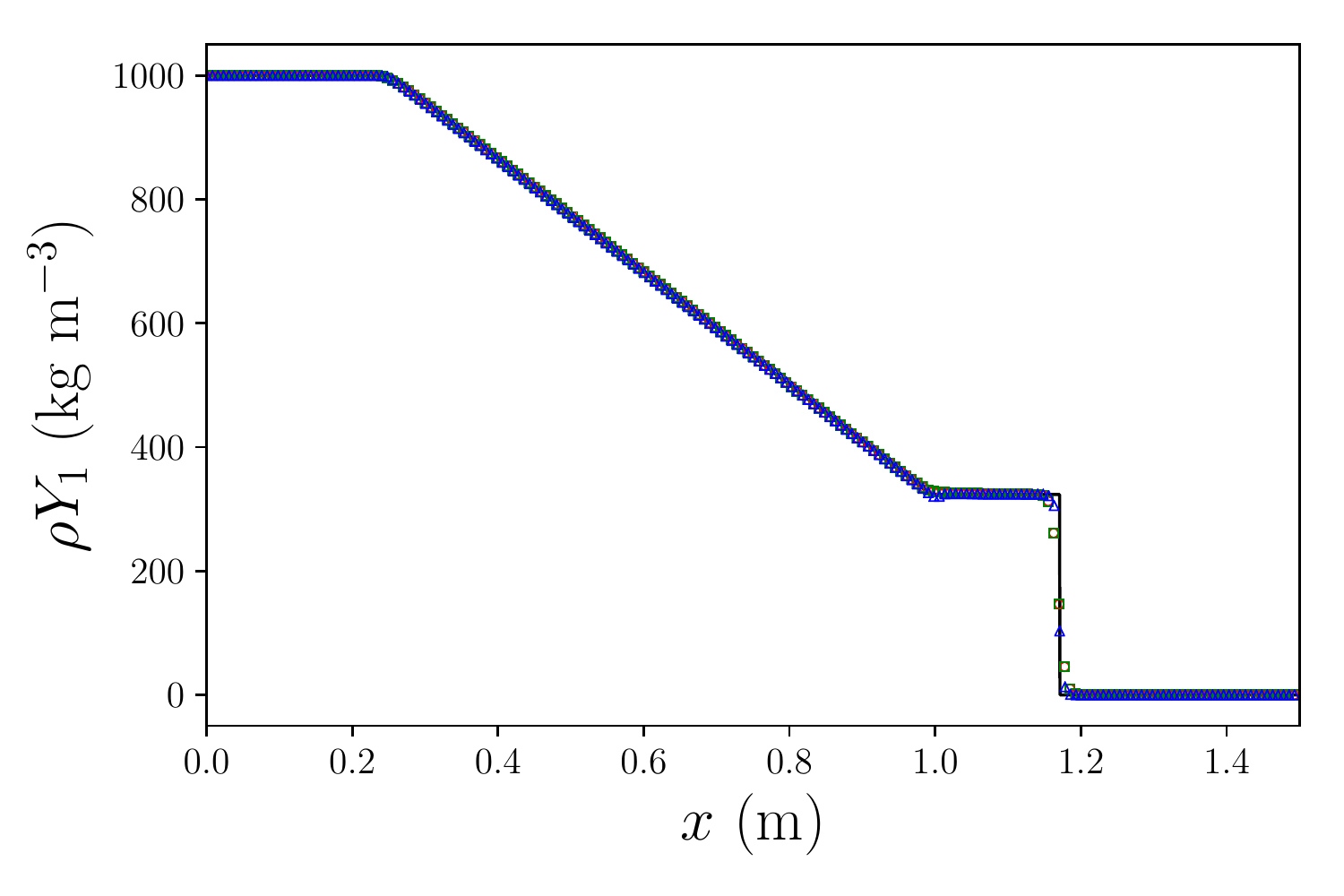}
\label{fig:compare_extreme_gas_liquid_rhoY1_global}}
\subfigure[Partial density of gas profile]{%
\includegraphics[width=0.45\textwidth]{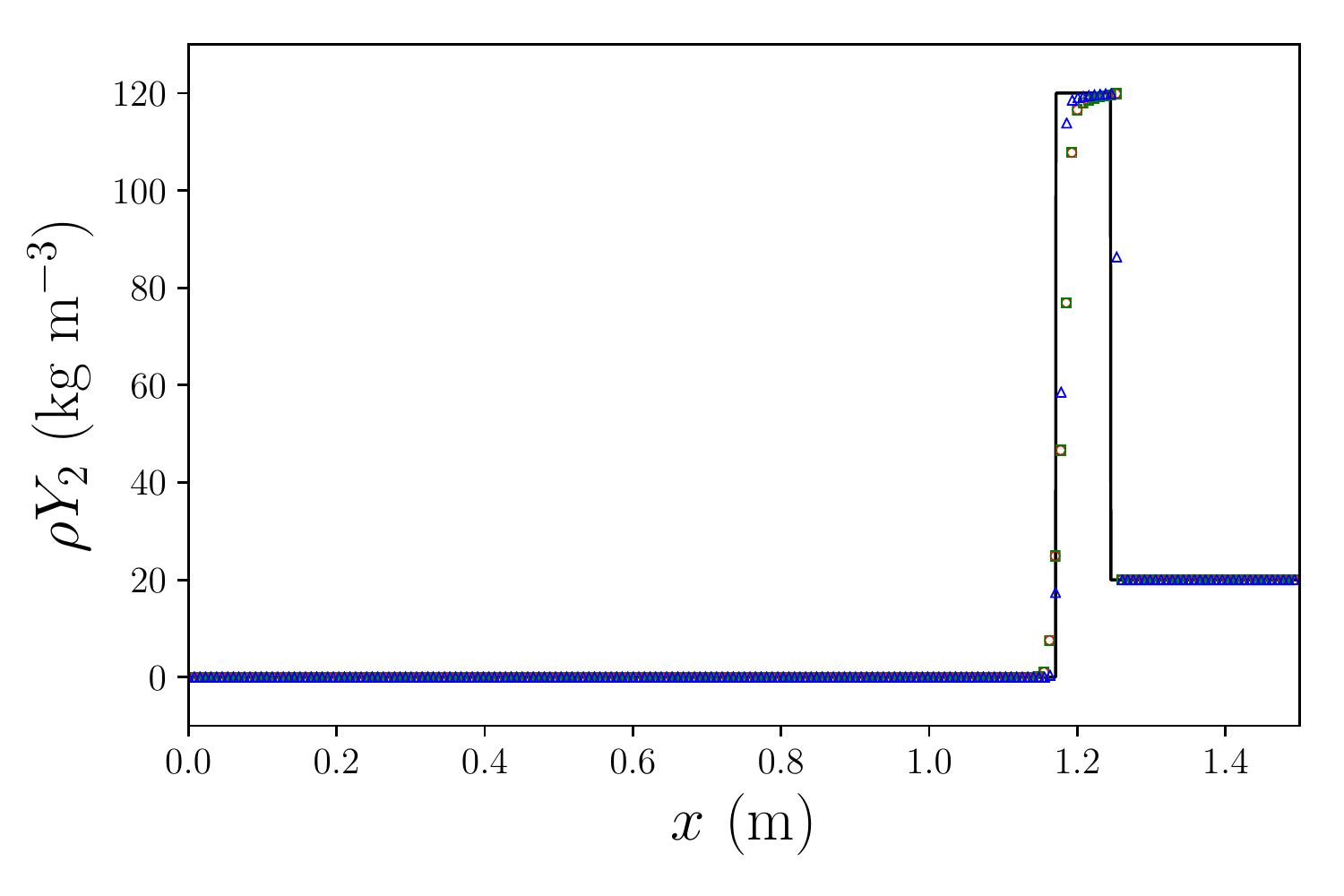}
\label{fig:compare_extreme_gas_liquid_rhoY2_global}}
\subfigure[Velocity profile]{%
\includegraphics[width=0.45\textwidth]{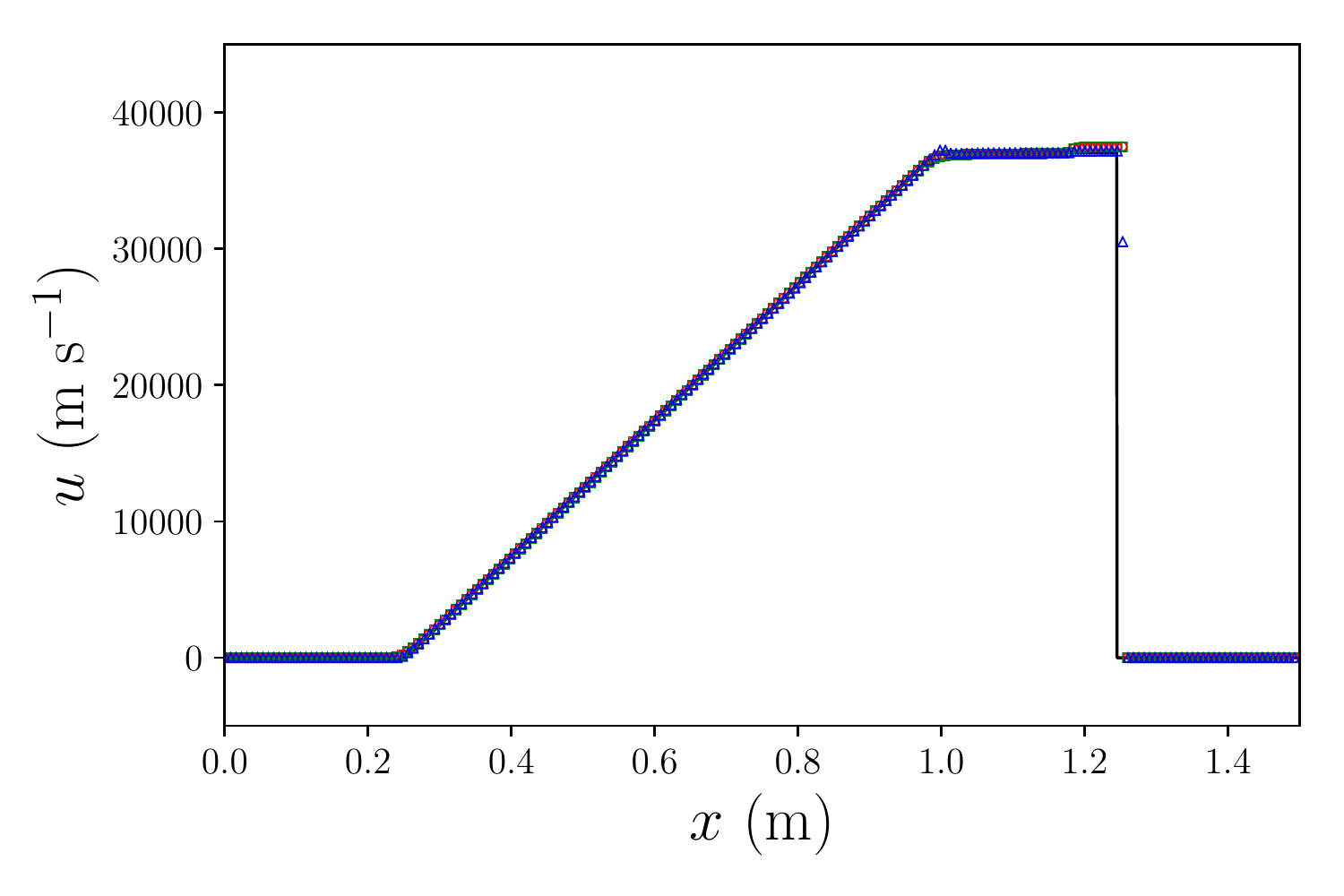}
\label{fig:compare_extreme_gas_liquid_u_global}}
\subfigure[Pressure profile]{%
\includegraphics[width=0.45\textwidth]{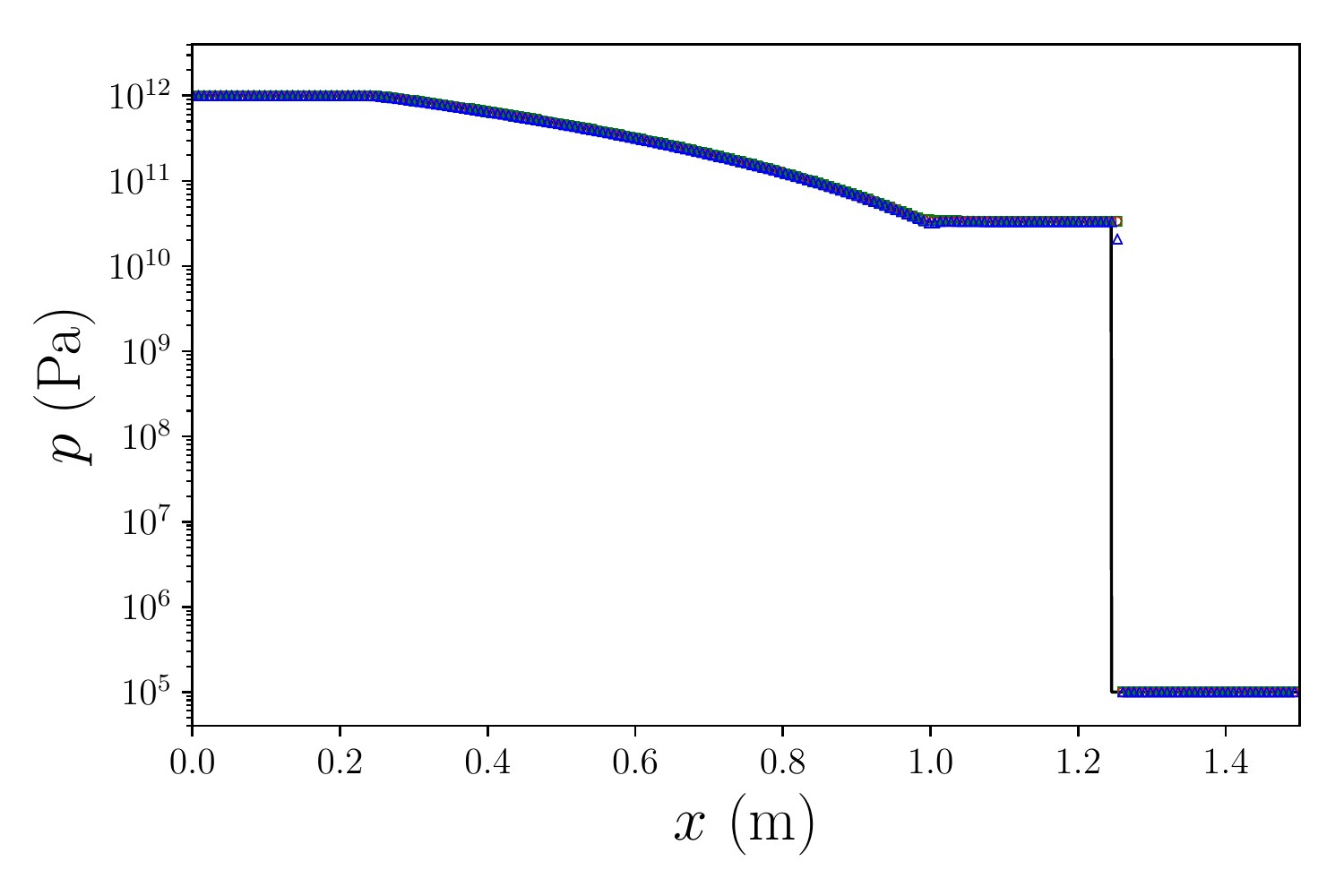}
\label{fig:compare_extreme_gas_liquid_p_global}}
\subfigure[Temperature profile]{%
\includegraphics[width=0.45\textwidth]{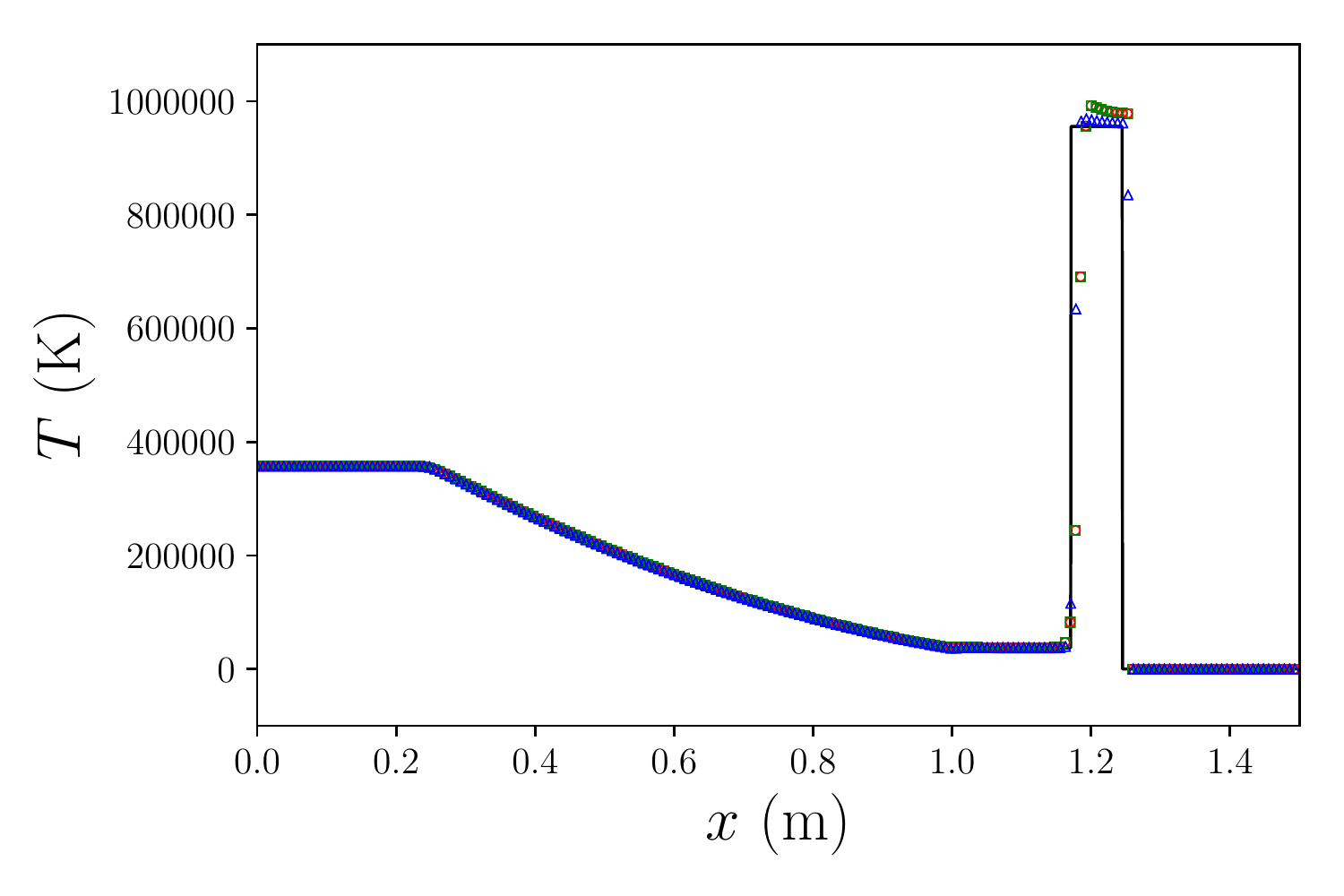}
\label{fig:compare_extreme_gas_liquid_T_global}}
\subfigure[Volume fraction of liquid profile]{%
\includegraphics[width=0.45\textwidth]{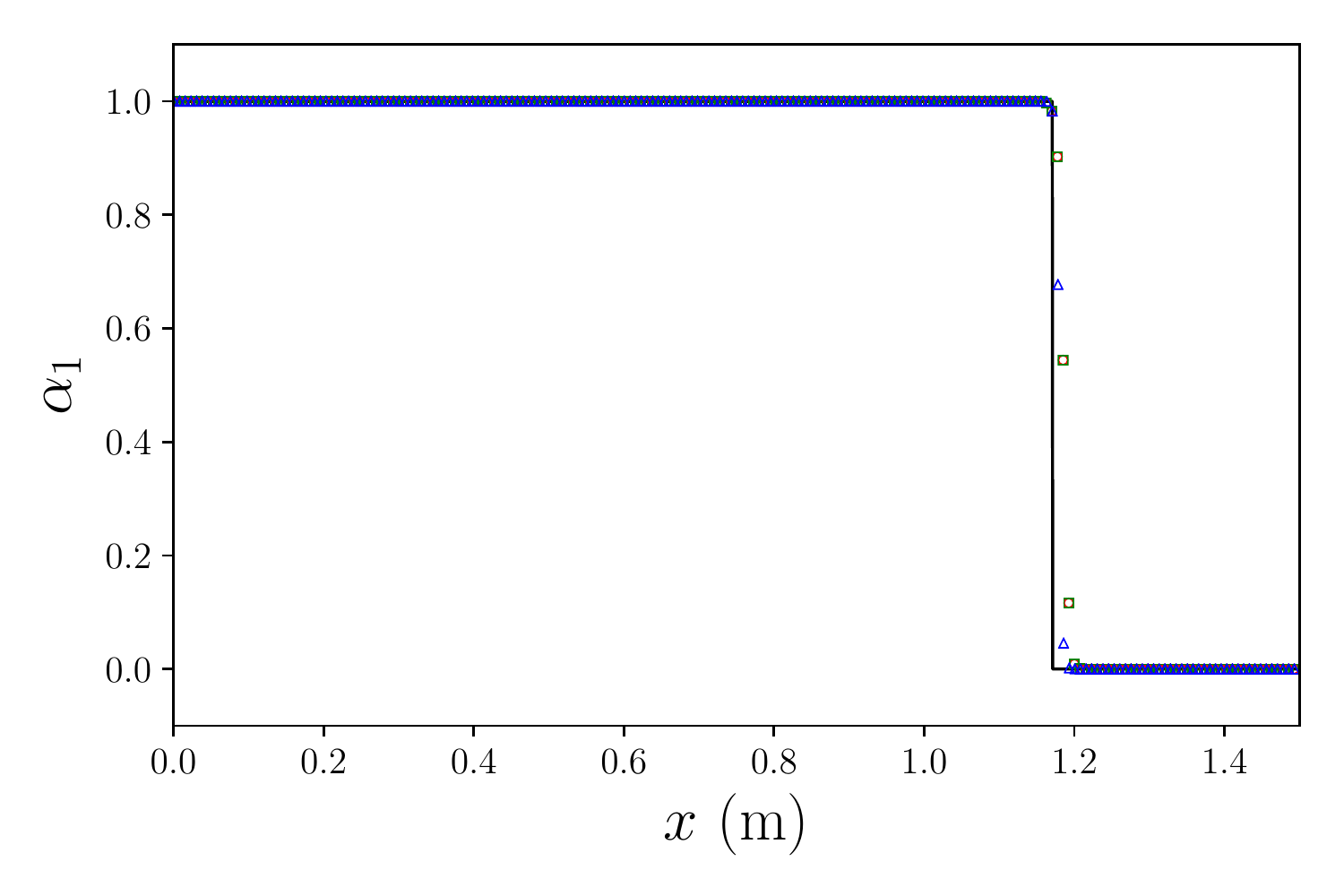}
\label{fig:compare_extreme_gas_liquid_vol_frac_global}}
\caption{Extreme gas/liquid shock tube problem at $t = 1\mathrm{e}{-5} \ \mathrm{s}$ using different combinations of numerical algorithms and schemes. Black solid line: exact; red circles: four-equation HRM with the first order HLLC; green squares: five-equation model with the numerical thermal relaxation using the first order HLLC; blue triangles: five-equation model with the numerical thermal relaxation using the PP-WCNS-IS. 1 out of 50 grid points and 1 out of 5 grid points are plotted for the first order HLLC and PP-WCNS-IS respectively.}
\label{fig:compare_extreme_gas_liquid}
\end{figure}

Figure~\ref{fig:compare_extreme_gas_liquid} compares the numerical solutions from the different combinations of numerical algorithms and schemes with the exact solutions.
From the solutions of different fields, it can be seen that the fractional algorithm with PP-WCNS-IS can capture the discontinuities and the rarefaction wave well without obvious undershoots and overshoots. With much fewer number of grid cells, the PP-WCNS-IS method can give solutions of similar quality as the first order methods, or even more accurate solutions in some of the fields such as the partial density of gas and the temperature. This can be seen in the solutions in the regions between the material interface (left discontinuity) and the shock (right discontinuity).

\begin{table}[!ht]
\small
  \begin{center}
    \begin{tabular}{@{}c | ccccc@{}}\toprule
    $(m)$ &
    \addstackgap{\stackanchor{$\alpha_1 \rho_1$}{$(\mathrm{kg\ m^{-3}})$}} &
    \stackanchor{$\alpha_2 \rho_2$}{$(\mathrm{kg\ m^{-3}})$} &
    \stackanchor{$u$}{$(\mathrm{m\ s^{-1}})$} &
    \stackanchor{$p$}{$(\mathrm{Pa})$} &
    $\alpha_1$ \\ \midrule
    \addstackgap{$x < 0.8$} & $9.9999998999999991\mathrm{e}{2}$ & $9.7235797802400965\mathrm{e}{-5}$ & 0 & $1.0\mathrm{e}{12}$ & $1 - 1.0\mathrm{e}{-8}$ \\
    \addstackgap{$x \geq 0.8$} & $1.7538240816326533\mathrm{e}{-4}$ & $1.9999999799999998\mathrm{e}{1}$ & 0 & $1.0\mathrm{e}{5}$ & $1.0\mathrm{e}{-8}$ \\ \bottomrule
    \end{tabular}
  \end{center}
  \caption{Initial conditions of the 1D extreme gas/liquid shock tube problem.}
  \label{table:IC_1D_extreme_gas_liquid_shock_tube_problem}
\end{table}

\subsection{One-dimensional extreme shock-interface interaction problem}

This is another 1D extreme test problem but with three species. The first species is liquid water while the second and third species are air and sulphur hexafluoride ($\mathrm{SF_6}$) respectively. In this problem, a strong shock is initiated in liquid water and interacts with a discontinuous material interface separating the liquid water and a gas mixture of air and $\mathrm{SF_6}$.
When the incident shock hits the material interface, a rarefaction wave is reflected and a shock is transmitted into the gas mixture.
The shock and the interface are initially located at $x = 0.15\ \mathrm{m}$ and $x = 0.3\ \mathrm{m}$ respectively. The initial pressure and temperature fields across the material interface are uniformly at $101325\ \mathrm{Pa}$ and $298\ \mathrm{K}$. The pressure of post-shock region in liquid water is $1.0\mathrm{e}{12}\ \mathrm{Pa}$. The gas mixture is initially composed of half air and half $\mathrm{SF_6}$ essentially in terms of volume fractions.
The initial conditions are given by table~\ref{table:IC_1D_extreme_shock_interface_interaction}.
Various simulations are conducted with the fractional algorithm composed of the five-equation model and the thermal relaxation using the first-order HLLC and PP-WCNS-IS methods.
A uniform grid with 10000 grid points and $\Delta t = 4.8\mathrm{e}{-10} \ \mathrm{s}$ are chosen for the first order HLLC method, while two uniform grids with 800 and 1600 grid cells are employed with $\Delta t = 6.0\mathrm{e}{-9} \ \mathrm{s}$ and $\Delta t = 3.0\mathrm{e}{-9} \ \mathrm{s}$ respectively for the PP-WCNS-IS method. Numerical failures are experienced when the positivity-preserving limiters are turned off for the PP-WCNS-IS method.

\begin{figure}[!ht]
\centering
\subfigure[Partial density of species 1 profile]{%
\includegraphics[width=0.45\textwidth]{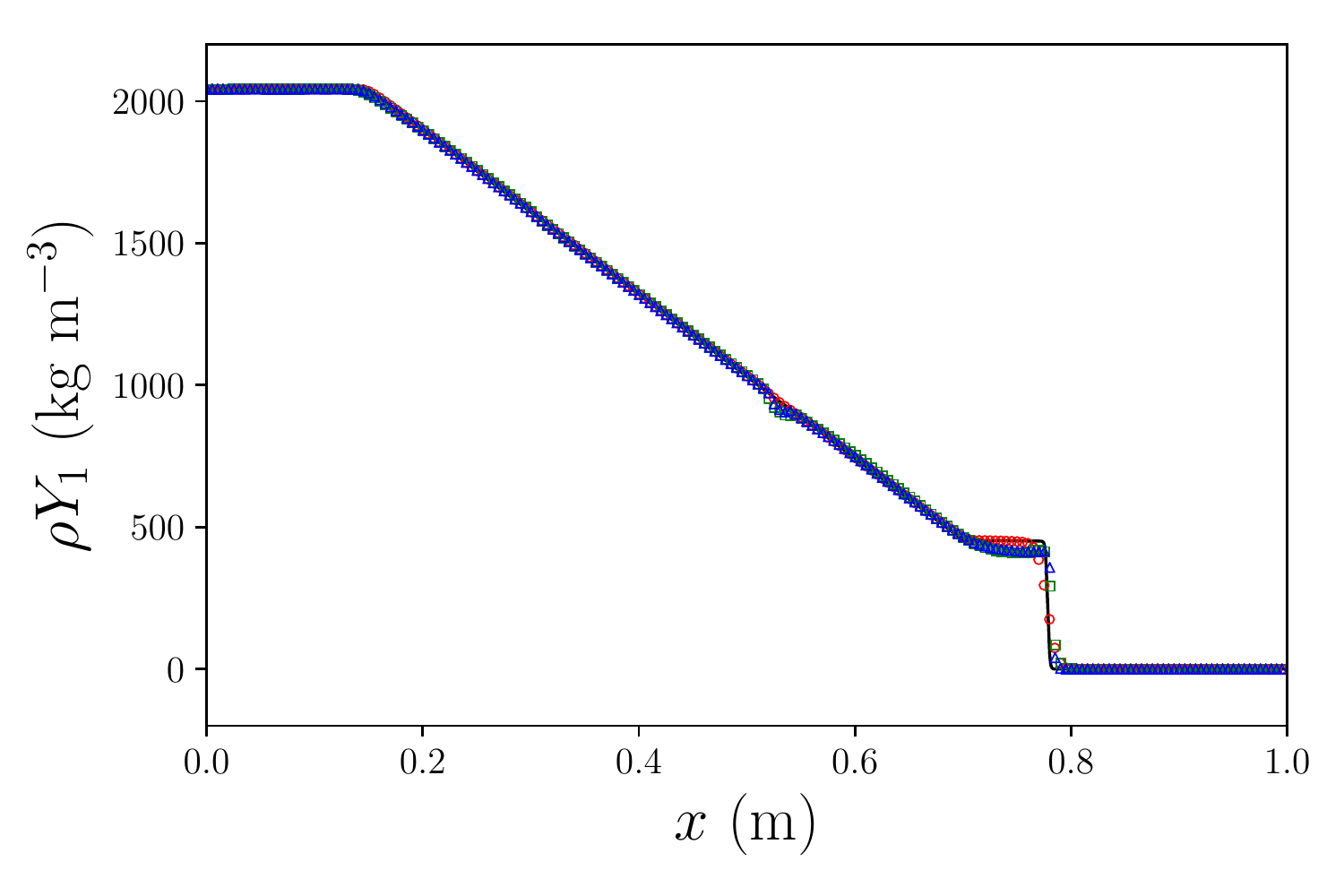}
\label{fig:compare_extreme_shock_interface_rhoY1_global}}
\subfigure[Partial density of species 2 profile]{%
\includegraphics[width=0.45\textwidth]{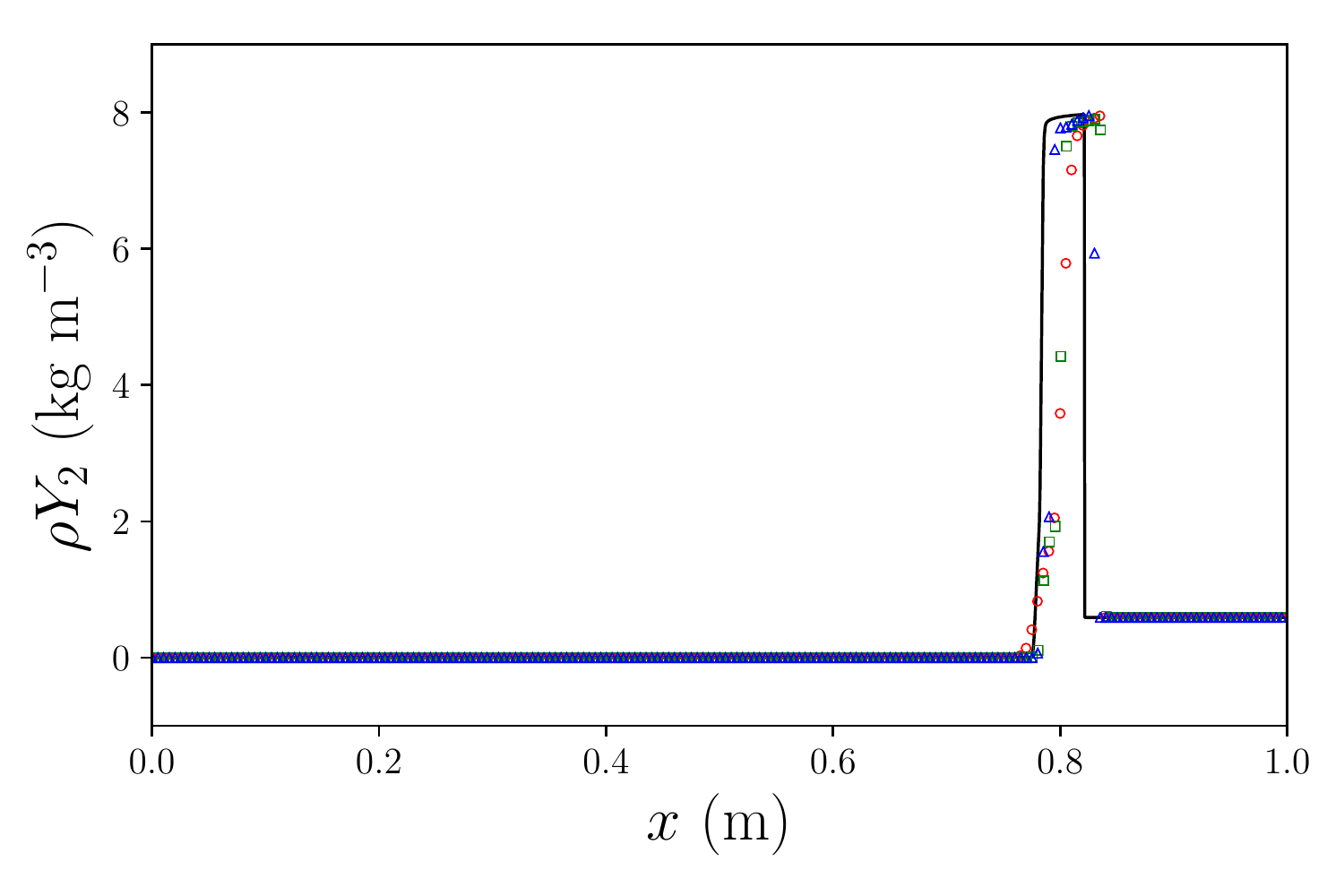}
\label{fig:compare_extreme_shock_interface_rhoY2_global}}
\subfigure[Partial density of species 3 profile]{%
\includegraphics[width=0.45\textwidth]{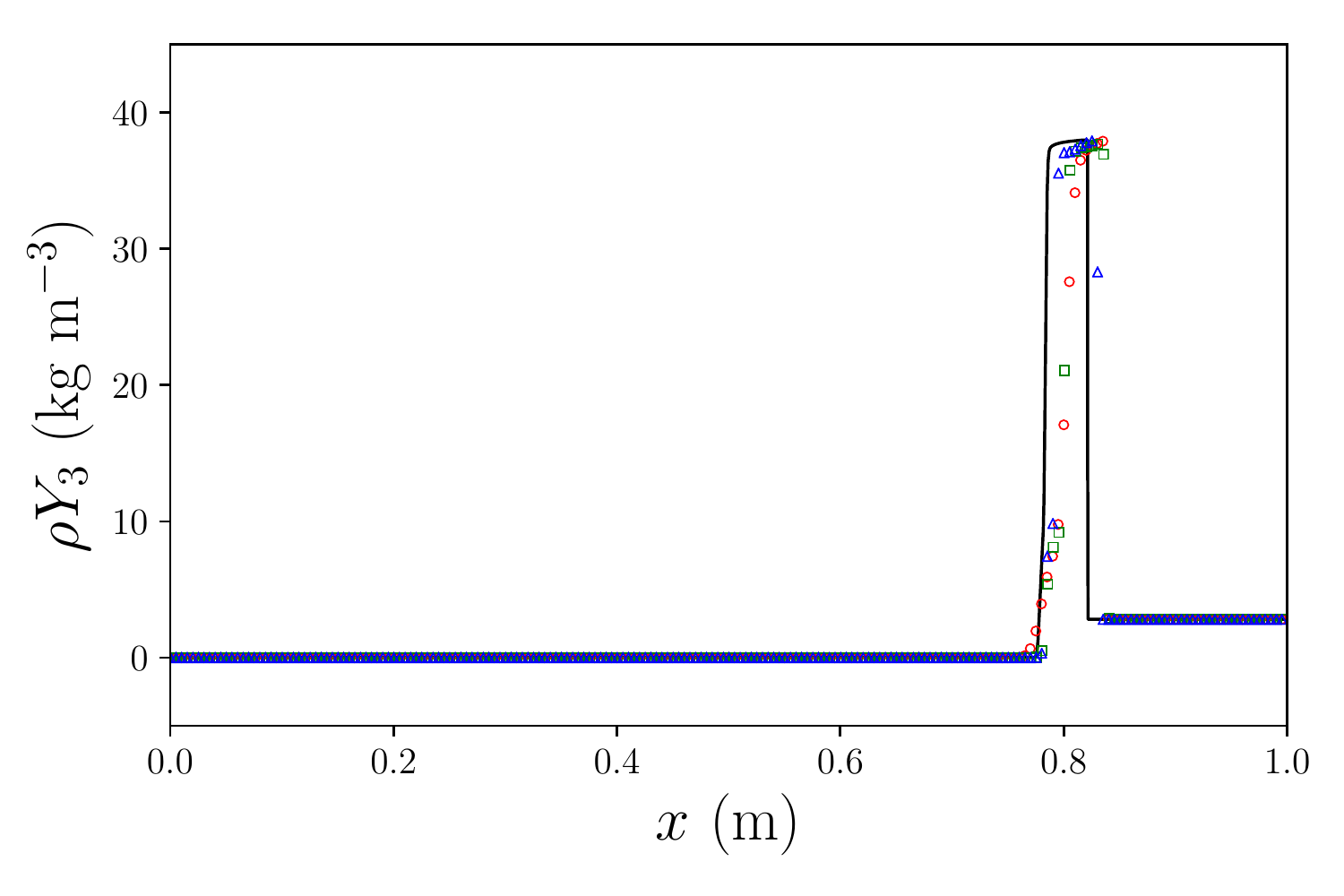}
\label{fig:compare_extreme_shock_interface_rhoY3_global}}
\subfigure[Velocity profile]{%
\includegraphics[width=0.45\textwidth]{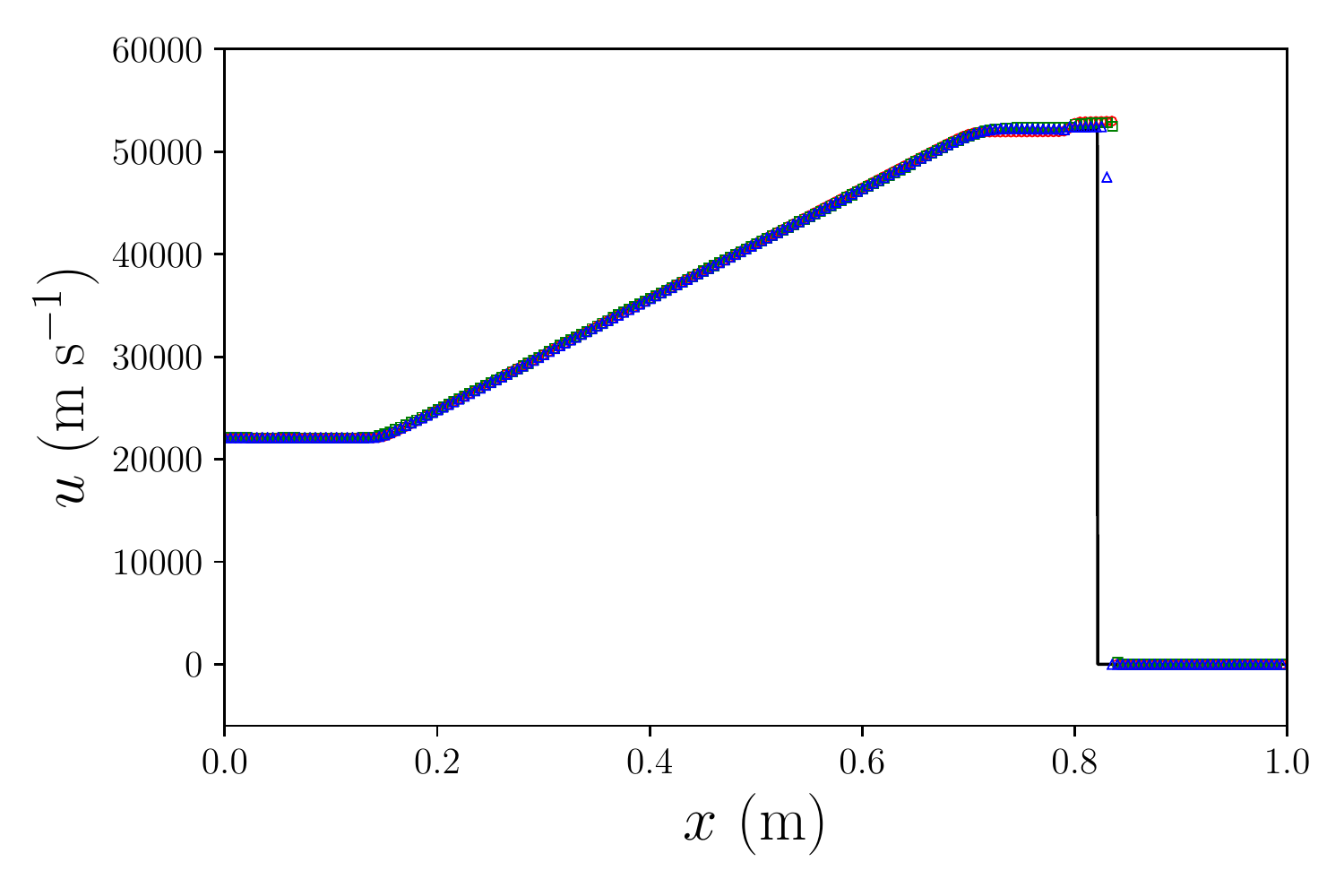}
\label{fig:compare_extreme_shock_interface_u_global}}
\subfigure[Pressure profile]{%
\includegraphics[width=0.45\textwidth]{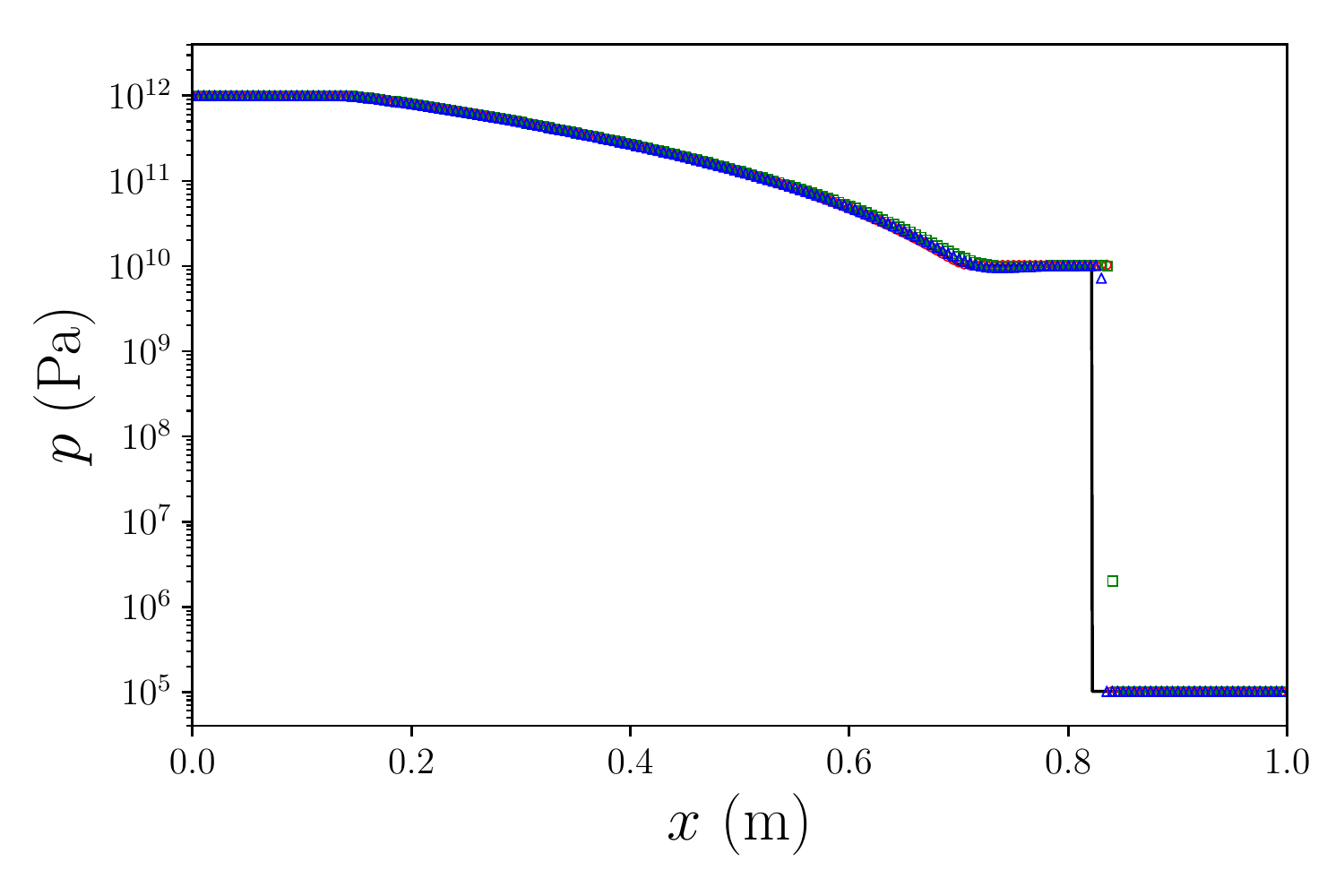}
\label{fig:compare_extreme_shock_interface_p_global}}
\subfigure[Temperature profile]{%
\includegraphics[width=0.45\textwidth]{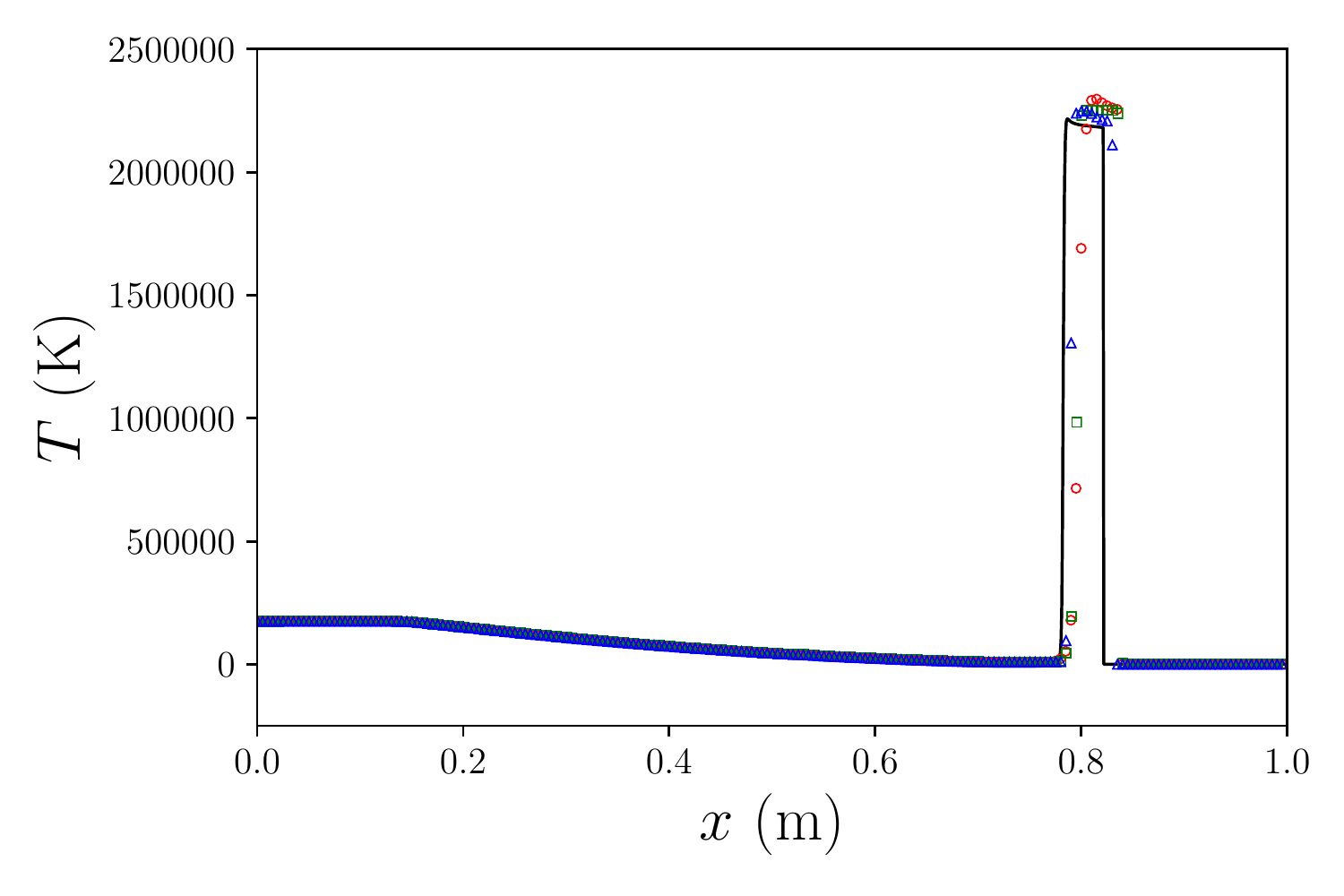}
\label{fig:compare_extreme_shock_interface_T_global}}
\caption{Extreme shock-interface interaction problem at $t = 1.26\mathrm{e}{-5} \ \mathrm{s}$ using the fractional algorithm with the five-equation model and the numerical thermal relaxation with different schemes. Black solid line: reference solutions (using the first order HLLC with 200000 points); red circles: first order HLLC with 10000 points; cyan diamonds: PP-WCNS-IS with 800 points; blue triangles: PP-WCNS-IS with 1600 points. 1 out of 50 grid points is plotted for the first order HLLC. 1 out of 4 grid points and 1 out of 8 grid points are plotted for the PP-WCNS-IS with 800 and 1600 points respectively.}
\label{fig:compare_extreme_shock_interface_interaction_1}
\end{figure}

\begin{figure}[!ht]
\centering
\subfigure[Volume fraction of species 1 profile]{%
\includegraphics[width=0.45\textwidth]{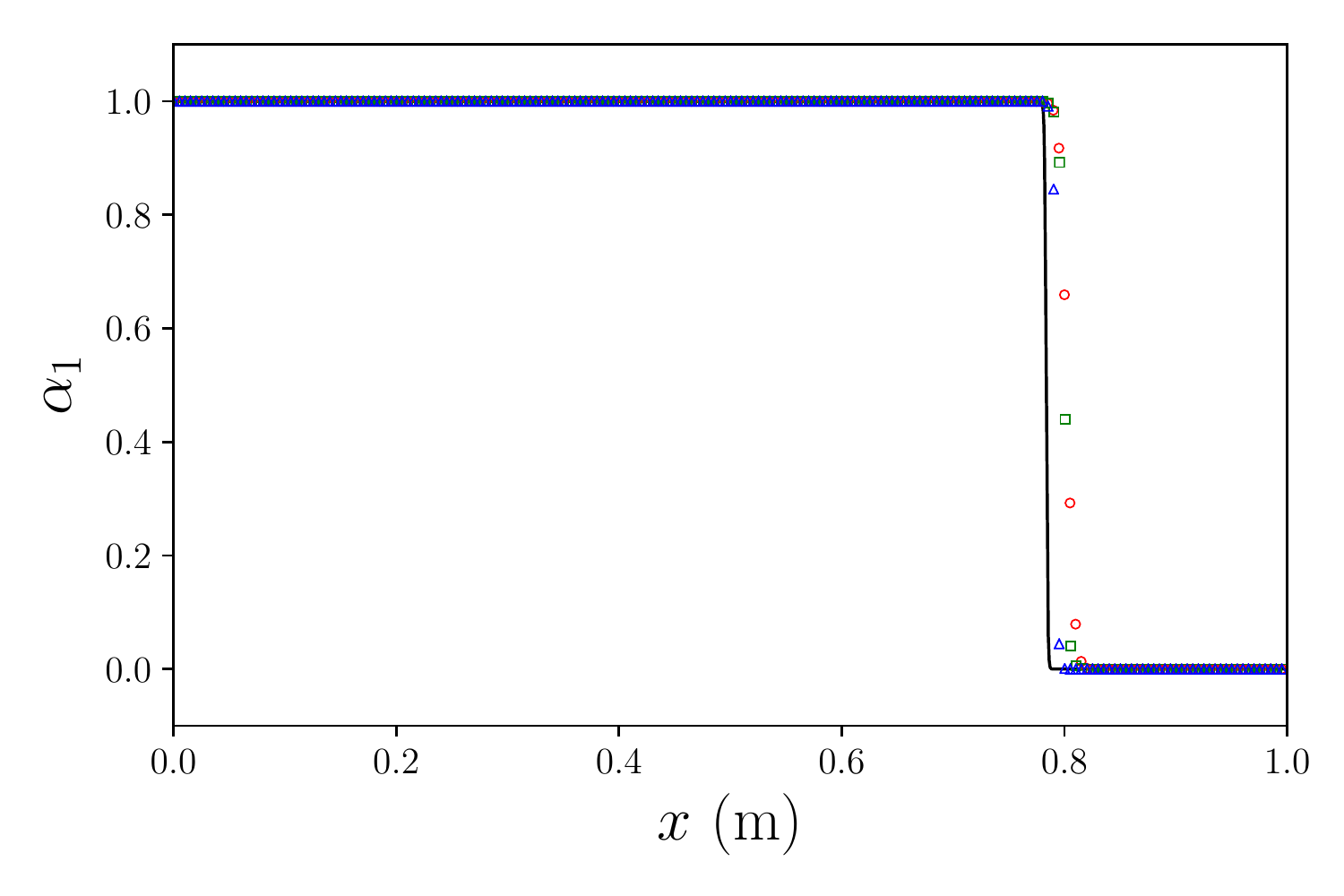}
\label{fig:compare_extreme_shock_interface_vol_frac_1_global}}
\subfigure[Volume fraction of species 2 profile]{%
\includegraphics[width=0.45\textwidth]{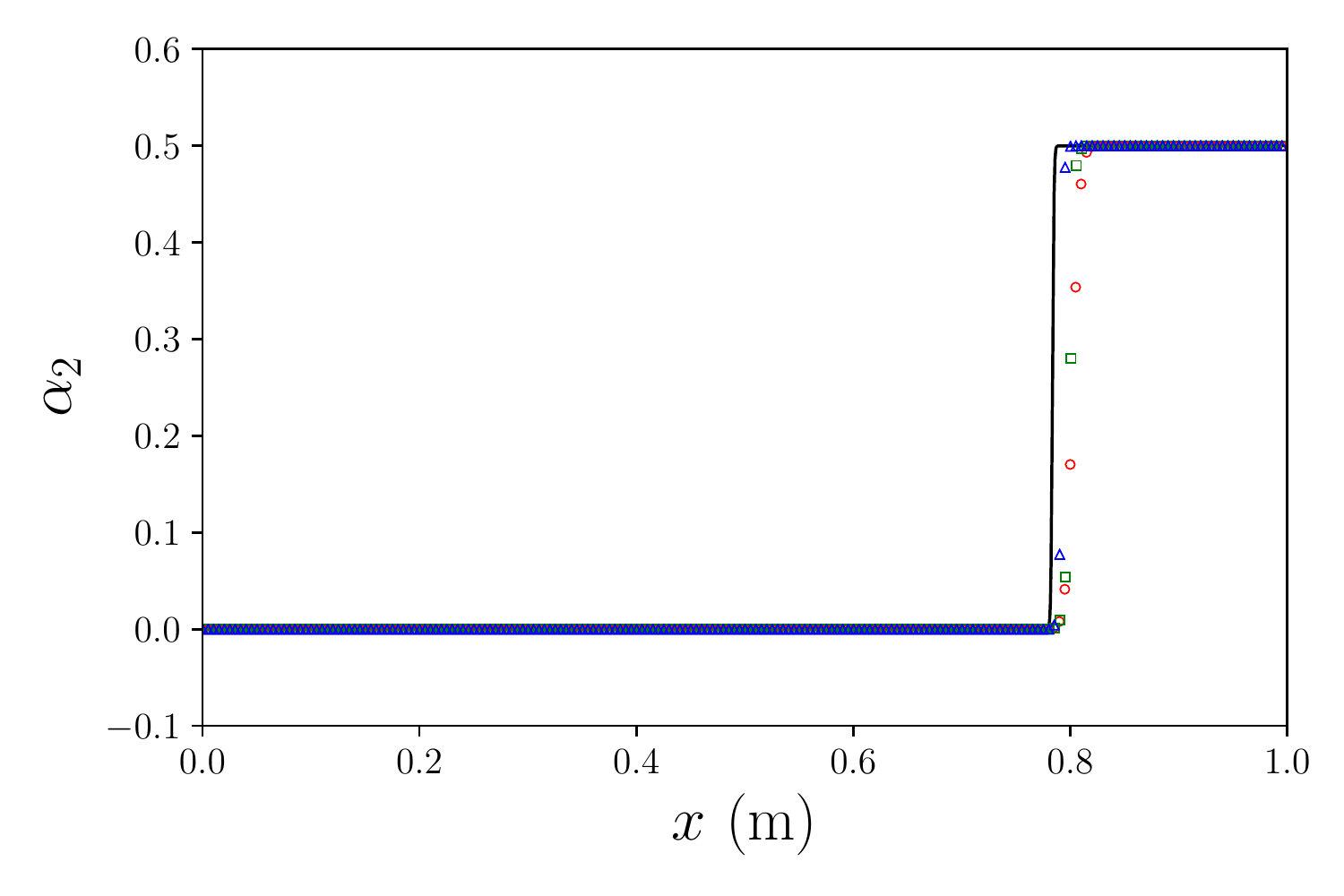}
\label{fig:compare_extreme_shock_interface_vol_frac_2_global}}
\caption{Extreme shock-interface interaction problem at $t = 1.26\mathrm{e}{-5} \ \mathrm{s}$ using the fractional algorithm with the five-equation model and the numerical thermal relaxation with different schemes. Black solid line: reference solutions (using the first order HLLC with 200000 points); red circles: first order HLLC with 10000 points; green squares: PP-WCNS-IS with 800 points; blue triangles: PP-WCNS-IS with 1600 points. 1 out of 50 grid points is plotted for the first order HLLC. 1 out of 4 grid points and 1 out of 8 grid points are plotted for the PP-WCNS-IS with 800 and 1600 points respectively.}
\label{fig:compare_extreme_shock_interface_interaction_2}
\end{figure}

Figures~\ref{fig:compare_extreme_shock_interface_interaction_1} and \ref{fig:compare_extreme_shock_interface_interaction_2} compare the numerical solutions from different cases with the reference solutions. 
The reference solutions are computed using the first order HLLC scheme with 200000 grid points and $\Delta t = 2.4\mathrm{e}{-11}\ \mathrm{s}$.
Again, it can be seen from the figures that PP-WCNS-IS method can capture the discontinuities and the rarefaction wave similarly well or even better with much fewer number of grid cells compared with the first order HLLC scheme. No obvious spurious oscillations are observed in the solutions obtained with the PP-WCNS-IS method. There are dips in the liquid partial density solutions computed with PP-WCNS-IS as seen in figure~\ref{fig:compare_extreme_shock_interface_rhoY1_global} due to the start up errors caused by the use of exact shock jump initial conditions. This is not obvious in the late-time solutions obtained with the first order scheme as it is heavily smeared out due to much larger numerical dissipation. Comparing the solutions given by PP-WCNS-IS using different levels of grid resolution, it can be seen that the solutions converge towards the reference solutions with increasing grid resolution.

\begin{table}[!ht]
\small
  \begin{center}
    \begin{tabular}{@{}c | c c c}\toprule
    $(m)$ &
    \addstackgap{\stackanchor{$\alpha_1 \rho_1$}{$(\mathrm{kg\ m^{-3}})$}} &
    \stackanchor{$\alpha_2 \rho_2$}{$(\mathrm{kg\ m^{-3}})$} &
    \stackanchor{$\alpha_3 \rho_3$}{$(\mathrm{kg\ m^{-3}})$}\\ \midrule
    \addstackgap{$x < 0.15$} & $2.0429330221481357\mathrm{e}{3}$ & $1.9864622524021913\mathrm{e}{-4}$ & $9.4681765791131904\mathrm{e}{-4}$ \\
    \addstackgap{$0.15 \leq x < 0.3 $} & $1.0227724208197188\mathrm{e}{3}$ & $1.1817862212832324\mathrm{e}{-8}$ & $5.6328080779493768\mathrm{e}{-8}$ \\
    \addstackgap{$x \geq 0.3 $} & $1.0227724412751677\mathrm{e}{-5}$ & $0.59089310473268508$ & $2.8164040108106478$ \\ \bottomrule
    \end{tabular}
    \begin{tabular}{@{}c | c c c c}\toprule
    $(m)$ &
    \stackanchor{$u$}{$(\mathrm{m\ s^{-1}})$} &
    \stackanchor{$p$}{$(\mathrm{Pa})$} &
    $\alpha_1$ &
    $\alpha_2/\alpha_3$ \\ \midrule
    \addstackgap{$x < 0.15$} & $2.2096204395563851\mathrm{e}{4}$ & $1.0\mathrm{e}{12}$ & $1 - 1.0\mathrm{e}{-8}$ & $1.0\mathrm{e}{-8}$ \\
    \addstackgap{$0.15 \leq x < 0.3 $} & 0 & $1.01325\mathrm{e}{5}$ & $1 - 1.0\mathrm{e}{-8}$ & $1.0\mathrm{e}{-8}$ \\
    \addstackgap{$x \geq 0.3 $} & 0 & $1.01325\mathrm{e}{5}$ & $1.0\mathrm{e}{-8}$ & $0.5 - 5.0\mathrm{e}{-9}$ \\ \bottomrule
    \end{tabular}
  \end{center}
  \caption{Initial conditions of the 1D extreme shock-interface interaction problem.}
  \label{table:IC_1D_extreme_shock_interface_interaction}
\end{table}

\subsection{Two-dimensional shock water cylinder interaction problem}

\begin{figure}[!ht]
  \centering
  \begin{tikzpicture}[thick,scale=0.8, every node/.style={transform shape}]
    \useasboundingbox (0cm,-1cm)  rectangle (8cm,7cm);
    \draw[black]        (-0.5cm,0.0cm) rectangle ++(9cm,6cm);
    \draw[black]        ( 4.0cm,2.9cm) circle (1.2cm);
    \draw[black, thick] ( 2.2cm,0.0cm) -- (2.2cm,6cm);

    \node[text width=3cm] at (1.3cm,4.75cm) {Post-shock air};
    \node[text width=3cm] at (4.5cm,4.75cm) {Pre-shock air};
    \node[text width=3cm] at (5.05cm,2.2cm) {Water};
    
    \draw (4.0cm, 3.0cm) node[cross] {};

    \draw[{Straight Barb[angle'=60,scale=3]}-{Straight Barb[angle'=60,scale=3]}] ( 7.0cm,0.0cm) -- (7.0cm,6cm);
    \draw[{Straight Barb[angle'=60,scale=3]}-{Straight Barb[angle'=60,scale=3]}] (-0.5cm,6.5cm) -- (8.5cm,6.5cm);
    \draw[{Straight Barb[angle'=60,scale=3]}-{Straight Barb[angle'=60,scale=3]}] ( 2.8cm,3.0cm) -- (5.2cm,3.0cm);

    \node[text width=3cm] at (5.2cm,6.8cm) {$20D$};
    \node[text width=3cm] at (8.7cm,3.0cm) {$10D$};
    \node[text width=3cm] at (5.35cm,2.6cm) {$D$};
    \node[text width=3cm] at (5.1cm,3.45cm) { $[0, \, 0]$ };

    \draw[-{Straight Barb[angle'=60,scale=3]}] (4.0cm,-0.5cm) -- (5.5cm,-0.5cm);
    \node[text width=3cm] at (7.1cm,-0.6cm) {$x$};
    \draw[black] (4.0cm,-0.75cm) -- (4.0cm,-0.25cm);
    \draw[-{Straight Barb[angle'=60,scale=3]}] (-1.0cm,3.0cm) -- (-1.0cm,4.5cm);
    \node[text width=1cm] at (-0.5cm,4.75cm) {$y$};
    \draw[black] (-1.25cm,3.0cm) -- (-0.75cm,3.0cm);
  \end{tikzpicture}
  \caption{Schematic diagram of the 2D Mach 1.47 shock water cylinder interaction problem.} \label{fig:schematic_2D_shock_water_cylinder}
\end{figure}
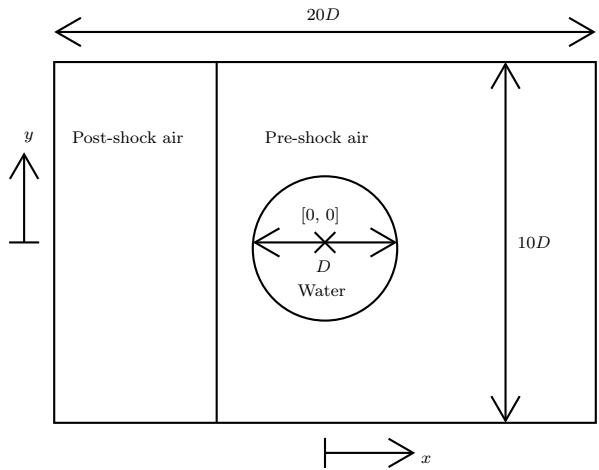

In this two-species 2D problem, a Mach 1.47 shock wave in air interacts with a water column with initial diameter $D=4.8\ \mathrm{mm}$. The pre-shock air and liquid water are initially at atmospheric conditions with $p = 101325 \ \mathrm{Pa}$ and $T = 298 \ \mathrm{K}$. The deformation and breakup of the water cylinder was first studied by~\citet{igra2001investigation,igra2001numerical} both experimentally and numerically. This problem was used for validation and verification in a number of research works~\cite{terashima2010front,meng2015numerical,aslani2018localized,kaiser2020investigation}.
The problem has a domain size of $\left[ -10D, 10D \right] \times \left[ -5D, 5D \right]$. The water cylinder is initially placed at the origin of the domain. The shock is launched from the left of the water cylinder such that the shock interacts with the water cylinder after $5\ \mu\mathrm{s}$. Constant extrapolation is used at all domain boundaries.
Figure~\ref{fig:schematic_2D_shock_water_cylinder} shows the schematic of the initial flow field and domain. 
The initial conditions are given by table~\ref{table:IC_2D_shock_water_cylinder}.
The accuracy of the fractional algorithm with the five-equation model by Allaire et al. and the thermal relaxation is analyzed with the computations performed on three different levels of mesh resolution:
$512 \times 256$, $1024 \times 512$, and $2048 \times 1024$,
using the PP-WCNS-IS method.

\begin{figure}[!ht]
\centering
\subfigure[Experiment]{%
\includegraphics[trim=-0.5cm -2.0cm -0.5cm -0.4cm,height=0.4\textwidth]{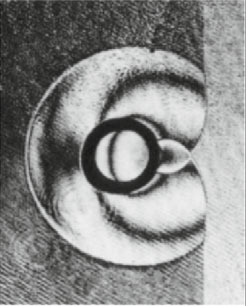}
\label{fig:compare_2D_shock_water_cylinder_t1_experiment}}
\subfigure[Simulation at $t = 18\ \mu\mathrm{s}$ ($t^*=0.0391$)]{%
\includegraphics[height=0.4\textwidth]{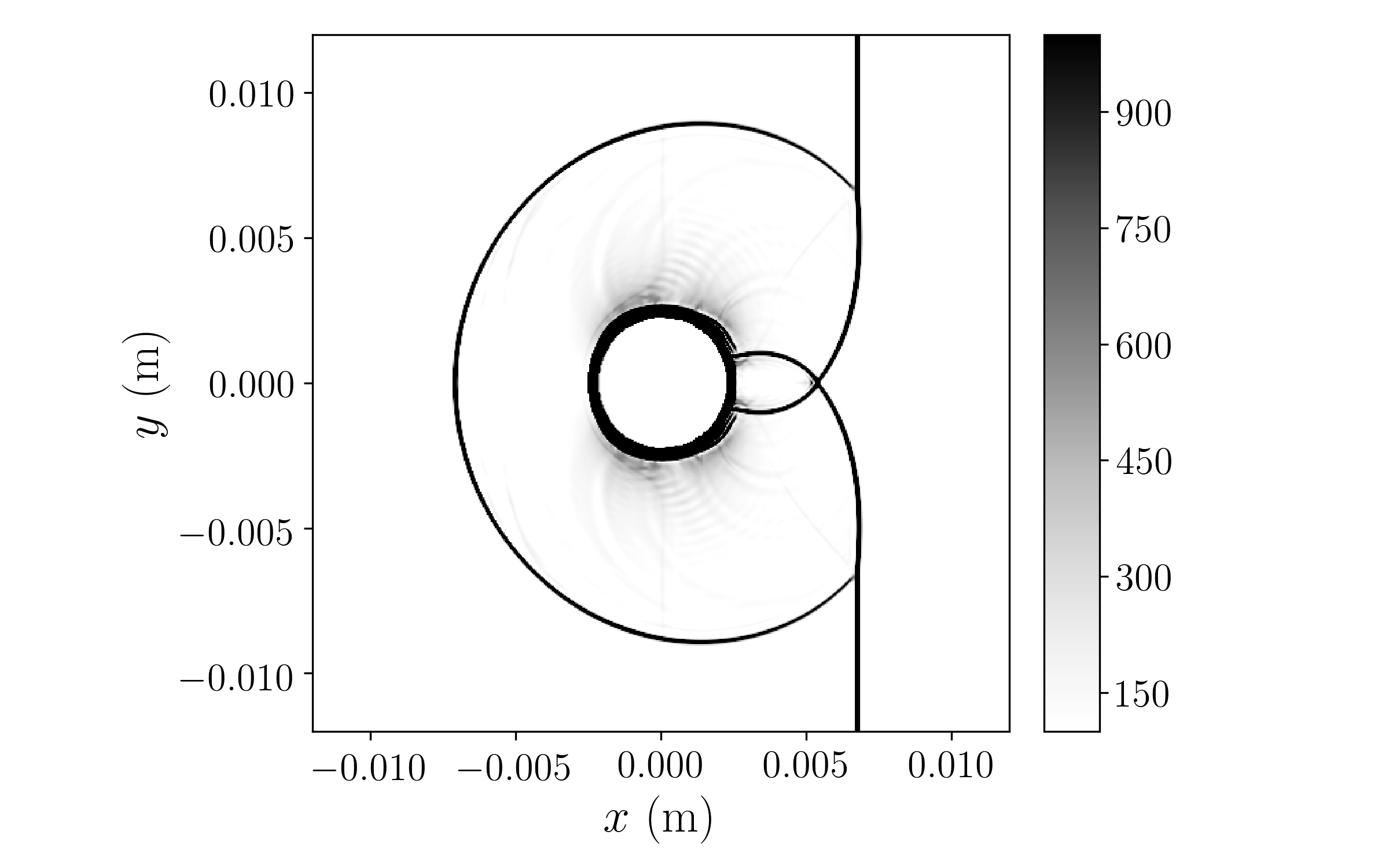}
\label{fig:compare_2D_shock_water_cylinder_t1_WCNS5_IS_PP}}
\newline
\subfigure[Experiment]{%
\includegraphics[trim=0 -2.0cm 0 -0.4cm,height=0.4\textwidth]{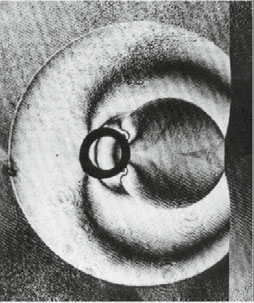}
\label{fig:compare_2D_shock_water_cylinder_t2_experiment}}
\subfigure[Simulation at $t = 32\ \mu\mathrm{s}$ ($t^*=0.0695$)]{%
\includegraphics[height=0.4\textwidth]{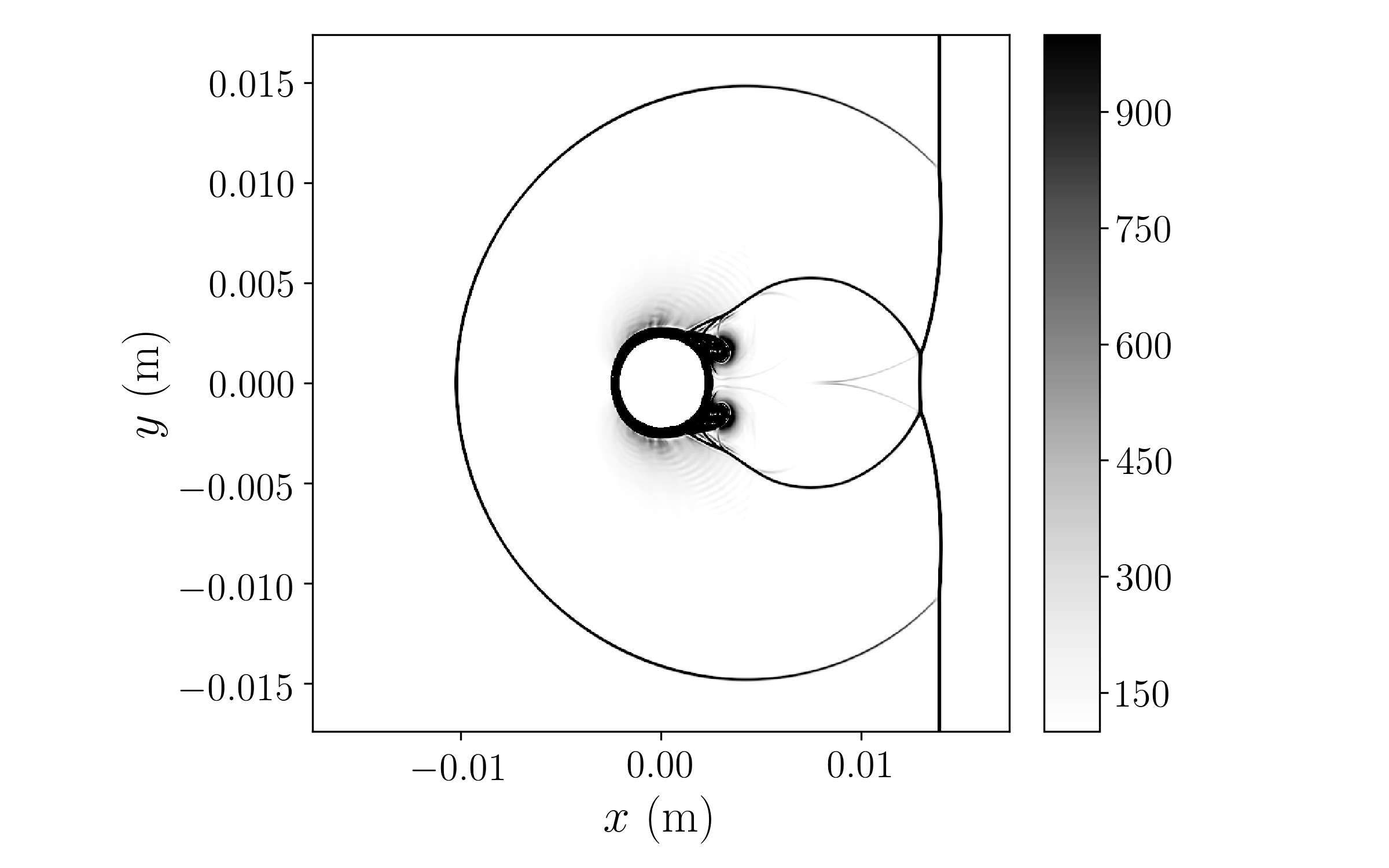}
\label{fig:compare_2D_shock_water_cylinder_t2_WCNS5_IS_PP}}
\caption{Comparison of the 2D shock water cylinder interaction problem by~\citet{igra2001numerical}. Left: holographic interferograms; right: density gradient magnitude, $\left| \nabla \rho \right|$ ($\mathrm{kg\ m^{-4}}$), from simulation using fractional algorithm with PP-WCNS-IS on a mesh with resolution $2048 \times 1024$. Reprinted from \cite{meng2015numerical} with permission from Springer.}
\label{fig:2D_shock_water_interaction_experiment_compare}
\end{figure}

\begin{figure}[!ht]
\centering
\subfigure[$t = 4\ \mu \mathrm{s}$ ($t^*=0.00869$)]{%
\includegraphics[trim=0 1.5cm 0.0cm 3.0cm,clip,height=0.22\textwidth]{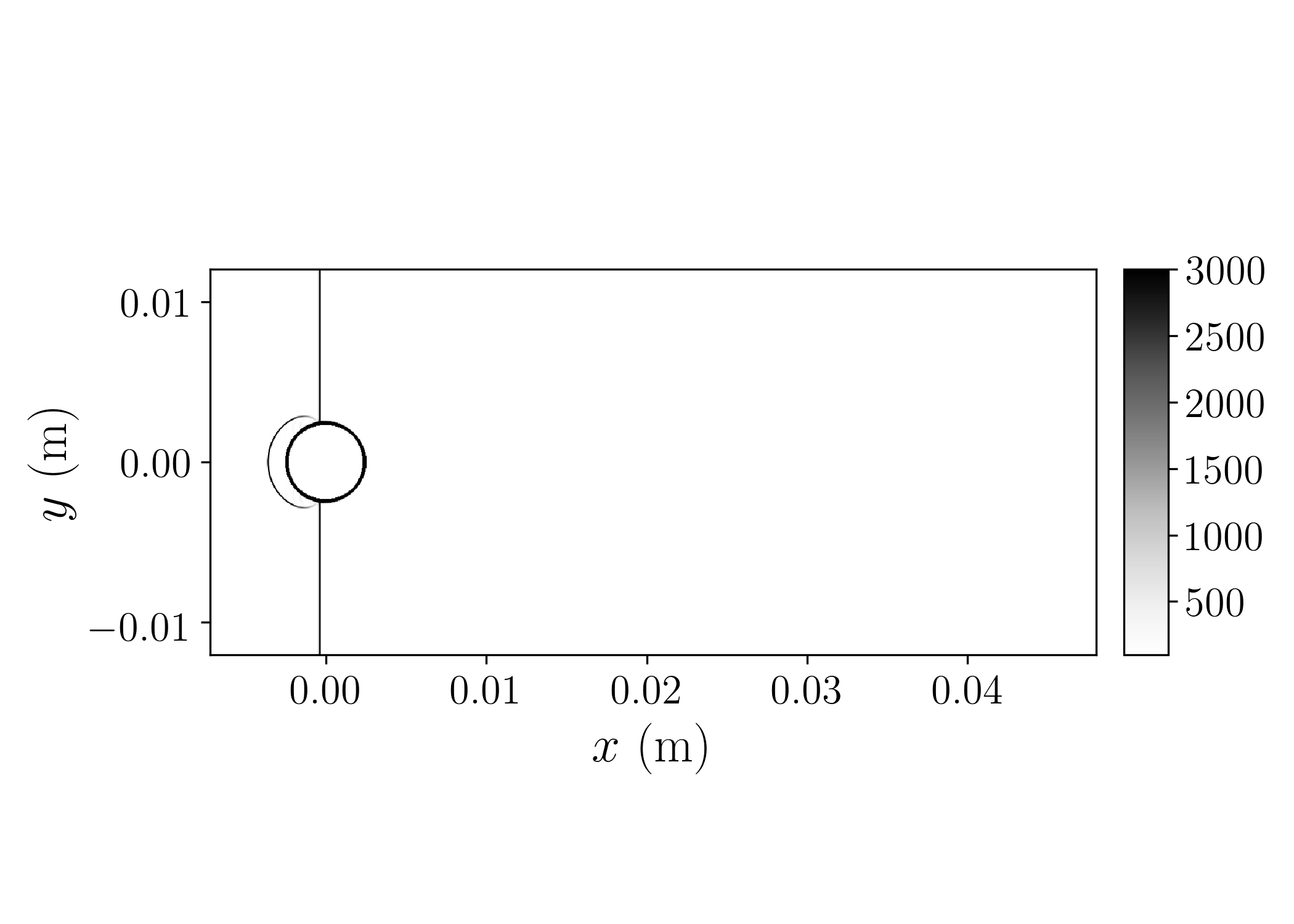}}
\subfigure[$t = 8\ \mu\mathrm{s}$ ($t^*=0.0174$)]{%
\includegraphics[trim=0 1.5cm 0 3.0cm,clip,height=0.22\textwidth]{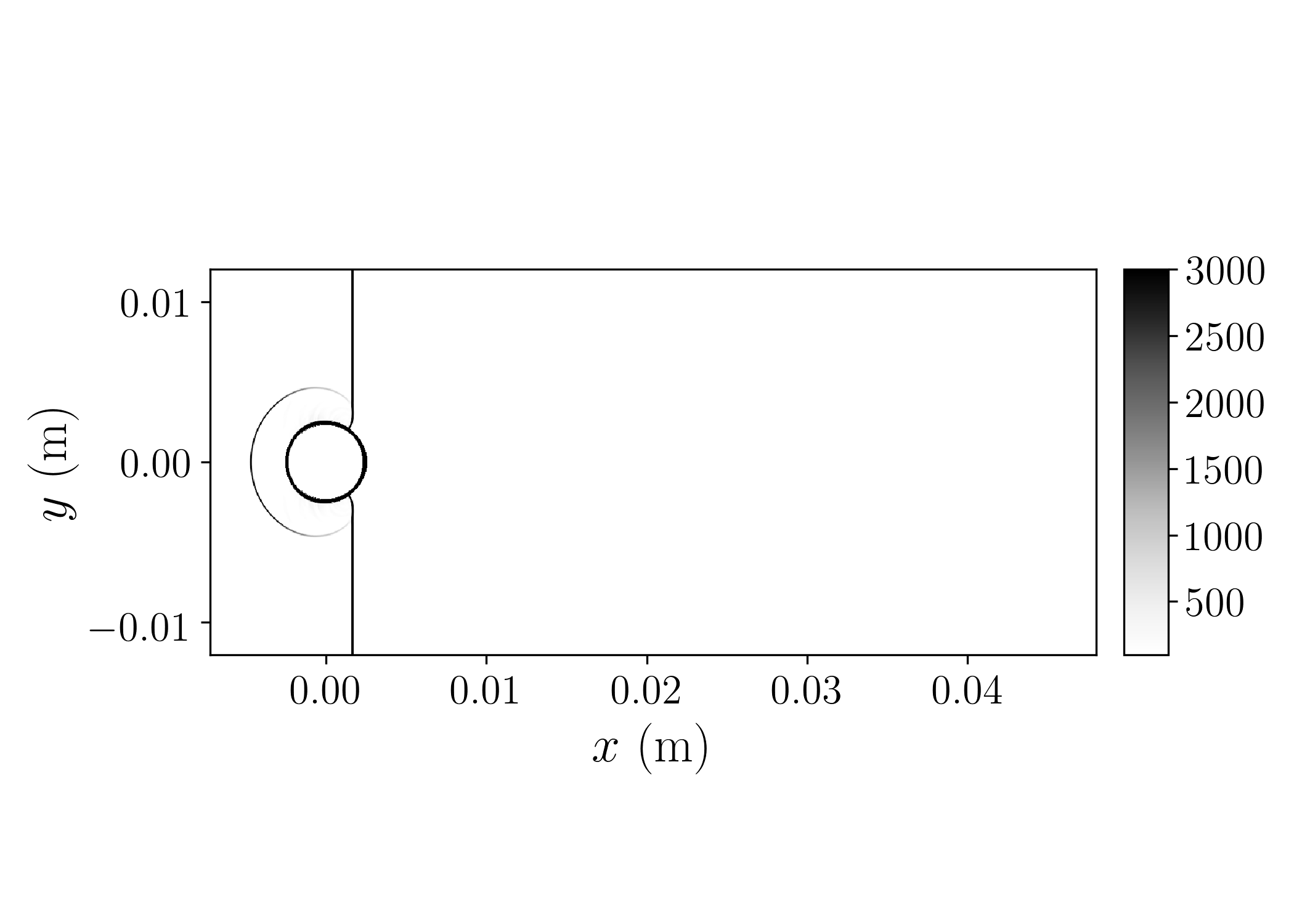}}
\subfigure[$t = 80\ \mu\mathrm{s}$ ($t^*=0.174$)]{%
\includegraphics[trim=0 1.5cm 0.0cm 3.0cm,clip,height=0.22\textwidth]{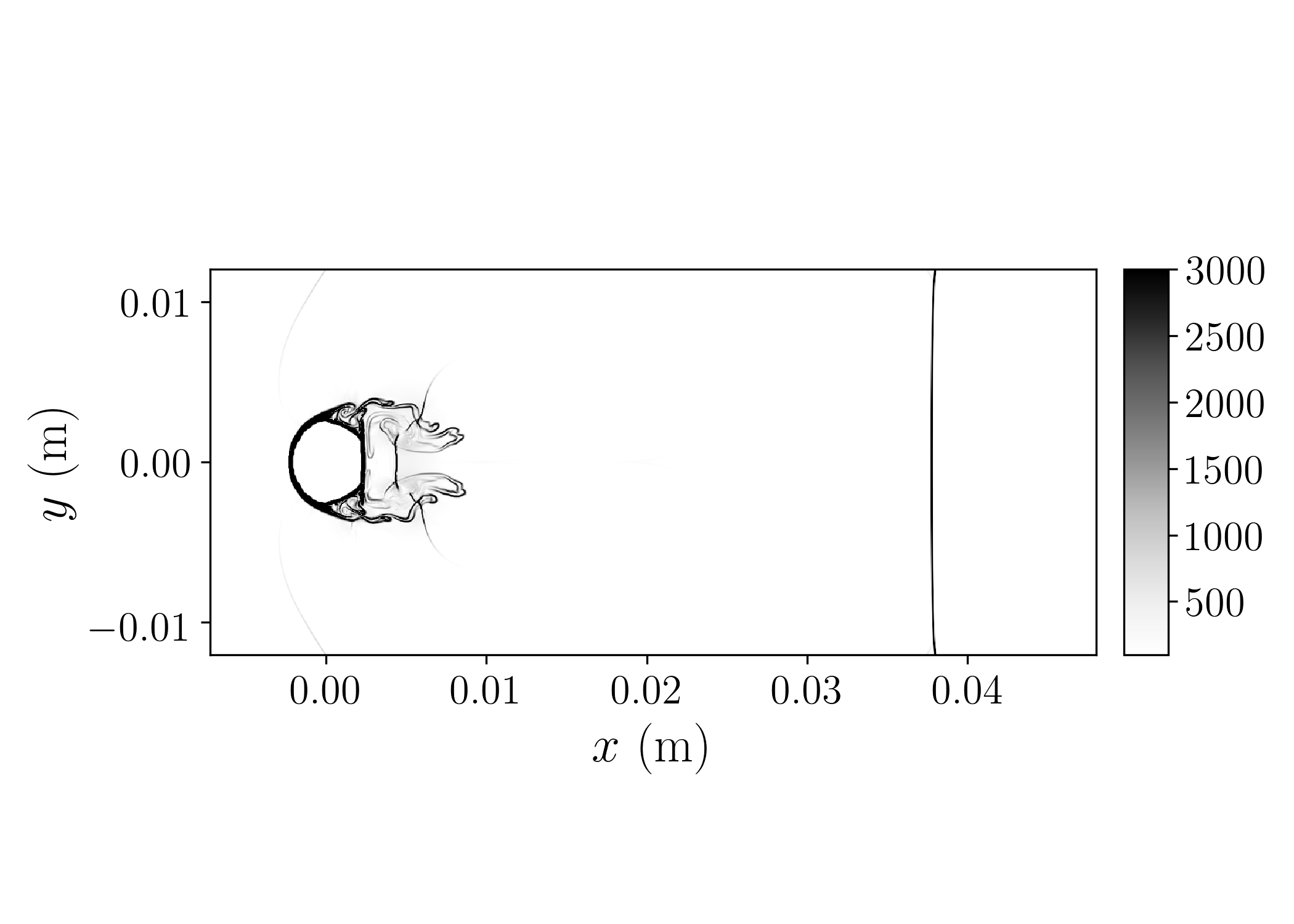}}
\subfigure[$t = 120\ \mu\mathrm{s}$ ($t^*=0.261$)]{%
\includegraphics[trim=0 1.5cm 0 3.0cm,clip,height=0.22\textwidth]{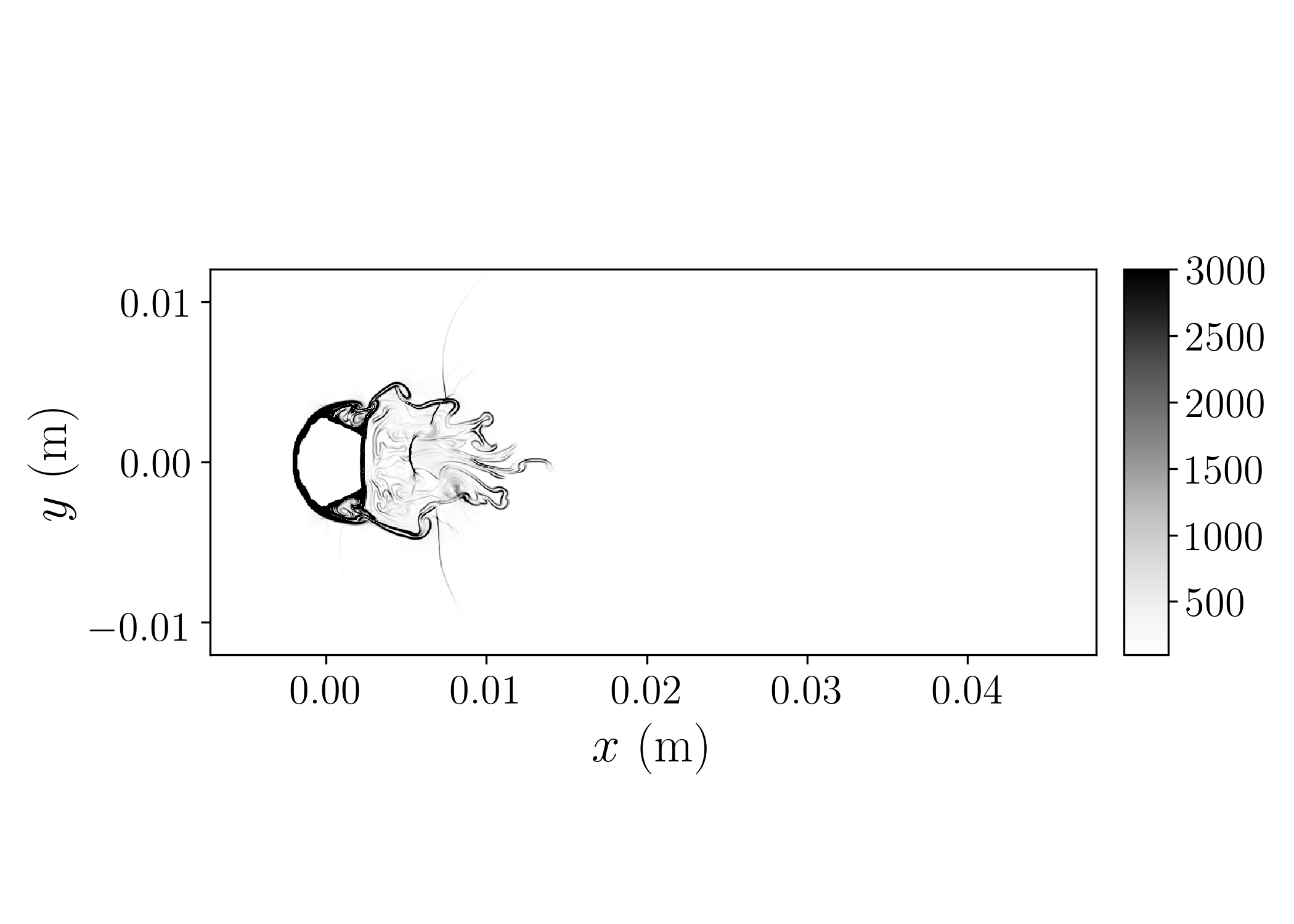}}
\subfigure[$t = 200\ \mu\mathrm{s}$ ($t^*=0.435$)]{%
\includegraphics[trim=0 1.5cm 0.0cm 3.0cm,clip,height=0.22\textwidth]{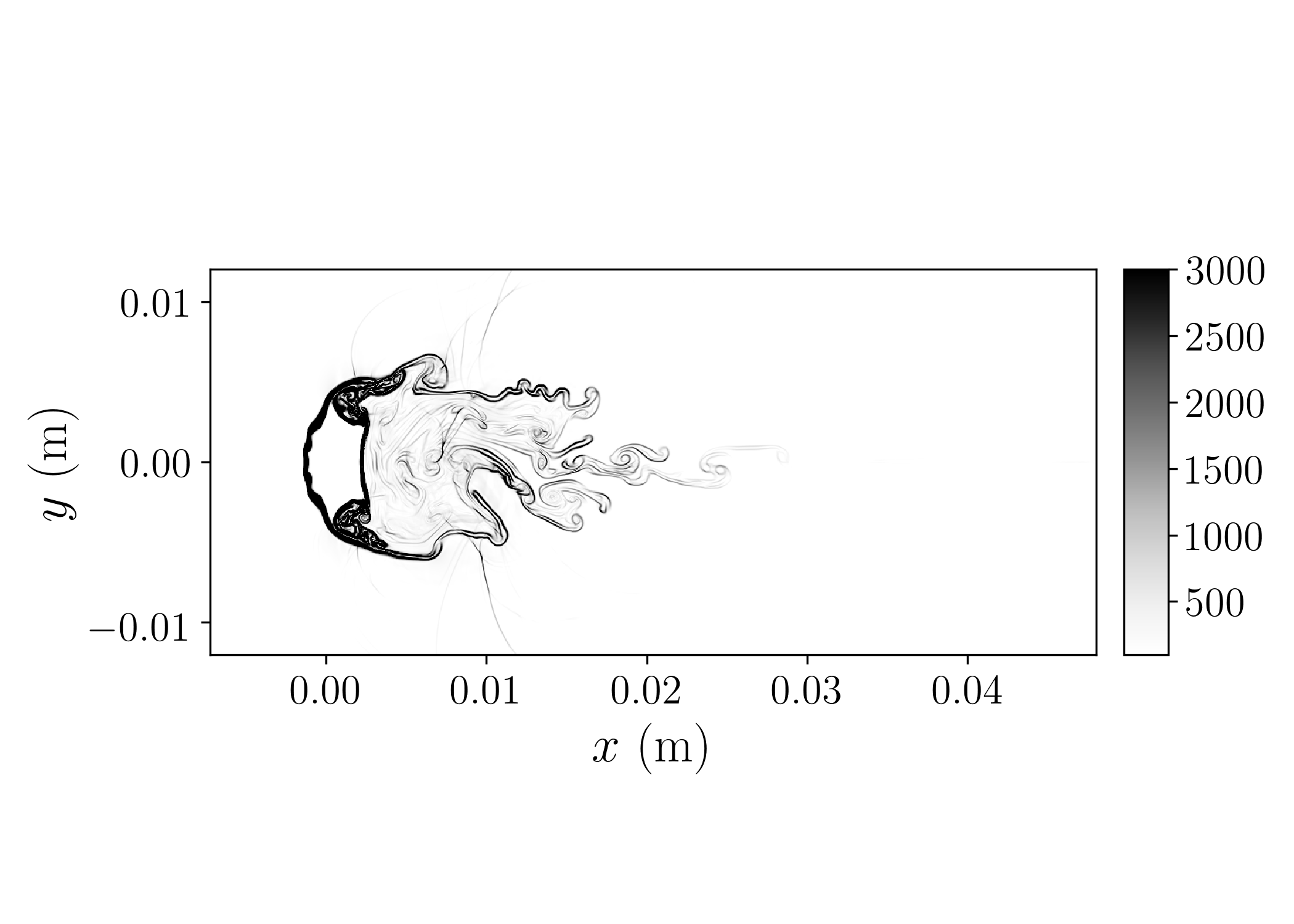}}
\subfigure[$t = 290\ \mu\mathrm{s}$ ($t^*=0.630$)]{%
\includegraphics[trim=0 1.5cm 0 3.0cm,clip,height=0.22\textwidth]{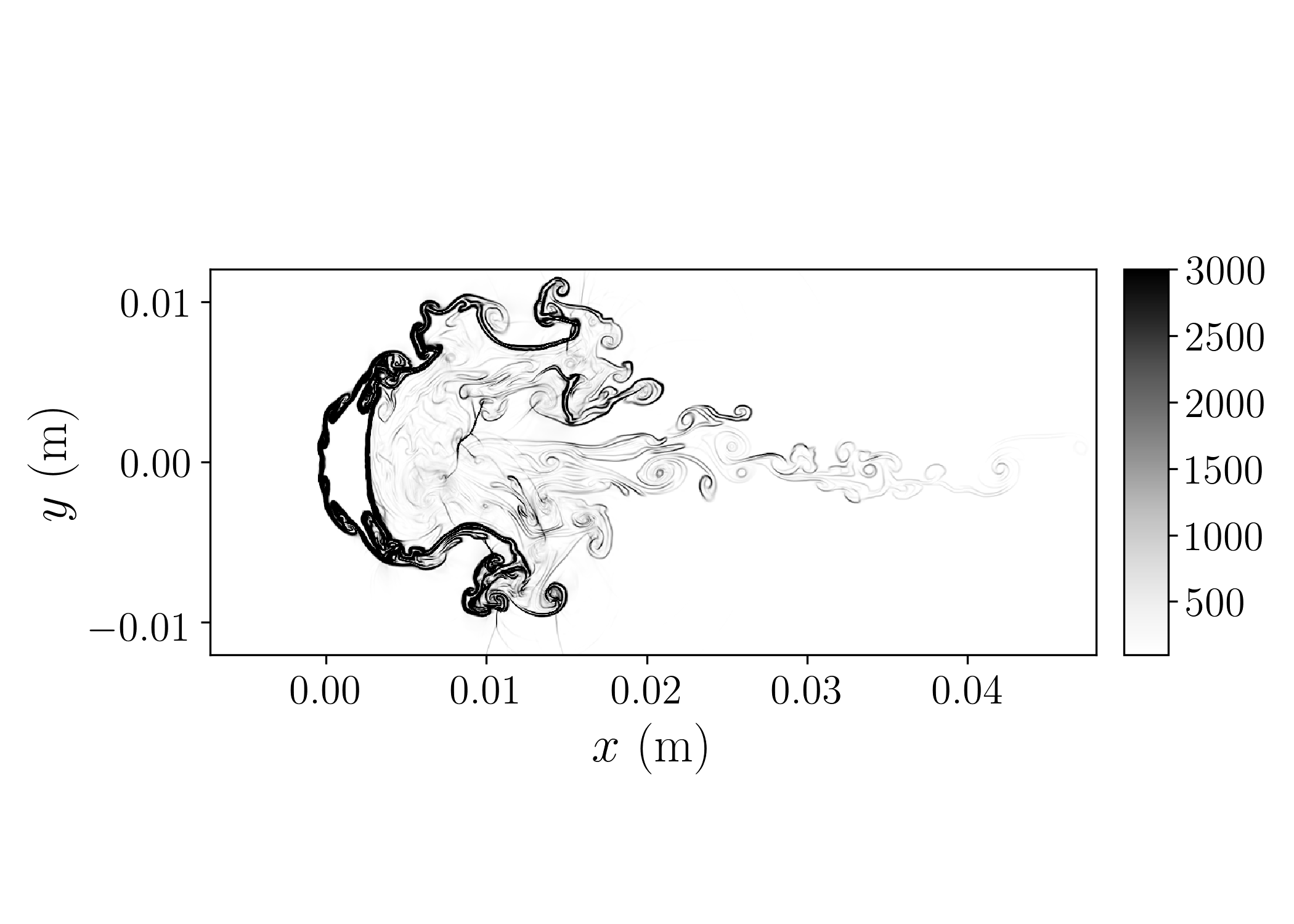}}
\caption{Numerical schlieren ($\left( \left| \nabla \rho \right| / \rho \right)$) of the 2D Mach 1.47 shock water cylinder interaction problem on a mesh with resolution $2048 \times 1024$.}
\label{fig:2D_shock_water_interaction_schl}
\end{figure}

\begin{figure}[!ht]
\centering
\subfigure[$t = 8\ \mu\mathrm{s}$ ($t^*=0.0174$), $\alpha_1$]{%
\includegraphics[trim=0 1.5cm 0.0cm 3.0cm,clip,height=0.22\textwidth]{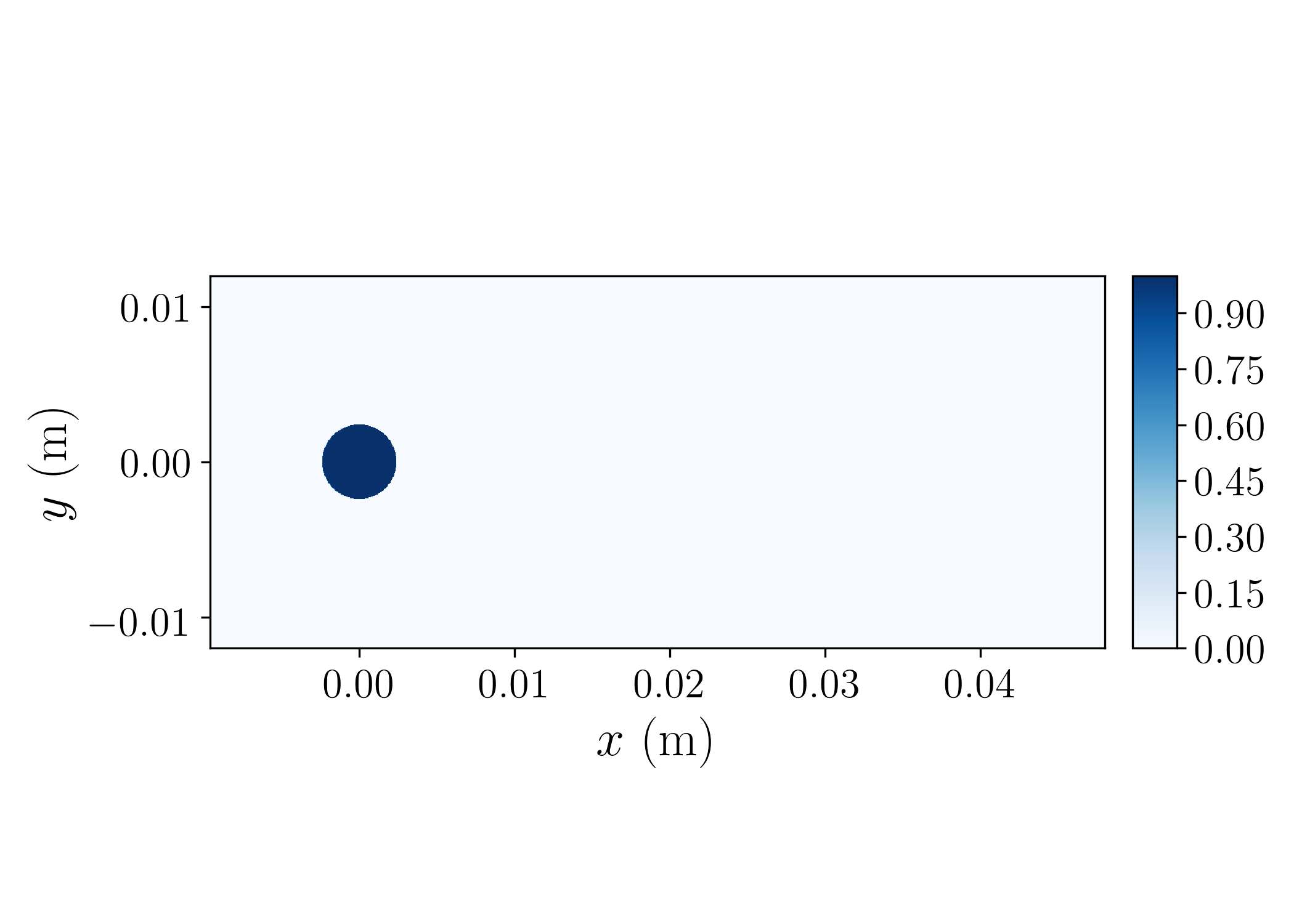}}
\subfigure[$t = 8\ \mu\mathrm{s}$ ($t^*=0.0174$), $T$ (K)]{%
\includegraphics[trim=0 1.5cm 0 3.0cm,clip,height=0.22\textwidth]{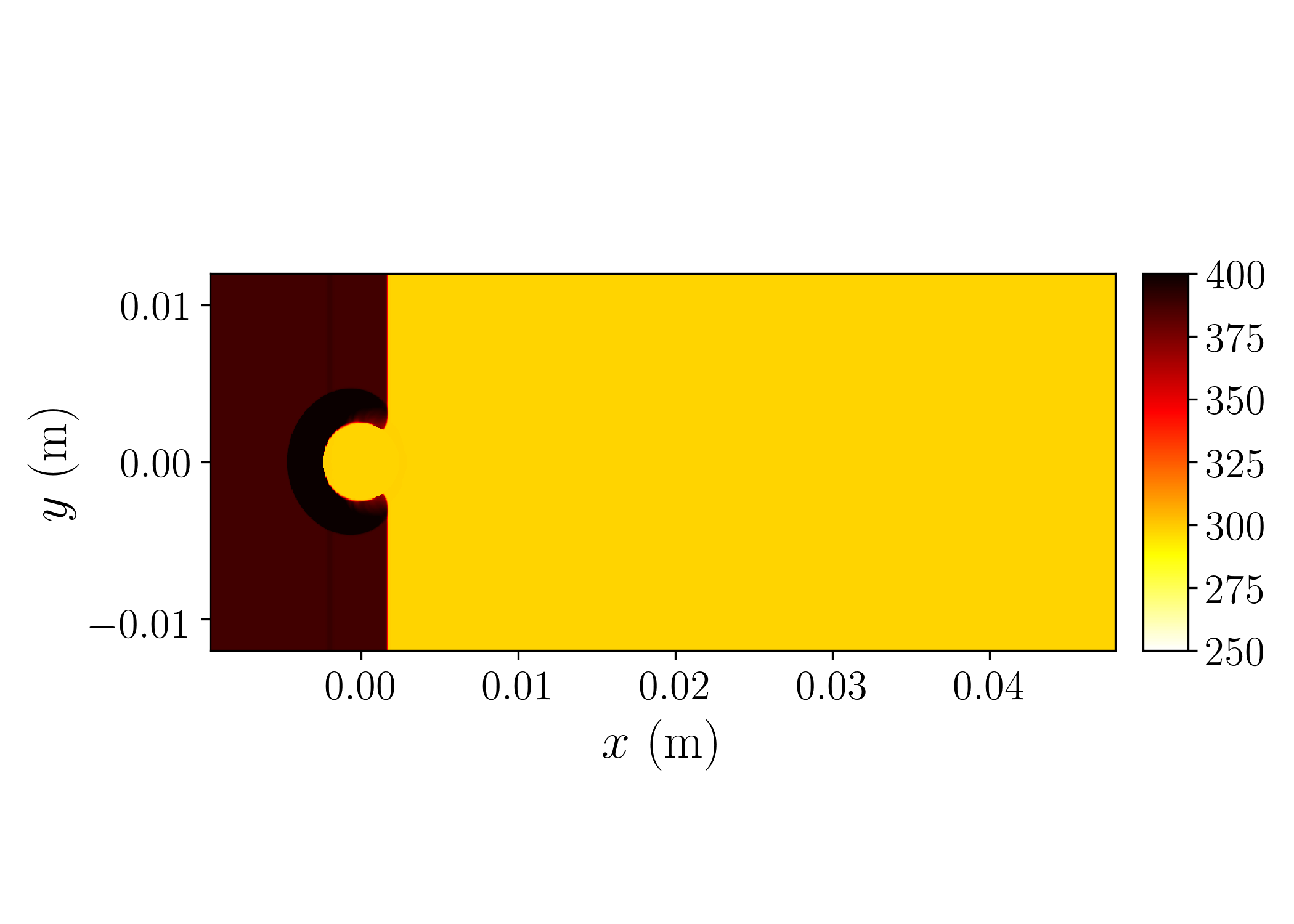}}
\subfigure[$t = 120\ \mu\mathrm{s}$ ($t^*=0.261$), $\alpha_1$]{%
\includegraphics[trim=0 1.5cm 0.0cm 3.0cm,clip,height=0.22\textwidth]{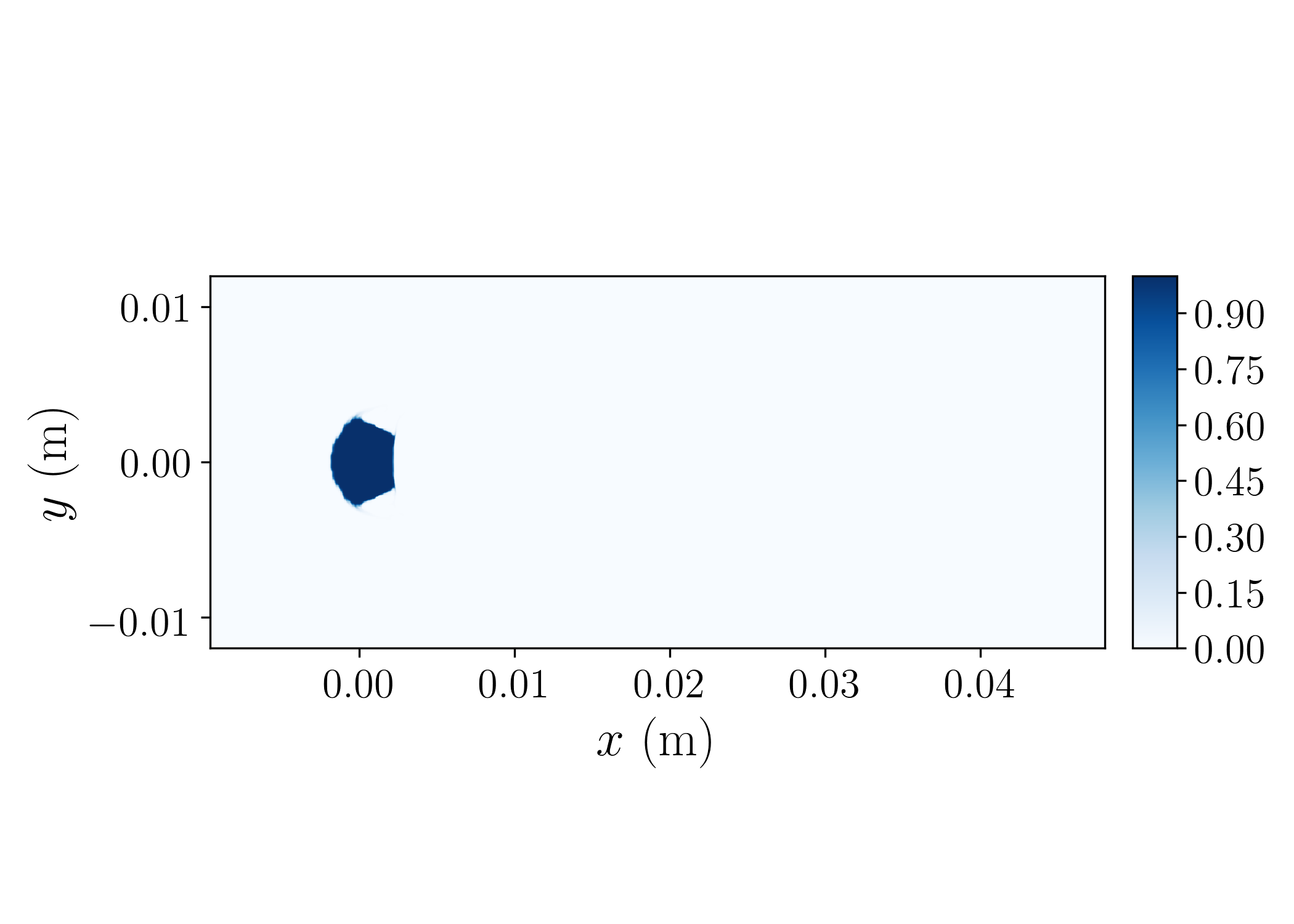}}
\subfigure[$t = 120\ \mu\mathrm{s}$ ($t^*=0.261$), $T$ (K)]{%
\includegraphics[trim=0 1.5cm 0 3.0cm,clip,height=0.22\textwidth]{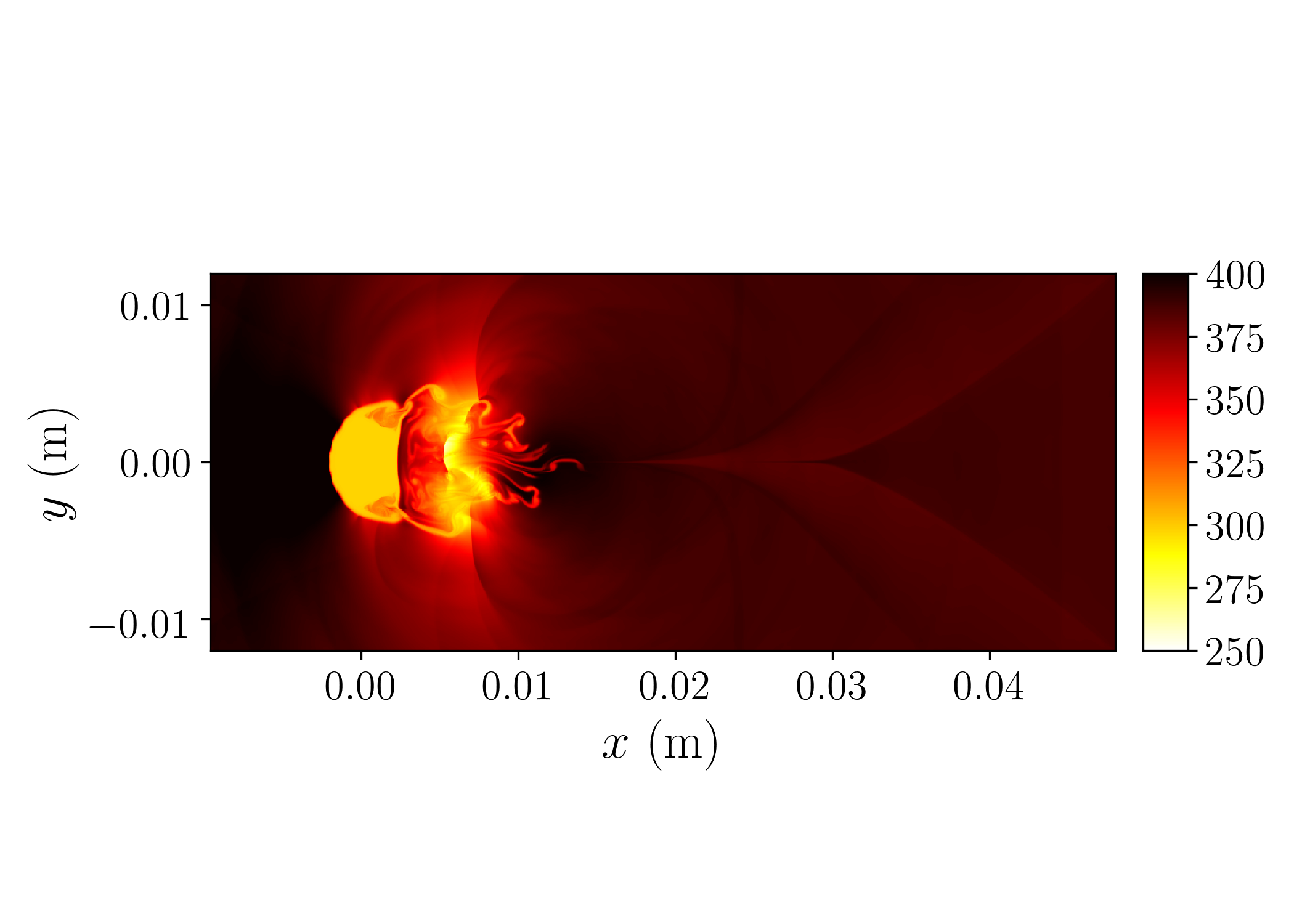}}
\subfigure[$t = 290\ \mu\mathrm{s}$ ($t^*=0.630$), $\alpha_1$]{%
\includegraphics[trim=0 1.5cm 0.0cm 3.0cm,clip,height=0.22\textwidth]{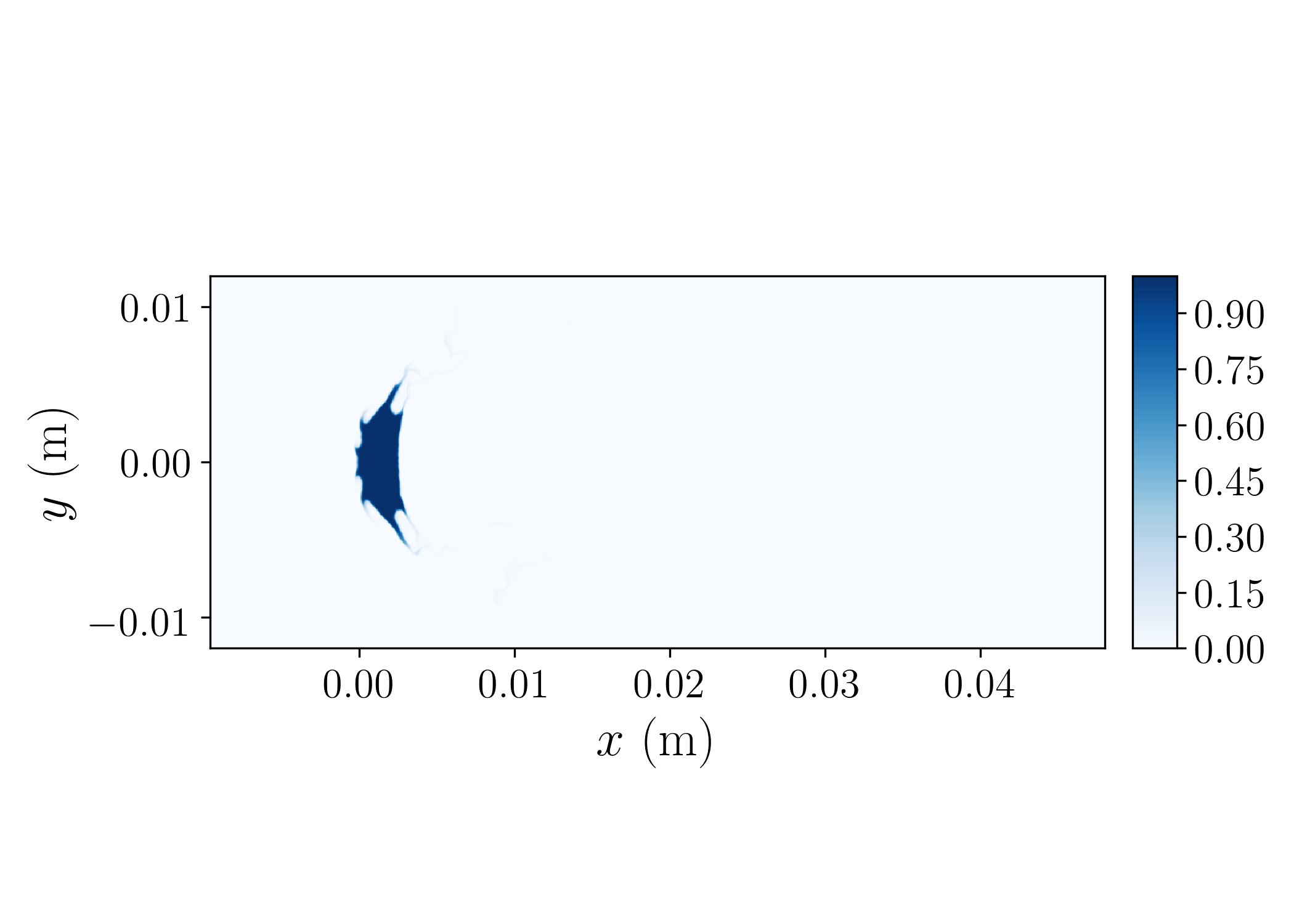}}
\subfigure[$t = 290\ \mu\mathrm{s}$ ($t^*=0.630$), $T$ (K)]{%
\includegraphics[trim=0 1.5cm 0 3.0cm,clip,height=0.22\textwidth]{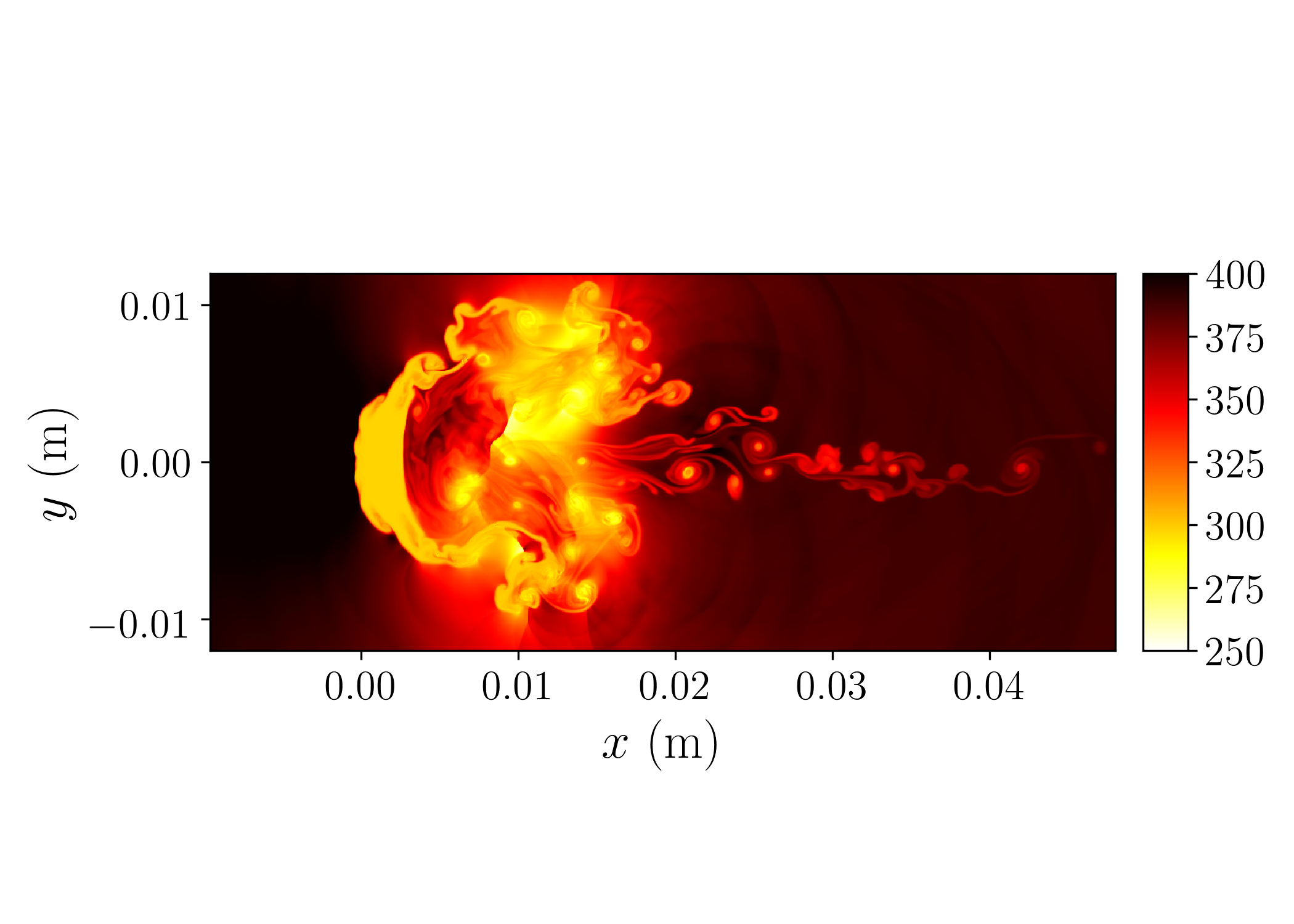}}
\caption{Volume fraction of water and temperature (K) of the 2D Mach 1.47 shock water cylinder interaction problem on a mesh with resolution $2048 \times 1024$. Left column: volume fraction of water; right column: temperature.}
\label{fig:plot_2D_shock_water_interaction}
\end{figure}

Figure~\ref{fig:2D_shock_water_interaction_experiment_compare} compares the density gradient of the highest resolution simulation with the holographic interferograms from the experiment at early times. With this mesh resolution, there are around 102 grid cells across the diameter of the water cylinder. At these two early times, the water cylinder appears rigid to the surrounding flow and does not deform much yet.
The density gradient compares qualitatively well with the experimental results, as important wave features such as the incident and reflected shocks are captured accurately, as well as the Mach stems on the sides of the cylinder. The numerical schlieren, the volume fraction of water, and temperature fields at different times are shown in figures~\ref{fig:2D_shock_water_interaction_schl} and \ref{fig:plot_2D_shock_water_interaction}. 
When the shock passes through the interface, baroclinic torque is generated due to the misalignment of the density and pressure gradients, which produces a large amount of vorticity that distorts the water cylinder over time, as seen from the figures.
At late times, vortices are shed downstream which form a chaotic wake. Most of the vortices are preserved for a long time as minimal dissipation is added by the high-order method.

\begin{figure}
    \centering
    \subfigure[Normalized centroid location]{%
    \includegraphics[width=0.48\textwidth]{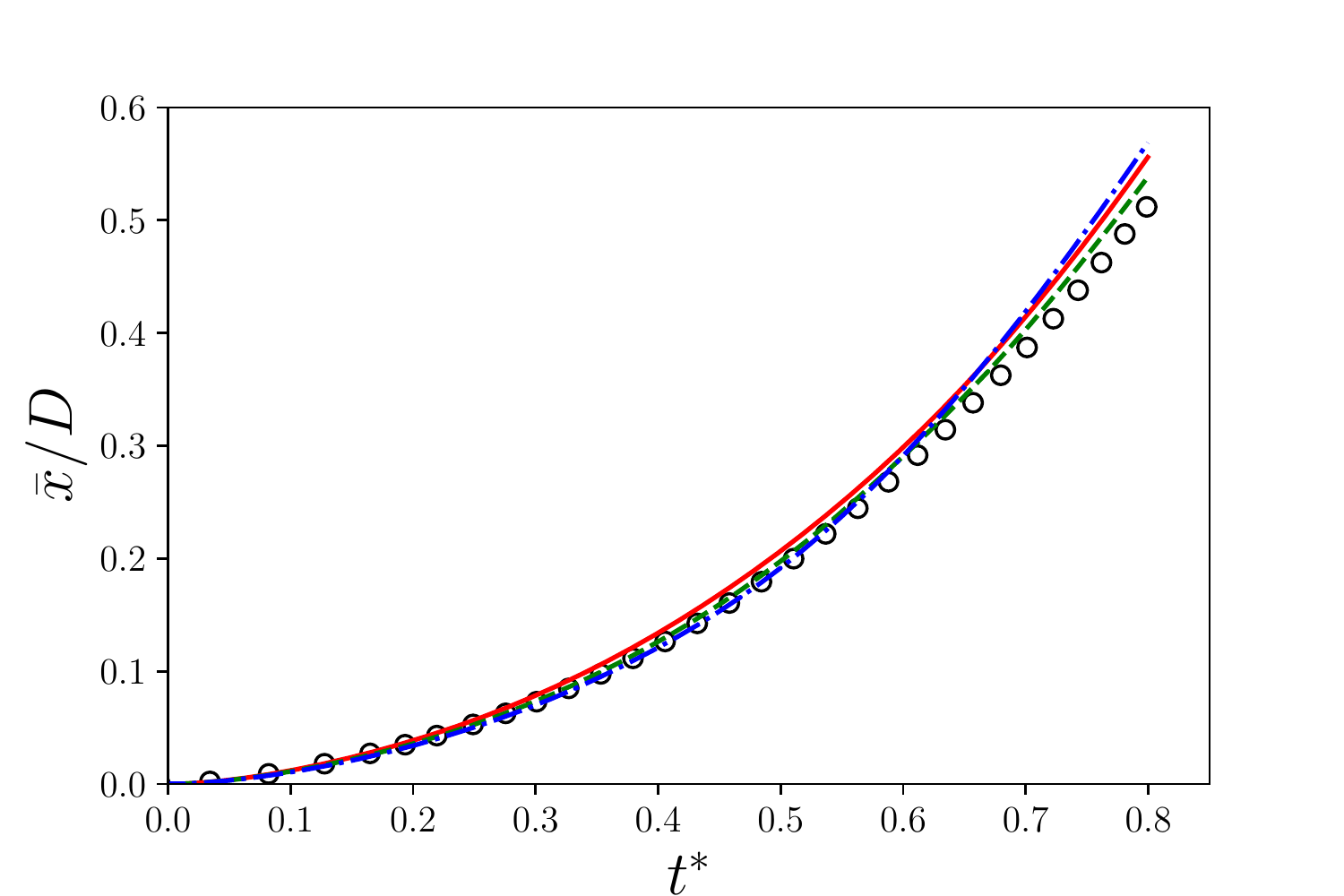}
    \label{fig:2D_shock_water_cylinder_centroid}}
    \subfigure[Drag coefficient]{%
    \includegraphics[width=0.48\textwidth]{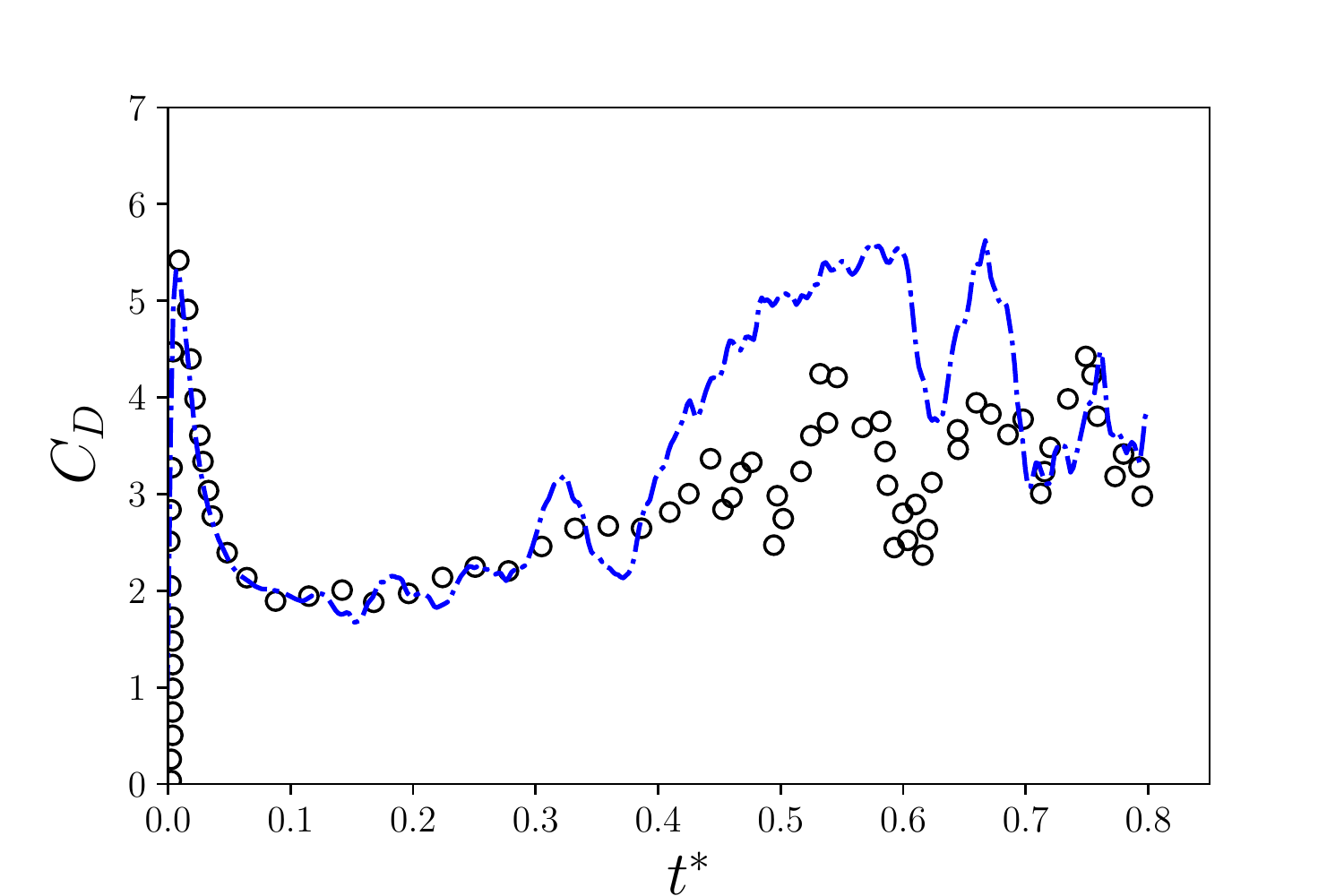}
    \label{fig:2D_shock_water_cylinder_drag_coeff}}
    \caption{Time history of the normalized centroid location and drag coefficient of the water cylinder in the 2D Mach 1.47 shock water cylinder interaction problem.
    Black circles: numerical results in~\citet{meng2015numerical};
    Red solid line: $512 \times 256$ mesh;
    green dashed line: $1024 \times 512$ mesh;
    blue dash-dotted line: $2048 \times 1024$ mesh.}
    \label{fig:2D_shock_water_cylinder_centroid_drag_coeff}
\end{figure}

In figure~\ref{fig:2D_shock_water_cylinder_centroid_drag_coeff}, the time evolution of the centroid location and the drag coefficient is compared with that in the numerical study by~\citet{meng2015numerical}, where the five-equation model by Allaire et al. with a WENO scheme was utilized.
The drag coefficient, $C_D$, is defined as:
\begin{equation}
    C_D = \frac{m \bar{a}}{\half \rho_g (u_g - \bar{u})^2 D},
\end{equation}
where $\rho_g$ and $u_g$ are the post-shock density and streamwise velocity of the air. The constant mass of the water cylinder, $m$, is computed with the initial liquid water density $\rho_l$ as:
\begin{equation}
    m = \rho_l \pi \left( \frac{D}{2} \right)^2 .
\end{equation}
The centroid streamwise velocity and acceleration of the water cylinder, $\bar{u}$ and $\bar{a}$ are computed with the liquid partial density $\alpha_1 \rho_1$:
\begin{align}
    \bar{u} &= \frac{\int \alpha_1 \rho_1 u dV}{\int \alpha_1 \rho_1 dV} , \\
    \bar{a} &= \frac{\frac{d}{dt} \int \alpha_1 \rho_1 u dV}{\int \alpha_1 \rho_1 dV} .
\end{align}
The time is normalized with the reference time scale $t_r$, defined as:
\begin{equation}
    t_r = \frac{D}{u_g} \sqrt{\frac{\rho_l}{\rho_g}} .
\end{equation}
From figure~\ref{fig:2D_shock_water_cylinder_centroid}, it can be seen that the normalized centroid location has insignificant grid sensitivity for the set of mesh resolution chosen throughout the course of the simulations. The centroid locations from our simulations have excellent agreement with the computational results by~\citet{meng2015numerical} until $t^*=0.4$. After that, slightly higher acceleration on the cylinder is observed in our results. The slightly larger acceleration of the water cylinder at late times can also be confirmed by the discrepancy in the drag coefficient obtained in this work and in \cite{meng2015numerical} after $t^*=0.4$, which is shown in figure~\ref{fig:2D_shock_water_cylinder_drag_coeff}. At late times, the flow becomes chaotic from laminar flow and the results are sensitive to the numerical methods and the problem set-up. 
Moreover, it has become more essential to run 3D simulation with subgrid-scale (SGS) model for more accurate prediction of the turbulent wake.
Note that in~\cite{meng2015numerical}, the flow is assumed to be symmetric across the cylinder's centerline and only half of the cylinder was considered in the simulations while it is clear that the flow fields at late times are not symmetric anymore as the flow is no longer laminar, which can be seen in figure~\ref{fig:2D_shock_water_interaction_schl} and other figures. The difference in the assumption on the flow can also lead to discrepancy between the results at late times.

\begin{table}[!ht]
\small
  \begin{center}
    \begin{tabular}{@{}c | cccccc@{}}\toprule
     &
    \addstackgap{\stackanchor{$\alpha_1 \rho_1$}{$(\mathrm{kg\ m^{-3}})$}} &
    \stackanchor{$\alpha_2 \rho_2$}{$(\mathrm{kg\ m^{-3}})$} &
    \stackanchor{$u$}{$(\mathrm{m\ s^{-1}})$} &
    \stackanchor{$v$}{$(\mathrm{m\ s^{-1}})$} &
    \stackanchor{$p$}{$(\mathrm{Pa})$} &
    $\alpha_1$ \\
    \midrule
    \addstackgap{\stackanchor{pre-shock}{air}} & $1.0227724412751677\mathrm{e}{-5}$ & $1.1817862094653702$ & 0 & 0 & $1.01325\mathrm{e}{5}$ & $1.0\mathrm{e}{-8}$ \\
    \addstackgap{\stackanchor{post-shock}{air}} & $7.8666528513009387\mathrm{e}{-6}$ & $2.1397213925906398$ & $u_{post}$ & 0 & $p_{post}$ & $1.0\mathrm{e}{-8}$ \\
    \addstackgap{\stackanchor{water}{cylinder}} & $1.0227724310474432\mathrm{e}{3}$ & $1.1817862212832324\mathrm{e}{-8}$ & 0 & 0 & $1.01325\mathrm{e}{5}$ & $1 - 1.0\mathrm{e}{-8}$ \\
    \bottomrule
    \end{tabular}
  \end{center}
  \caption{Initial conditions of the 2D Mach 1.47 shock water cylinder interaction problem. The post-shock air velocity and pressure are  $u_{post}=228.00747932015310\ \mathrm{m\ s^{-1}}$ and $p_{post}=2.3855789125\mathrm{e}{5}\ \mathrm{Pa}$ respectively.}
  \label{table:IC_2D_shock_water_cylinder}
\end{table}

\subsection{Two-dimensional shock bubble collapse problem}

In contrast to the previous test with a water column surrounded by air, an air bubble (column) is placed in liquid water and interacts with a 1.9 GPa incident shock in this 2D problem. This problem was first experimentally investigated by~\citet{bourne1992shock} and was considered in many previous numerical works~\cite{ball2000shock, terashima2010front, lauer2012numerical, nourgaliev2006adaptive, nourgaliev2007high, shukla2010interface, shukla2014nonlinear, haimovich2017numerical}. 

The pre-shock water and air cylinder are initially at atmospheric conditions with $p = 101325 \ \mathrm{Pa}$ and $T = 298 \ \mathrm{K}$.
The domain has a size of $\left[ 0, 24L \right] \times \left[ -7.5L, 7.5L \right]$, where  $L = 1\ \mathrm{mm}$ is chosen. The air cylinder with radius $R_b = 3L$ is initially placed at location $\left[ 12L, 0 \right]$. The initial position of the shock is at $5.4L$ to the left of the center of the air cavity.
The initial conditions are given by table~\ref{table:IC_1_9_GPa_shock_bubble_collapse_problem}.
The accuracy of the fractional algorithm with the five-equation model by Allaire et al. and the thermal relaxation is examined with the simulations performed on three different levels of mesh resolution:
$512 \times 320$, $1024 \times 640$, and $2048 \times 1280$,
using the PP-WCNS-IS method.

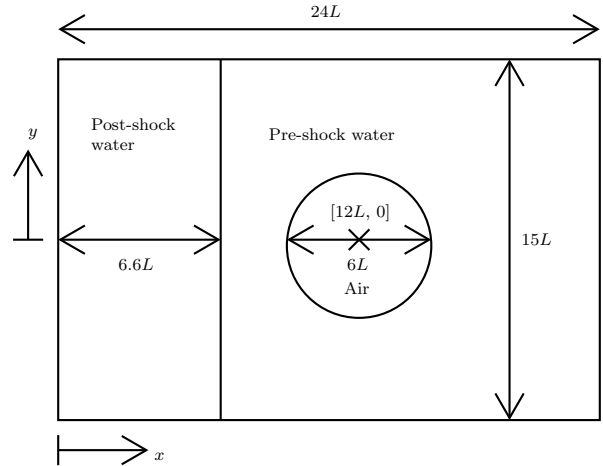
\begin{figure}[!ht]
  \centering
  \begin{tikzpicture}[thick,scale=0.8, every node/.style={transform shape}]
    \useasboundingbox (0cm,-1cm)  rectangle (8cm,7cm);
    \draw[black]        (-0.5cm,0.0cm) rectangle ++(9cm,6cm);
    \draw[black]        ( 4.5cm,2.9cm) circle (1.2cm);
    \draw[black, thick] ( 2.2cm,0.0cm) -- (2.2cm,6cm);

    \node[text width=3cm] at (1.55cm,4.75cm) {Post-shock \\ water};
    \node[text width=3cm] at (4.5cm,4.75cm) {Pre-shock water};
    \node[text width=3cm] at (5.75cm,2.2cm) {Air};
    
    \draw (4.5cm, 3.0cm) node[cross] {};

    \draw[{Straight Barb[angle'=60,scale=3]}-{Straight Barb[angle'=60,scale=3]}] ( 7.0cm,0.0cm) -- (7.0cm,6cm);
    \draw[{Straight Barb[angle'=60,scale=3]}-{Straight Barb[angle'=60,scale=3]}] (-0.5cm,6.5cm) -- (8.5cm,6.5cm);
    \draw[{Straight Barb[angle'=60,scale=3]}-{Straight Barb[angle'=60,scale=3]}] ( 3.3cm,3.0cm) -- (5.7cm,3.0cm);
    \draw[{Straight Barb[angle'=60,scale=3]}-{Straight Barb[angle'=60,scale=3]}] (-0.5cm,3.0cm) -- (2.2cm,3.0cm);
    
    \node[text width=3cm] at (5.2cm,6.8cm) {$24L$};
    \node[text width=3cm] at (8.7cm,3.0cm) {$15L$};
    \node[text width=3cm] at (5.8cm,2.6cm) {$6L$};
    \node[text width=3cm] at (2.0cm,2.6cm) {$6.6L$};
    \node[text width=3cm] at (5.52cm,3.45cm) { $[12L, \, 0]$ };

    \draw[-{Straight Barb[angle'=60,scale=3]}] (-0.5cm,-0.5cm) -- (1.0cm,-0.5cm);
    \node[text width=3cm] at (2.6cm,-0.6cm) {$x$};
    \draw[black] (-0.5cm,-0.75cm) -- (-0.5cm,-0.25cm);
    \draw[-{Straight Barb[angle'=60,scale=3]}] (-1.0cm,3.0cm) -- (-1.0cm,4.5cm);
    \node[text width=1cm] at (-0.5cm,4.75cm) {$y$};
    \draw[black] (-1.25cm,3.0cm) -- (-0.75cm,3.0cm);
  \end{tikzpicture}
  \caption{Schematic diagram of the 2D 1.9 GPa shock bubble collapse problem.} \label{fig:schematic_2D_1_9_GPa_shock_bubble_collapse_problem}
\end{figure}

\begin{table}[!ht]
\small
  \begin{center}
    \begin{tabular}{@{}c | cccccc@{}}\toprule
     &
    \addstackgap{\stackanchor{$\alpha_1 \rho_1$}{$(\mathrm{kg\ m^{-3}})$}} &
    \stackanchor{$\alpha_2 \rho_2$}{$(\mathrm{kg\ m^{-3}})$} &
    \stackanchor{$u$}{$(\mathrm{m\ s^{-1}})$} &
    \stackanchor{$v$}{$(\mathrm{m\ s^{-1}})$} &
    \stackanchor{$p$}{$(\mathrm{Pa})$} &
    $\alpha_1$ \\
    \midrule
    \addstackgap{\stackanchor{pre-shock}{water}} & $1.0227724310474432\mathrm{e}{3}$ & $1.1817862212832324\mathrm{e}{-8}$ & 0 & 0 & $1.01325\mathrm{e}{5}$ & $1 - 1.0\mathrm{e}{-8}$ \\
    \addstackgap{\stackanchor{post-shock}{water}} & $1.4584493961445507\mathrm{e}{3}$ & $9.7946301317627371\mathrm{e}{-5}$ & $u_{post}$ & 0 & $1.9\mathrm{e}{9}$ & $1 - 1.0\mathrm{e}{-8}$ \\
    \addstackgap{\stackanchor{air}{cylinder}} & $1.0227724412751677\mathrm{e}{-5}$ & $1.1817862094653702$ & 0 & 0 & $1.01325\mathrm{e}{5}$ & $1.0\mathrm{e}{-8}$ \\
    \bottomrule
    \end{tabular}
  \end{center}
  \caption{Initial conditions of the 2D 1.9 GPa shock bubble collapse problem. The post-shock water velocity is  $u_{post}=744.92462003762398\ \mathrm{m\ s^{-1}}$.}
  \label{table:IC_1_9_GPa_shock_bubble_collapse_problem}
\end{table}

\begin{figure}[!ht]
\centering
\subfigure[$t = 0.638\ \mu \mathrm{s}$]{%
\includegraphics[trim=0 1.5cm 0 1.0cm,clip,width=0.4\textwidth]{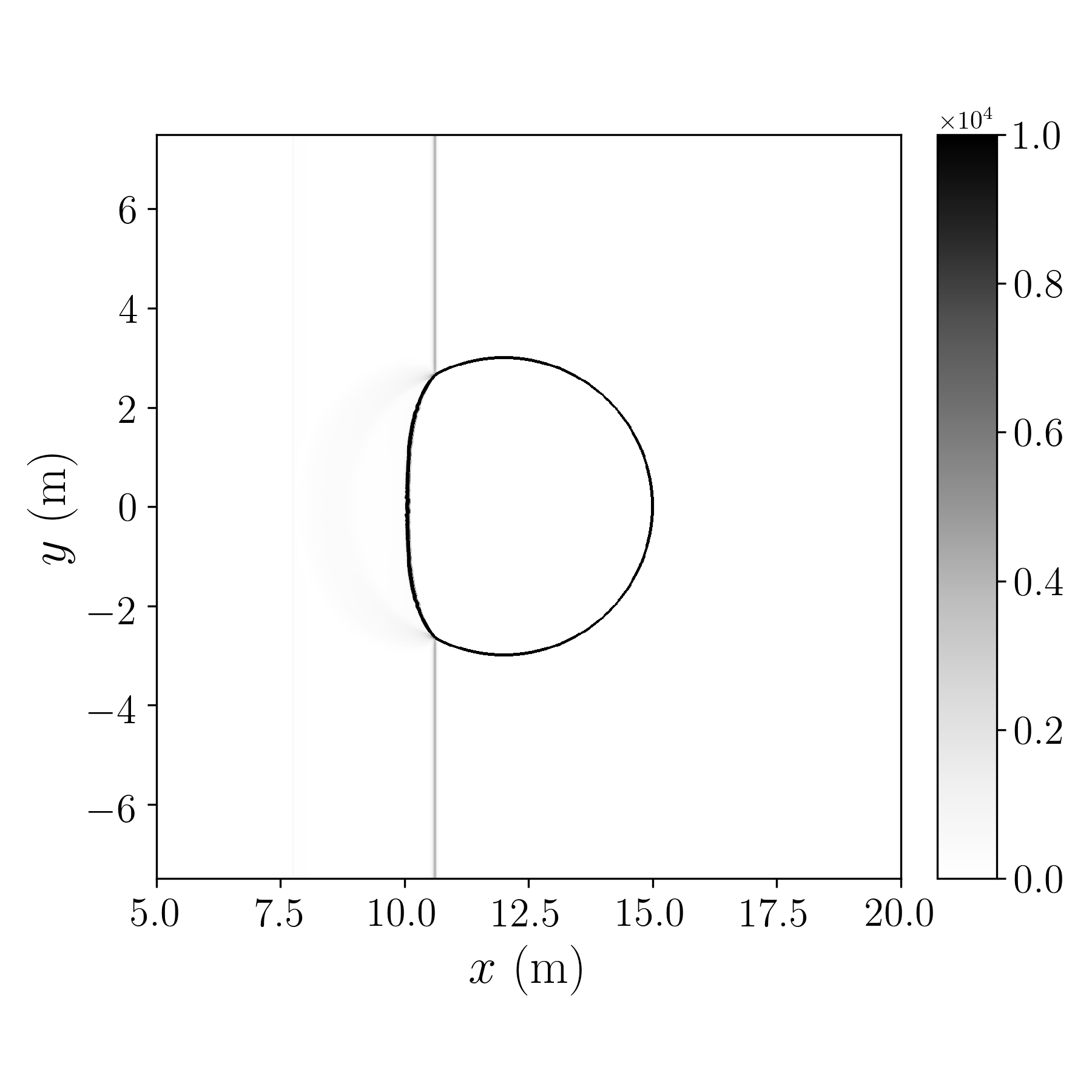}}
\subfigure[$t = 2.24\ \mu \mathrm{s}$]{%
\includegraphics[trim=0 1.5cm 0 1.0cm,clip,width=0.4\textwidth]{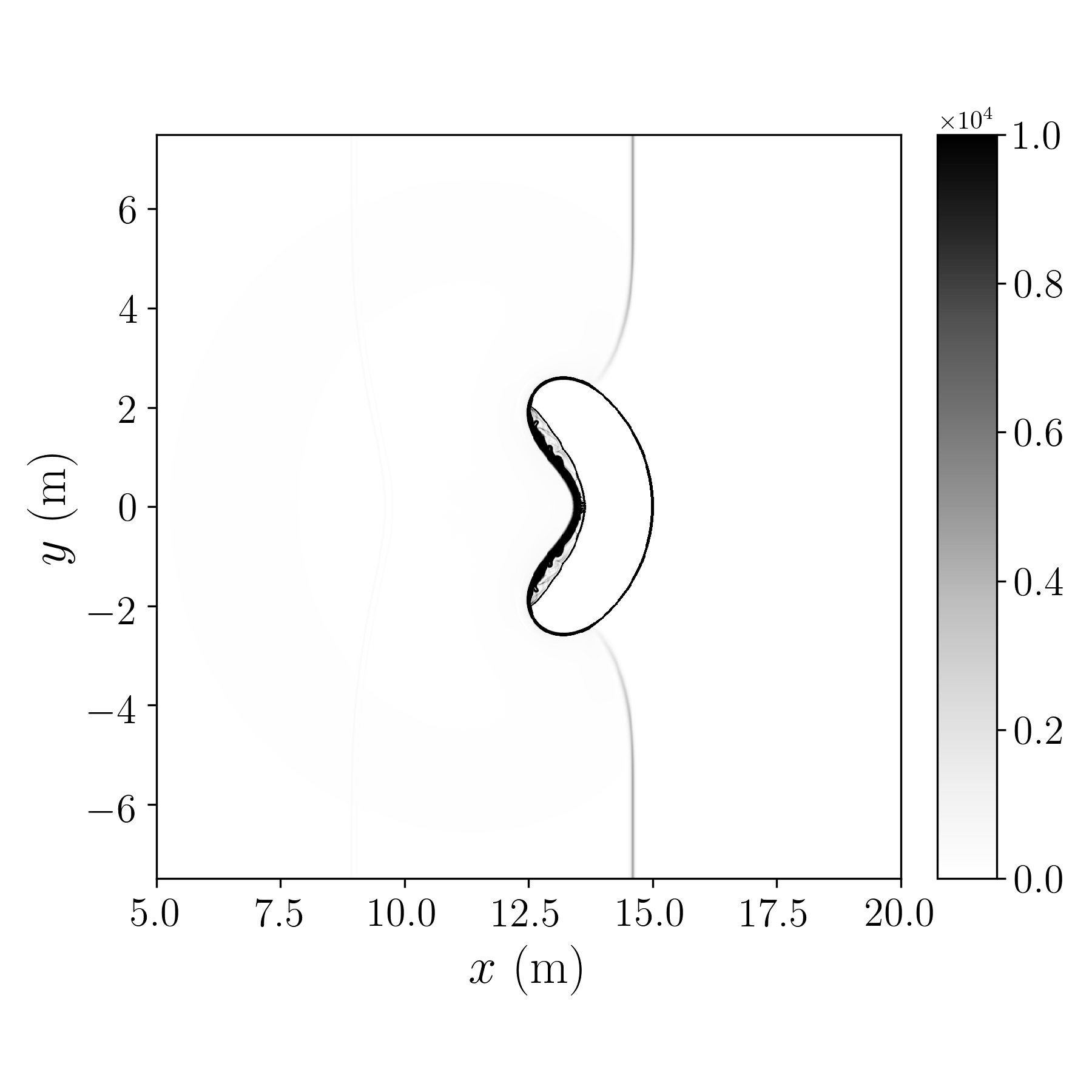}}
\subfigure[$t = 2.80\ \mu \mathrm{s}$]{%
\includegraphics[trim=0 1.5cm 0 1.0cm,clip,width=0.4\textwidth]{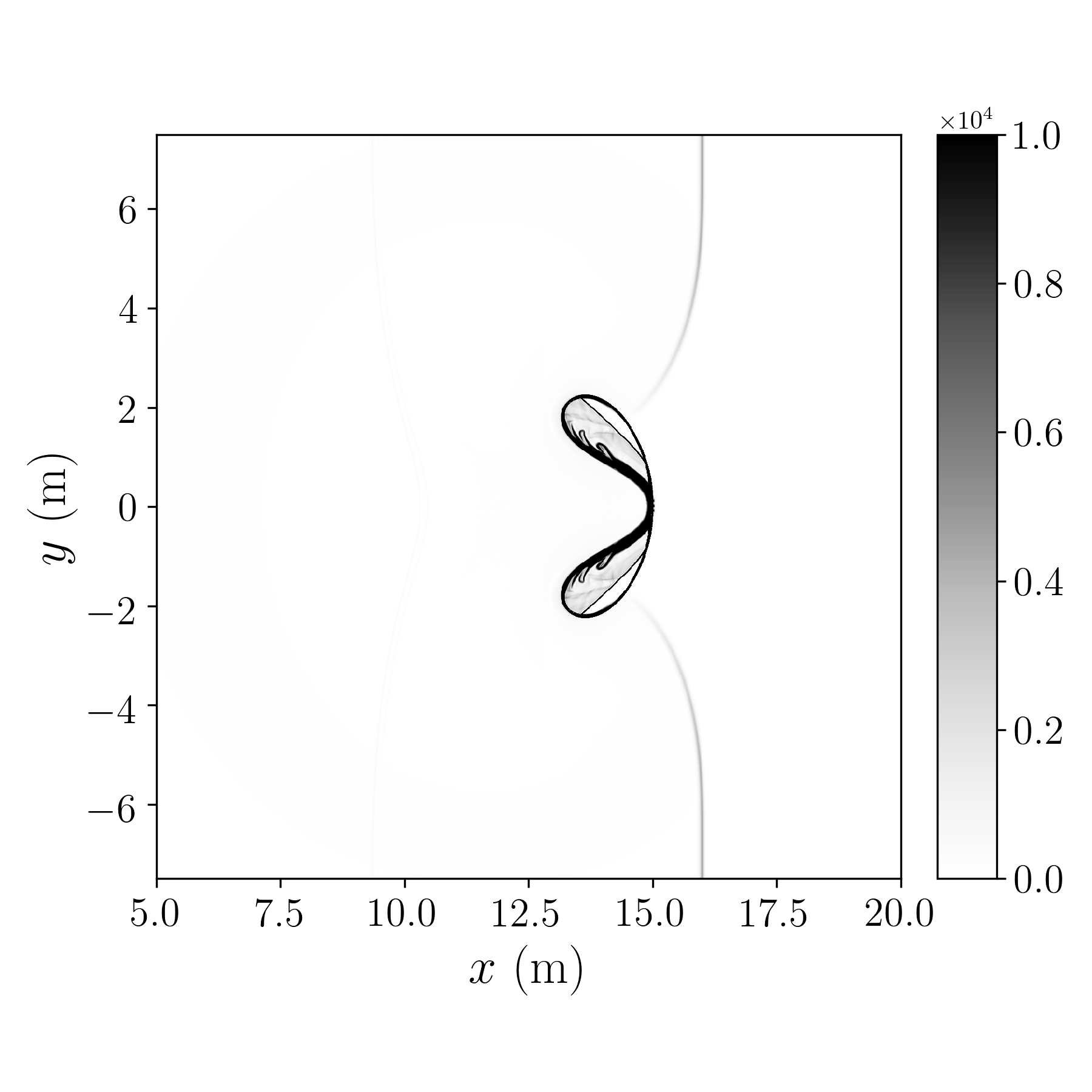}}
\subfigure[$t = 3.44\ \mu \mathrm{s}$]{%
\includegraphics[trim=0 1.5cm 0 1.0cm,clip,width=0.4\textwidth]{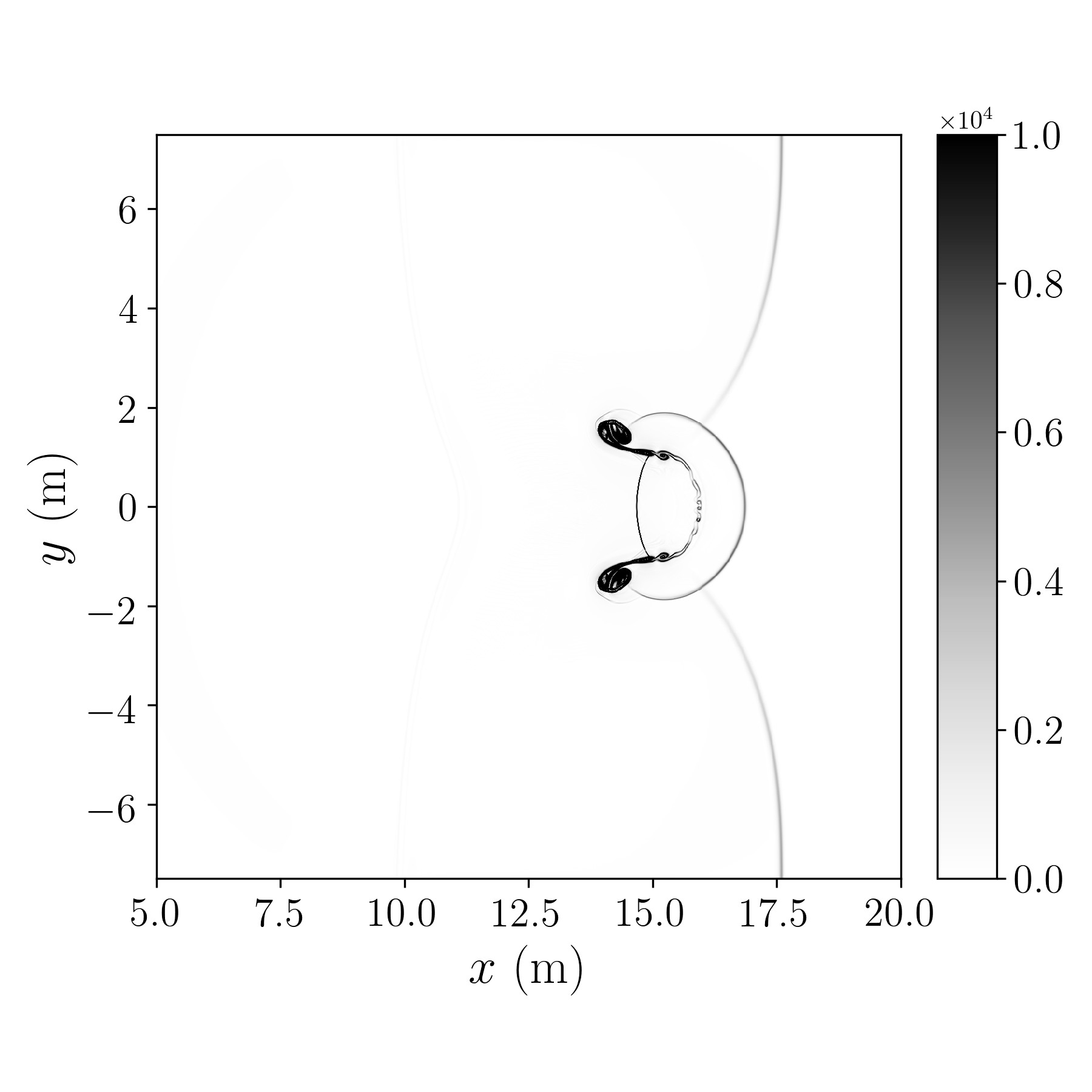}}
\caption{Numerical schlieren ($\left( \left| \nabla \rho \right| / \rho \right)$) of the 2D 1.9 GPa shock bubble collapse problem on a mesh with resolution $2048 \times 1280$.}
\label{fig:2D_1_9_GPa_shock_bubble_collapse_problem_schl}
\end{figure}

\begin{figure}[!ht]
\centering
\subfigure[$t = 0.638\ \mu \mathrm{s}$]{%
\includegraphics[trim=0 1.5cm 0 1.0cm,clip,width=0.4\textwidth]{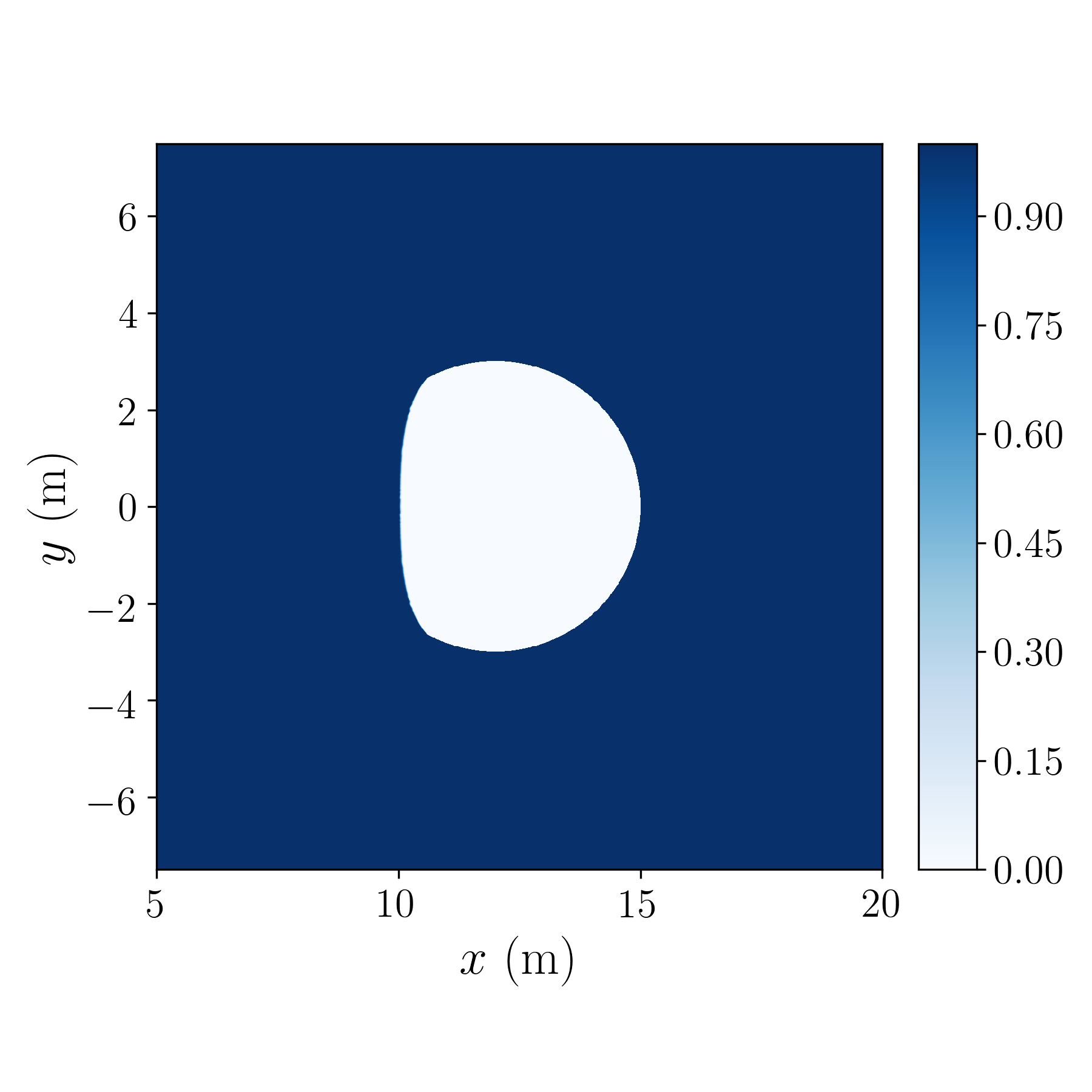}}
\subfigure[$t = 2.24\ \mu \mathrm{s}$]{%
\includegraphics[trim=0 1.5cm 0 1.0cm,clip,width=0.4\textwidth]{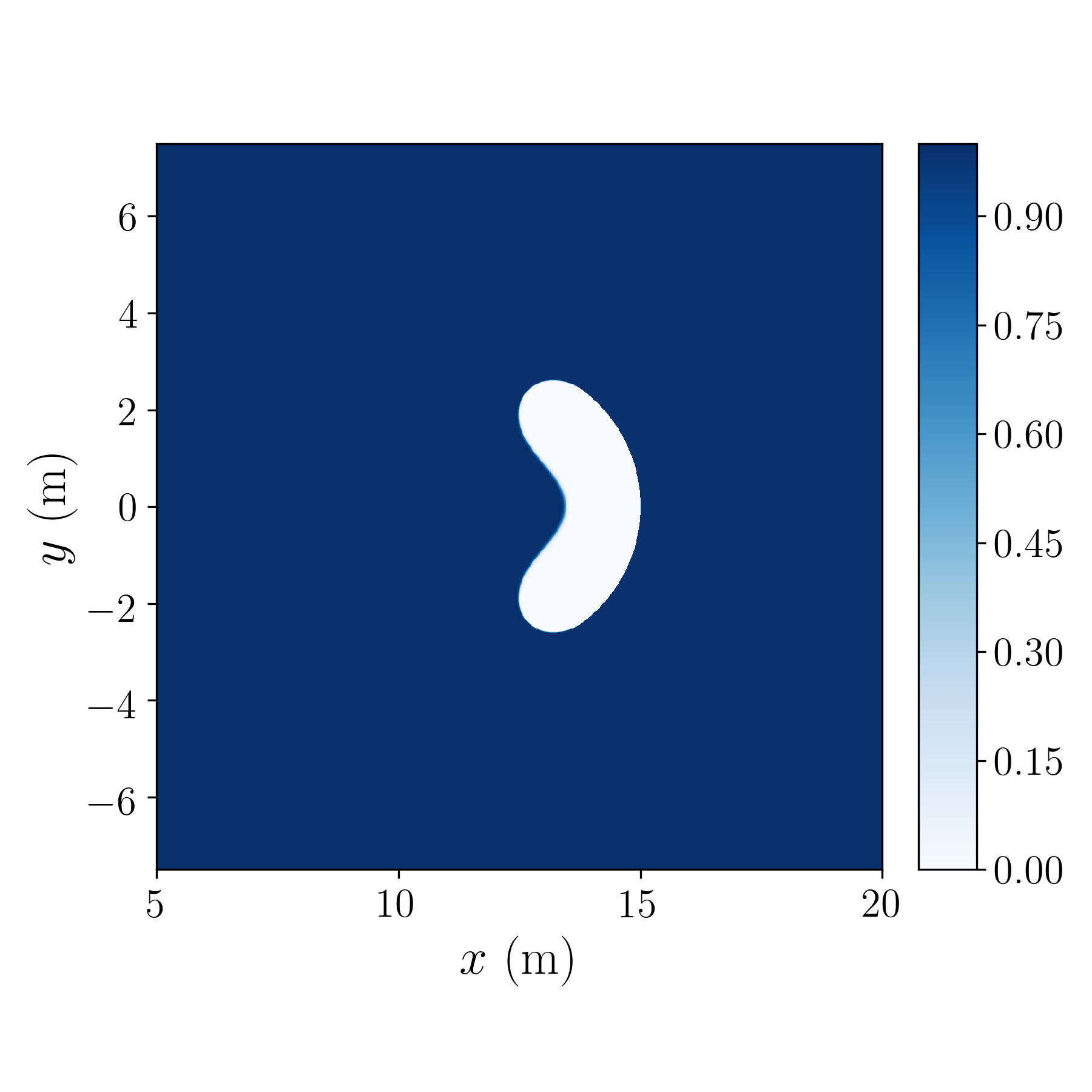}}
\subfigure[$t = 2.80\ \mu \mathrm{s}$]{%
\includegraphics[trim=0 1.5cm 0 1.0cm,clip,width=0.4\textwidth]{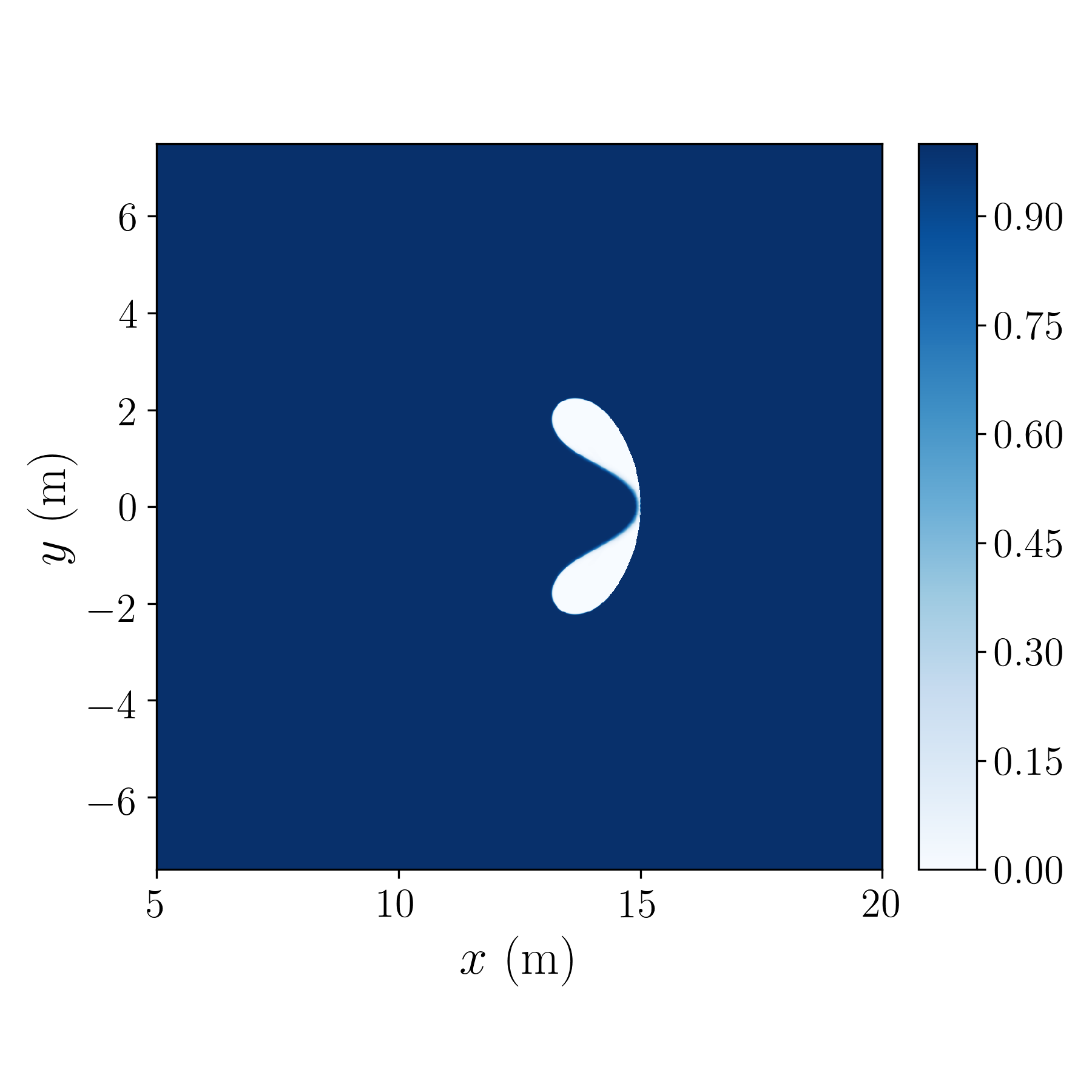}}
\subfigure[$t =3.44\ \mu \mathrm{s}$]{%
\includegraphics[trim=0 1.5cm 0 1.0cm,clip,width=0.4\textwidth]{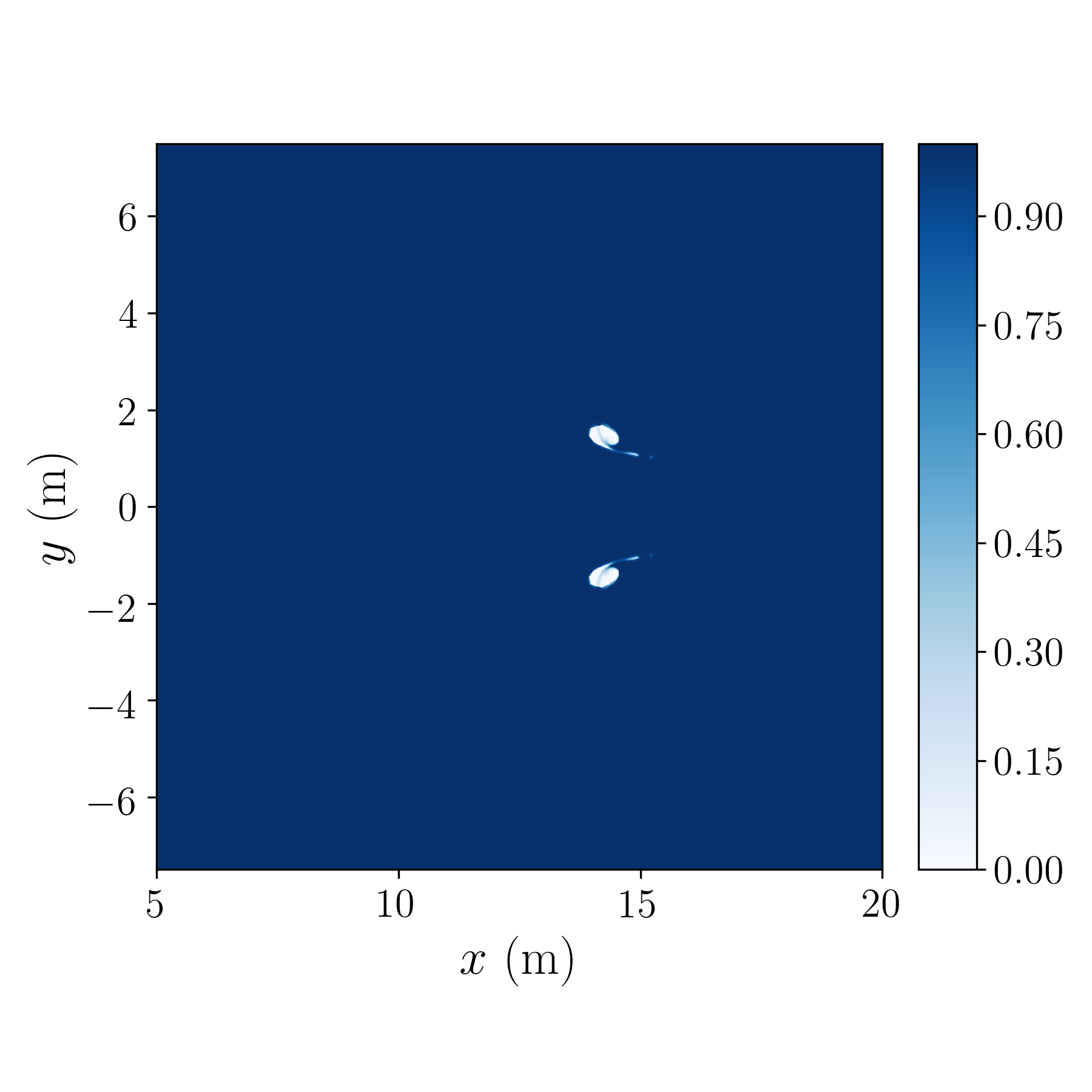}}
\caption{Volume fraction of water of the 2D 1.9 GPa shock bubble collapse problem on a mesh with resolution $2048 \times 1280$.}
\label{fig:2D_1_9_GPa_shock_bubble_collapse_problem_volume_fractions}
\end{figure}

\begin{figure}[!ht]
\centering
\subfigure[$t = 0.638\ \mu \mathrm{s}$]{%
\includegraphics[trim=0 1.5cm 0 1.0cm,clip,width=0.4\textwidth]{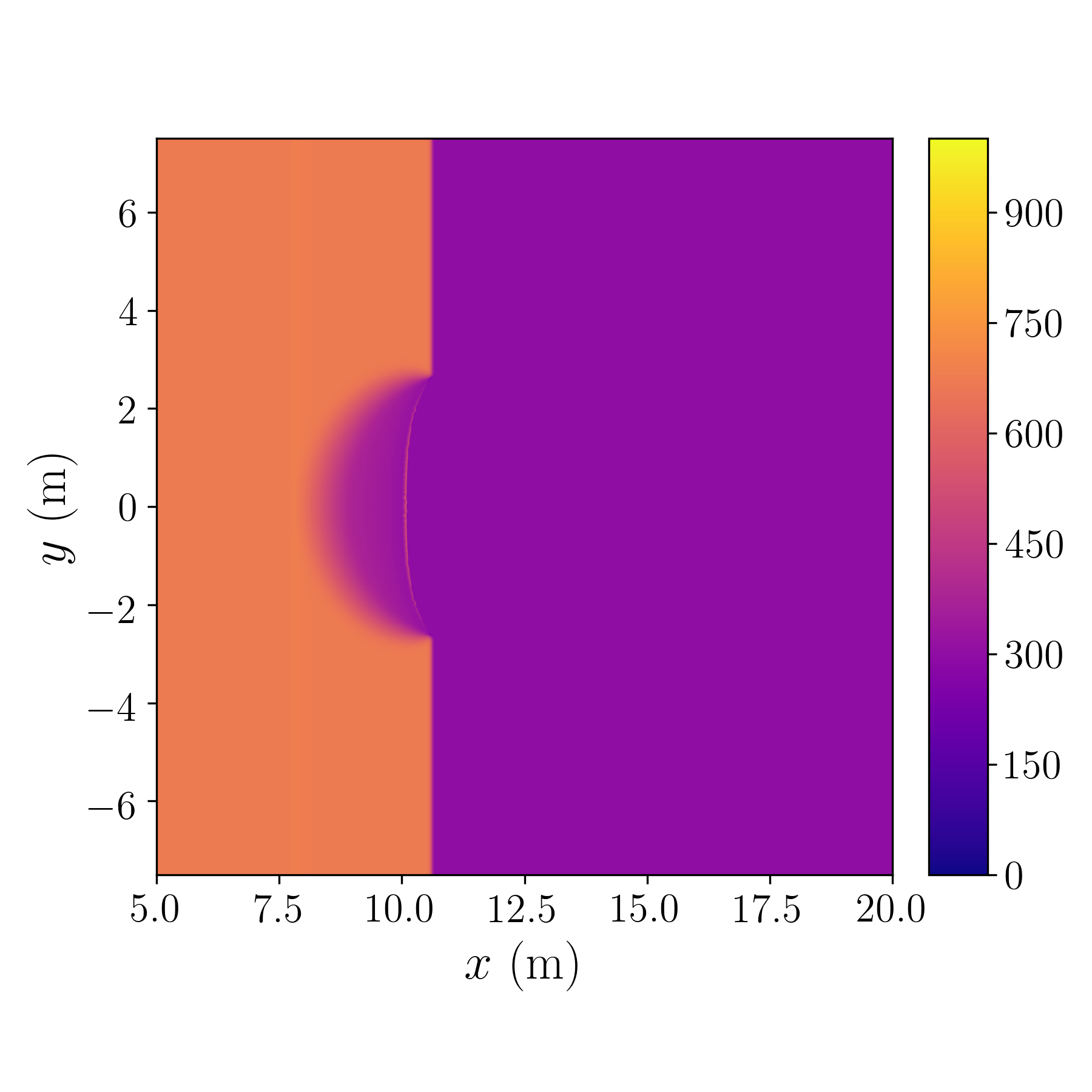}}
\subfigure[$t = 2.24\ \mu \mathrm{s}$]{%
\includegraphics[trim=0 1.5cm 0 1.0cm,clip,width=0.4\textwidth]{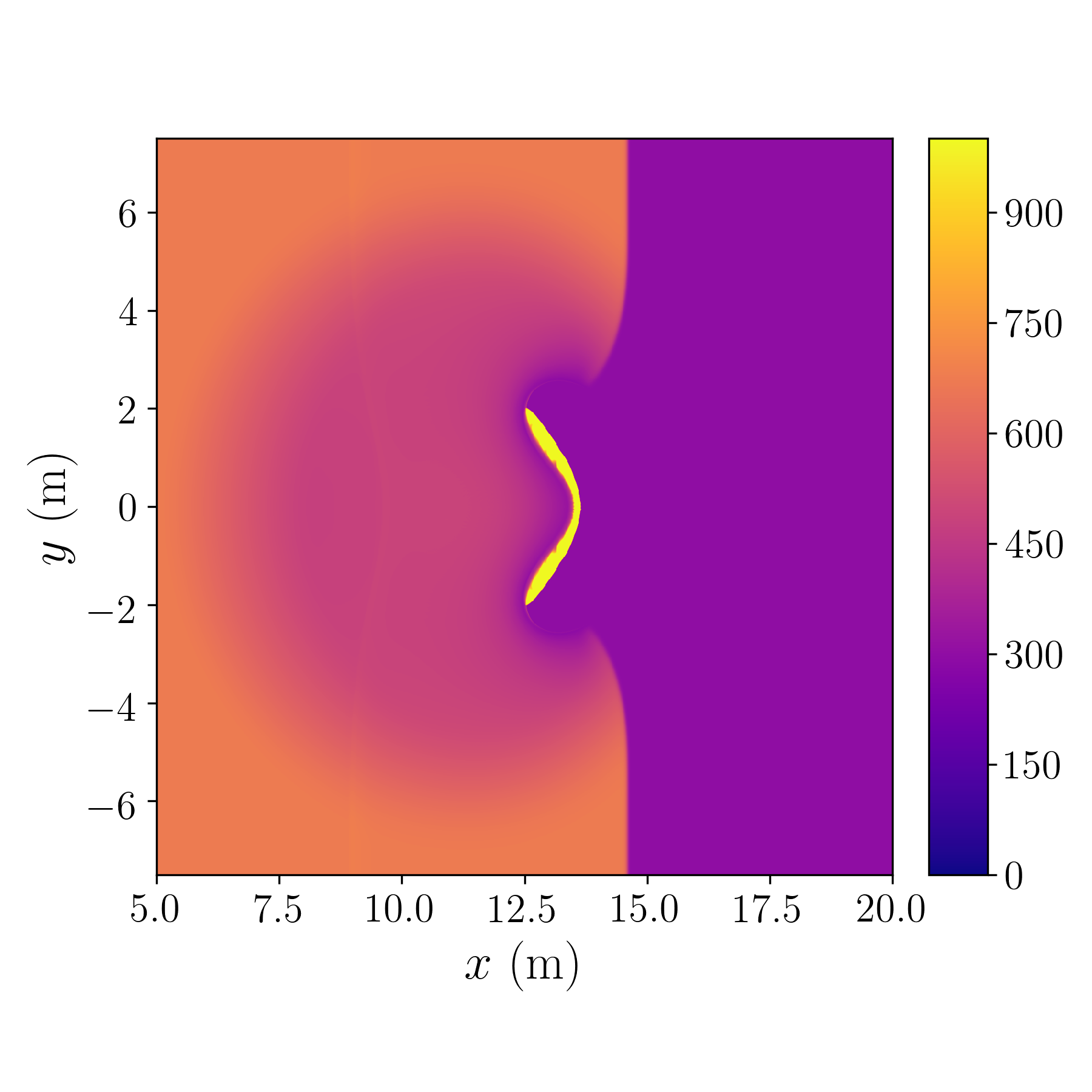}}
\subfigure[$t = 2.80\ \mu \mathrm{s}$]{%
\includegraphics[trim=0 1.5cm 0 1.0cm,clip,width=0.4\textwidth]{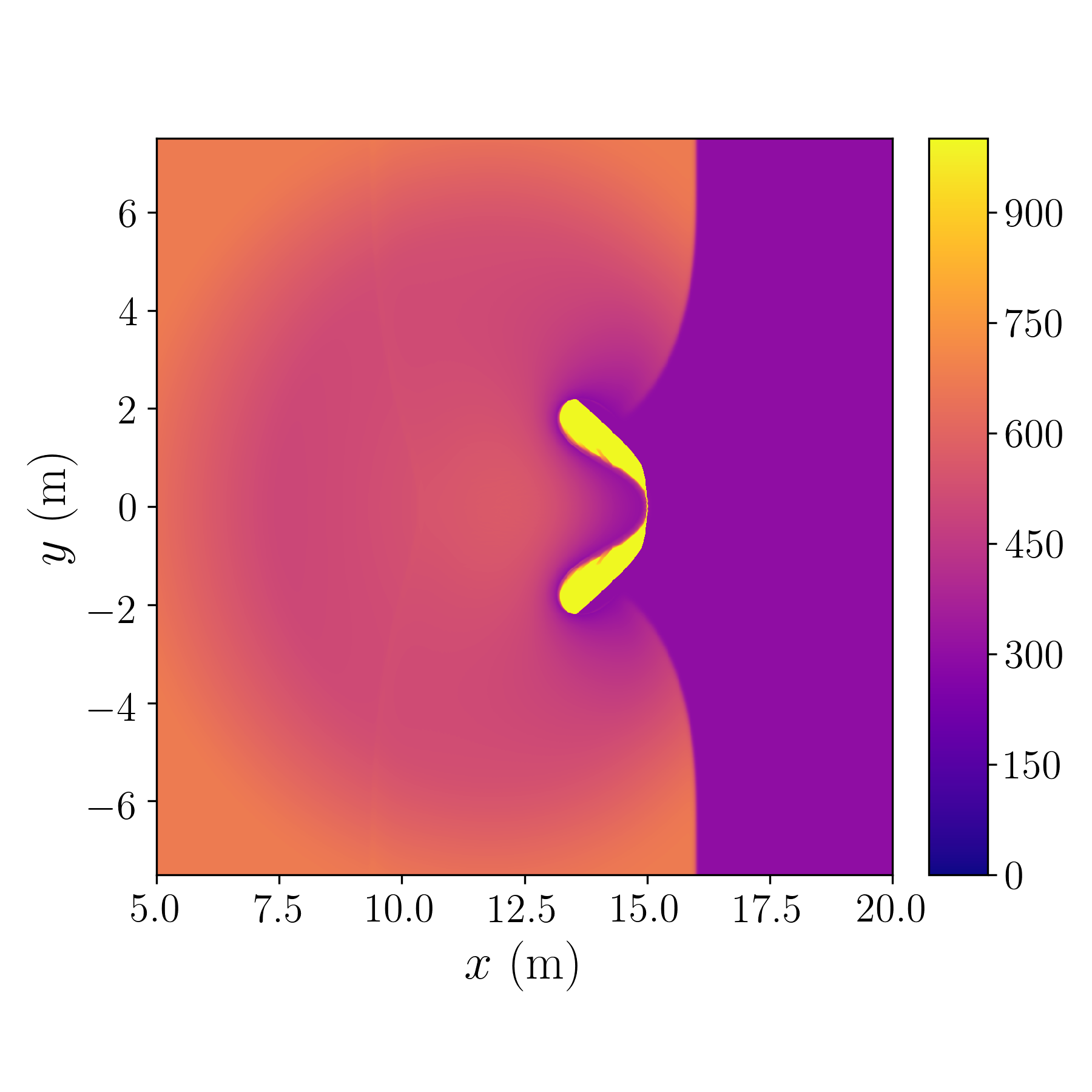}}
\subfigure[$t =3.44\ \mu \mathrm{s}$]{%
\includegraphics[trim=0 1.5cm 0 1.0cm,clip,width=0.4\textwidth]{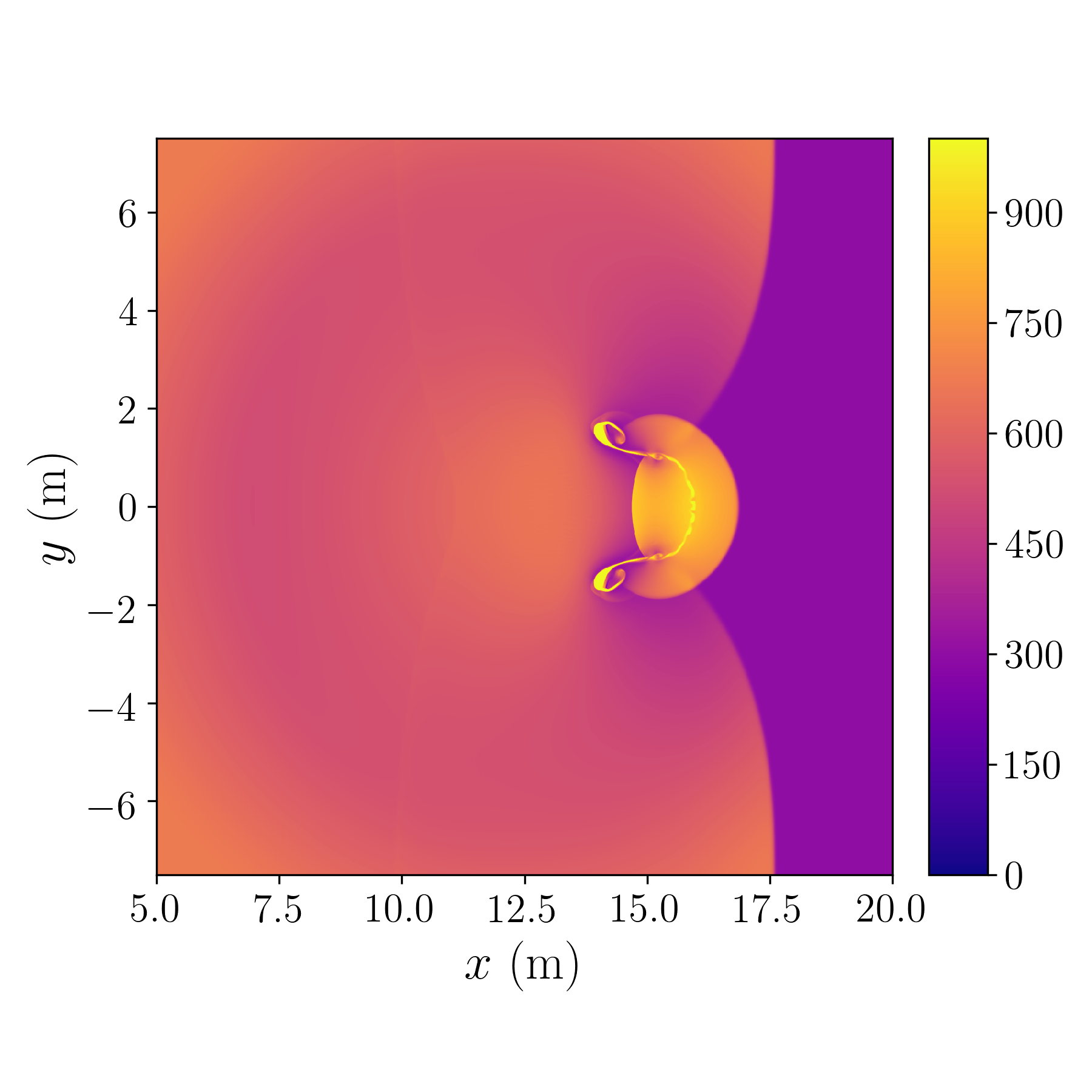}}
\caption{Temperature (K) of the 2D 1.9 GPa shock bubble collapse problem on a mesh with resolution $2048 \times 1280$.}
\label{fig:2D_1_9_GPa_shock_bubble_collapse_problem_temperature}
\end{figure}

The simulation starts at $t = -0.962436637988\ \mu \mathrm{s}$ such that $t = 0$ represents the time when the shock first hits the air bubble. Figures~\ref{fig:2D_1_9_GPa_shock_bubble_collapse_problem_schl}, \ref{fig:2D_1_9_GPa_shock_bubble_collapse_problem_volume_fractions}, and \ref{fig:2D_1_9_GPa_shock_bubble_collapse_problem_temperature} shows the numerical schlieren, volume fraction of the water, and the temperature at different times during the bubble collapse in the highest resolution case. From the figures, it can be seen that the shocks and interfaces are well captured with only a small amount of numerical diffusion.
In the simulation, the water jet reaches the right bubble wall at around $t = 2.80\ \mu\mathrm{s}$. This collapse time is comparable to the values reported by~\citet{ball2000shock} and \citet{lauer2012numerical}, where the collapse times were reported to be around $t = 3.1\ \mu\mathrm{s}$. In those works, instead of using the stiffened gas equation of state, the Tait equation of state is chosen for the water. Besides, Lagrangian and level set approaches are used respectively in the two works, instead of an Eulerian diffuse interface method. Note that observing the schlieren images of the cavity collapse shown in~\cite{bourne1992shock}, it is estimated in~\cite{ball2000shock} that the collapse time is significantly larger than $t = 1.5\ \mu\mathrm{s}$ but before $t = 3.5\ \mu\mathrm{s}$.

\begin{figure}
    \centering
    \includegraphics[width=0.6\textwidth]{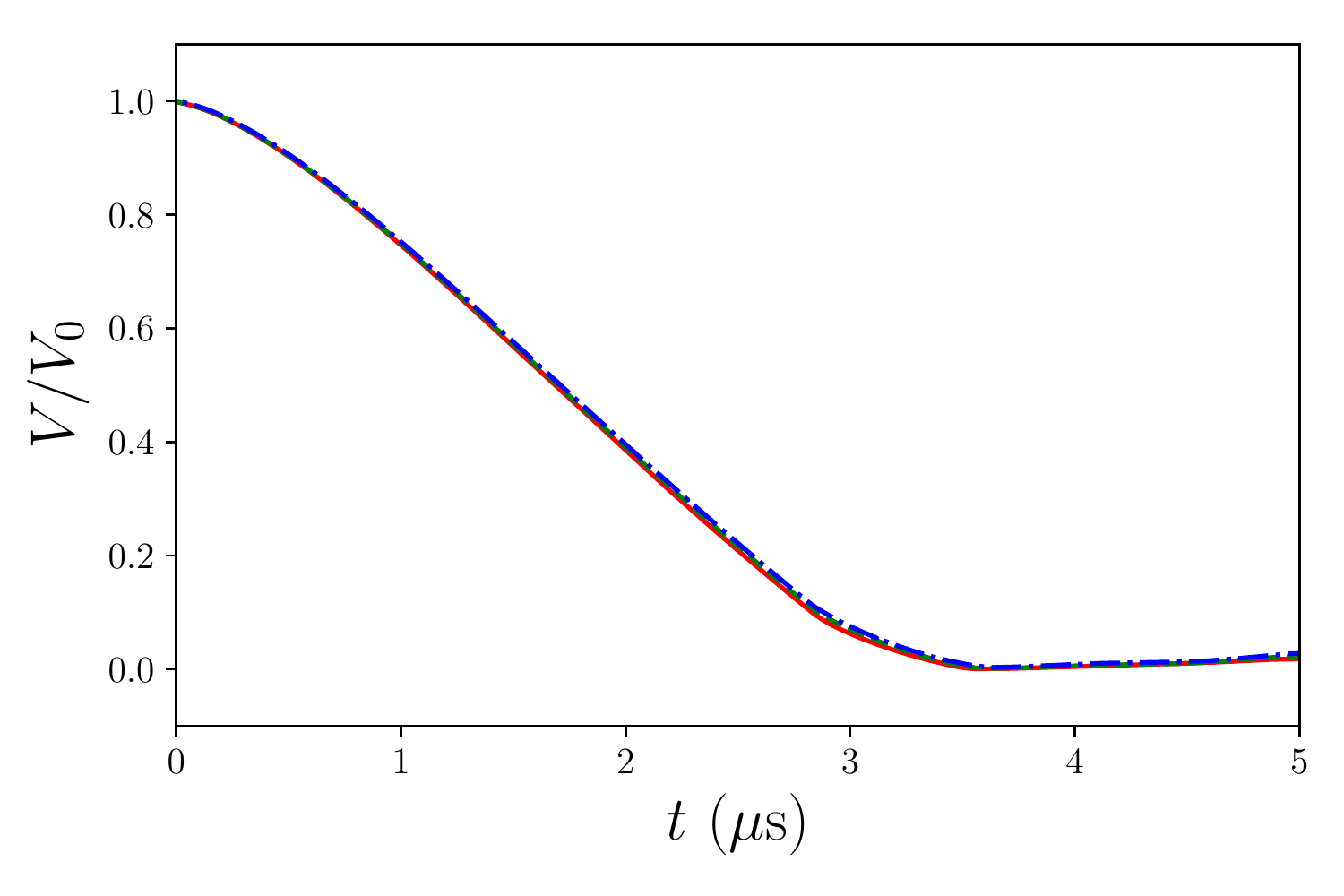}
    \caption{Time history of the normalized air cavity volume of the 2D 1.9 GPa shock bubble collapse problem.
    Red solid line: $512 \times 320$ mesh;
    green dashed line: $1024 \times 640$ mesh;
    blue dash-dotted line: $2048 \times 1280$ mesh.}
    \label{fig:2D_1_9_GPa_shock_bubble_collapse_problem_cavity_volume}
\end{figure}

Figure~\ref{fig:2D_1_9_GPa_shock_bubble_collapse_problem_cavity_volume} shows the time history of the air cavity volume $V$ normalized by the initial volume $V_0 = \pi R_b^2$ from different levels of mesh resolution. The quantity is well converged with the levels of mesh resolution chosen. As seen from the figure, there is a linear decay of the cavity volume from $t = 1.0\ \mu\mathrm{s}$ to $t = 2.8\ \mu\mathrm{s}$. This is consistent with the experimental observation of \citet{bourne1992shock} and computational results of \citet{ball2000shock} and \citet{lauer2012numerical}.

\subsection{Two-dimensional Mach 100 water jet problem}

This two-species test case is proposed in our previous work~\cite{wong2021positivity} and is adapted with the species in mechanical and thermal equilibria. In this intense 2D problem, a Mach 100 water jet enters a domain filled with essentially pure air with initial pressure and temperature fields uniformly at $101325\ \mathrm{Pa}$ and $298\ \mathrm{K}$ respectively.
The domain size is $\left[ 0, L \right] \times \left[ -0.25L, 0.25L \right]$, where $L = 1\ \mathrm{m}$ is chosen. The initial conditions of the ambient air are given by table~\ref{table:IC_2D_Mach_100_water_jet}. Constant extrapolation is used at top, bottom, and right boundaries. The left boundary is described by Dirichlet boundary conditions given by table~\ref{table:BC_2D_Mach_100_water_jet}. The speed of the jet is $1.5\mathrm{e}{5}\ \mathrm{m\ s^{-1}}$, which is around Mach 100 with respect to the sound speed of the water jet. The robustness of the fractional algorithm with the five-equation model by Allaire et al. and the thermal relaxation is examined with a simulation performed on a $1024 \times 512$ mesh using the PP-WCNS-IS method. Note that numerical failures are experienced when the positivity-preserving limiters are turned off.

\begin{table}[!ht]
\small
  \begin{center}
    \begin{tabular}{@{}cccccc@{}}\toprule
    \addstackgap{\stackanchor{$\alpha_1 \rho_1$}{$(\mathrm{kg\ m^{-3}})$}} &
    \stackanchor{$\alpha_2 \rho_2$}{$(\mathrm{kg\ m^{-3}})$} &
    \stackanchor{$u$}{$(\mathrm{m\ s^{-1}})$} &
    \stackanchor{$v$}{$(\mathrm{m\ s^{-1}})$} &
    \stackanchor{$p$}{$(\mathrm{Pa})$} &
    $\alpha_1$ \\ \midrule
    $1.0227724412751677\mathrm{e}{-5}$ & 1.1817862094653702 & 0 & 0 & $1.01325\mathrm{e}{5}$ & $1.0\mathrm{e}{-8}$ \\
    \bottomrule
    \end{tabular}
  \end{center}
  \caption{Initial conditions of the 2D Mach 100 water jet problem.}
  \label{table:IC_2D_Mach_100_water_jet}
\end{table}

\begin{figure}[!ht]
\centering
\subfigure[$t = 1\ \mu\mathrm{s}$, $\alpha_1$]{%
\includegraphics[width=0.45\textwidth]{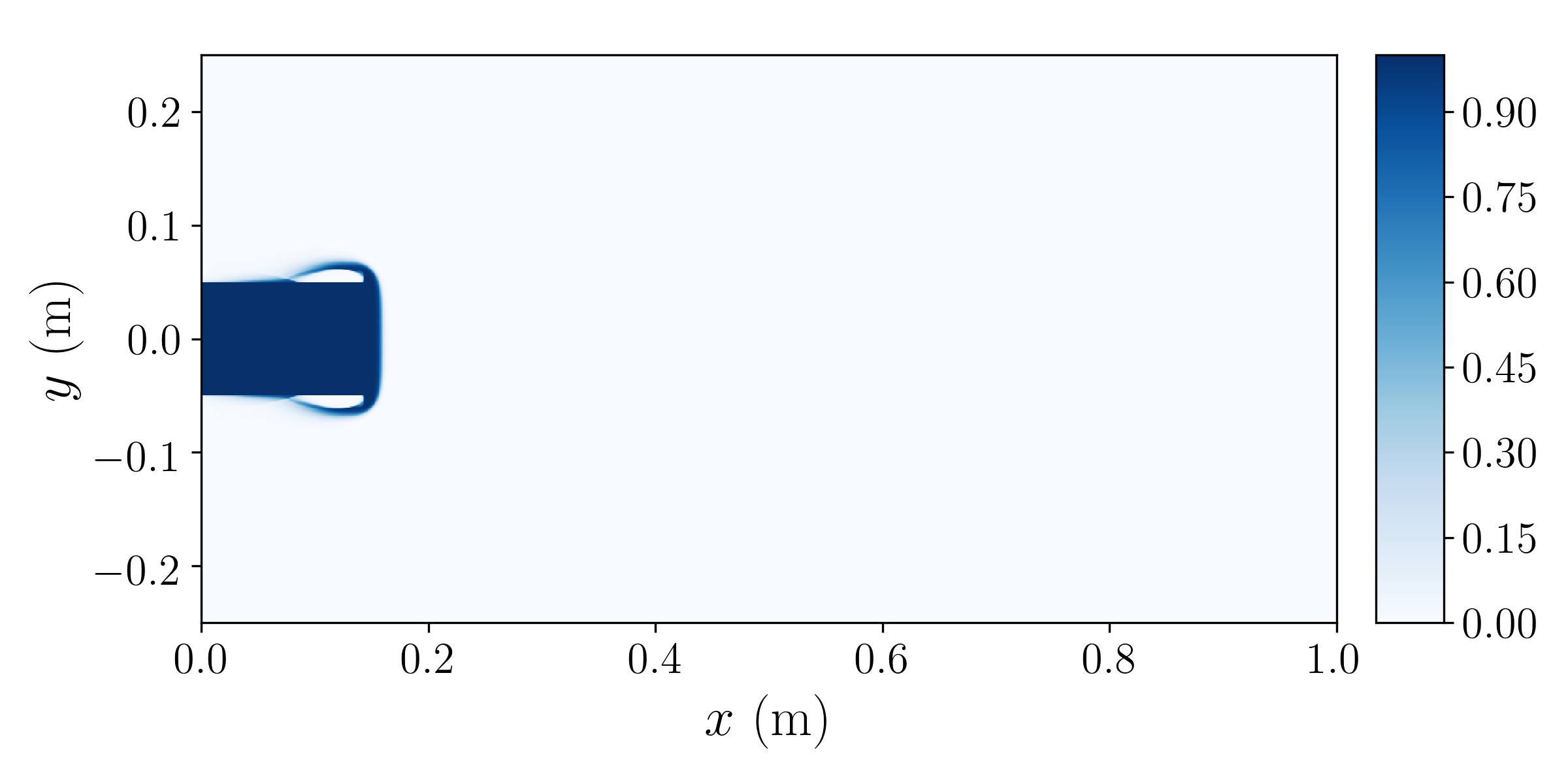}
\label{fig:plot_2D_Mach_100_water_jet_vol_frac_t1}}
\subfigure[$t = 1\ \mu\mathrm{s}$, $\left( \left| \nabla \rho \right| / \rho \right)$]{%
\includegraphics[width=0.45\textwidth]{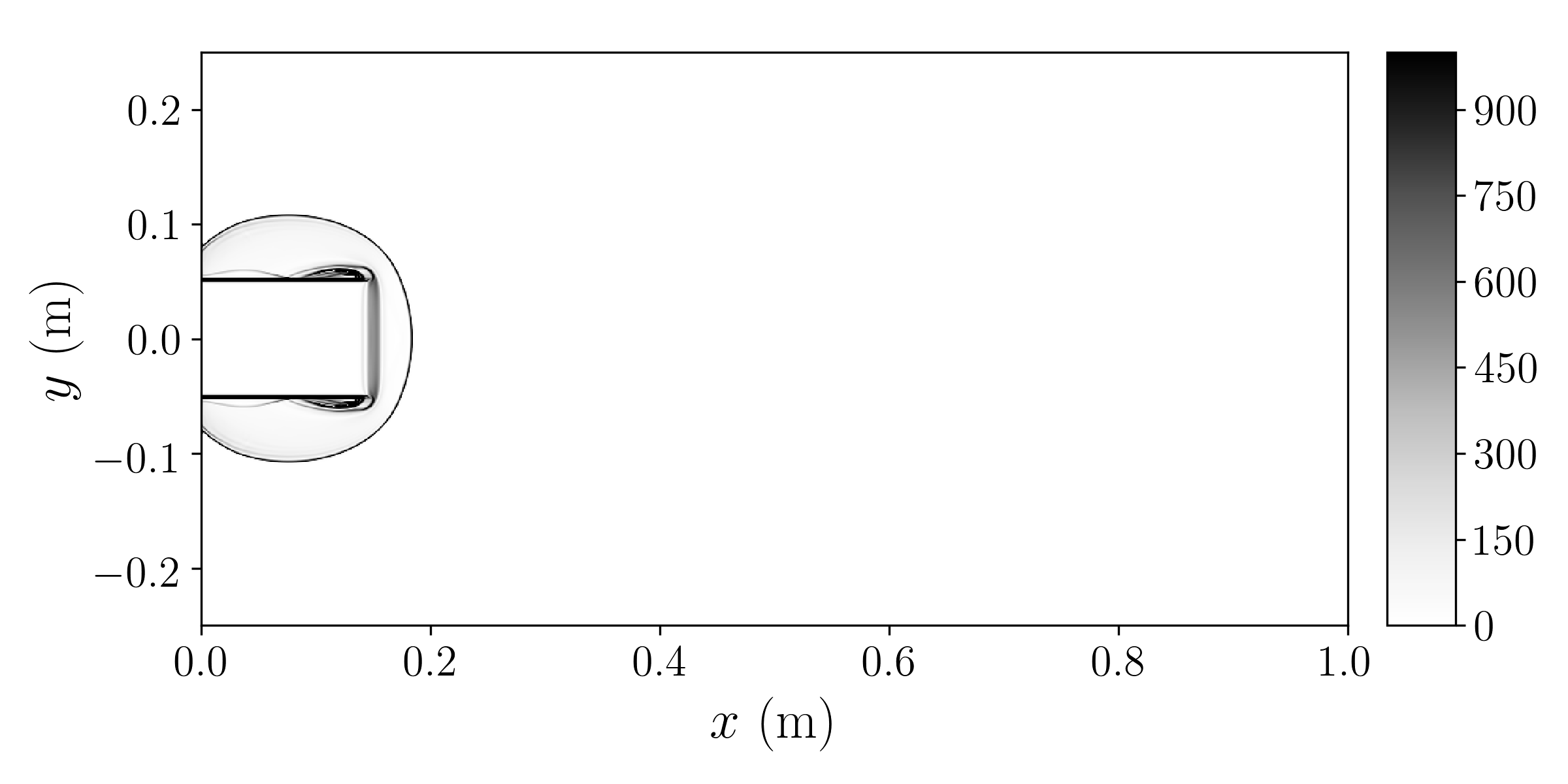}
\label{fig:plot_2D_Mach_100_water_jet_schl_t1}}
\subfigure[$t = 2\ \mu\mathrm{s}$, $\alpha_1$]{%
\includegraphics[width=0.45\textwidth]{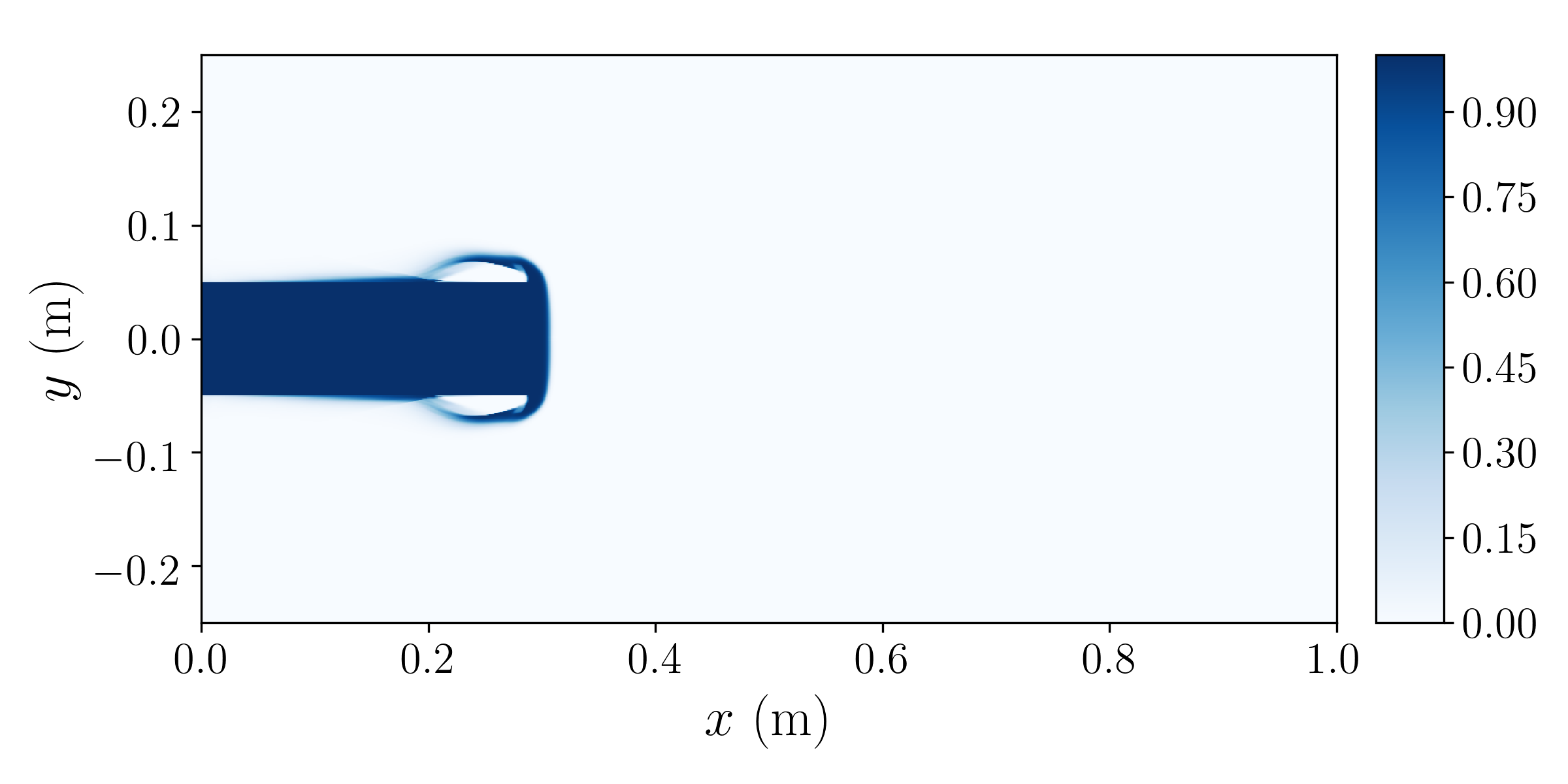}
\label{fig:plot_2D_Mach_100_water_jet_vol_frac_t2}}
\subfigure[$t = 2\ \mu\mathrm{s}$, $\left( \left| \nabla \rho \right| / \rho \right)$]{%
\includegraphics[width=0.45\textwidth]{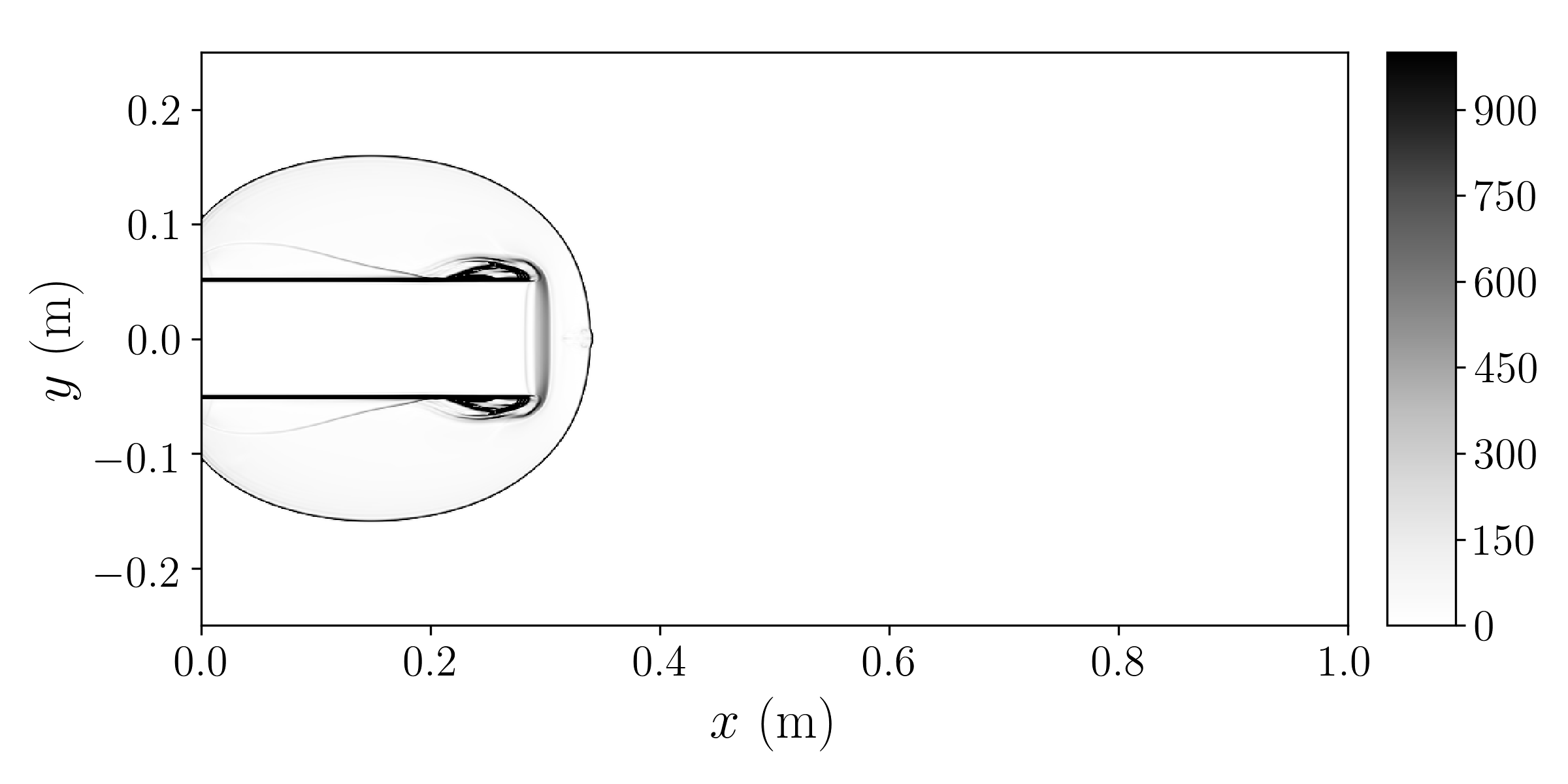}
\label{fig:plot_2D_Mach_100_water_jet_schl_t2}}
\subfigure[$t = 4\ \mu\mathrm{s}$, $\alpha_1$]{%
\includegraphics[width=0.45\textwidth]{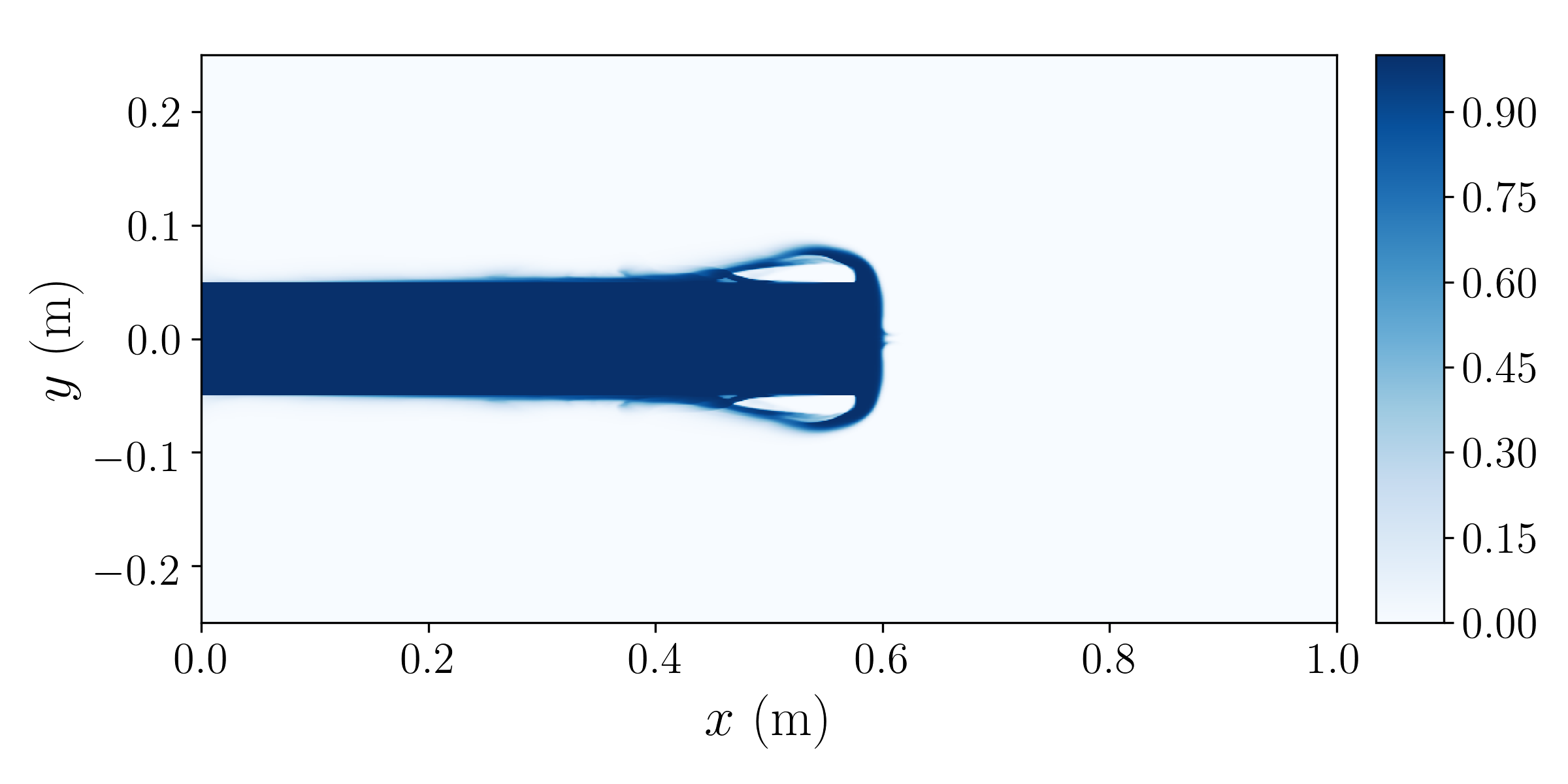}
\label{fig:plot_2D_Mach_100_water_jet_vol_frac_t3}}
\subfigure[$t = 4\ \mu\mathrm{s}$, $\left( \left| \nabla \rho \right| / \rho \right)$]{%
\includegraphics[width=0.45\textwidth]{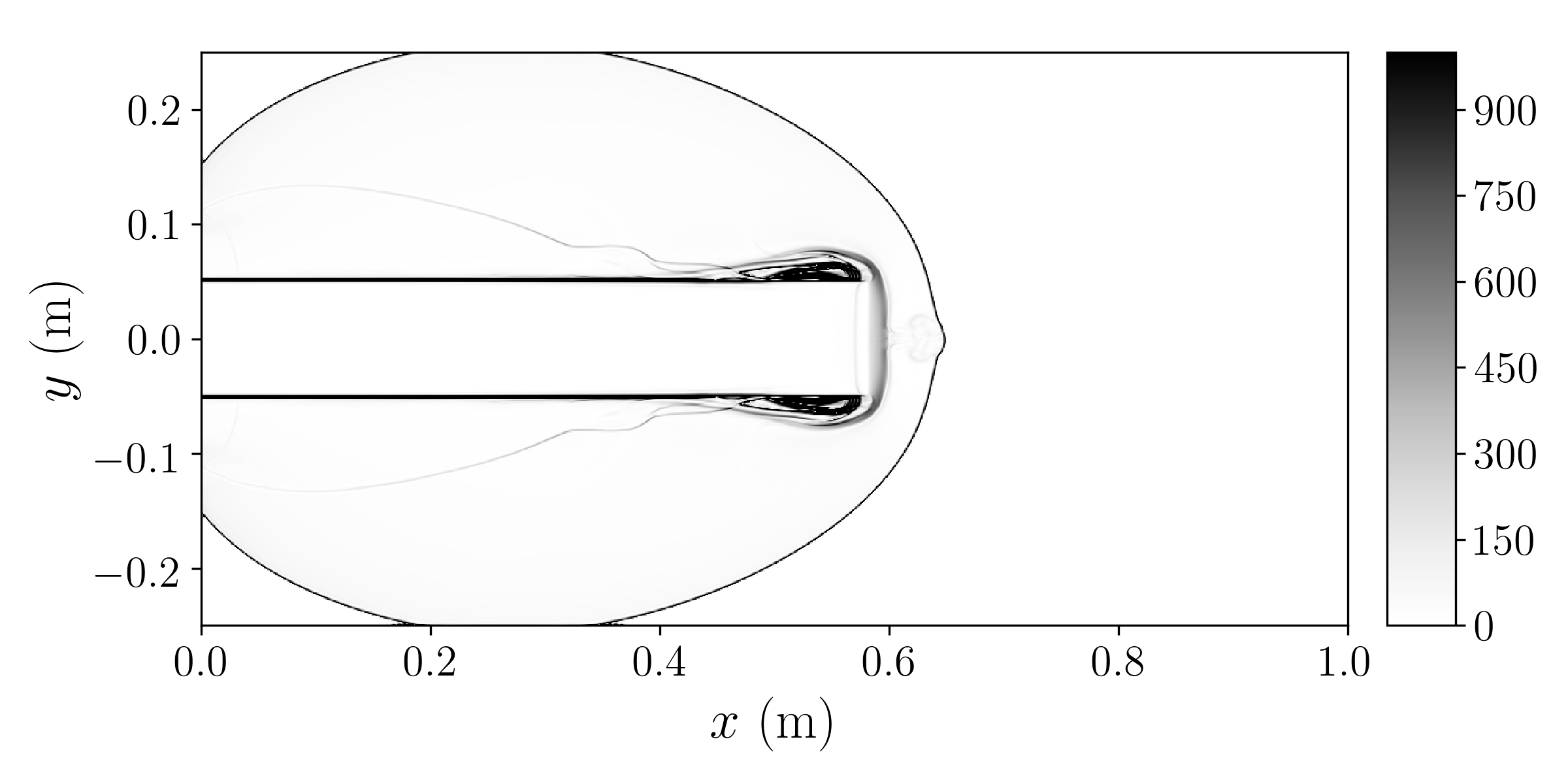}
\label{fig:plot_2D_Mach_100_water_jet_schl_t3}}
\subfigure[$t = 6\ \mu\mathrm{s}$, $\alpha_1$]{%
\includegraphics[width=0.45\textwidth]{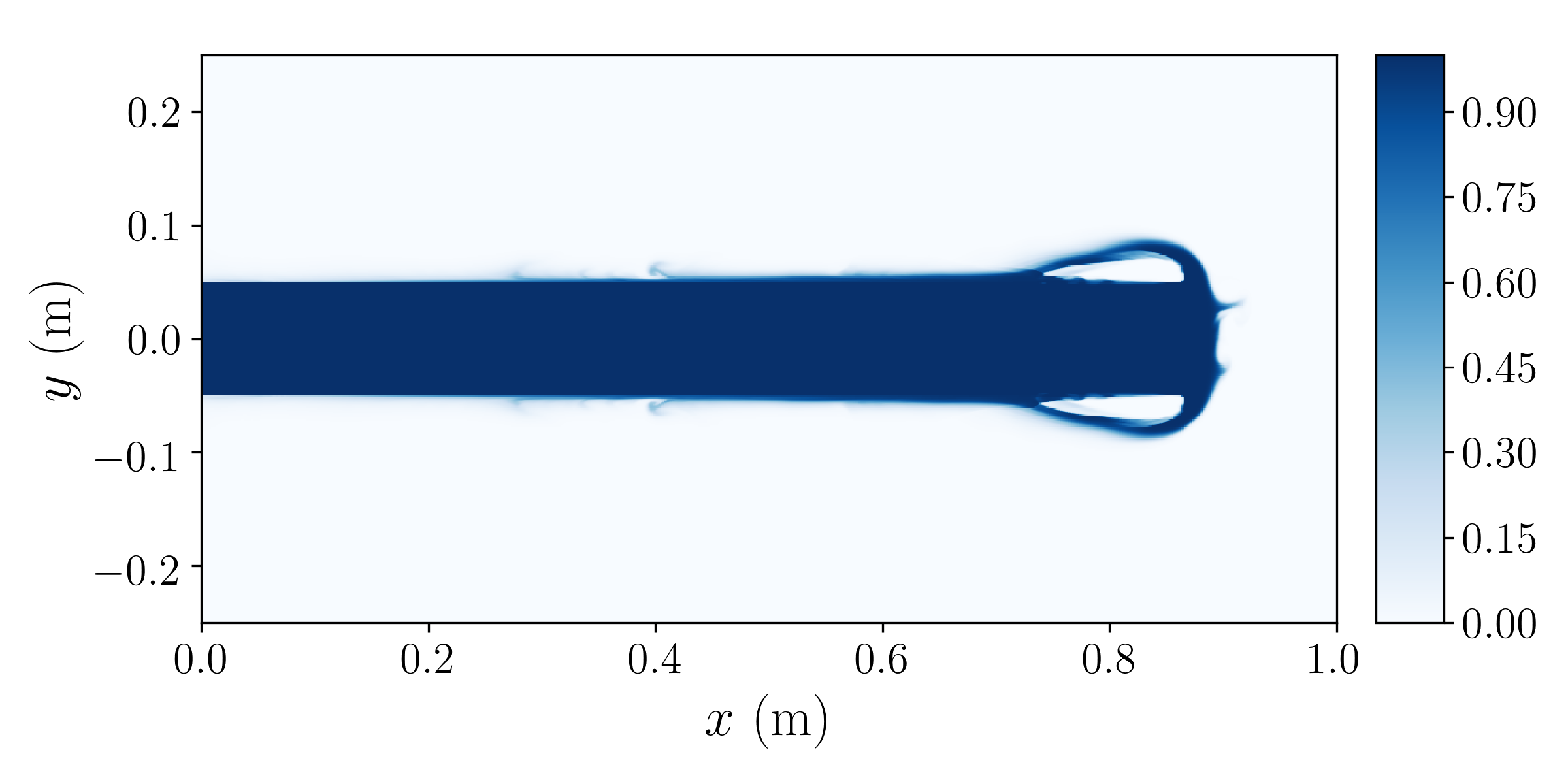}
\label{fig:plot_2D_Mach_100_water_jet_vol_frac_t4}}
\subfigure[$t = 6\ \mu\mathrm{s}$, $\left( \left| \nabla \rho \right| / \rho \right)$]{%
\includegraphics[width=0.45\textwidth]{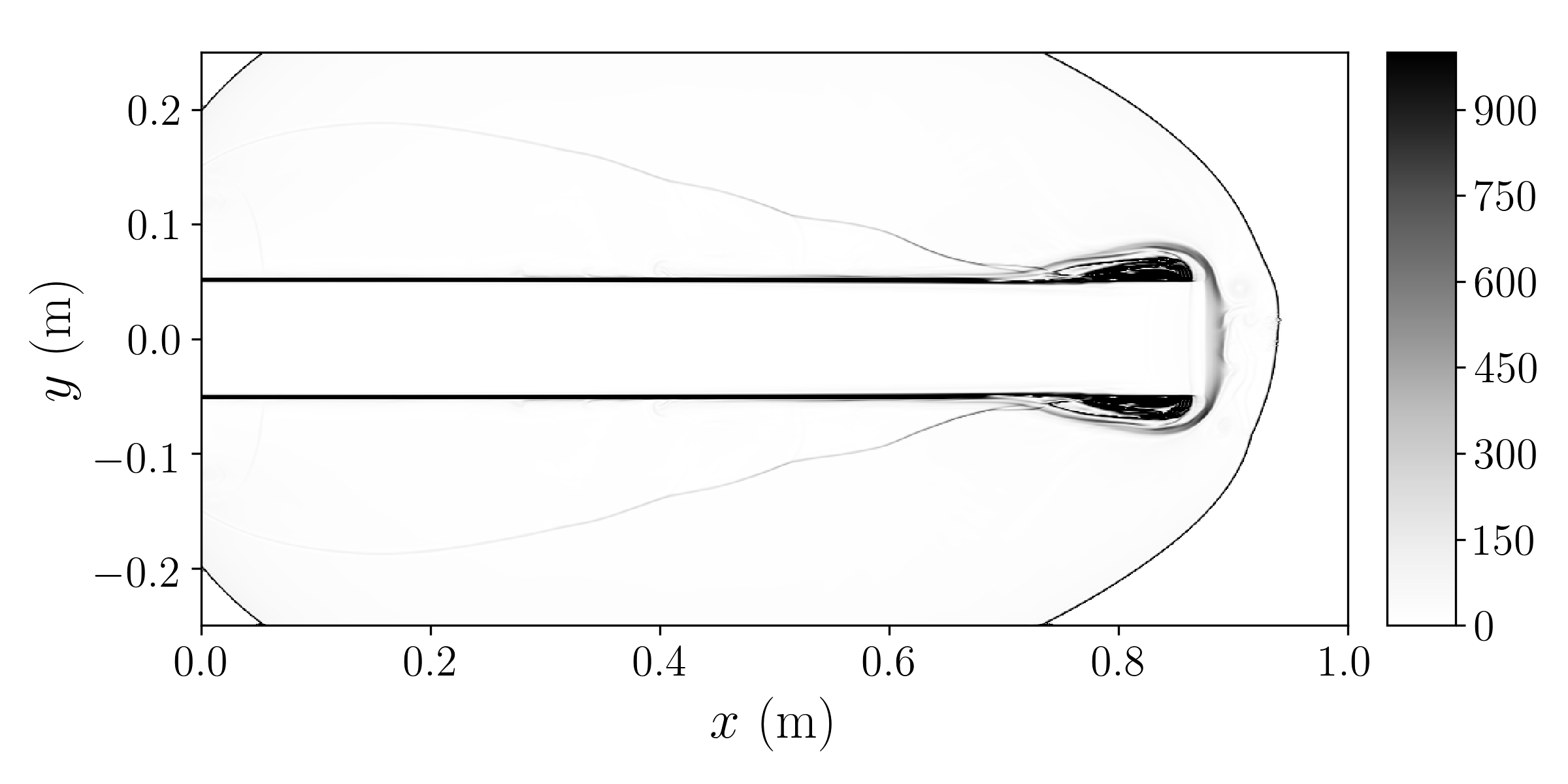}
\label{fig:plot_2D_Mach_100_water_jet_schl_t4}}
\caption{Volume fraction of water and numerical schlieren ($\left( \left| \nabla \rho \right| / \rho \right)$) of the 2D Mach 100 water jet problem. Left column: volume fraction of water; right column: numerical schlieren.}
\label{fig:plot_2D_Mach_100_water_jet}
\end{figure}

\begin{figure}[!ht]
\centering
\subfigure[$t = 1\ \mu\mathrm{s}$, five-equation model]{%
\includegraphics[width=0.45\textwidth]{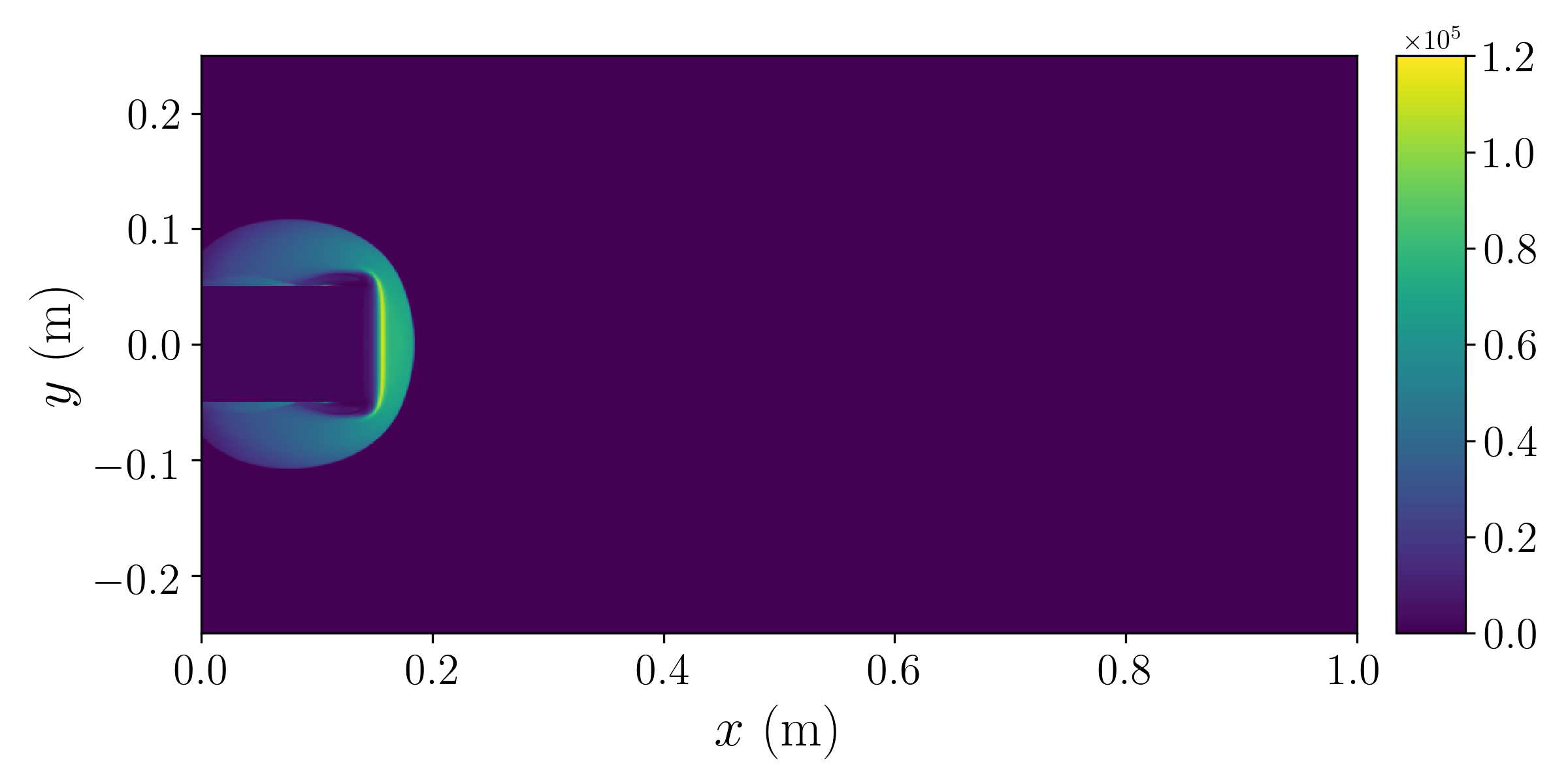}
\label{fig:compare_2D_Mach_100_water_jet_sos_t1_5_eqn}}
\subfigure[$t = 1\ \mu\mathrm{s}$, four-equation model]{%
\includegraphics[width=0.45\textwidth]{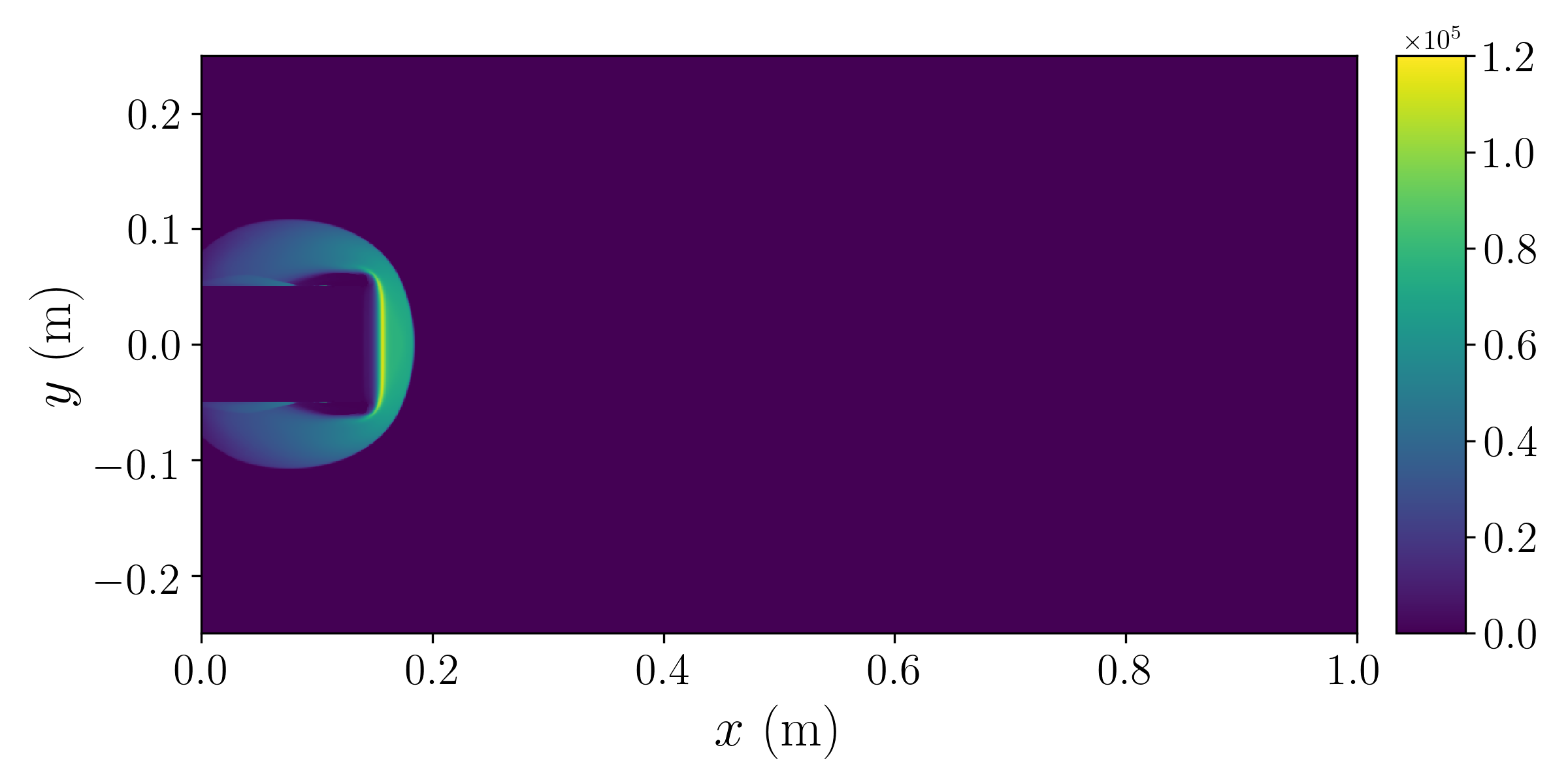}
\label{fig:compare_2D_Mach_100_water_jet_sos_t1_4_eqn}}
\subfigure[$t = 2\ \mu\mathrm{s}$, five-equation model]{%
\includegraphics[width=0.45\textwidth]{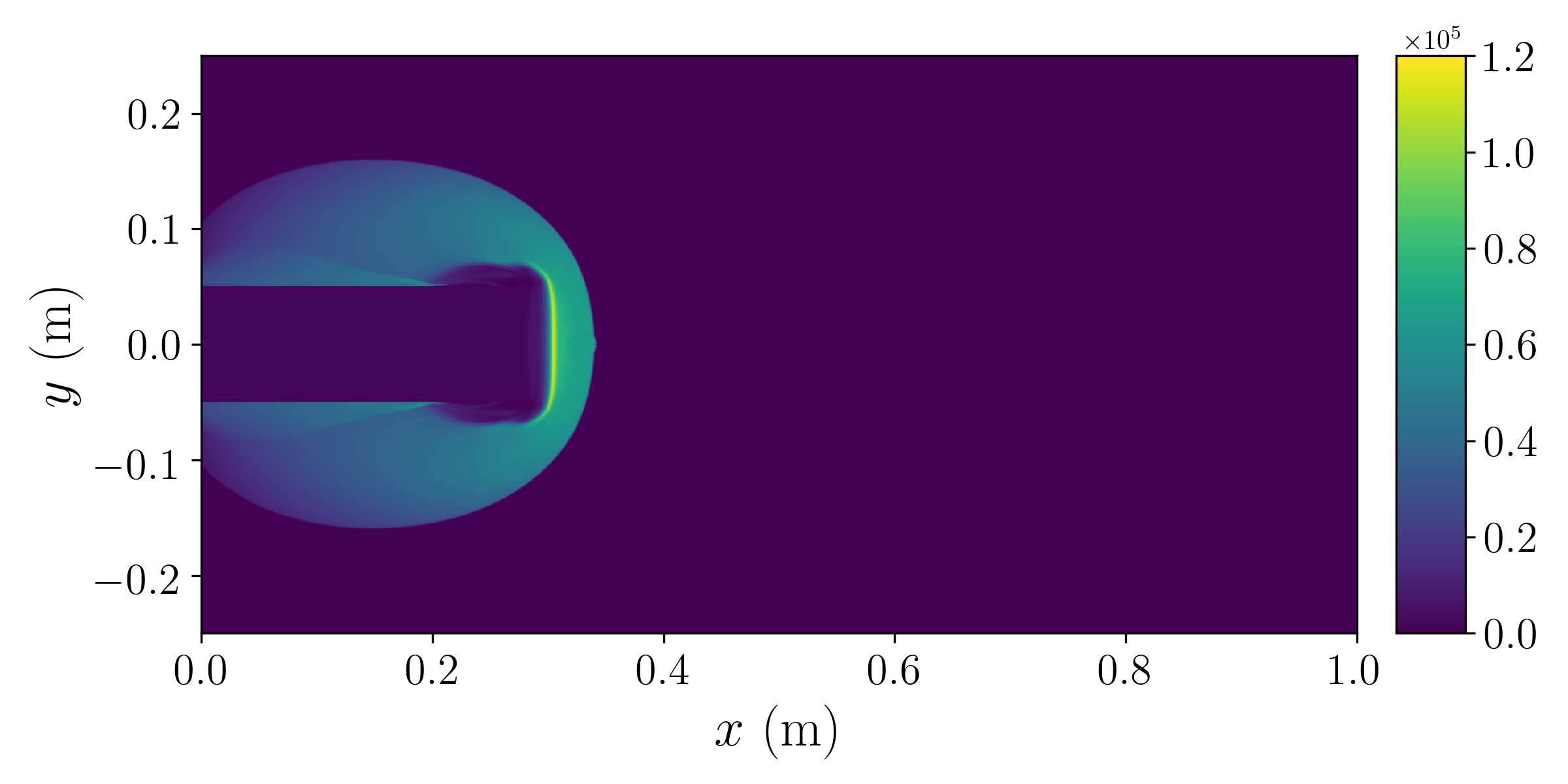}
\label{fig:compare_2D_Mach_100_water_jet_sos_t2_5_eqn}}
\subfigure[$t = 2\ \mu\mathrm{s}$, four-equation model]{%
\includegraphics[width=0.45\textwidth]{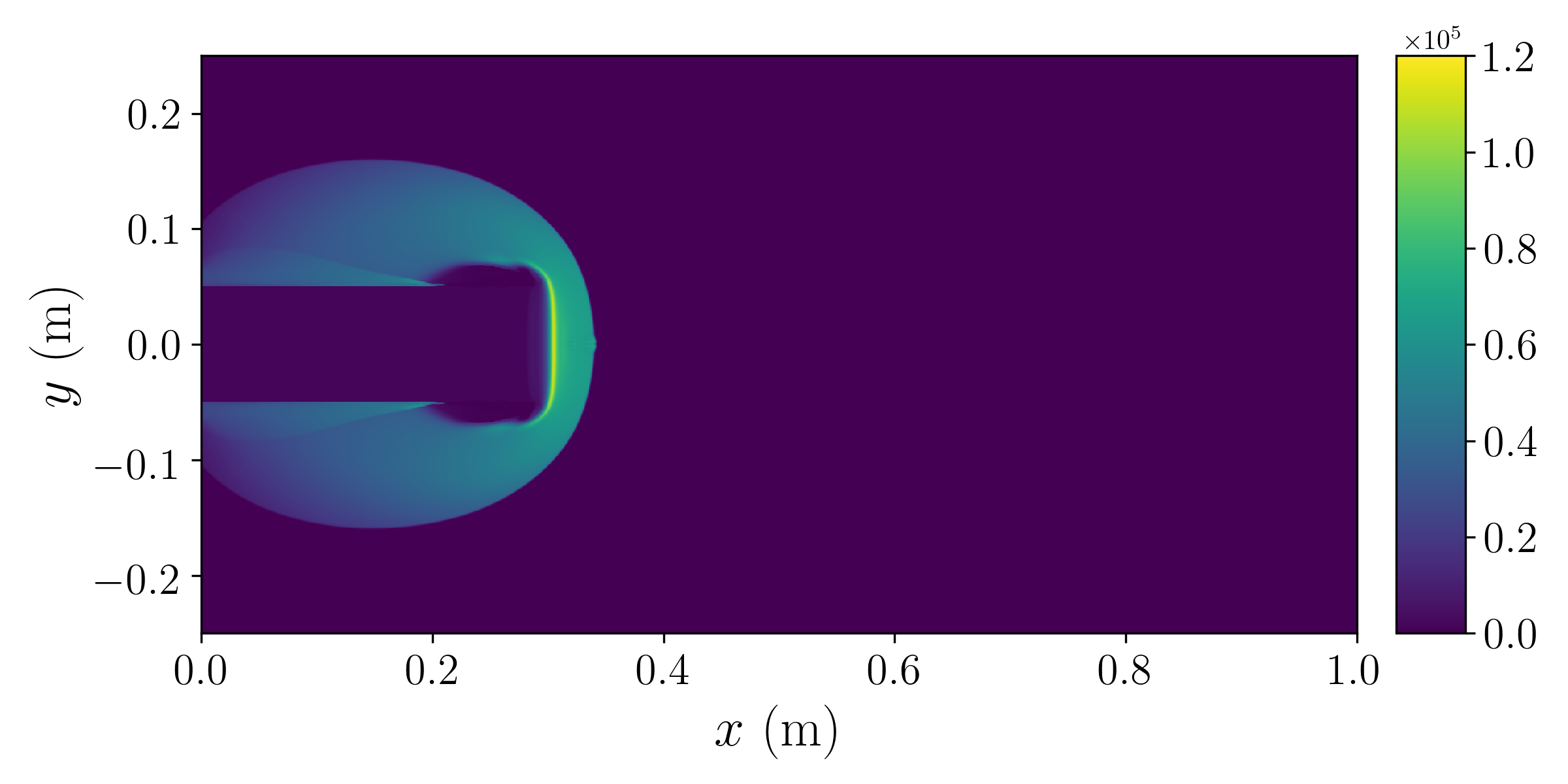}
\label{fig:compare_2D_Mach_100_water_jet_sos_t2_4_eqn}}
\subfigure[$t = 4\ \mu\mathrm{s}$, five-equation model]{%
\includegraphics[width=0.45\textwidth]{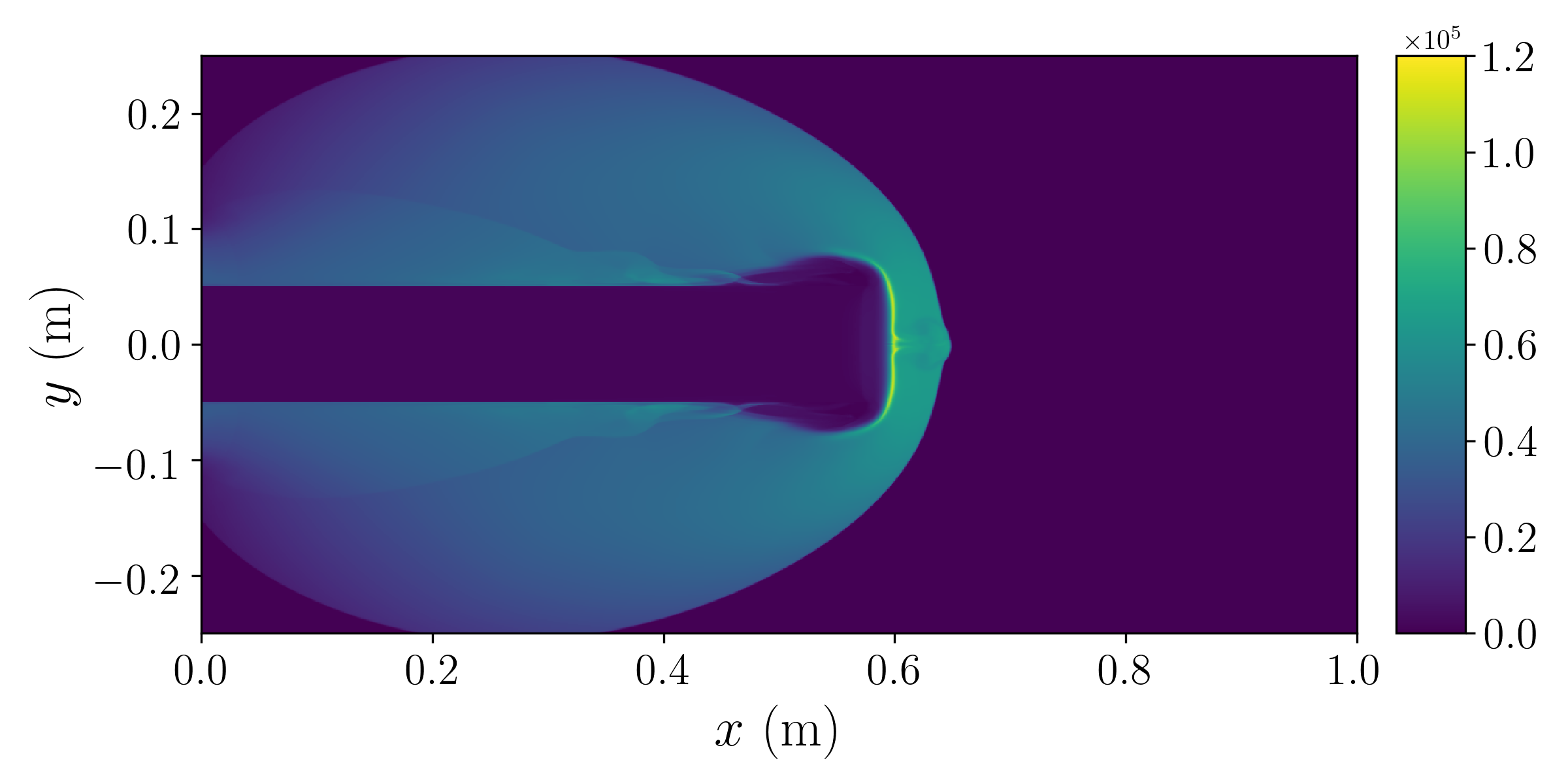}
\label{fig:compare_2D_Mach_100_water_jet_sos_t3_5_eqn}}
\subfigure[$t = 4\ \mu\mathrm{s}$, four-equation model]{%
\includegraphics[width=0.45\textwidth]{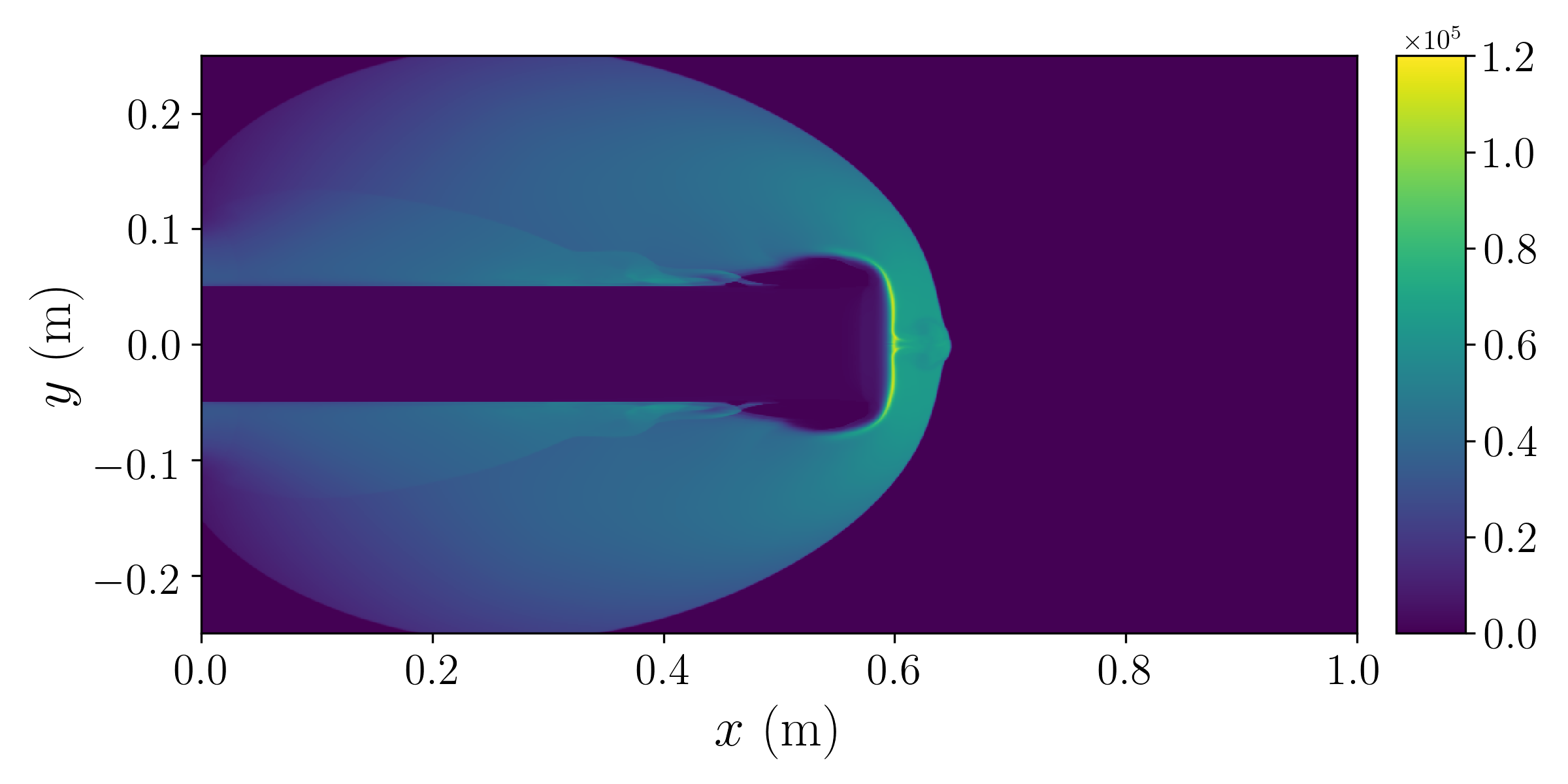}
\label{fig:compare_2D_Mach_100_water_jet_sos_t3_4_eqn}}
\subfigure[$t = 6\ \mu\mathrm{s}$, five-equation model]{%
\includegraphics[width=0.45\textwidth]{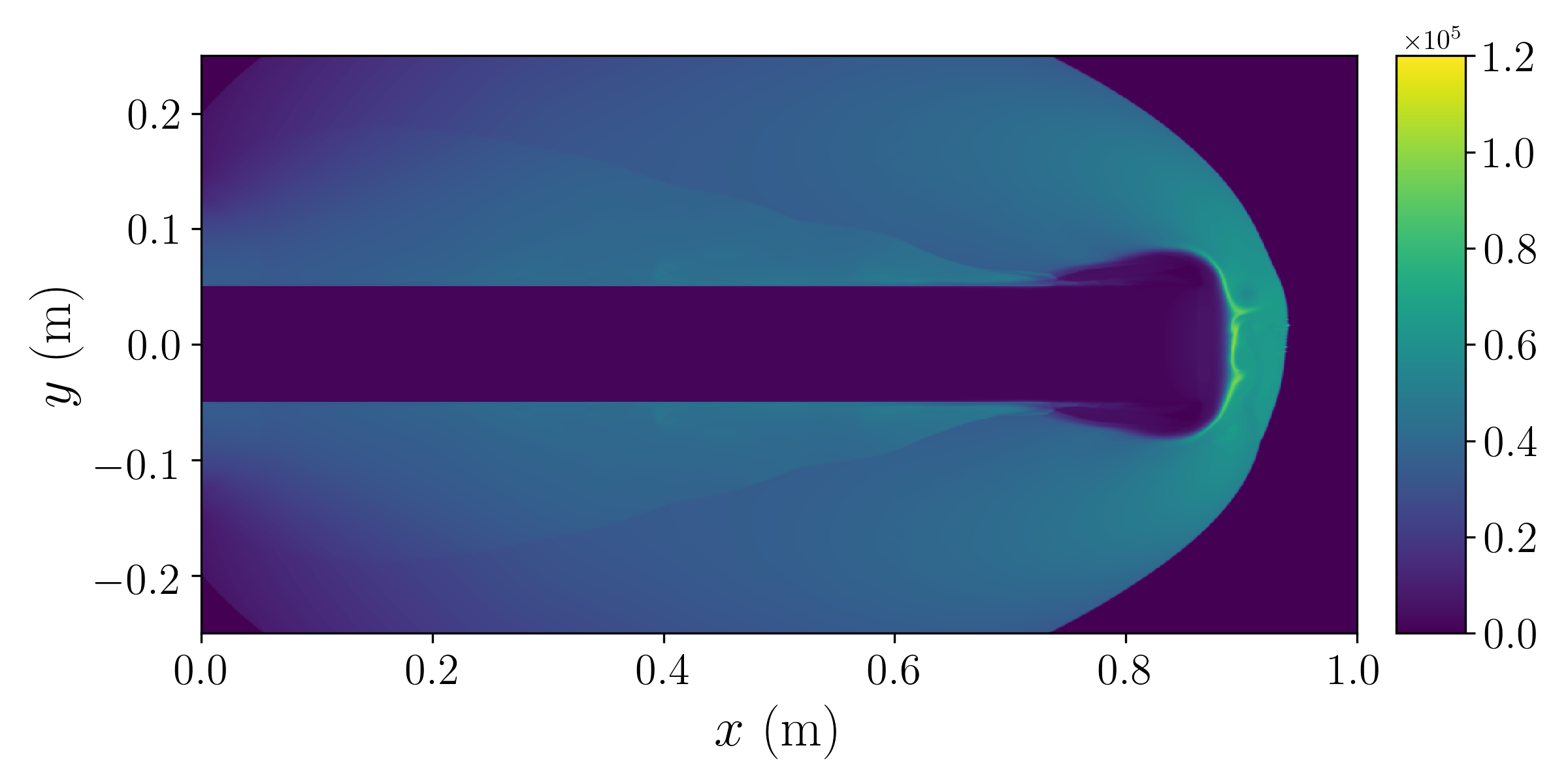}
\label{fig:compare_2D_Mach_100_water_jet_sos_t4_5_eqn}}
\subfigure[$t = 6\ \mu\mathrm{s}$, four-equation model]{%
\includegraphics[width=0.45\textwidth]{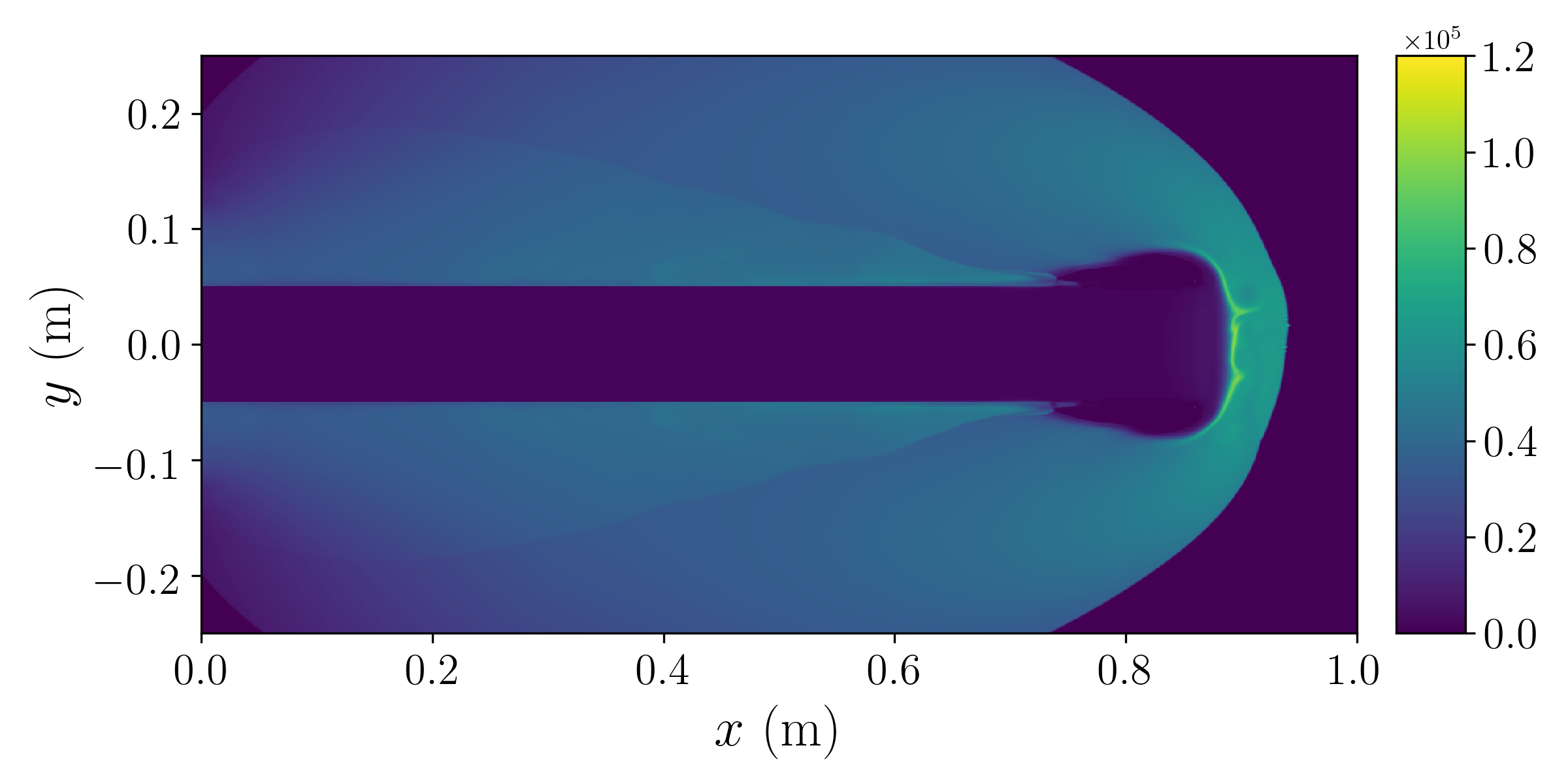}
\label{fig:compare_2D_Mach_100_water_jet_sos_t4_4_eqn}}
\caption{Speed of sound ($\mathrm{m\ s^{-1}}$) of the 2D Mach 100 water jet problem. Left column: sound speed of five-equation model by Allaire et al.; right column: sound speed of four-equation HRM.}
\label{fig:compare_2D_Mach_100_water_jet_sos}
\end{figure}

The volume fraction of water and numerical schlieren of this test case at different times are shown in figure~\ref{fig:plot_2D_Mach_100_water_jet}. From the volume fraction plots, it can be seen that the air-water interfaces are accurately captured with only very few grid cells across them. While numerical dissipation is added at the interfaces, some roll-ups can still be produced along the sides of the jet at late times by the Kelvin--Helmholtz instability. Minimally diffused interfaces can also be observed in the numerical schlieren. In those schlieren, a strong bow shock ahead of the water jet can be visualized as the jet penetrates into the ambient air. Similar to the interfaces, the shock is also well captured by the numeral method at this mesh resolution with very few grid cells. The speed of sound computed using the formulae for the five-equation model and the four-equation HRM are compared in figure~\ref{fig:compare_2D_Mach_100_water_jet_sos}. While the correct sound speed given by the fractional algorithm should be the one for the the four-equation HRM, the plots only show small difference for the two definitions of sound speed as the interfaces are reasonably resolved in this test. The discrepancy mainly appears in the regions around the two entrained air bubbles, which are near the front of the jet.

\begin{table}[!ht]
\small
  \begin{center}
    \begin{tabular}{@{}c | cccccc@{}} \toprule
     &
    \addstackgap{\stackanchor{$\alpha_1 \rho_1$}{$(\mathrm{kg\ m^{-3}})$}} &
    \stackanchor{$\alpha_2 \rho_2$}{$(\mathrm{kg\ m^{-3}})$} &
    \stackanchor{$u$}{$(\mathrm{m\ s^{-1}})$} &
    \stackanchor{$v$}{$(\mathrm{m\ s^{-1}})$} &
    \stackanchor{$p$}{$(\mathrm{Pa})$} &
    $\alpha_1$ \\ \midrule
    \addstackgap{$\left| y \right| \leq 0.05L$}  & $1.0227724310474432\mathrm{e}{3}$ & $1.1817862272214237\mathrm{e}{-8}$ & $1.5\mathrm{e}{5}$ & 0 & $1.01325\mathrm{e}{5}$ & $1 - 1.0\mathrm{e}{-8}$ \\
    \addstackgap{otherwise} & $1.0227724412751677\mathrm{e}{-5}$ & 1.1817862094653702 & 0 & 0 & $1.01325\mathrm{e}{5}$ & $1.0\mathrm{e}{-8}$ \\ \bottomrule
    \end{tabular}
  \end{center}
  \caption{Left boundary conditions of the 2D Mach 100 water jet problem.}
  \label{table:BC_2D_Mach_100_water_jet}
\end{table}

\subsection{Two-dimensional three-species extreme underwater explosion (UNDEX) problem}

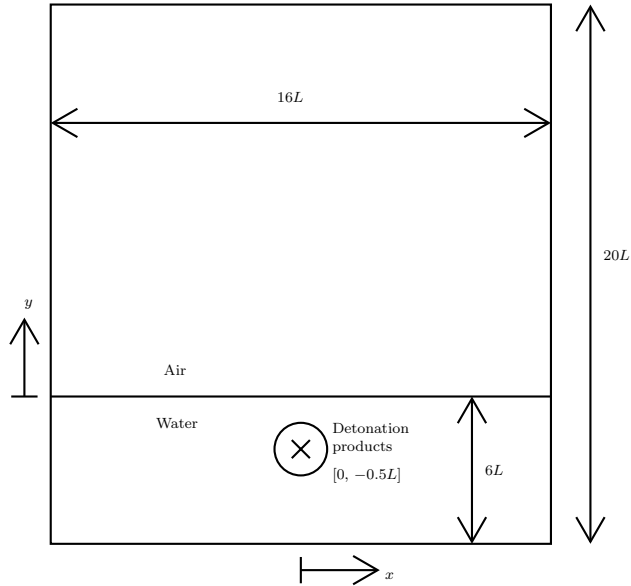
\begin{figure}[!ht]
  \centering
  \begin{tikzpicture}[thick,scale=0.7, every node/.style={transform shape}]
    \useasboundingbox (-1cm,-2cm)  rectangle (10cm,10cm);
    \draw[black]        (-0.5cm,-1.0cm) rectangle ++(9.5cm,10.25cm);
    \draw[black]        ( 4.25cm,0.8cm) circle (0.5cm);
    \draw[black, thick] (-0.5cm,1.8cm) -- (9.0cm,1.8cm);

    \node[text width=3cm] at (3.15cm,2.3cm) {Air};
    \node[text width=3cm] at (3.0cm,1.3cm) {Water};
    \node[text width=3cm] at (6.35cm,1.0cm) {Detonation \\ products};

    \draw (4.25cm, 0.8cm) node[cross] {};
    
    \draw[{Straight Barb[angle'=60,scale=3]}-{Straight Barb[angle'=60,scale=3]}] ( 9.75cm,-1.0cm) -- (9.75cm,9.25cm);
    \draw[{Straight Barb[angle'=60,scale=3]}-{Straight Barb[angle'=60,scale=3]}] (-0.5cm,7.0cm) -- (9.0cm,7.0cm);
    \draw[{Straight Barb[angle'=60,scale=3]}-{Straight Barb[angle'=60,scale=3]}] ( 7.5cm,-1.0cm) -- (7.5cm,1.8cm);

    \node[text width=3cm] at ( 5.3cm,7.5cm) {$16L$};
    \node[text width=3cm] at (11.5cm,4.5cm) {$20L$};
    \node[text width=3cm] at (9.25cm,0.4cm) {$6L$};
    \node[text width=3cm] at (6.35cm,0.3cm) { $[0, \, -0.5L]$ };

    \draw[-{Straight Barb[angle'=60,scale=3]}] (4.25cm,-1.5cm) -- (5.75cm,-1.5cm);
    \node[text width=3cm] at (7.35cm,-1.6cm) {$x$};
    \draw[black] (4.25cm,-1.75cm) -- (4.25cm,-1.25cm);
    \draw[-{Straight Barb[angle'=60,scale=3]}] (-1.0cm,1.8cm) -- (-1.0cm,3.3cm);
    \node[text width=1cm] at (-0.5cm,3.55cm) {$y$};
    \draw[black] (-1.25cm,1.8cm) -- (-0.75cm,1.8cm);
  \end{tikzpicture}
  \caption{Schematic diagram of the 2D three-species extreme UNDEX problem.} \label{fig:schematic_2D_three_species_extreme_UNDEX_problem}
\end{figure}

This is an intensive 2D underwater explosion (UNDEX) problem with liquid water and two ideal gases which are air and detonation products. 
Early theoretical analyses, experimental, and computational investigations of this problem can be found in~\cite{cole1948under,keller1956damping,holt1977underwater}. This is a popular test problem in many other works on compressible multi-phase numerical methods~\cite{shyue1998efficient, luo2004computation, farhat2008higher, miller2013pressure, shyue2014eulerian, chiapolino2017sharpening}.
Initially, there is an ultra-high pressure and density detonation gas bubble, with $p=1.0\mathrm{e}{12}\ \mathrm{Pa}$ and $\rho_3=1250\ \mathrm{kg\ m^{-3}}$, settled underwater that is close to the air-water interface in a domain of $\left[ -8L, 8L \right] \times \left[ 0, 20L \right]$, where $L = 1\ \mathrm{m}$ is chosen.
Figure~\ref{fig:schematic_2D_three_species_extreme_UNDEX_problem} shows the schematic of the initial flow field and domain.
The detonation products at high pressure and temperature are initially located at $\left[ 0, -0.5L \right]$ with radius $0.3L$ and are treated as an ideal gas, with gas properties $\gamma = 1.25$, $c_p = 1000\ \mathrm{J\ kg^{-1} K^{-1}}$.
The air and water surrounding the gas bubble are at atmospheric conditions, i.e. $p = 101325 \ \mathrm{Pa}$ and $T = 298 \ \mathrm{K}$. The initial conditions are detailed in table~\ref{table:IC_2D_three_species_extreme_UNDEX_problem}.
Constant extraploation is used at all domain boundaries. The computation is performed with the fractional algorithm using the PP-WCNS-IS on a $1536 \times 1920$ mesh until $t = 0.25\ \mathrm{ms}$.
Numerical failures are encountered when the positivity-preserving limiters are turned off.

\begin{table}[!ht]
\small
  \begin{center}
    \begin{tabular}{@{}c | c c c c}\toprule
     &
    \addstackgap{\stackanchor{$\alpha_1 \rho_1$}{$(\mathrm{kg\ m^{-3}})$}} &
    \stackanchor{$\alpha_2 \rho_2$}{$(\mathrm{kg\ m^{-3}})$} &
    \stackanchor{$\alpha_3 \rho_3$}{$(\mathrm{kg\ m^{-3}})$} \\
    \midrule
    \addstackgap{water} & $1.0227724208197188\mathrm{e}{3}$ & $1.1817862212832324\mathrm{e}{-8}$ & $1.7000838926174497\mathrm{e}{-8}$ \\
    \addstackgap{air} & $1.0227724412751677\mathrm{e}{-5}$ & $1.1817861976475079$ & $1.7000838926174497\mathrm{e}{-8}$ \\
    \addstackgap{\stackanchor{detonation}{products}} & $8.9361901785714267\mathrm{e}{-7}$ & $8.6891757696127126\mathrm{e}{-6}$ & $1.2499999749999999\mathrm{e}{3}$ \\
    \bottomrule
    \end{tabular}
    \begin{tabular}{@{}c | c c c c c c }\toprule
     &
    \addstackgap{\stackanchor{$u$}{$(\mathrm{m\ s^{-1}})$}} &
    \stackanchor{$v$}{$(\mathrm{m\ s^{-1}})$} &
    \stackanchor{$p$}{$(\mathrm{Pa})$} &
    $\alpha_1$ &
    $\alpha_2$ \\
    \midrule
    \addstackgap{water} & $0$& $0$ & $1.01325\mathrm{e}{5}$ & $1 - 2.0\mathrm{e}{-8}$ & $1.0\mathrm{e}{-8}$ \\
    \addstackgap{air} & $0$ & $0$ & $1.01325\mathrm{e}{5}$ & $1.0\mathrm{e}{-8}$ & $1 - 2.0\mathrm{e}{-8}$ \\
    \addstackgap{\stackanchor{detonation}{products}} & $0$ & $0$ & $1.0\mathrm{e}{12}$ & $1.0\mathrm{e}{-8}$ & $1.0\mathrm{e}{-8}$ \\
    \bottomrule
    \end{tabular}
  \end{center}
  \caption{Initial conditions of the 2D three-species extreme UNDEX problem. The initial velocity is $u = v = 0$.}
  \label{table:IC_2D_three_species_extreme_UNDEX_problem}
\end{table}

\begin{figure}[!ht]
\centering
\subfigure[$t = 0.1\ \mathrm{ms}, \ \alpha_1$]{%
\includegraphics[width=0.32\textwidth]{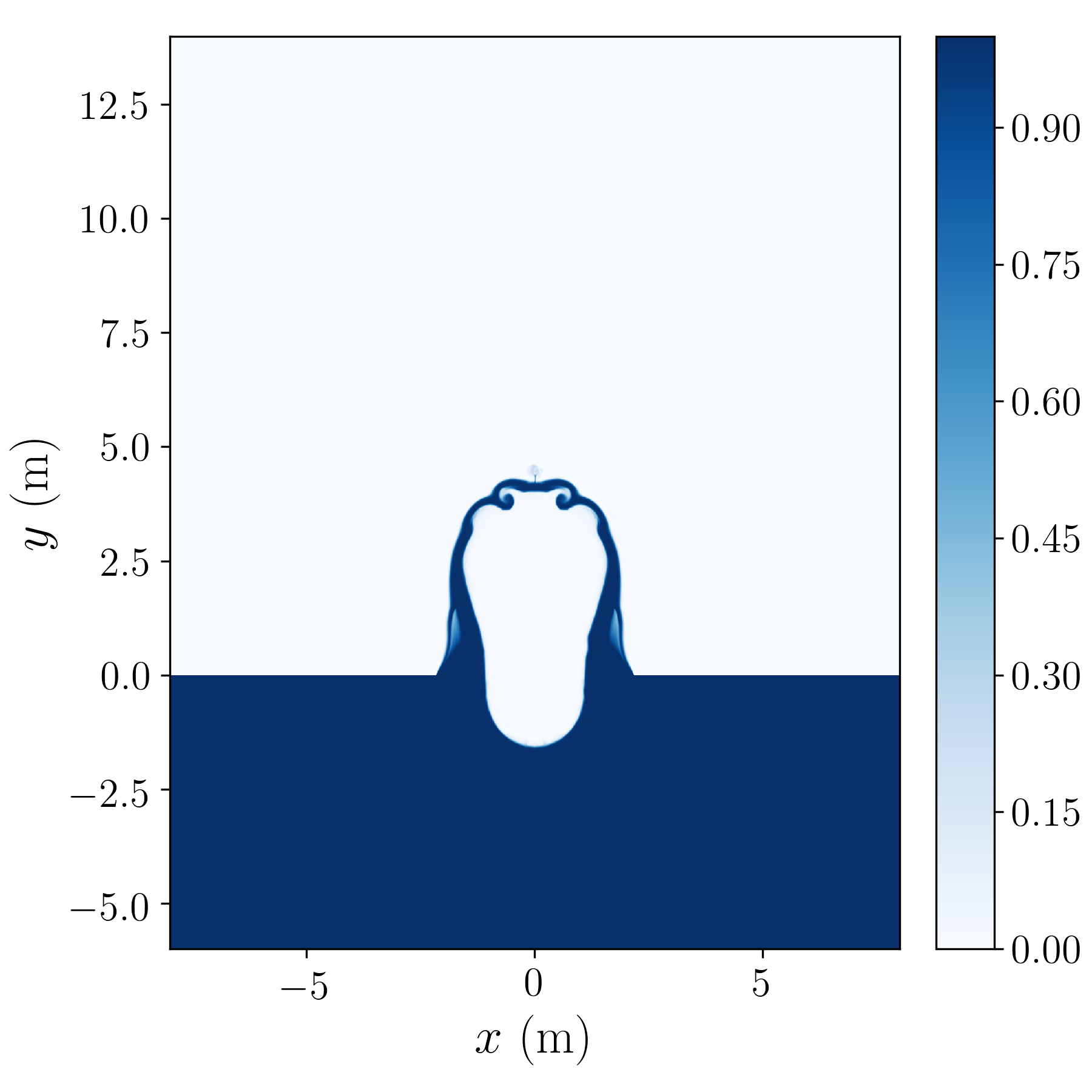}}
\subfigure[$t = 0.1\ \mathrm{ms}, \ \alpha_2$]{%
\includegraphics[width=0.32\textwidth]{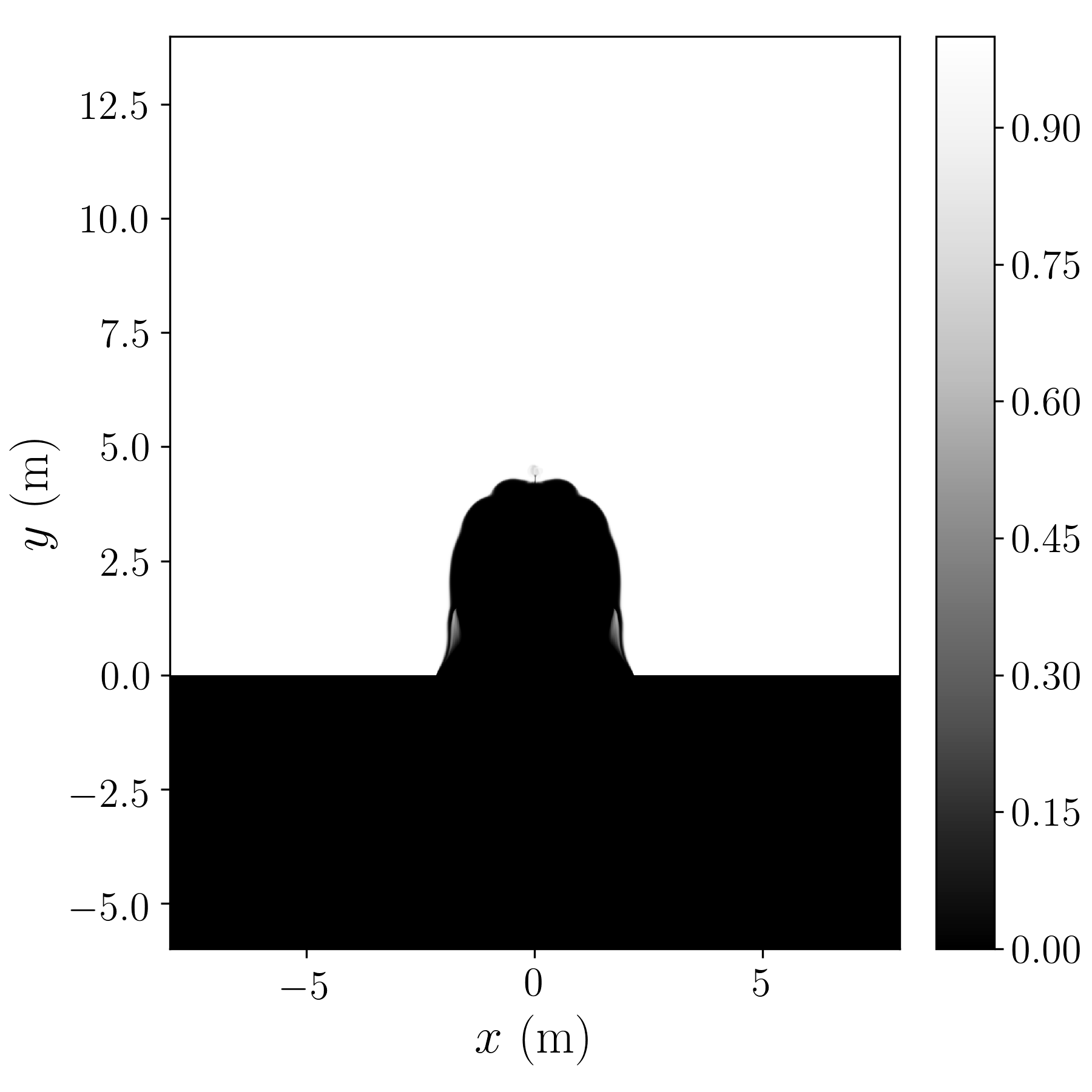}}
\subfigure[$t = 0.1\ \mathrm{ms}, \ \alpha_3$]{%
\includegraphics[width=0.32\textwidth]{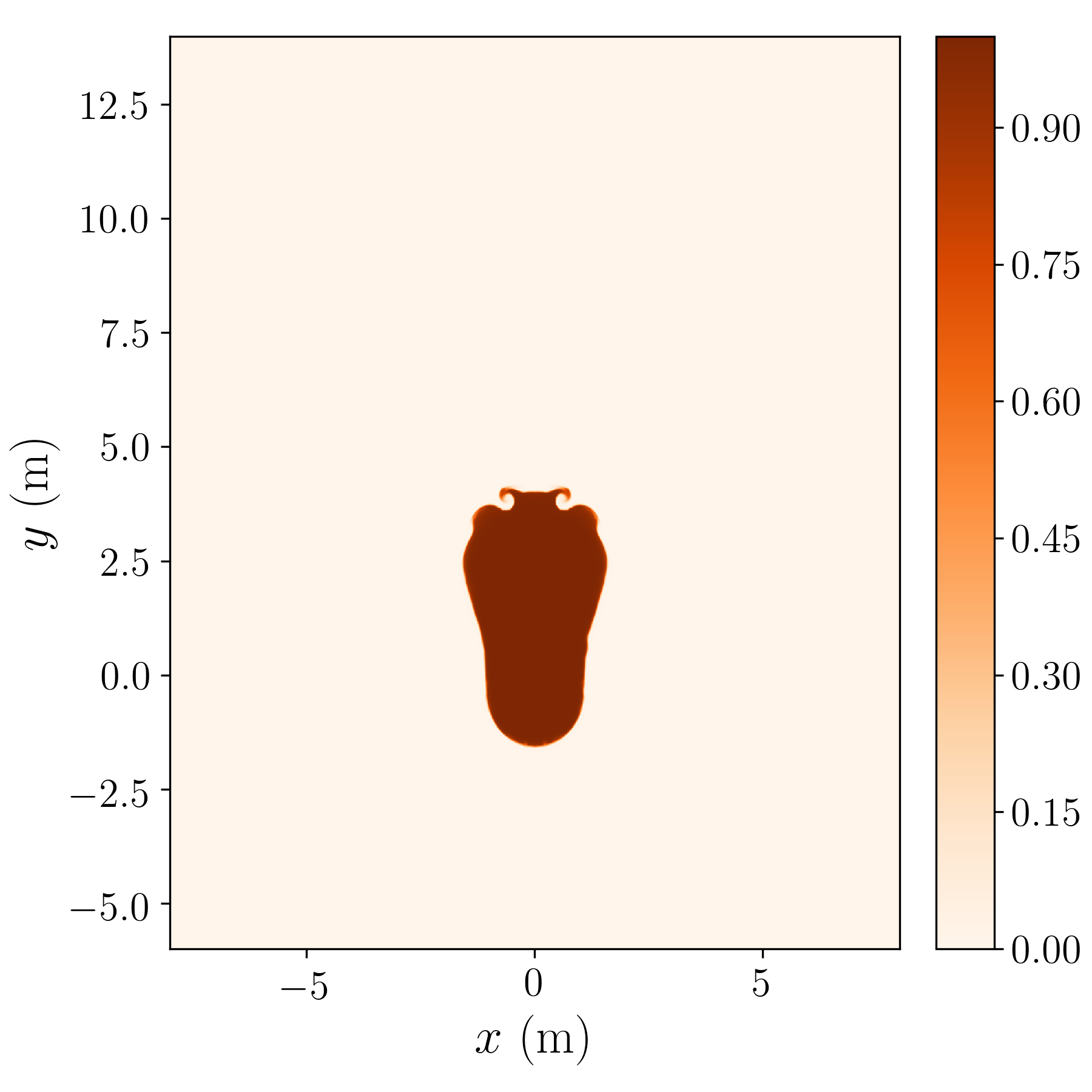}}
\subfigure[$t = 0.2\ \mathrm{ms}, \ \alpha_1$]{%
\includegraphics[width=0.32\textwidth]{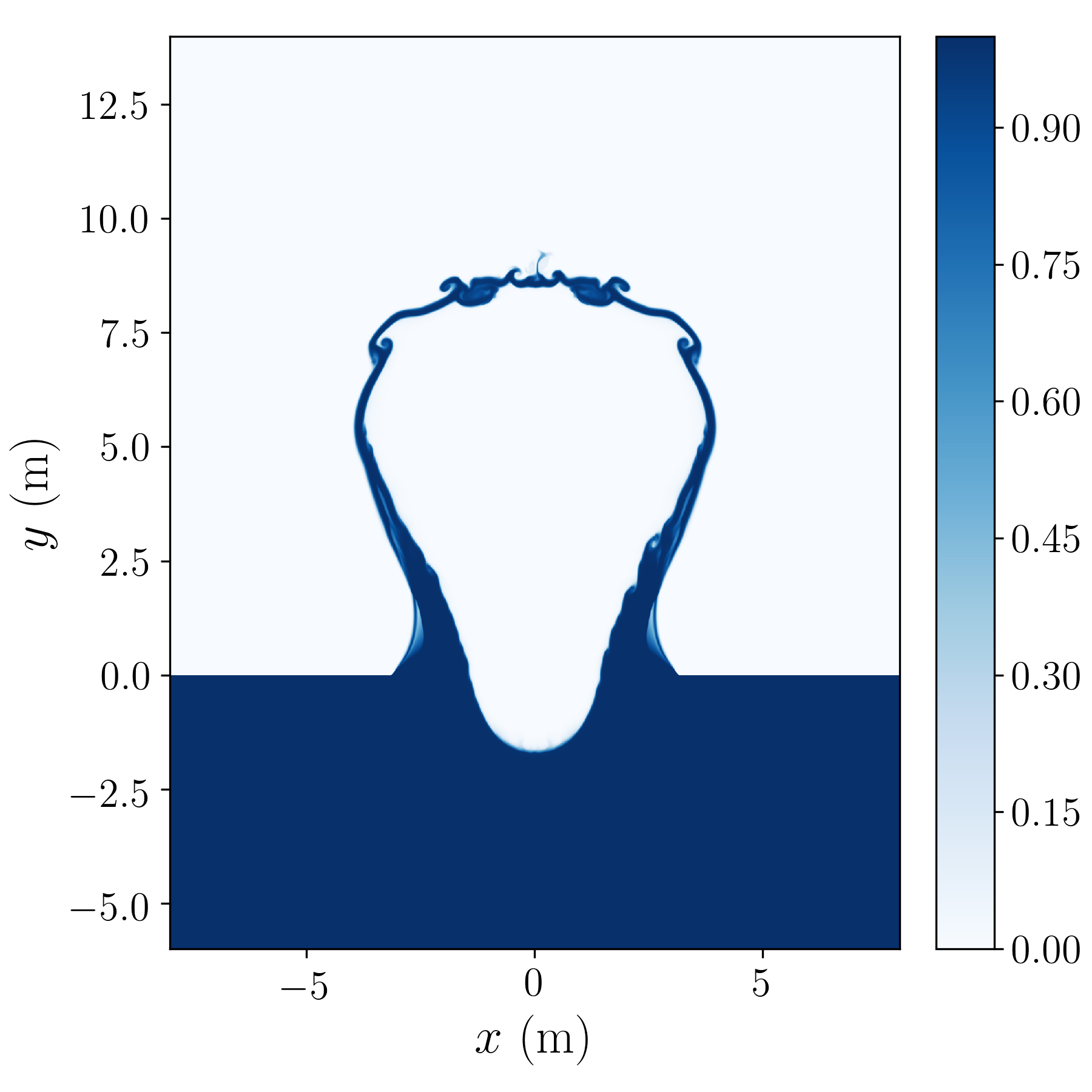}}
\subfigure[$t = 0.2\ \mathrm{ms}, \ \alpha_2$]{%
\includegraphics[width=0.32\textwidth]{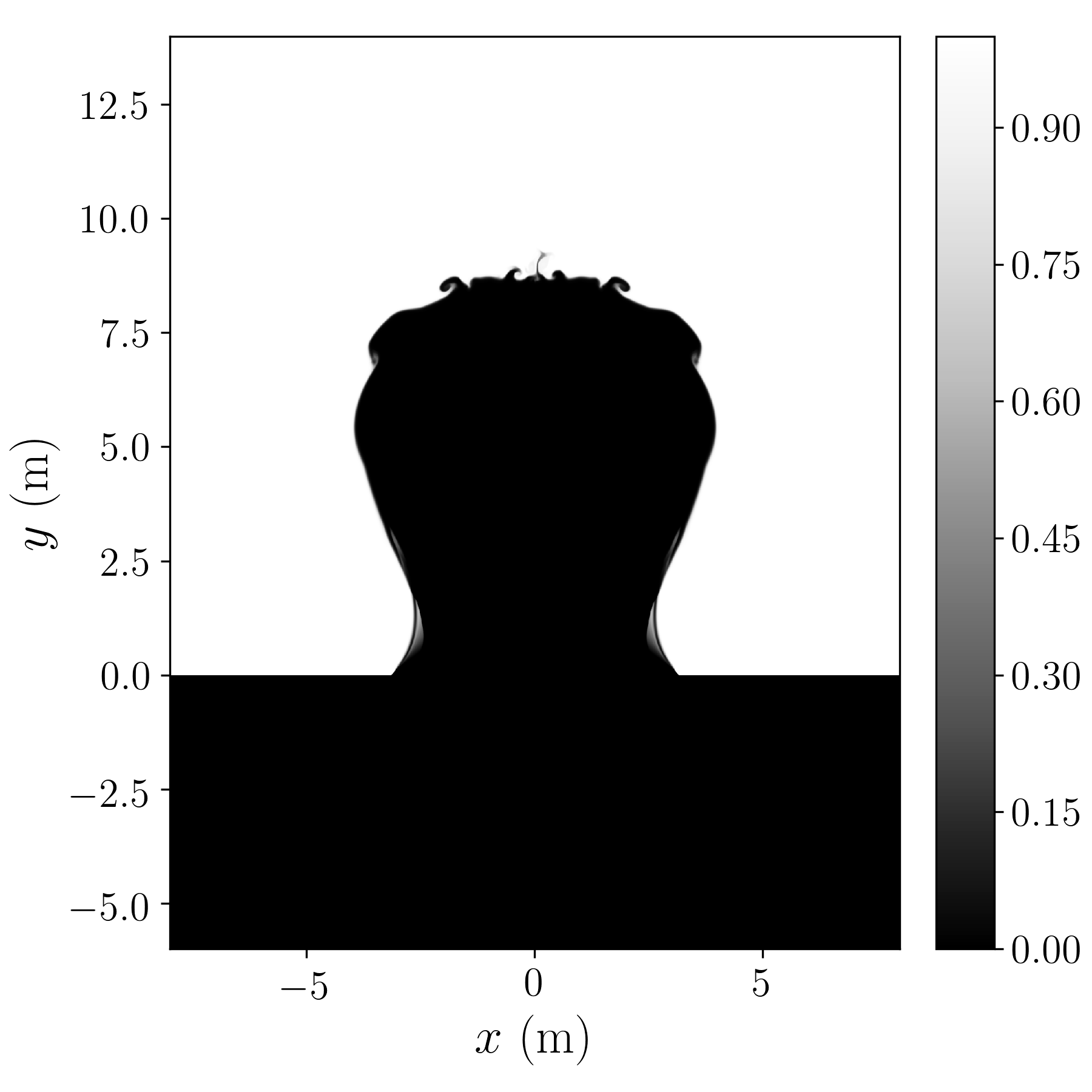}}
\subfigure[$t = 0.2\ \mathrm{ms}, \ \alpha_3$]{%
\includegraphics[width=0.32\textwidth]{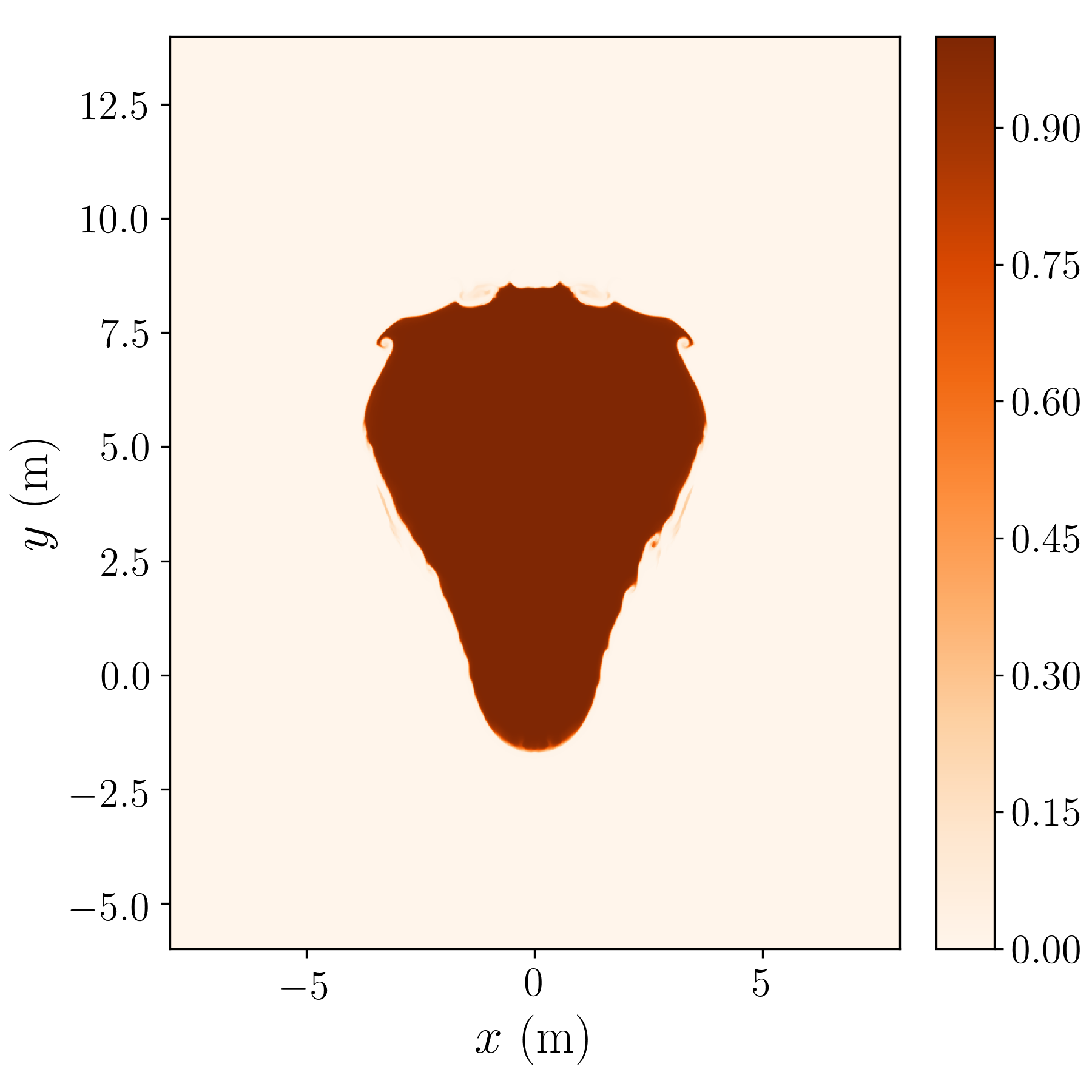}}
\subfigure[$t = 0.25\ \mathrm{ms}, \ \alpha_1$]{%
\includegraphics[width=0.32\textwidth]{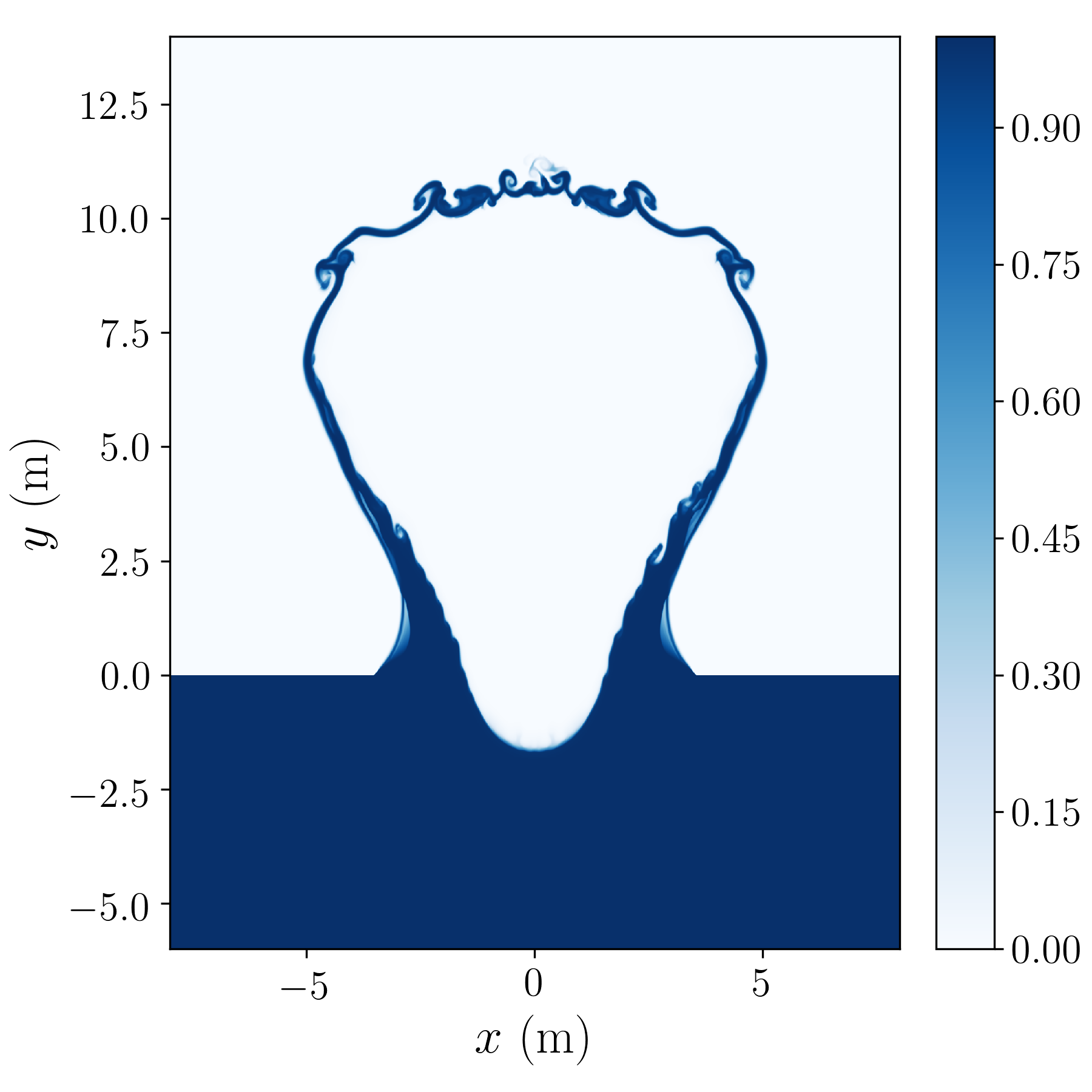}}
\subfigure[$t = 0.25\ \mathrm{ms}, \ \alpha_2$]{%
\includegraphics[width=0.32\textwidth]{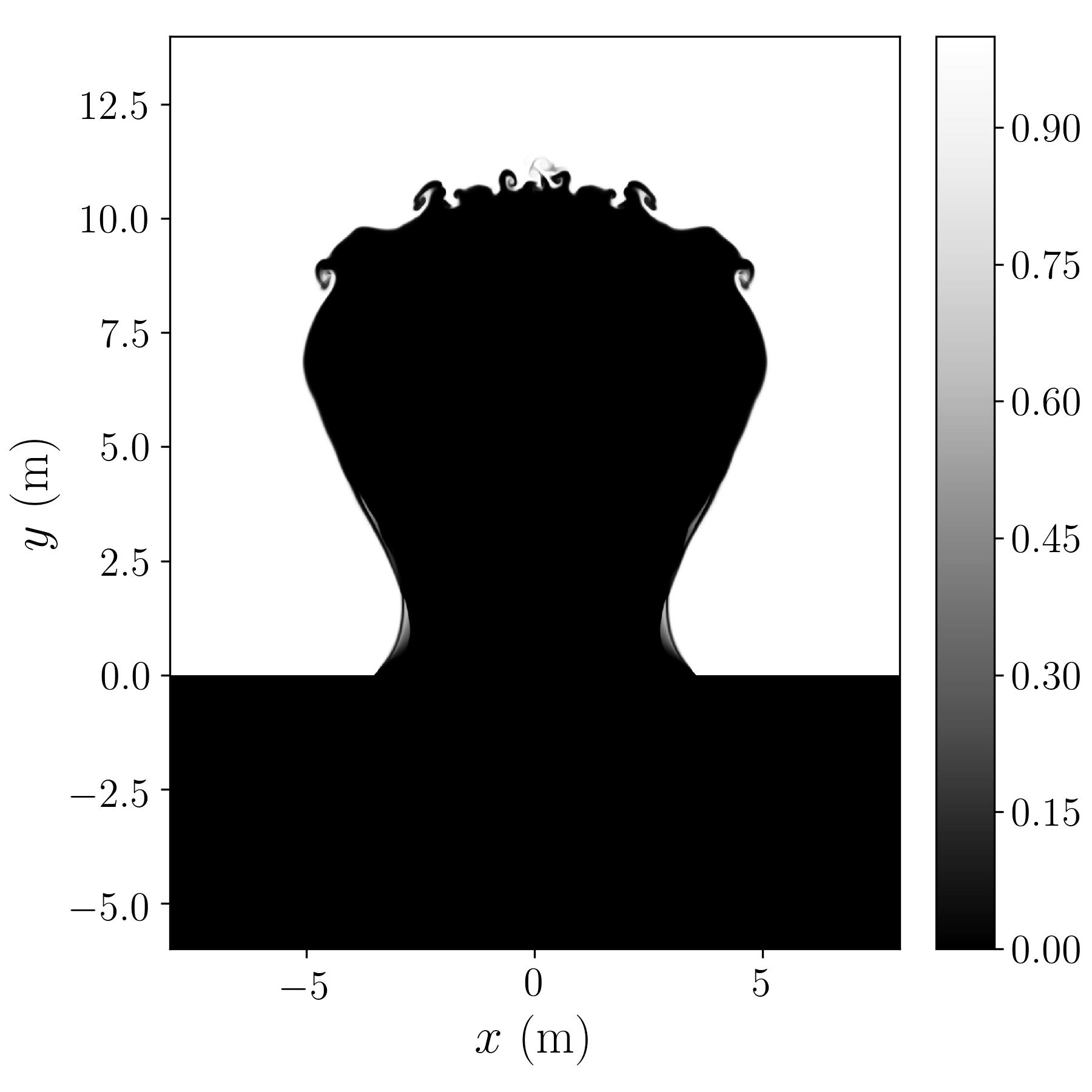}}
\subfigure[$t = 0.25\ \mathrm{ms}, \ \alpha_3$]{%
\includegraphics[width=0.32\textwidth]{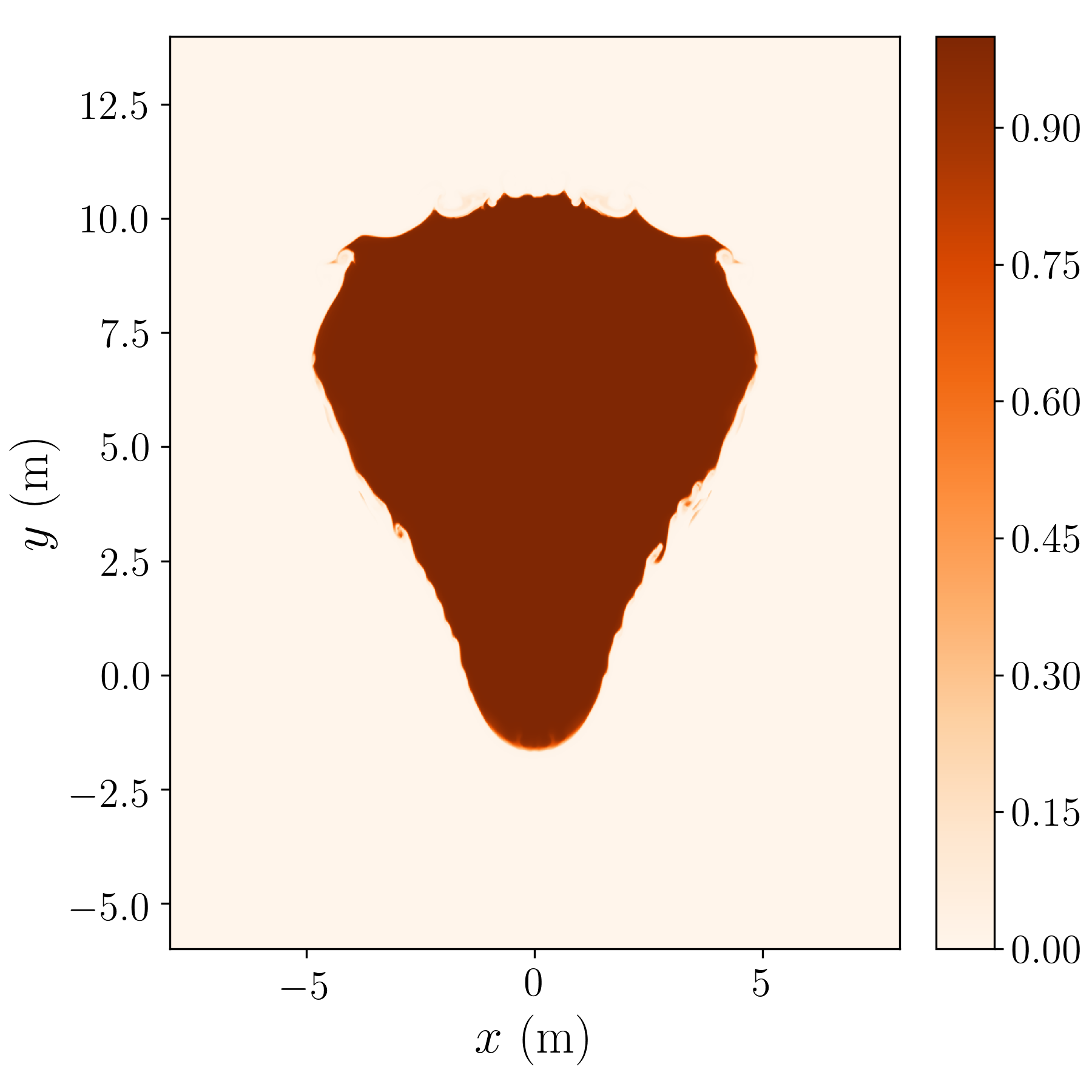}}
\caption{Volume fractions of the 2D three-species extreme UNDEX problem.}
\label{fig:three_species_extreme_UNDEX_problem_volume_fractions}
\end{figure}

\begin{figure}[!ht]
\centering
\subfigure[$t = 0.1\ \mathrm{ms}$]{%
\includegraphics[width=0.32\textwidth]{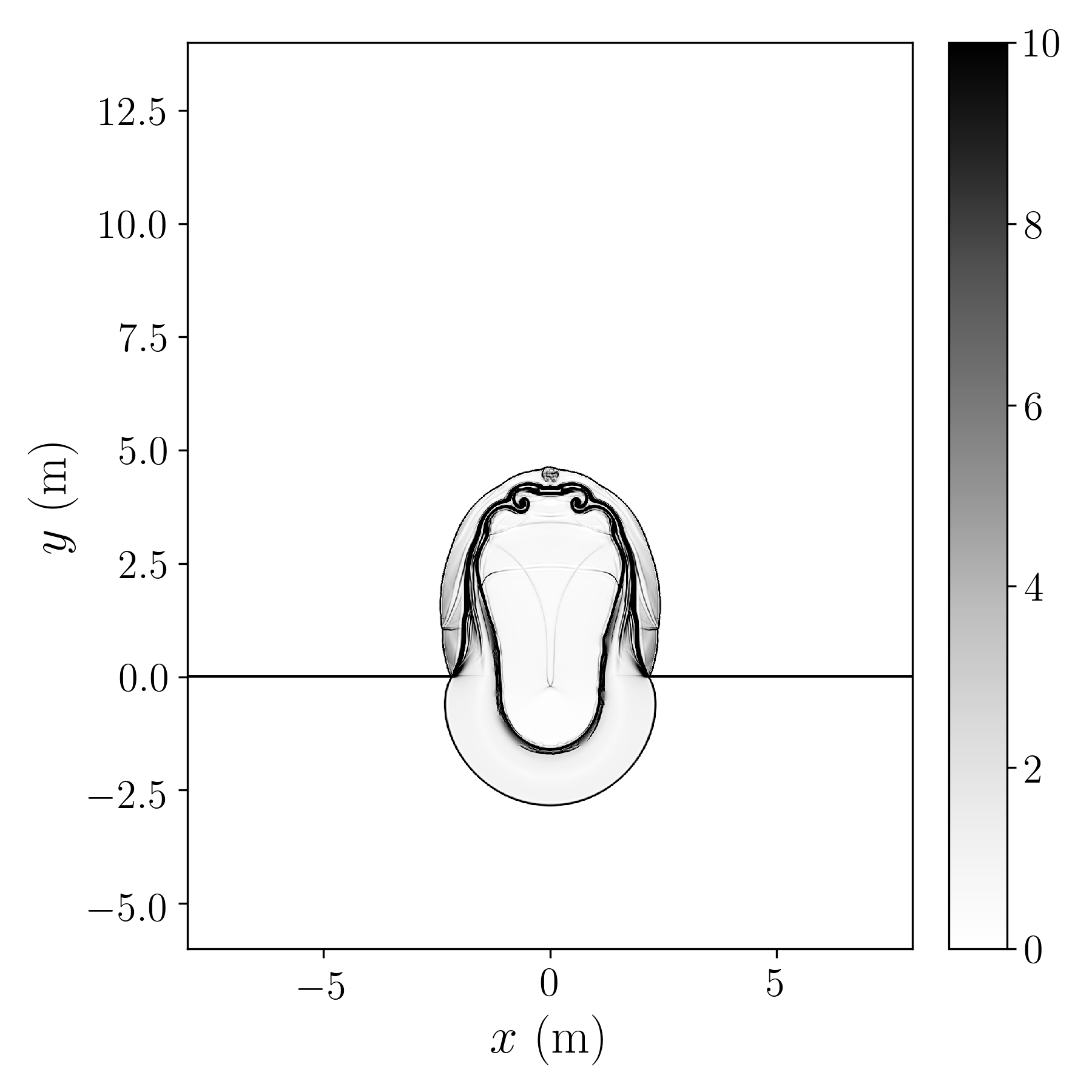}}
\subfigure[$t = 0.2\ \mathrm{ms}$]{%
\includegraphics[width=0.32\textwidth]{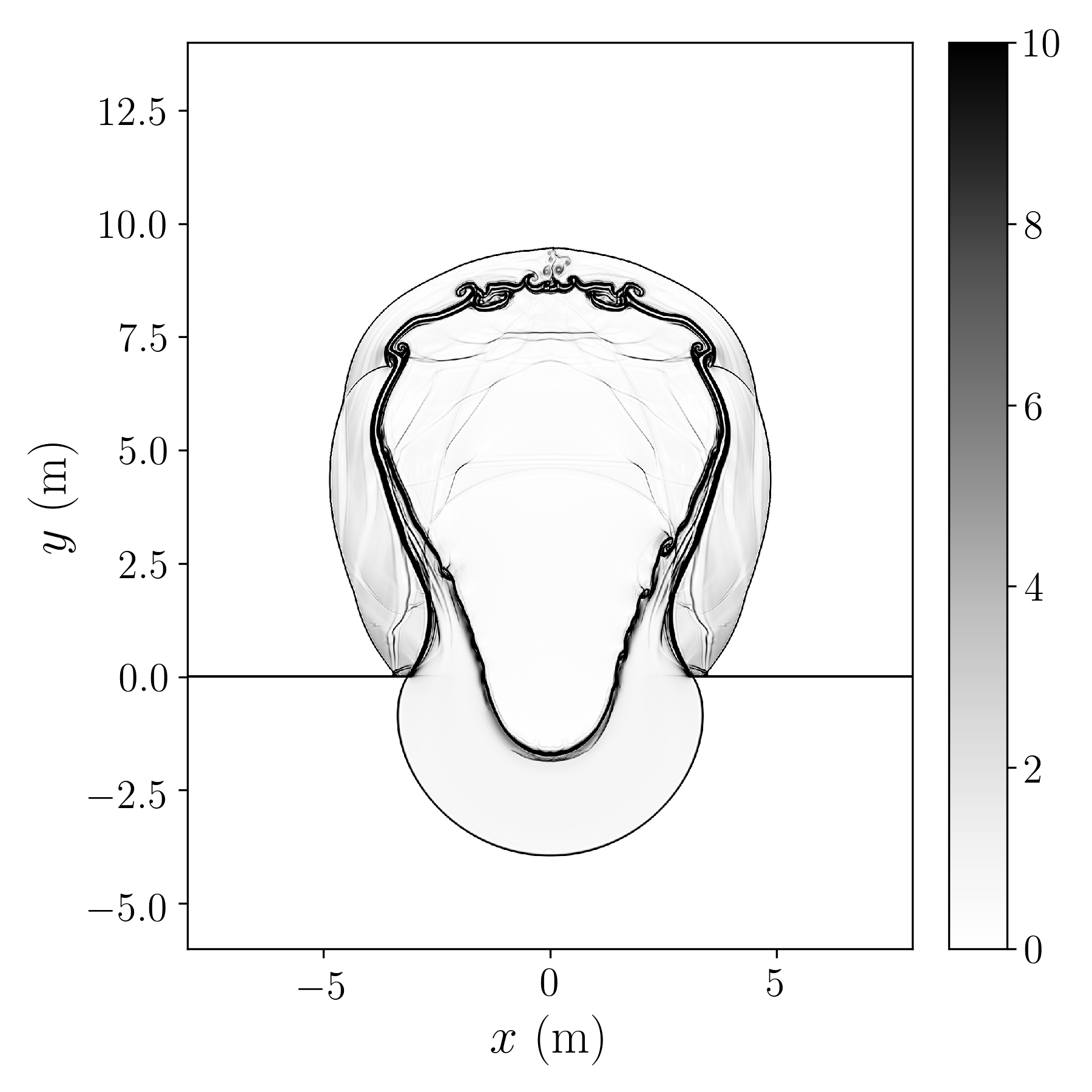}}
\subfigure[$t = 0.25\ \mathrm{ms}$]{%
\includegraphics[width=0.32\textwidth]{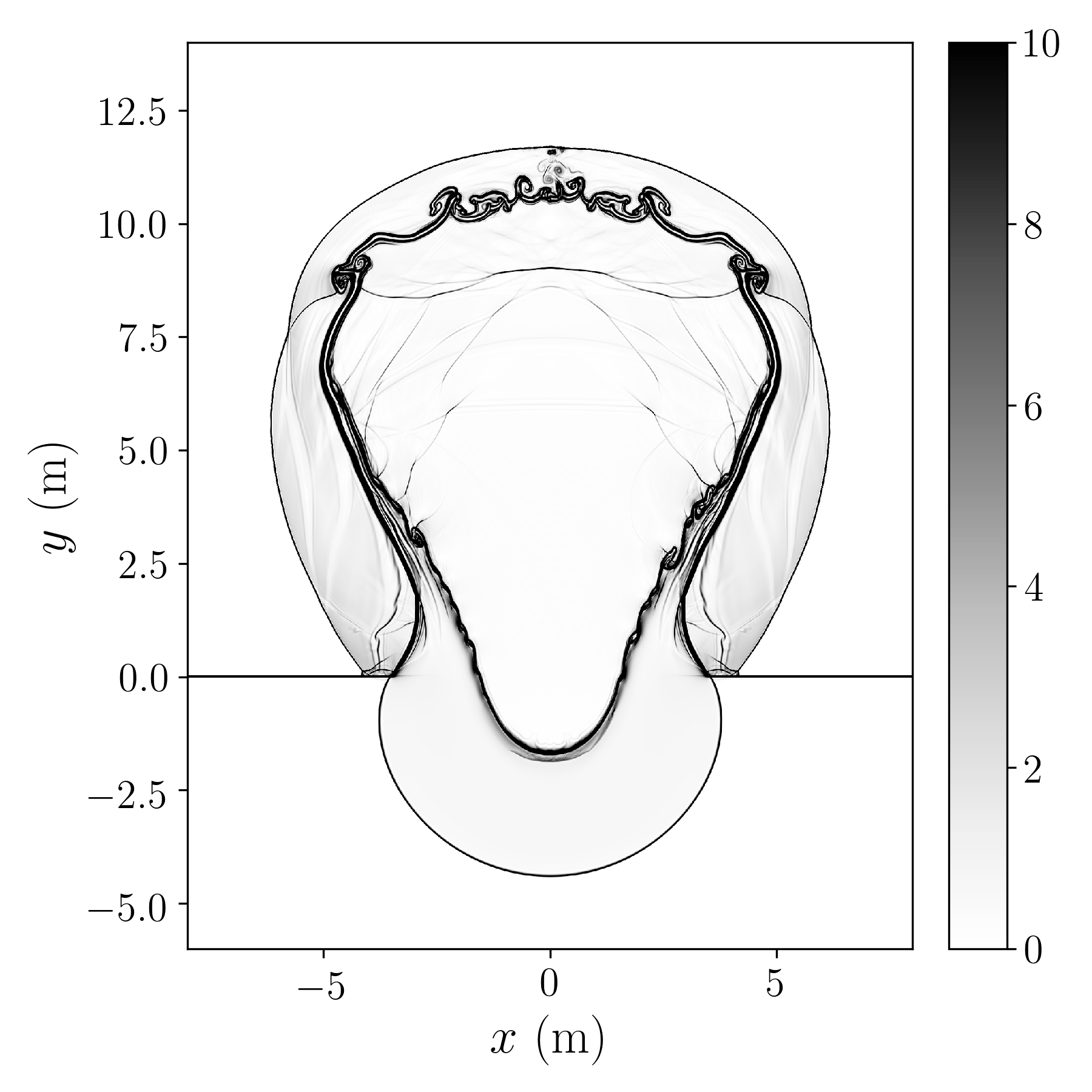}}
\caption{Numerical schlieren ($\left( \left| \nabla \rho \right| / \rho \right)$) of the 2D three-species extreme UNDEX problem.}
\label{fig:three_species_extreme_UNDEX_problem_schl}
\end{figure}

\begin{figure}[!ht]
\centering
\subfigure[$t = 0.1\ \mathrm{ms}$]{%
\includegraphics[width=0.32\textwidth]{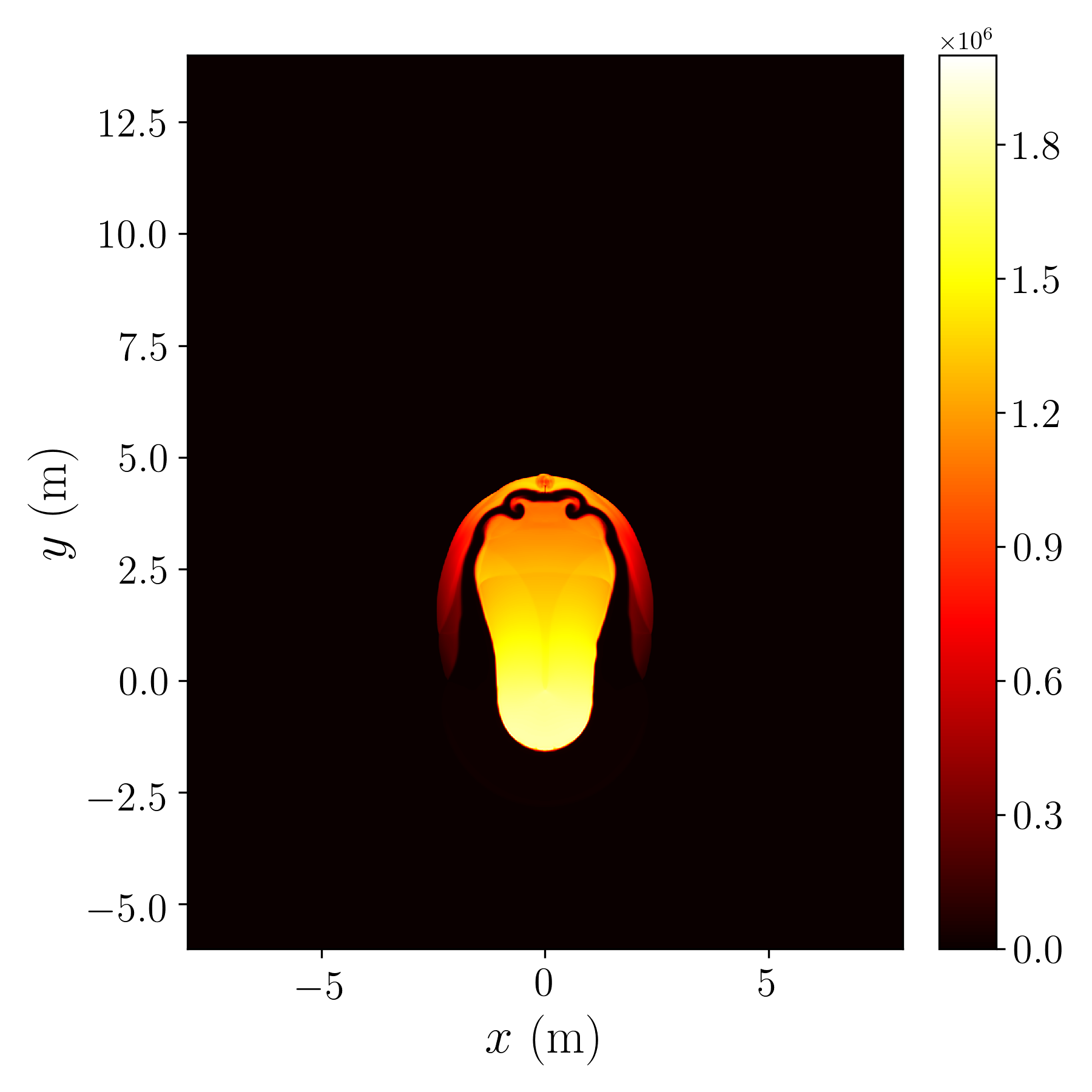}}
\subfigure[$t = 0.2\ \mathrm{ms}$]{%
\includegraphics[width=0.32\textwidth]{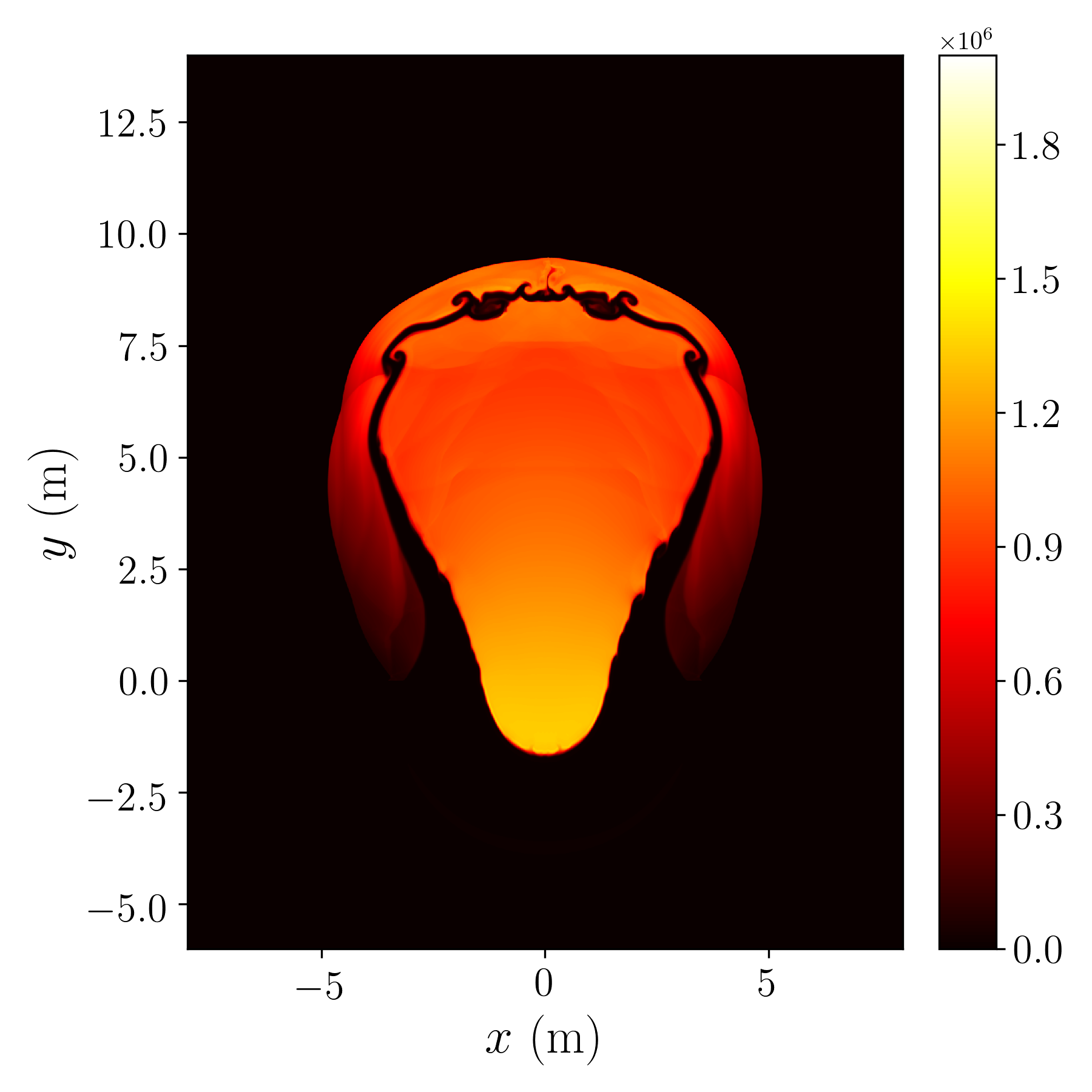}}
\subfigure[$t = 0.25\ \mathrm{ms}$]{%
\includegraphics[width=0.32\textwidth]{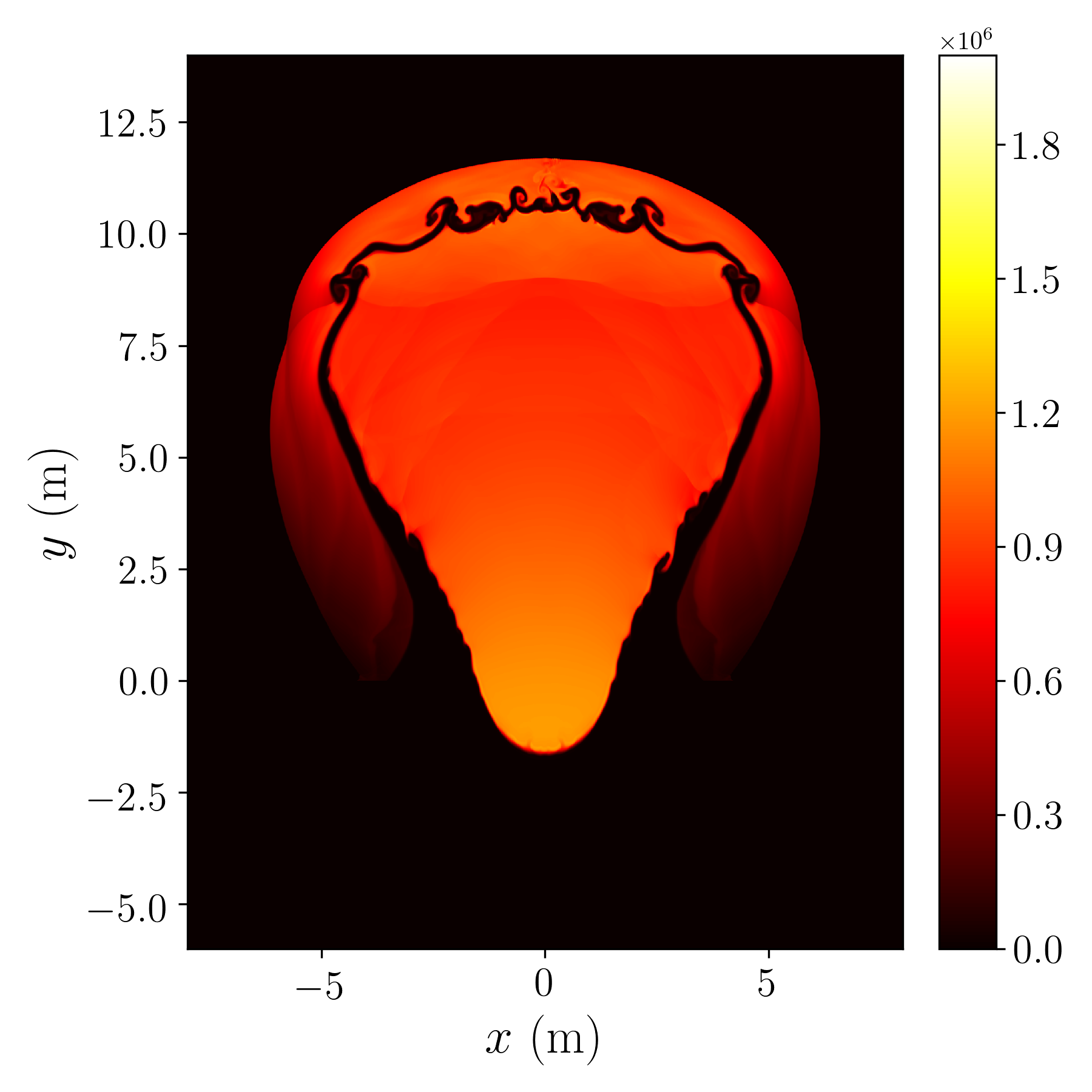}}
\caption{Temperature (K) of the 2D three-species extreme UNDEX problem.}
\label{fig:three_species_extreme_UNDEX_problem_temperature}
\end{figure}

\begin{figure}[!ht]
\centering
\subfigure[$t = 0.1\ \mathrm{ms}$, five-equation model]{%
\includegraphics[width=0.32\textwidth]{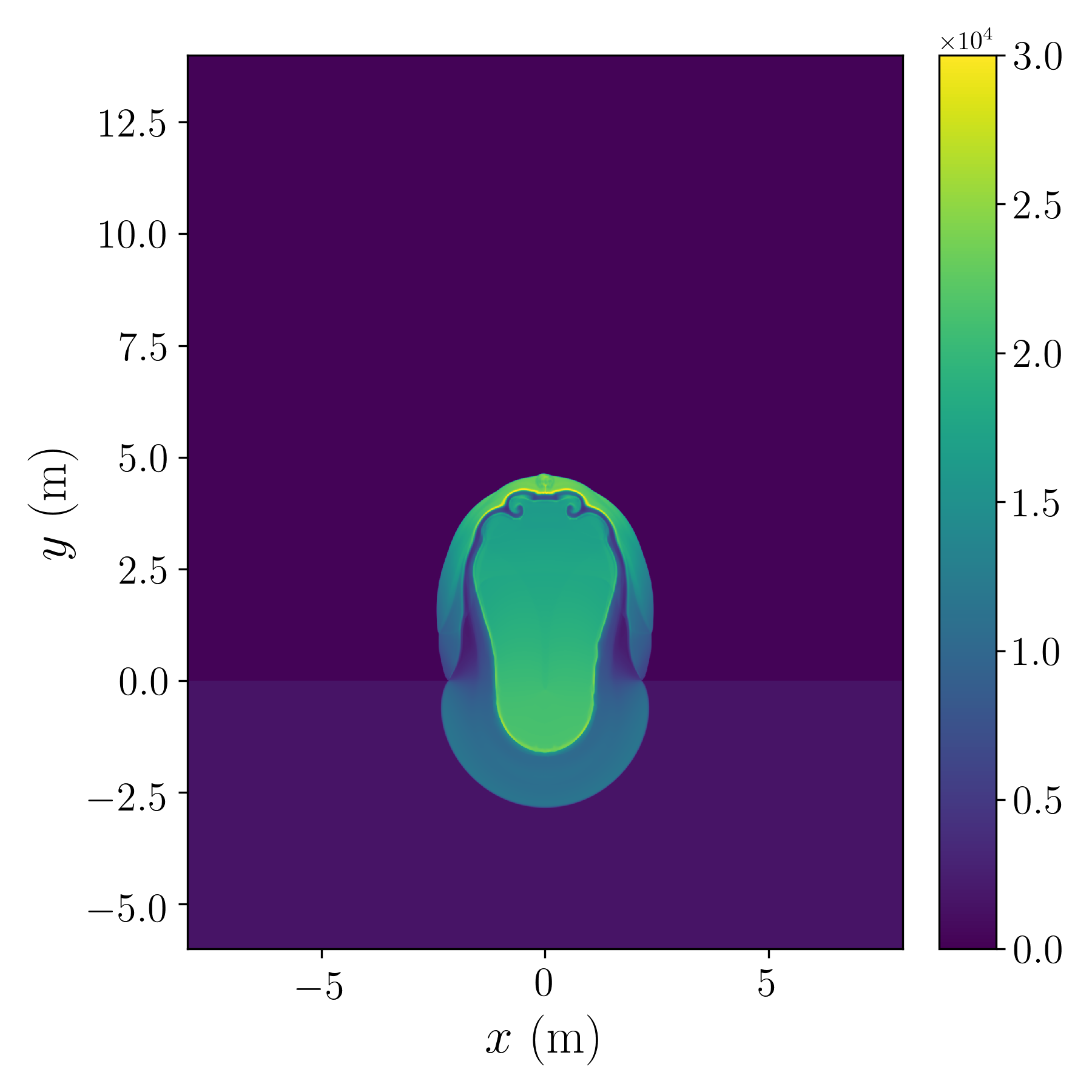}}
\subfigure[$t = 0.2\ \mathrm{ms}$, five-equation model]{%
\includegraphics[width=0.32\textwidth]{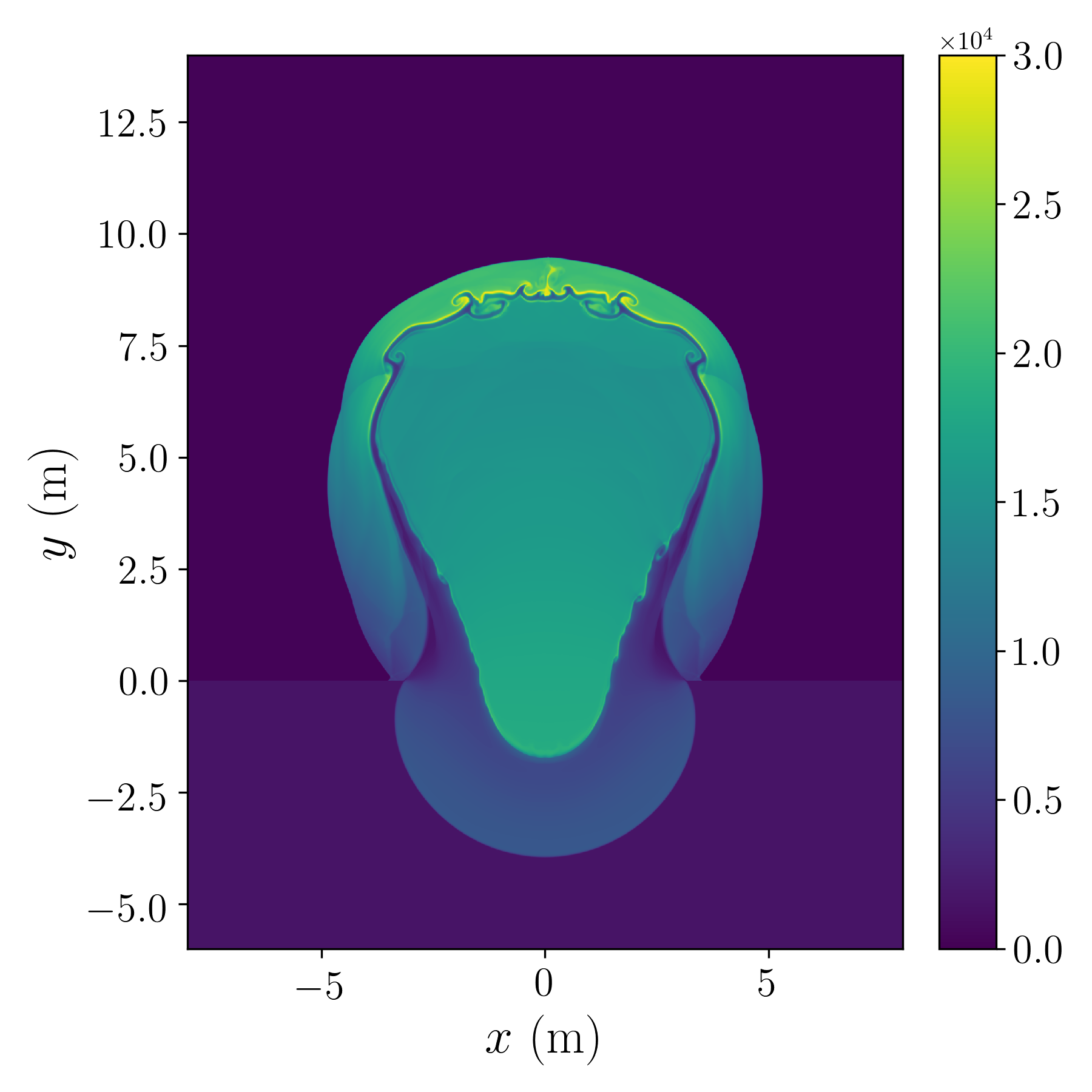}}
\subfigure[$t = 0.25\ \mathrm{ms}$, five-equation model]{%
\includegraphics[width=0.32\textwidth]{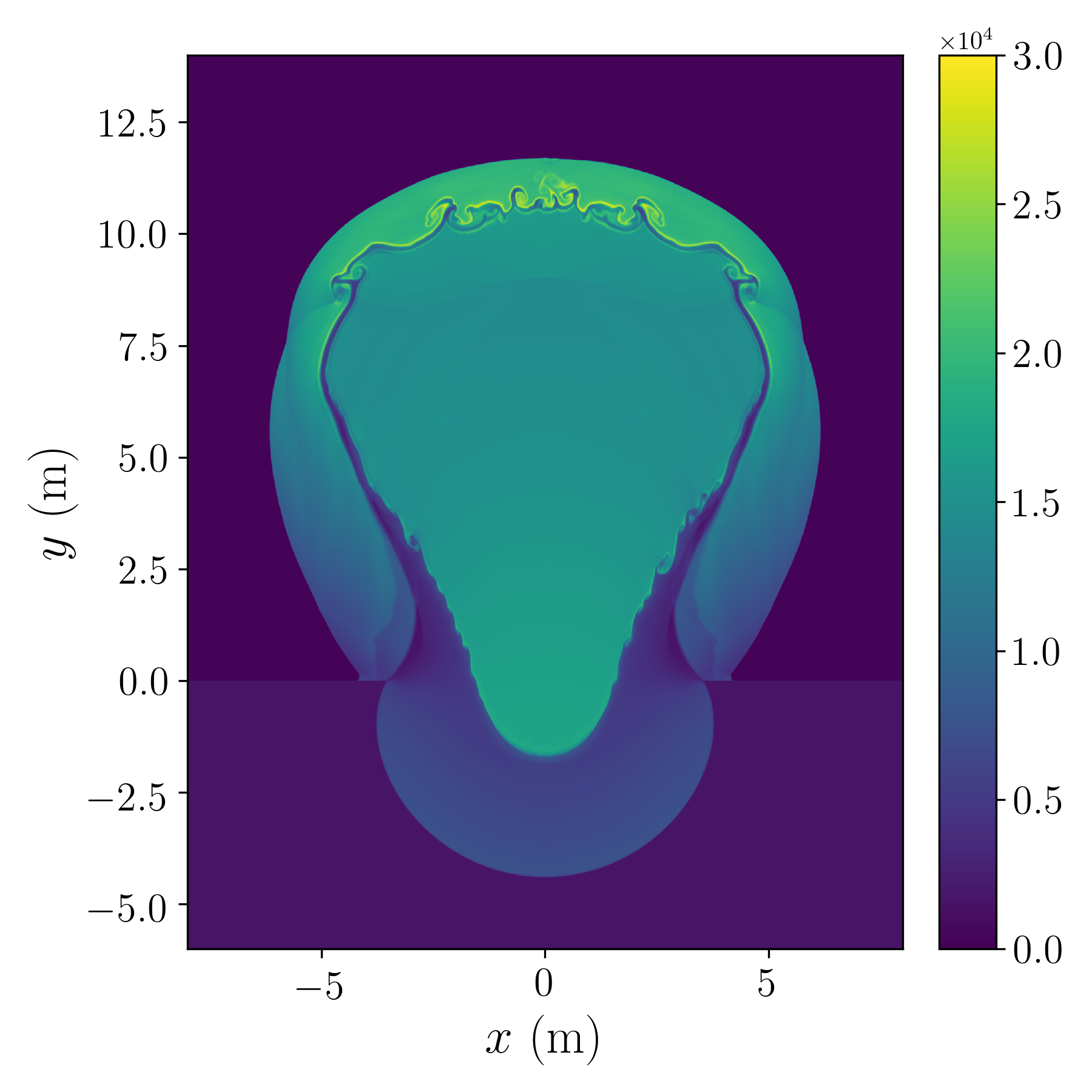}}
\subfigure[$t = 0.1\ \mathrm{ms}$, four-equation model]{%
\includegraphics[width=0.32\textwidth]{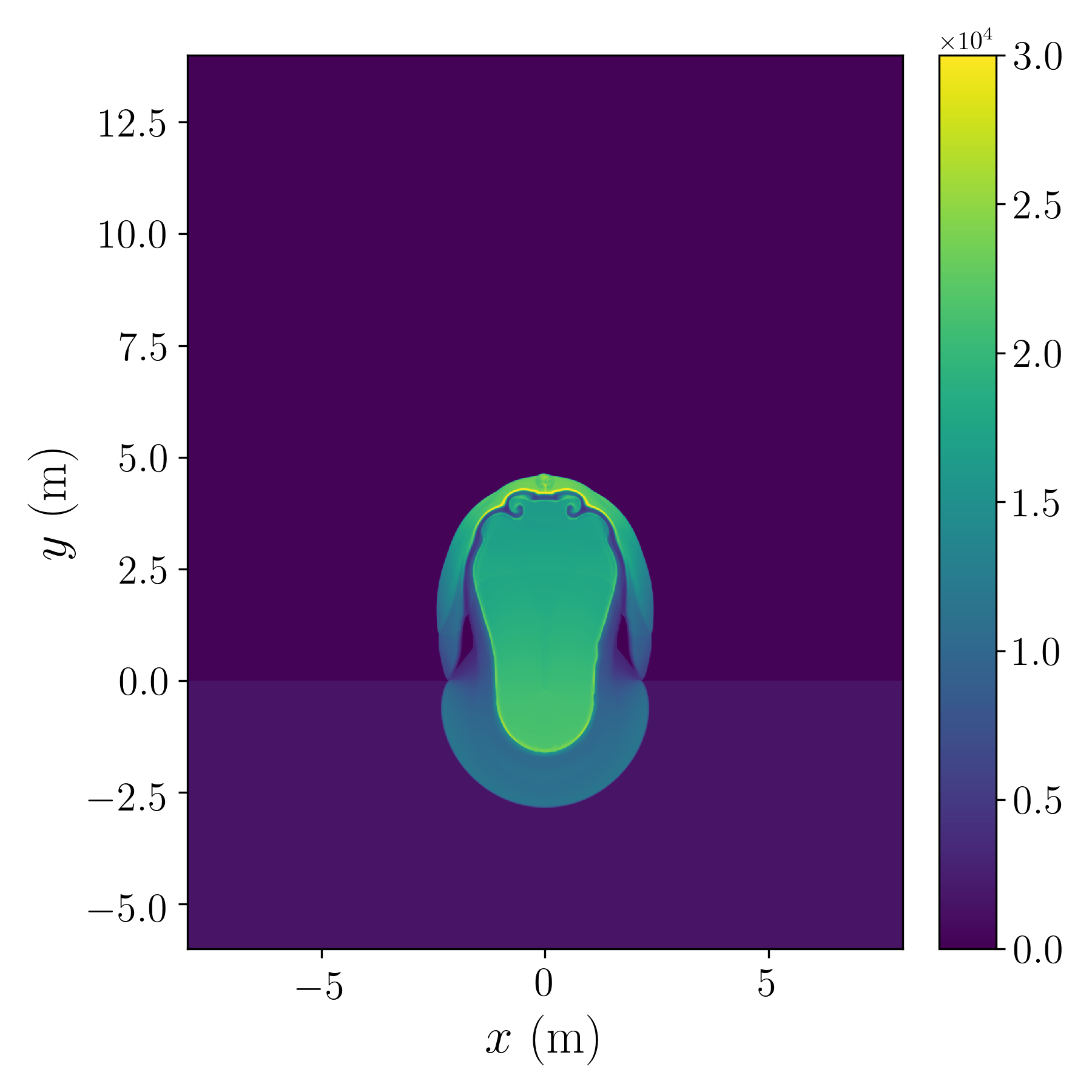}}
\subfigure[$t = 0.2\ \mathrm{ms}$, four-equation model]{%
\includegraphics[width=0.32\textwidth]{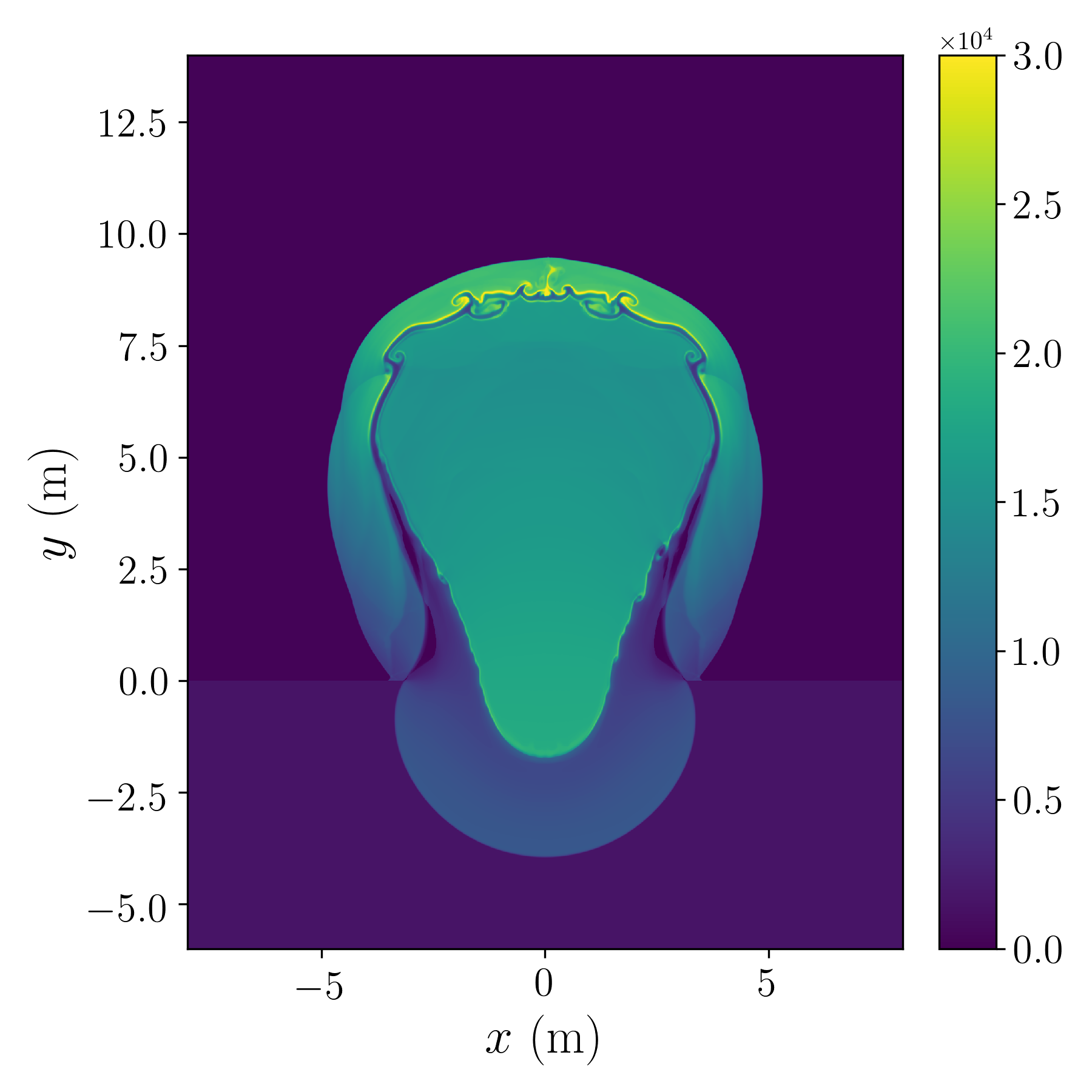}}
\subfigure[$t = 0.25\ \mathrm{ms}$, four-equation model]{%
\includegraphics[width=0.32\textwidth]{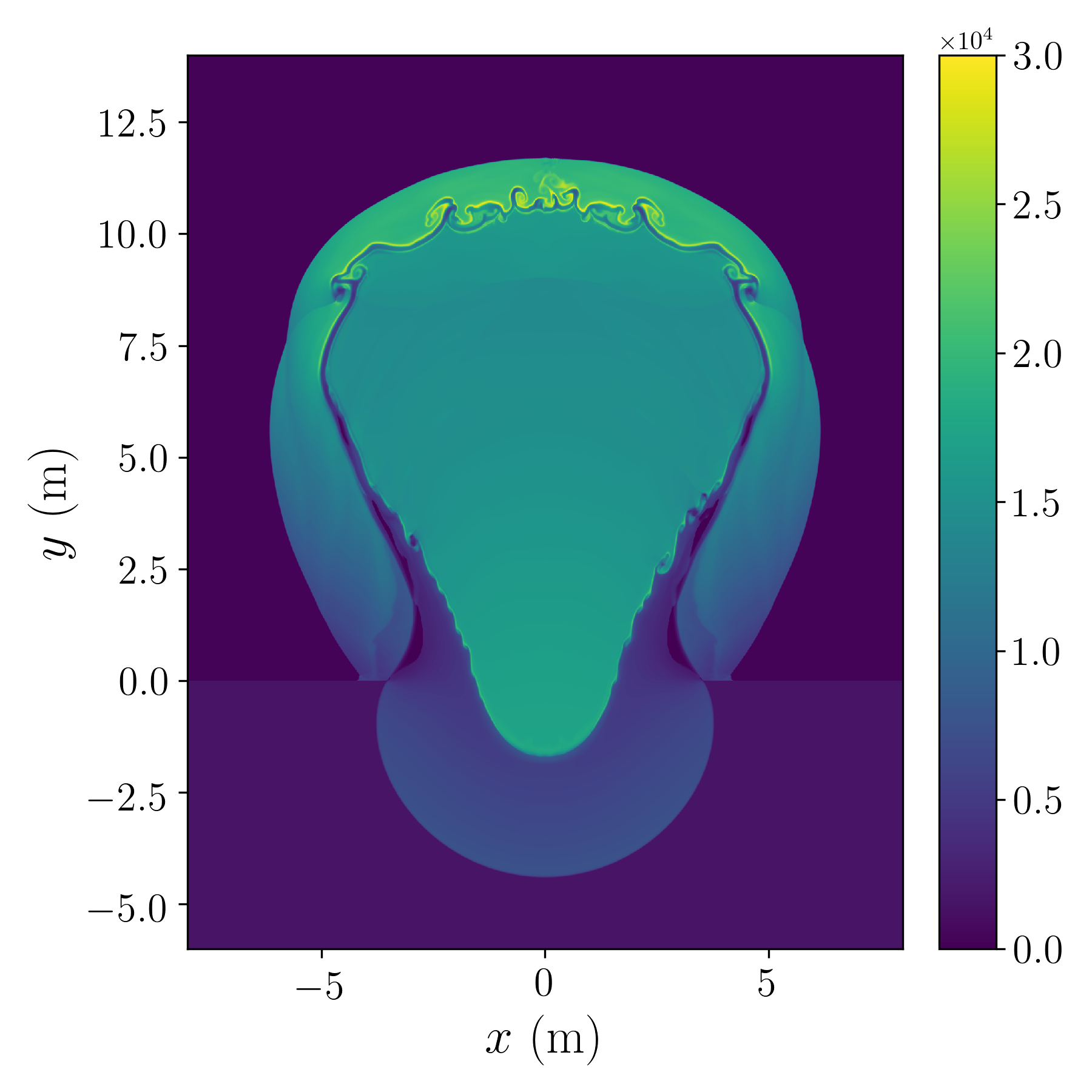}}
\caption{Speed of sound ($\mathrm{m\ s^{-1}}$) of the 2D three-species extreme UNDEX problem. Top row: sound speed of five-equation model by Allaire et al.; bottom row: sound speed of four-equation HRM.}
\label{fig:three_species_extreme_UNDEX_problem_sos}
\end{figure}

The volume fractions of the three different species: liquid water, air, and the detonation products at different times are shown in figure~\ref{fig:three_species_extreme_UNDEX_problem_volume_fractions}. Due to the large initial pressure and temperature differences between the detonation products and the liquid water, a very strong blast wave is generated which propagates into the water. When the shock reaches the water-air surface, baroclinic torque due to the difference between the pressure and density gradients produces vorticity at the interface, thus the interface is distorted over time. The distortion of the interface can be visualized more clearly in figure~\ref{fig:three_species_extreme_UNDEX_problem_schl}, together with the shocks. The large temperature gradient can be seen in figure~\ref{fig:three_species_extreme_UNDEX_problem_temperature}. Phase transition is not modeled in this case. Otherwise, cavitation will occur with the production of water vapor at the water-air surface due to the strong rarefaction wave generated when the shock waves travel from high acoustic impedance water to low impedance air~\cite{wang2014investigation}. Although this extreme test problem is artificial and unrealistic, the simulation does not fail due to negative squared speed of sound, and the pressure and temperature remain positive, as the outcomes of the positivity-preserving procedures.
The speed of sound computed using the formulae for the five-equation model and the four-equation HRM are compared in figure~\ref{fig:three_species_extreme_UNDEX_problem_sos}. Only minor difference between the two sound speeds can be seen in the plots and the difference mainly locates around the vortices along the interfaces.

\section{Concluding remarks} \label{sec:conclustion}

A positivity-preserving Eulerian diffuse interface method was proposed in this work for the simulations of compressible flows with a liquid and an arbitrary number of ideal gases, where the liquid is described by the stiffened gas equation of state. The method is a fractional algorithm that consists of a hyperbolic step of the five-equation model by Allaire et al. and an infinitely fast algebraic thermal relaxation step. Through mathematical proof, it was shown that the five-equation model with infinite thermal relaxation rate is reduced to the popular four-equation HRM. The speed of sound of the numerical model was also verified to converge to the analytical mixture sound speed of the four-equation HRM with a numerical experiment.
Quasi-positivity-preserving interpolation and positivity-preserving flux limiters were extended from our previous work for the use of the fractional algorithm with any high-order shock-capturing Cartesian finite difference or finite volume methods in a general form for robust simulations of multi-phase flows composed of a liquid and an arbitrary number of gases. The overall positivity-preserving fractional algorithm can maintain the positivity of the partial densities and the squared speed of sound, and the boundedness of the volume fractions for any SSP-RK time stepping schemes with proper CFL number. The high accuracy of the numerical method with the fifth order finite difference scheme, WCNS-IS, was demonstrated with different 1D benchmark problems and 2D experimental problems
where solutions of similar or even better accuracy can be obtained with much fewer grid cells compared to the first order accurate algorithms.
The robustness of the method with WCNS-IS was also evinced using various extreme problems with one stiffened gas and one or more ideal gases. The algorithm can be extended with additional relaxation steps, e.g. the thermo-chemical step for applications with phase transition, such as the simulations of space vehicle launches with water-based acoustic suppression systems~\cite{vu2013multiphase,kiris2016computational}. The capability of the diffuse interface method with phase transition for the acoustic prediction of launch environment with the water-based sound suppression systems has already been demonstrated in our previous work~\cite{lavaSciTech2023} and further analysis on the use of the algorithm for the space launch environment simulations and other related engineering applications will be addressed in a future work.


\begin{acknowledgements}
\noindent This work was partially supported by the NASA Exploration Ground Systems (EGS) program and the NASA Engineering and Safety Center (NESC). Computer time has been provided by the NASA Advanced Supercomputing (NAS) facility at NASA Ames Research Center.
We also gratefully acknowledge Dr. Michael F. Barad and Dr. Jeffrey A. Housman for valuable discussions.
\end{acknowledgements}

{
\footnotesize
\noindent \textbf{Data availability} The datasets generated and/or analyzed during the current study are available from the corresponding author on reasonable request.
}%

\section*{Declarations}

{
\noindent \footnotesize
\textbf{Conflict of interest} The authors have no relevant financial or non-financial interests to disclose.
}%


\clearpage

\section*{Appendices}

\appendix

\section{One-dimensional material interface advection \label{appendix:1D_material_advection}}

A 1D multi-phase problem with the advection of two material interfaces is considered here. The settings of this problem are similar to those in \cite{coralic2014finite,wong2017high,aslani2018localized} but the mixture in the entire domain including the material interfaces is also at thermal equilibrium initially, instead of only at mechanical equilibrium. The initial pressure and temperature fields are uniformly at $101325\ \mathrm{Pa}$ and $298\ \mathrm{K}$ respectively. The advection velocity is $100\ \mathrm{m\ s^{-1}}$. The domain size is $x \in \left[0, 1 \right) \ \mathrm{m}$ and the fluid in the middle of the domain, $0.25 \ \mathrm{m} \leq x < 0.75 \ \mathrm{m}$, is essentially pure liquid water surrounded by essentially pure ideal air. The initial conditions are given by table~\ref{table:IC_1D_material_interface_advection_no_temp_diff}. Periodic conditions are applied at both boundaries. The final time is set at $t = 0.01 \ \mathrm{s}$.
Simulations are evolved with constant time steps. When the first order HLLC scheme is used, $\Delta t = 5.0\mathrm{e}{-8} \ \mathrm{s}$ on a uniform grid with 5000 grid points is chosen. As for the PP-WCNS-IS, $\Delta t = 6.25\mathrm{e}{-7} \ \mathrm{s}$ on a uniform grid with 500 grid points is chosen.
The two material interfaces have exactly advected one period at the end of the simulations. Thus, the exact solutions are given by the initial conditions.

\begin{figure}[!ht]
\centering
\subfigure[Density field]{%
\includegraphics[width=0.45\textwidth]{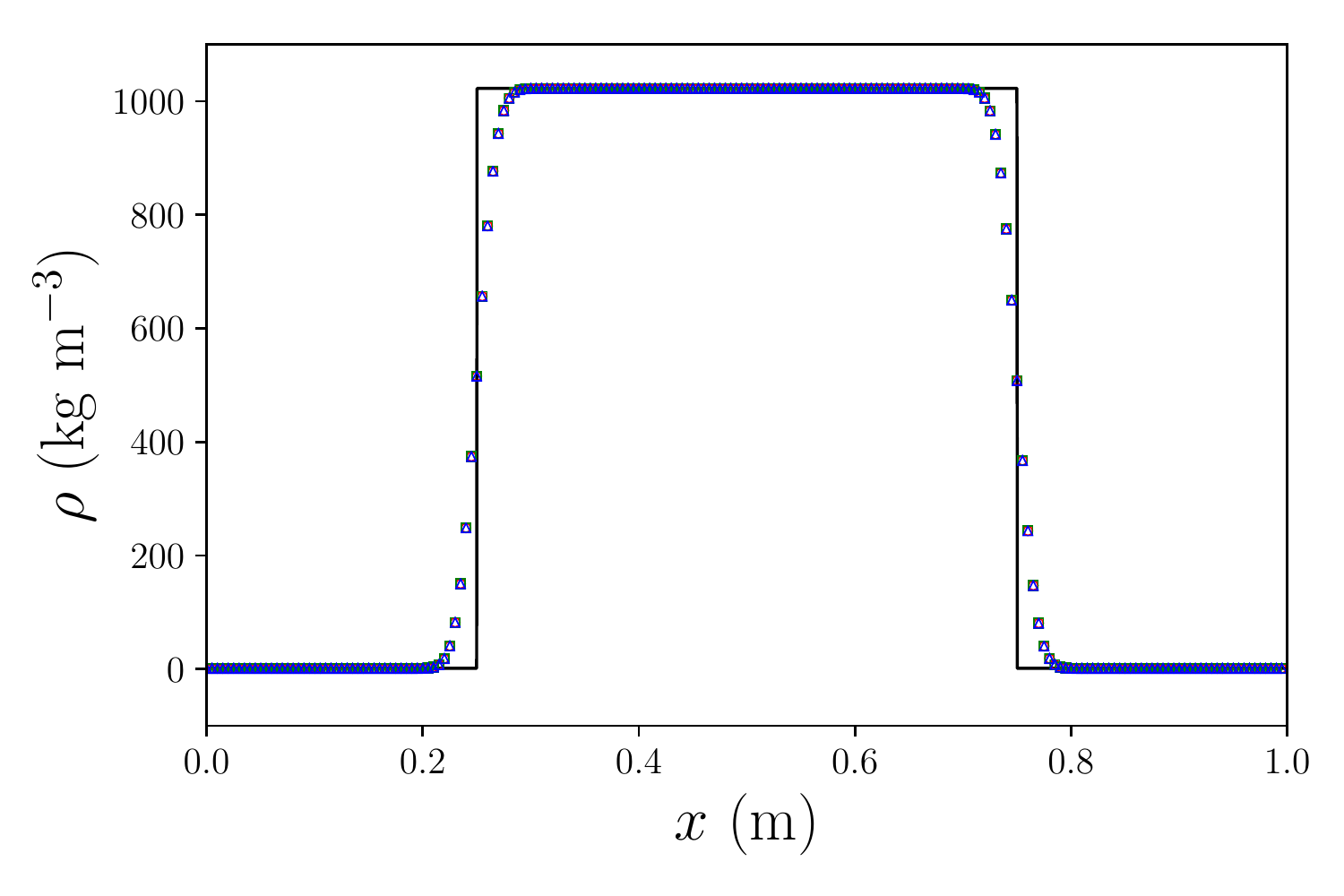}
\label{fig:compare_material_interface_advection_no_temp_diff_rho_HLLC}}
\subfigure[Velocity relative error]{%
\includegraphics[width=0.45\textwidth]{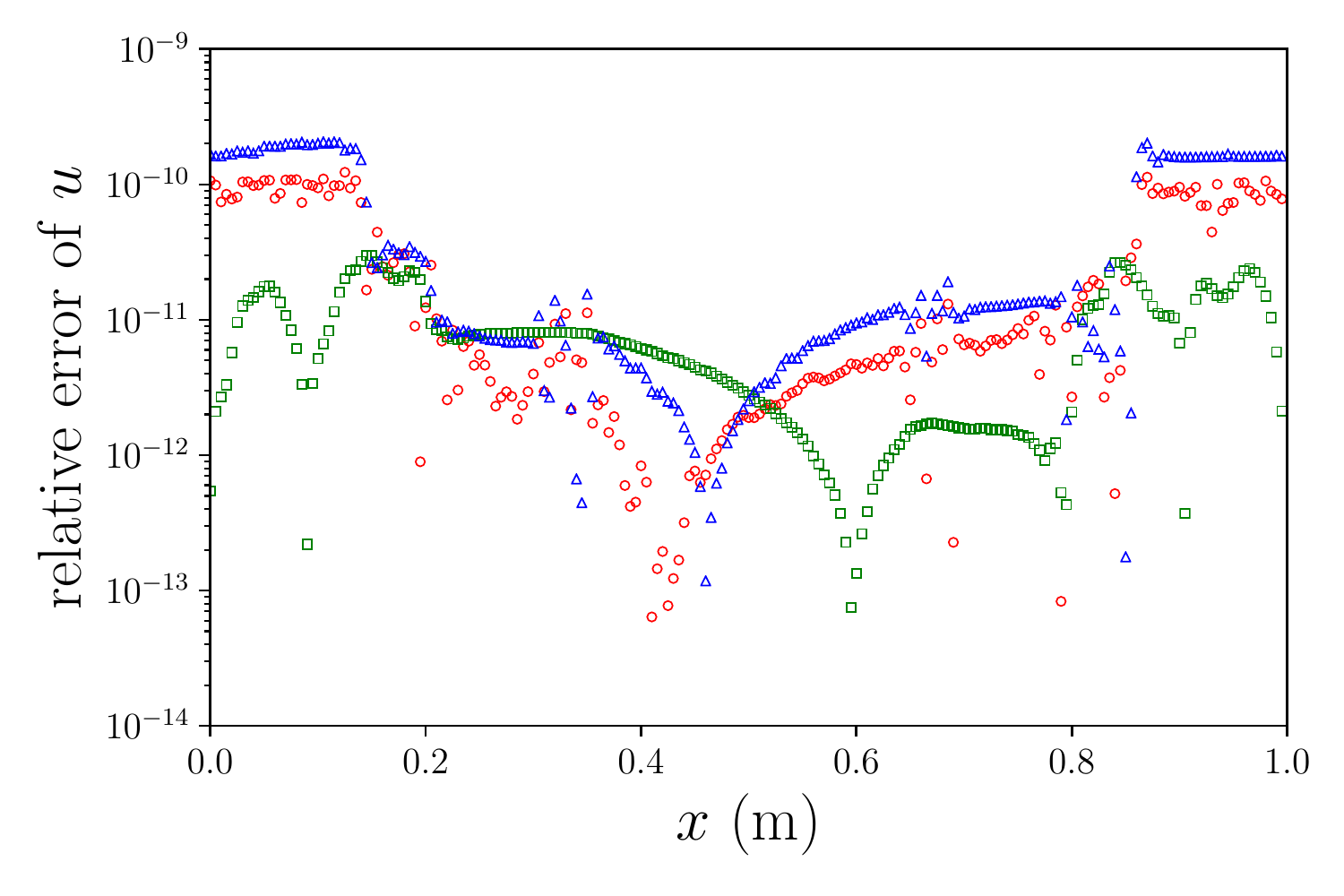}
\label{fig:compare_material_interface_advection_no_temp_diff_rel_error_u_HLLC}}
\subfigure[Pressure relative error]{%
\includegraphics[width=0.45\textwidth]{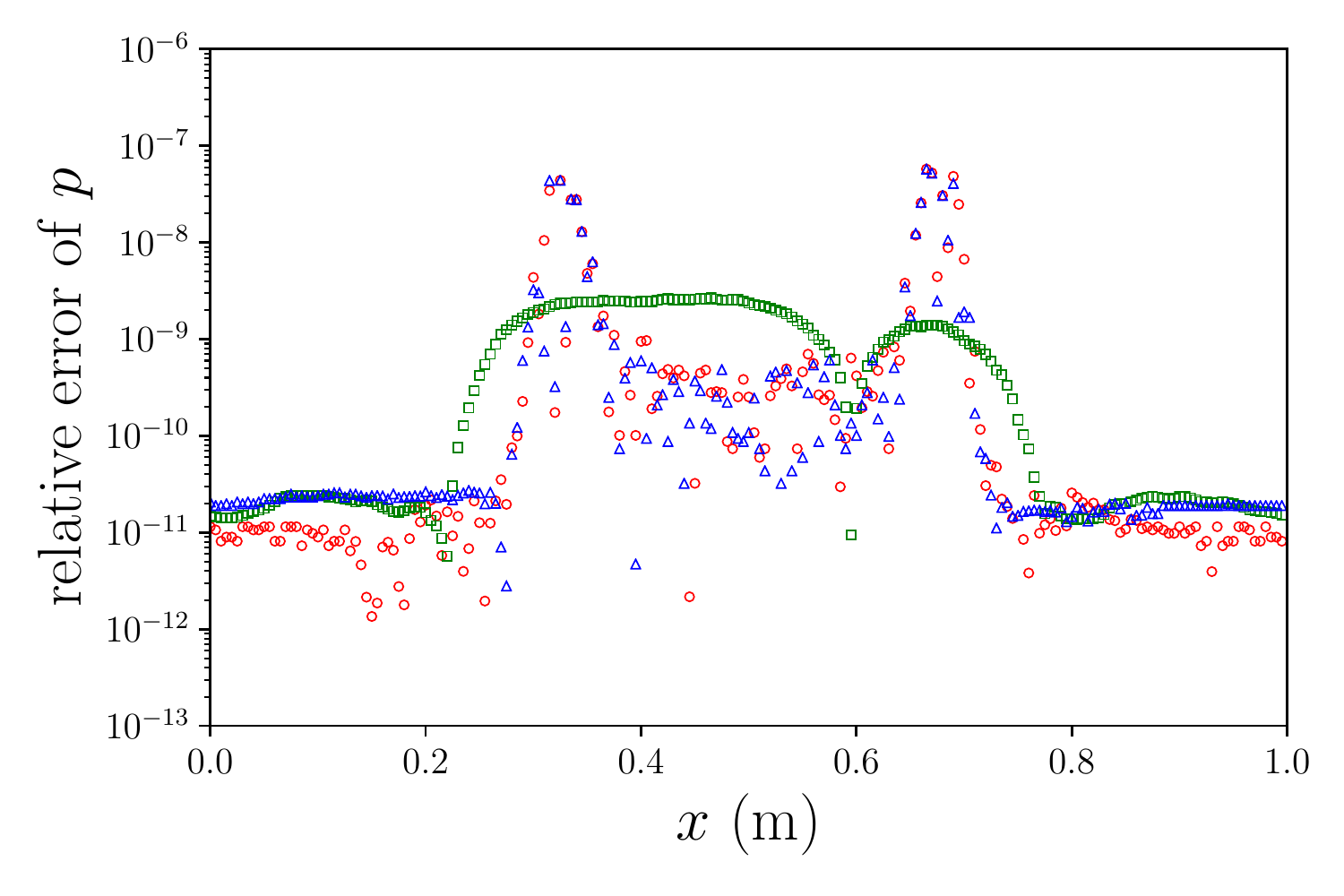}
\label{fig:compare_material_interface_advection_no_temp_diff_rel_err_p_HLLC}}
\subfigure[Temperature relative error]{%
\includegraphics[width=0.45\textwidth]{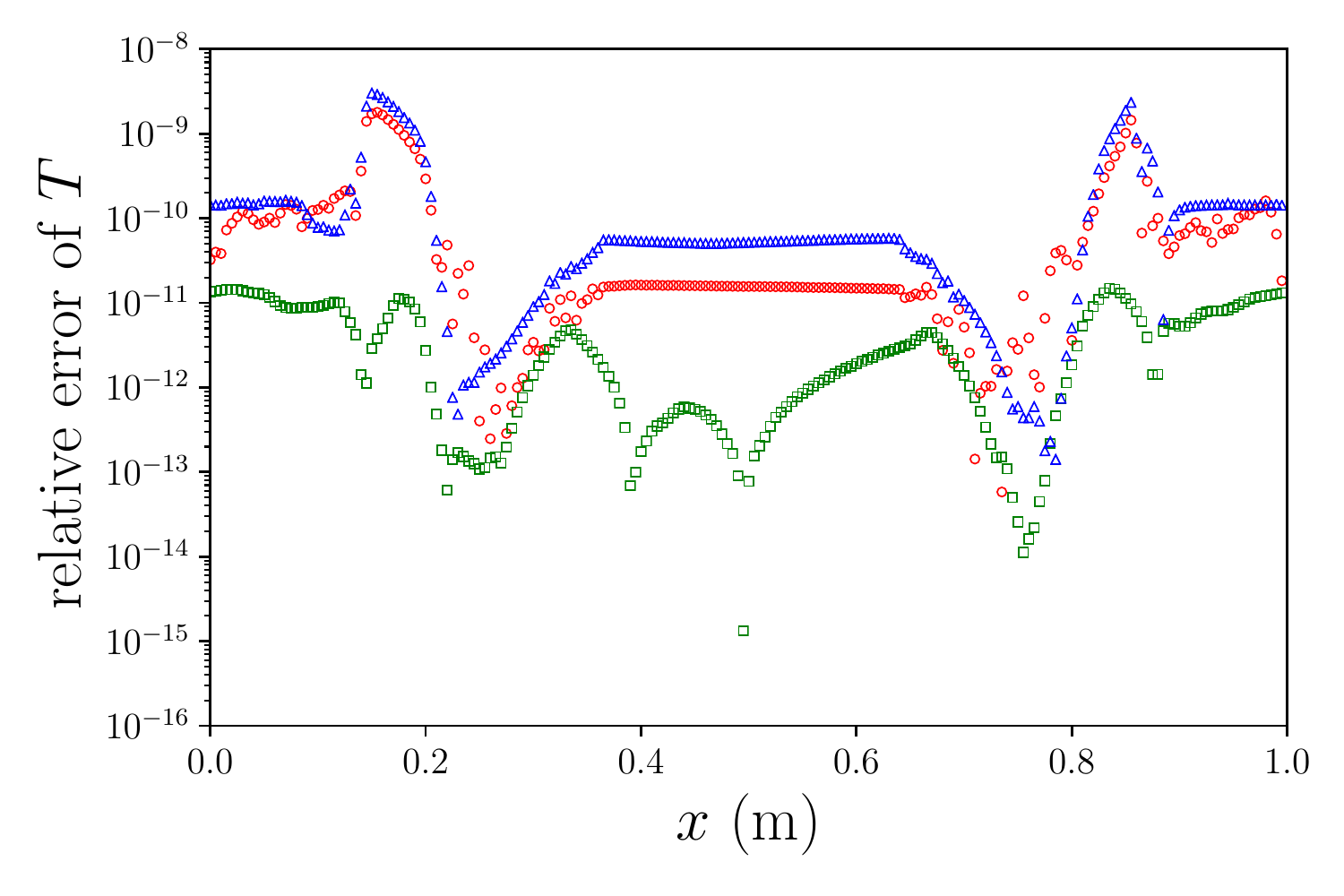}
\label{fig:compare_material_interface_advection_no_temp_diff_rel_err_T_HLLC}}
\caption{Material interface advection problem at $t = 0.01 \ \mathrm{s}$ using different flow models with the first order HLLC scheme. Black solid line: exact; red circles: four-equation HRM; green squares: five-equation model; blue triangles: five-equation model with the numerical thermal relaxation. 1 out of 25 grid points are plotted for each case.}
\label{fig:compare_material_interface_advection_no_temp_diff_HLLC}
\end{figure}

In figure~\ref{fig:compare_material_interface_advection_no_temp_diff_HLLC}, the density profiles and the relative errors in absolute values from three different numerical algorithms using the first order HLLC scheme are compared. The numerical algorithms are (1) four-equation HRM, (2) five-equation model without thermal relaxation, and (3) fractional algorithm composed of the hyperbolic time step of five-equation model and the numerical thermal relaxation step.
From the figure, it can be seen that all models capture the density discontinuities accurately. All algorithms can preserve pressure, temperature, and velocity equilibria well across the material interfaces where the relative errors in the three fields are below $1\mathrm{e}{-7}$, $1\mathrm{e}{-8}$, and $1\mathrm{e}{-9}$ respectively. 
The four-equation HRM and the fractional algorithm have larger relative errors in those three fields compared to the five-equation model without thermal relaxation as those two methods have larger overshoots and undershoots in those fields around the material interfaces.
The fractional algorithm has similar errors as the four-equation model for this advection problem since analytically the five-equation model with infinitely fast thermal relaxation is reduced to the four-equation model.

\begin{figure}[!ht]
\centering
\subfigure[Density field]{%
\includegraphics[width=0.45\textwidth]{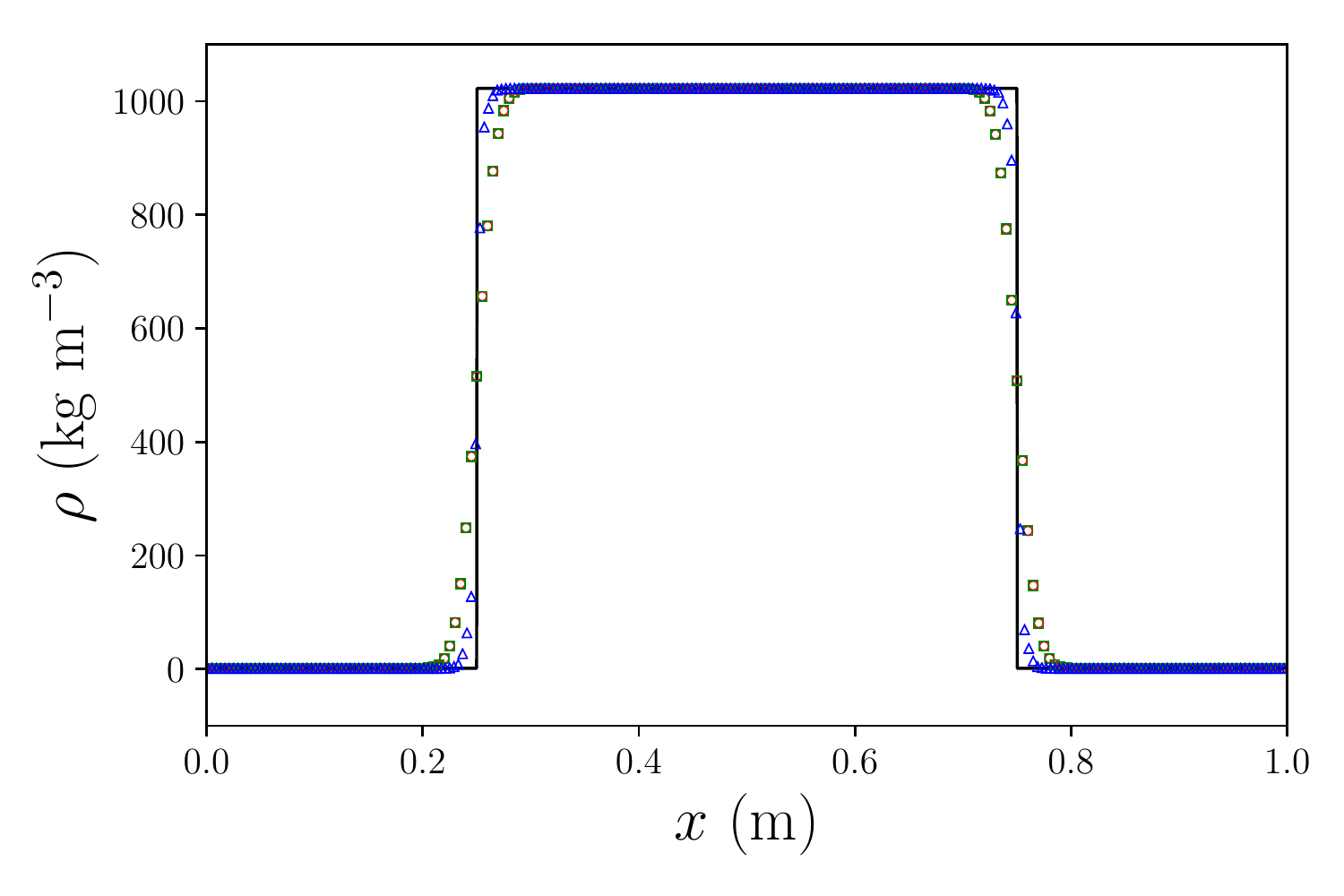}
\label{fig:compare_material_interface_advection_no_temp_diff_rho}}
\subfigure[Velocity relative error]{%
\includegraphics[width=0.45\textwidth]{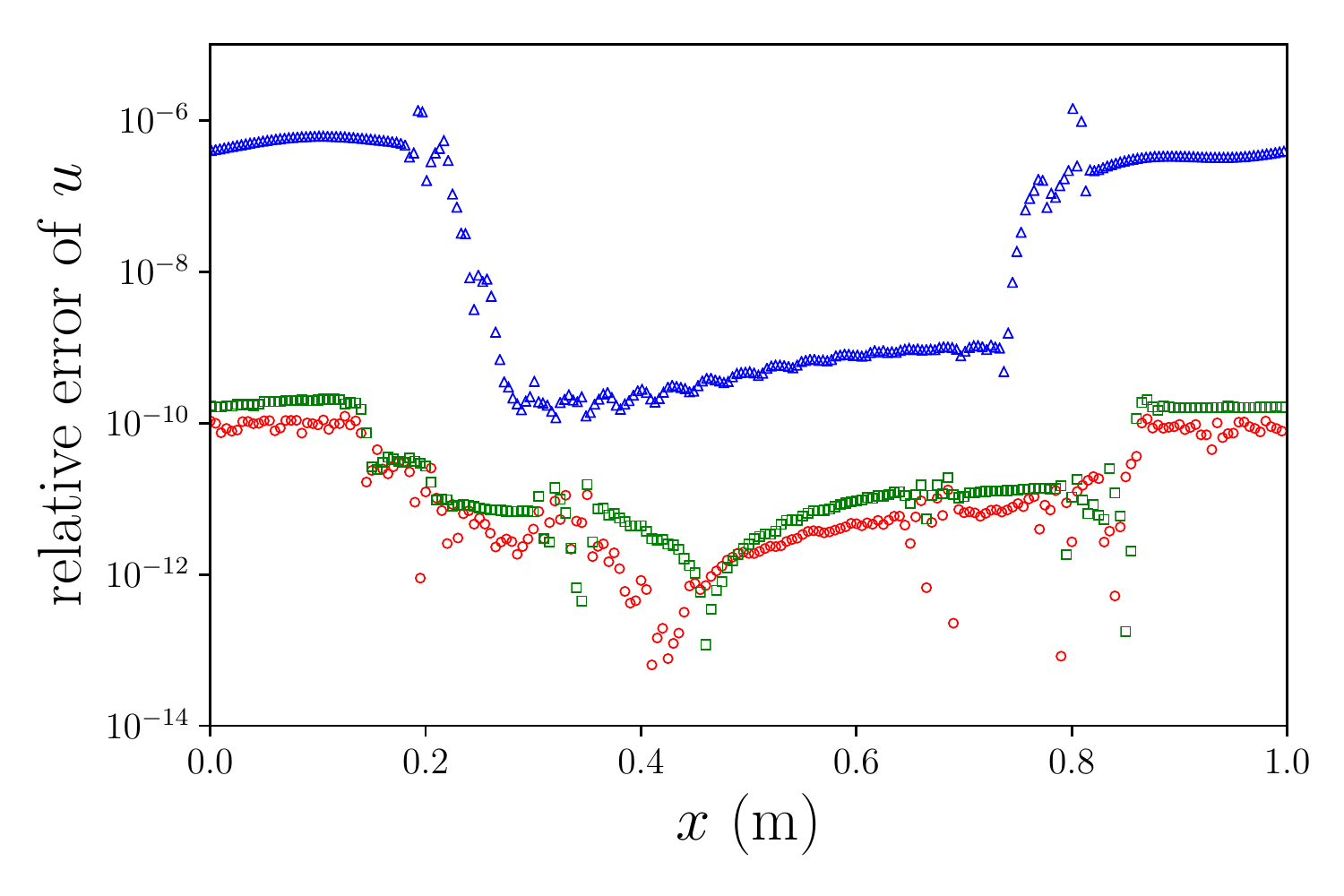}
\label{fig:compare_material_interface_advection_no_temp_diff_rel_err_u}}
\subfigure[Pressure relative error]{%
\includegraphics[width=0.45\textwidth]{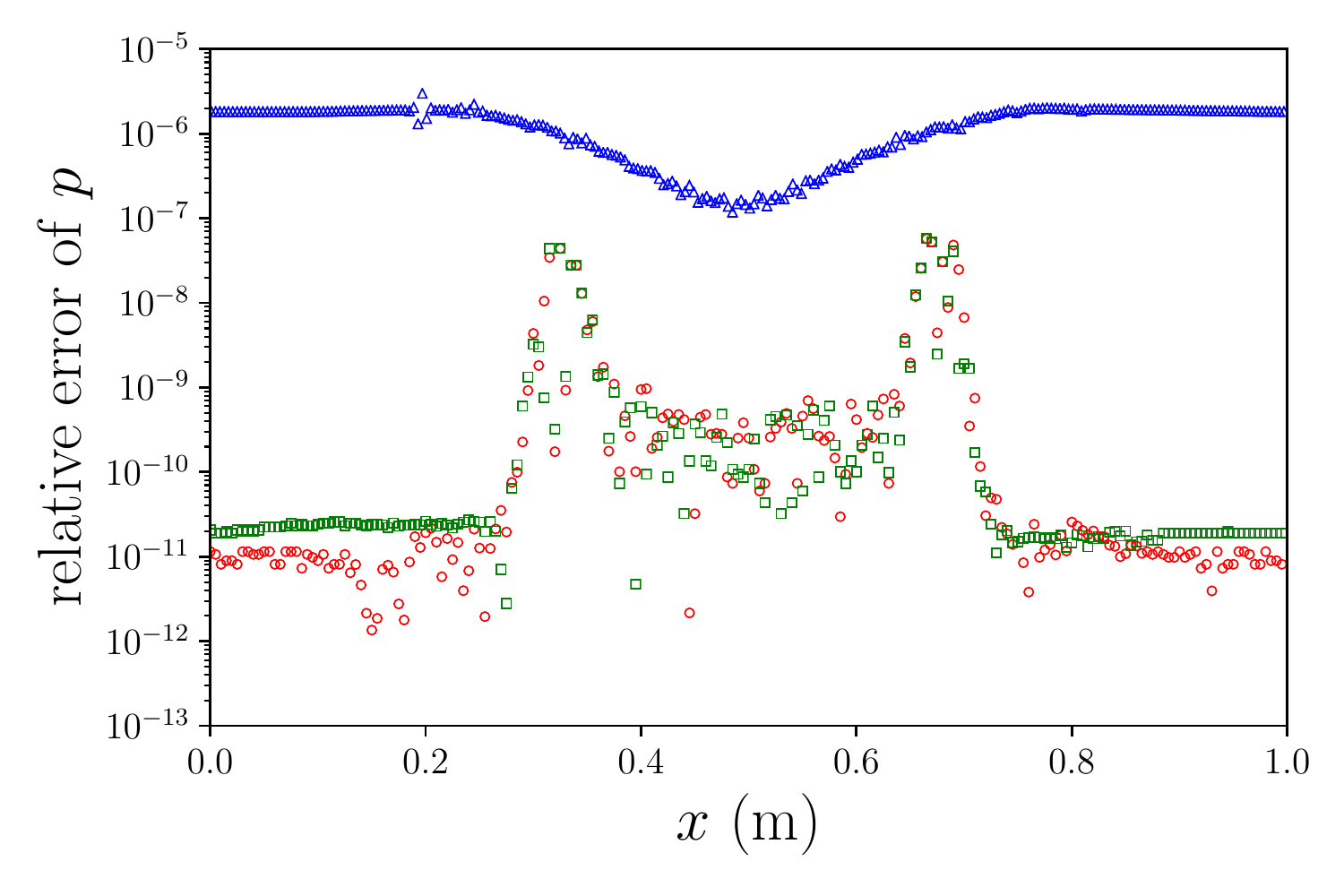}
\label{fig:compare_material_interface_advection_no_temp_diff_rel_err_p}}
\subfigure[Temperature relative error]{%
\includegraphics[width=0.45\textwidth]{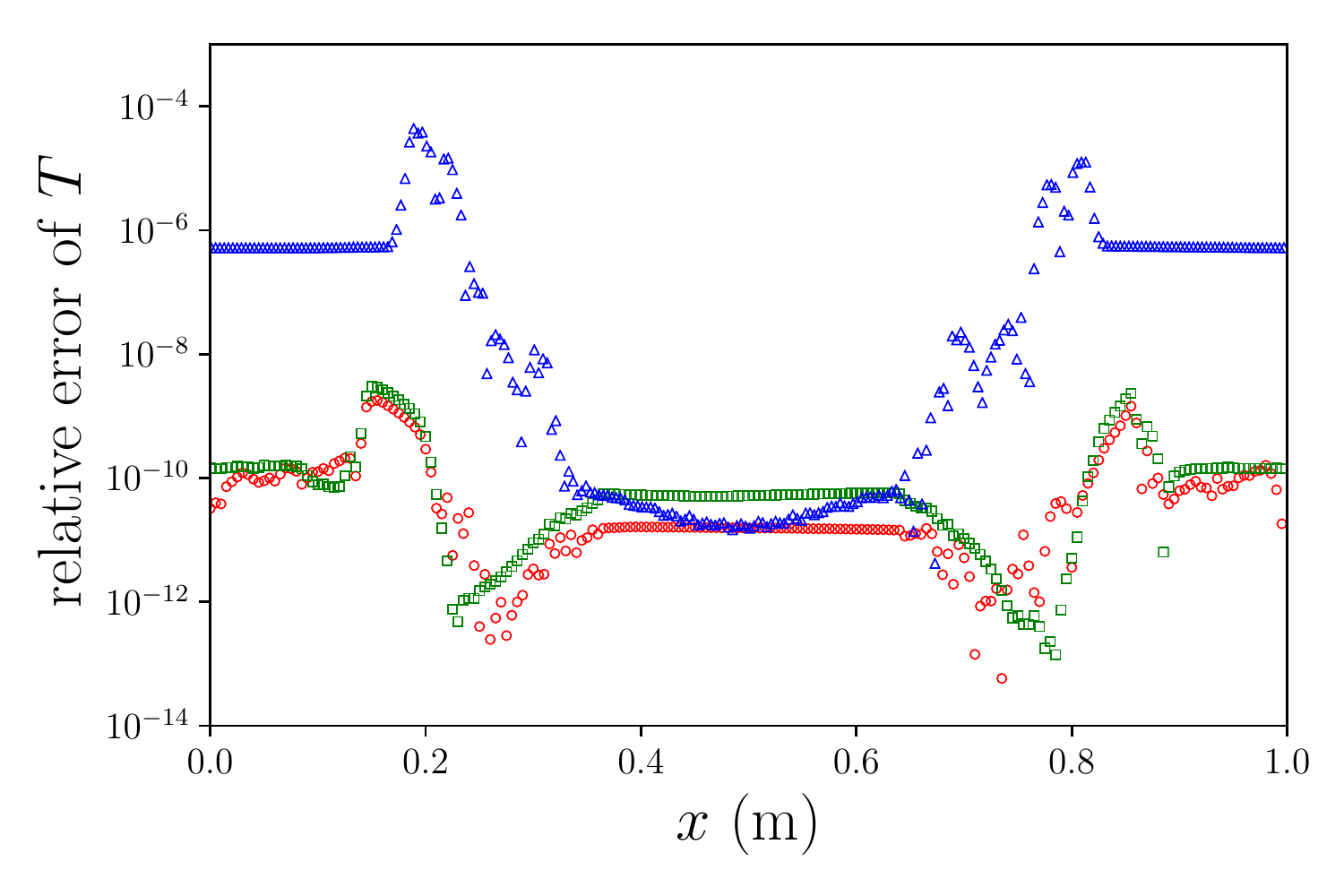}
\label{fig:compare_material_interface_advection_no_temp_diff_rel_err_T}}
\caption{Material interface advection problem at $t = 0.01 \ \mathrm{s}$ using different combinations of numerical algorithms and schemes. Black solid line: exact; red circles: four-equation HRM with the first order HLLC; green squares: five-equation model with the numerical thermal relaxation using the first order HLLC; blue triangles: five-equation model with the numerical thermal relaxation using the PP-WCNS-IS. 1 out of 25 grid points and 1 out of 2 grid points are plotted for the first order HLLC and PP-WCNS-IS respectively.}
\label{fig:compare_material_interface_advection_no_temp_diff}
\end{figure}

Figure~\ref{fig:compare_material_interface_advection_no_temp_diff} shows the comparison between the fractional algorithm with five-equation model and thermal relaxation using the PP-WCNS-IS and HLLC, and the four-equation model using HLLC. It can be seen from figure~\ref{fig:compare_material_interface_advection_no_temp_diff_rho} that the PP-WCNS-IS method can capture the density discontinuities more accurately using one order of magnitude fewer total grid points compared to the models using HLLC scheme. It is also found that there are larger relative errors in the pressure, temperature, and velocity fields when high order PP-WCNS-IS method is used. Nevertheless, the corresponding relative errors are still small which are in general around the orders of magnitude of $1\mathrm{e}{-5}$ to $1\mathrm{e}{-6}$.

\begin{table}[!ht]
  \begin{center}
    \begin{tabular}{@{}c | ccccc@{}}\toprule
    $(m)$ &
    \addstackgap{\stackanchor{$\alpha_1 \rho_1$}{$(\mathrm{kg\ m^{-3}})$}} &
    \stackanchor{$\alpha_2 \rho_2$}{$(\mathrm{kg\ m^{-3}})$} &
    \stackanchor{$u$}{$(\mathrm{m\ s^{-1}})$} &
    \stackanchor{$p$}{$(\mathrm{Pa})$} &
    $\alpha_1$ \\ \midrule
    \addstackgap{$0.25 \leq x < 0.75$} & $1.0227724310474432 \mathrm{e}{3}$ & $1.1817862272214237\mathrm{e}{-8}$ & 100 & 101325 & $1 - 1.0\mathrm{e}{-8}$ \\
    \addstackgap{otherwise} & $1.0227724412751677\mathrm{e}{-5}$ & 1.1817862094653702 & 100 & 101325 & $1.0\mathrm{e}{-8}$ \\ \bottomrule
    \end{tabular}
  \end{center}
  \caption{Initial conditions of 1D material interface advection problem.}
  \label{table:IC_1D_material_interface_advection_no_temp_diff}
\end{table}

As discussed in previous works~\cite{kawai2011high,nonomura2012numerical}, the errors and overshooting/undershooting around the material interfaces can be reduced for the conservative four-equation HRM by smoothing the initial discontinuities. The effects of the smoothing of the initial discontinuities in the 1D material advection problem discussed are analyzed here.
Volume fractions jumps are chosen to be smoothed in the initial conditions while the uniform pressure and temperature fields are maintained. The smoothing function is given by:
\begin{equation}
\begin{aligned}
    f_{s} &= \half \left[ \tanh \left( \frac{x - 0.25}{C_s \Delta x} \right) -  \tanh \left( \frac{x - 0.75}{C_s \Delta x} \right) \right], \\
    \alpha_1 &= \left( \alpha_{1,\mathrm{middle}} - \alpha_{1,\mathrm{side}} \right) f_{s} + \alpha_{1,\mathrm{side}}, \\
\end{aligned}
\end{equation}
where the subscripts ``middle" and ``side" refer to the region $0.25 \leq x < 0.75$ and the remaining regions respectively.
Test simulations are performed using the fractional algorithm with PP-WCNS-IS and different values for $C_s$ on a grid with 500 grid points.

\begin{figure}[!ht]
\centering
\subfigure[Density field]{%
\includegraphics[width=0.45\textwidth]{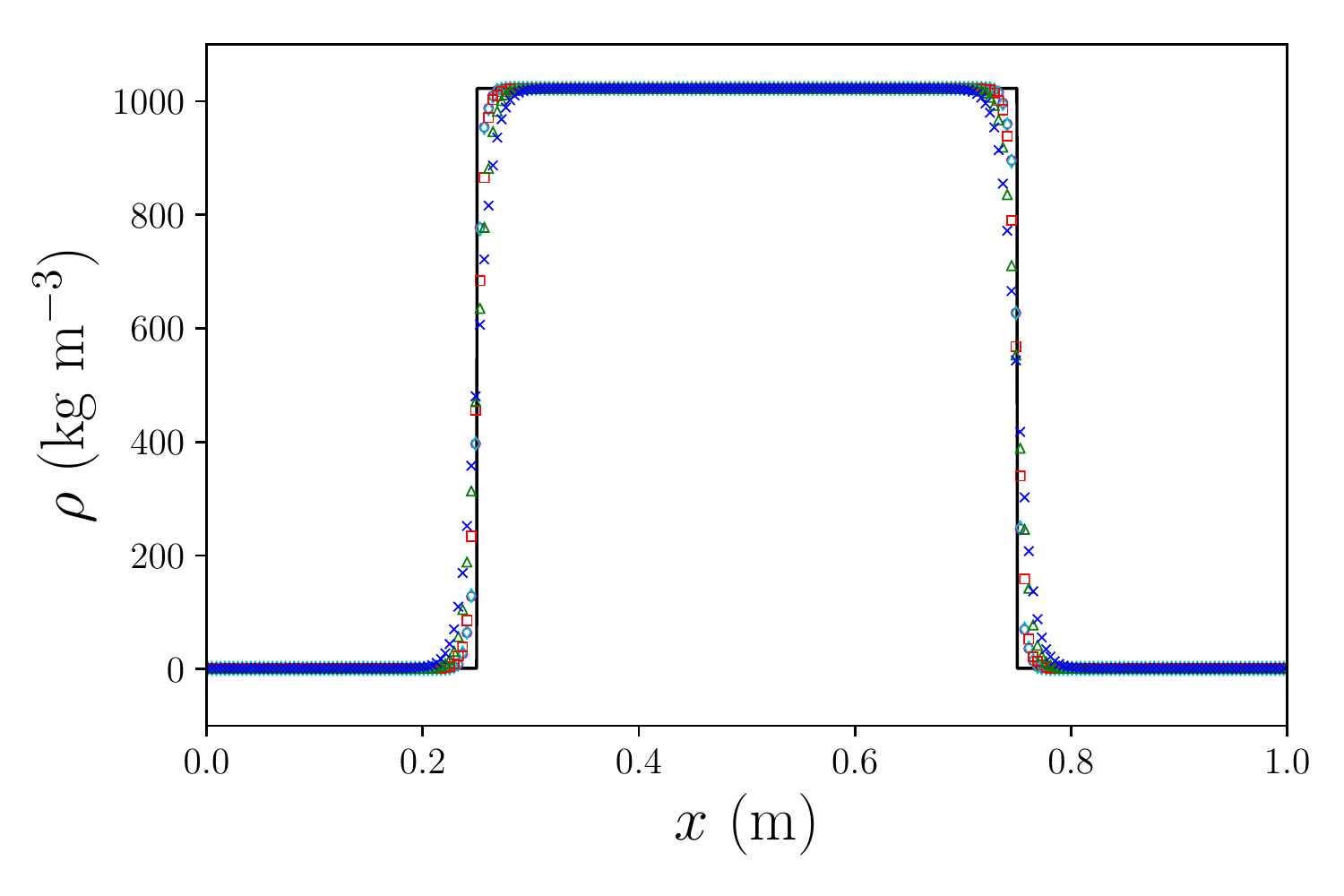}
\label{fig:compare_smoothed_material_interface_advection_no_temp_diff_rho}}
\subfigure[Velocity relative error]{%
\includegraphics[width=0.45\textwidth]{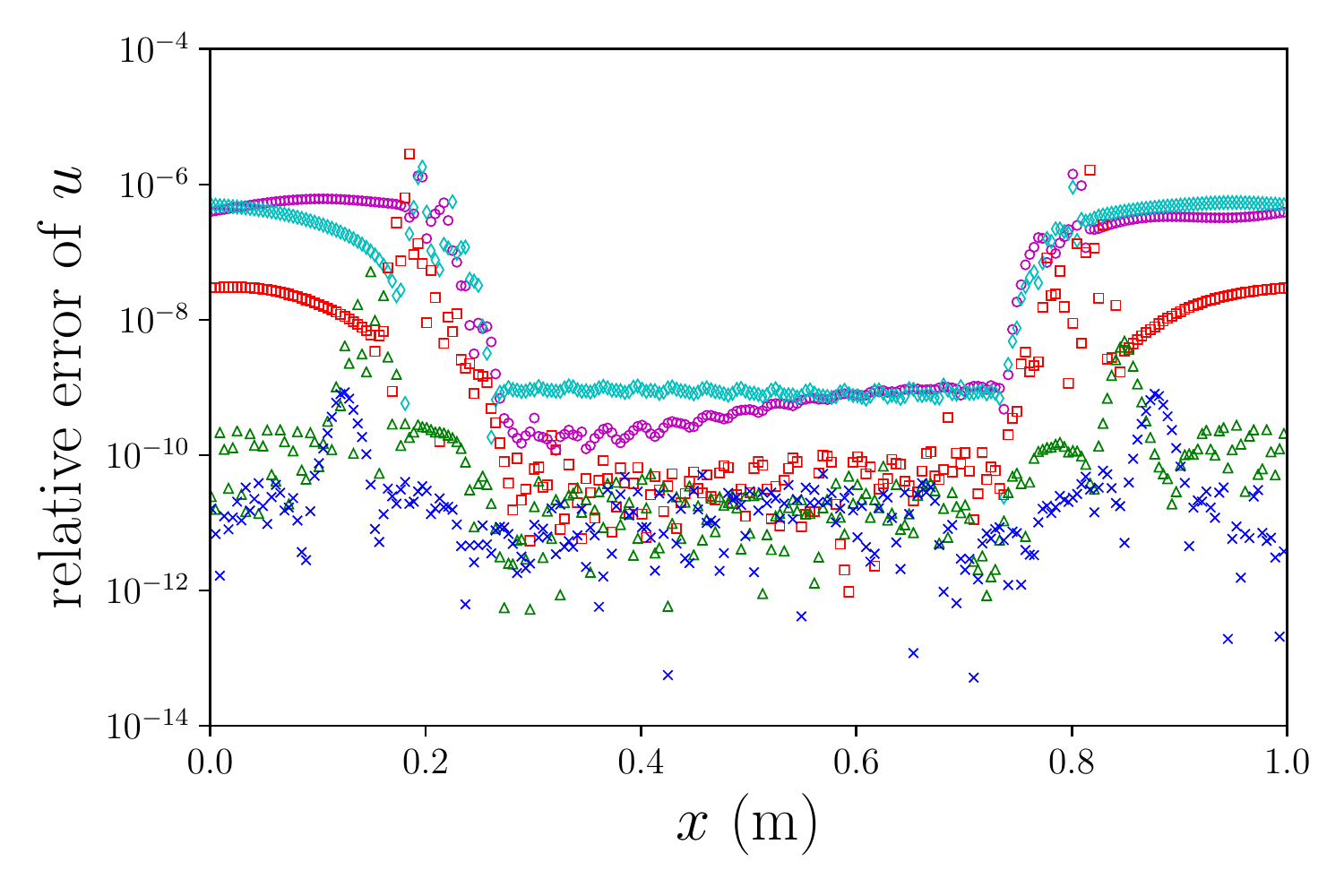}
\label{fig:compare_smoothed_material_interface_advection_no_temp_diff_rel_err_u}}
\subfigure[Pressure relative error]{%
\includegraphics[width=0.45\textwidth]{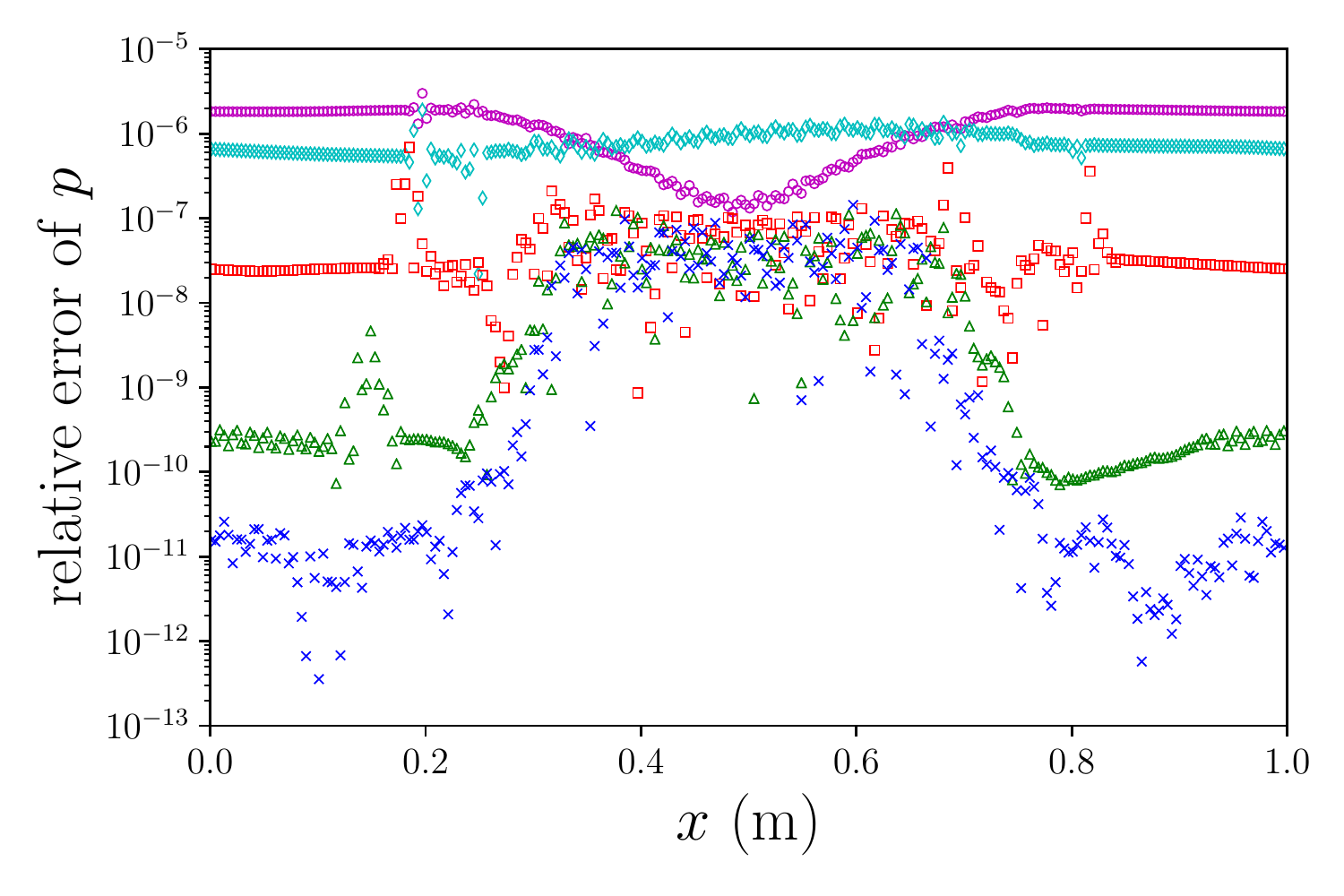}
\label{fig:compare_smoothed_material_interface_advection_no_temp_diff_rel_err_p}}
\subfigure[Temperature relative error]{%
\includegraphics[width=0.45\textwidth]{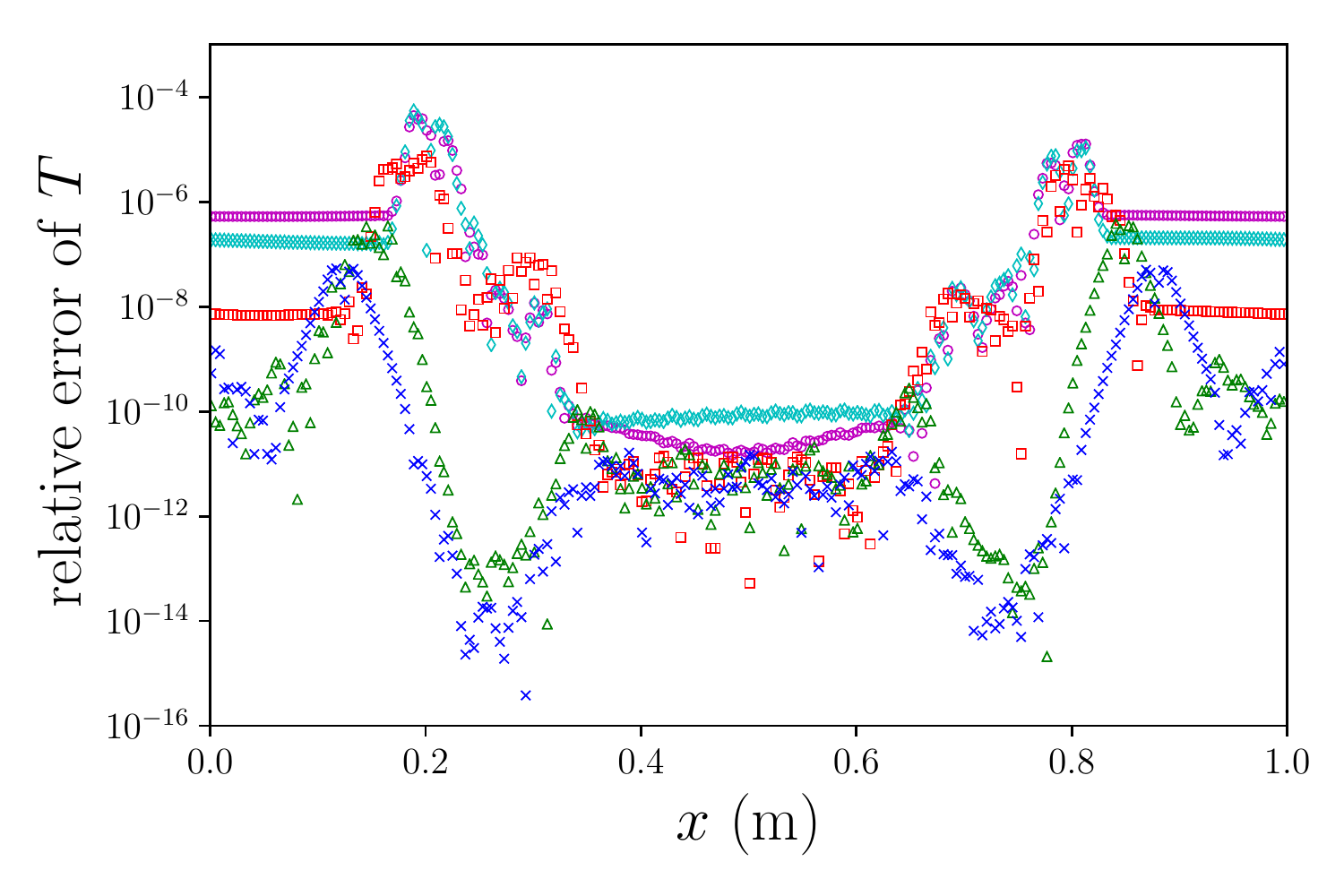}
\label{fig:compare_smoothed_material_interface_advection_no_temp_diff_rel_err_T}}
\caption{Material interface advection problem at $t = 0.01 \ \mathrm{s}$ using the PP-WCNS-IS with different initial thickness. Black solid line: exact; magenta circles: no initial smoothing; cyan diamonds: $C_s=2$; red squares: $C_s=4$; green triangles: $C_s=6$; blue crosses: $C_s=8$. 1 out of 2 grid points are plotted.}
\label{fig:compare_smoothed_material_interface_advection_no_temp_diff}
\end{figure}

Figure~\ref{fig:compare_smoothed_material_interface_advection_no_temp_diff} compares the density profiles and relative errors in absolute values of different fields after one period with different values for $C_s$.
As shown from the figure, when the initial interface is smoothed out more by increasing the user-defined parameter $C_s$, the errors in the velocity, pressure, and temperature fields given by the PP-WCNS-IS are decreasing. The relative errors in the those fields decrease by one or two orders of magnitudes when $C_s=6$ is used compared to the case with no initial smoothing. As expected, it can been seen from figure~\ref{fig:compare_smoothed_material_interface_advection_no_temp_diff_rho} that increasing $C_s$ smears the density discontinuities more. However, the smoothing effect is very mild even with $C_s=8$ compared with the case without smoothing since the shock-capturing scheme also adds dissipation to diffuse the interfaces during the simulations.

\section{Effects of flux blending in the extreme gas/liquid shock tube problem}

The blending between second order and high-order reconstructed fluxes and velocities given by equations~\eqref{eq:flux_blending} and \eqref{eq:velocity_blending} is used in all test problems for improving the accuracy and robustness near shocks. The effects of the blending are analyzed here with the extreme gas/liquid shock tube problem discussed in section~\ref{sec:1D_extreme_gas_liquid_Sod}. Two simulations using the fractional algorithm with PP-WCNS-IS are conducted with and without flux/velocity blending, where $\Gxhv_{i+1/2,j} = \hat{\mathbf{G}}^{\mathrm{HCS},x}_{i+1/2,j}$ and $\hat{u}_{i+1/2,j} = \hat{u}^{\mathrm{HCS}}_{i+1/2,j}$ are implied for the latter case. Following the previous section, $\Delta t = 8.0\mathrm{e}{-9} \ \mathrm{s}$ and a uniform grid with 1000 grid points are used. Thanks to the positivity-preserving interpolation and quasi-positivity-preserving flux limiters, no numerical failure is experience when flux/velocity blending is turned off in PP-WCNS-IS.

\begin{figure}[!ht]
\centering
\subfigure[Partial density of liquid profile]{%
\includegraphics[width=0.45\textwidth]{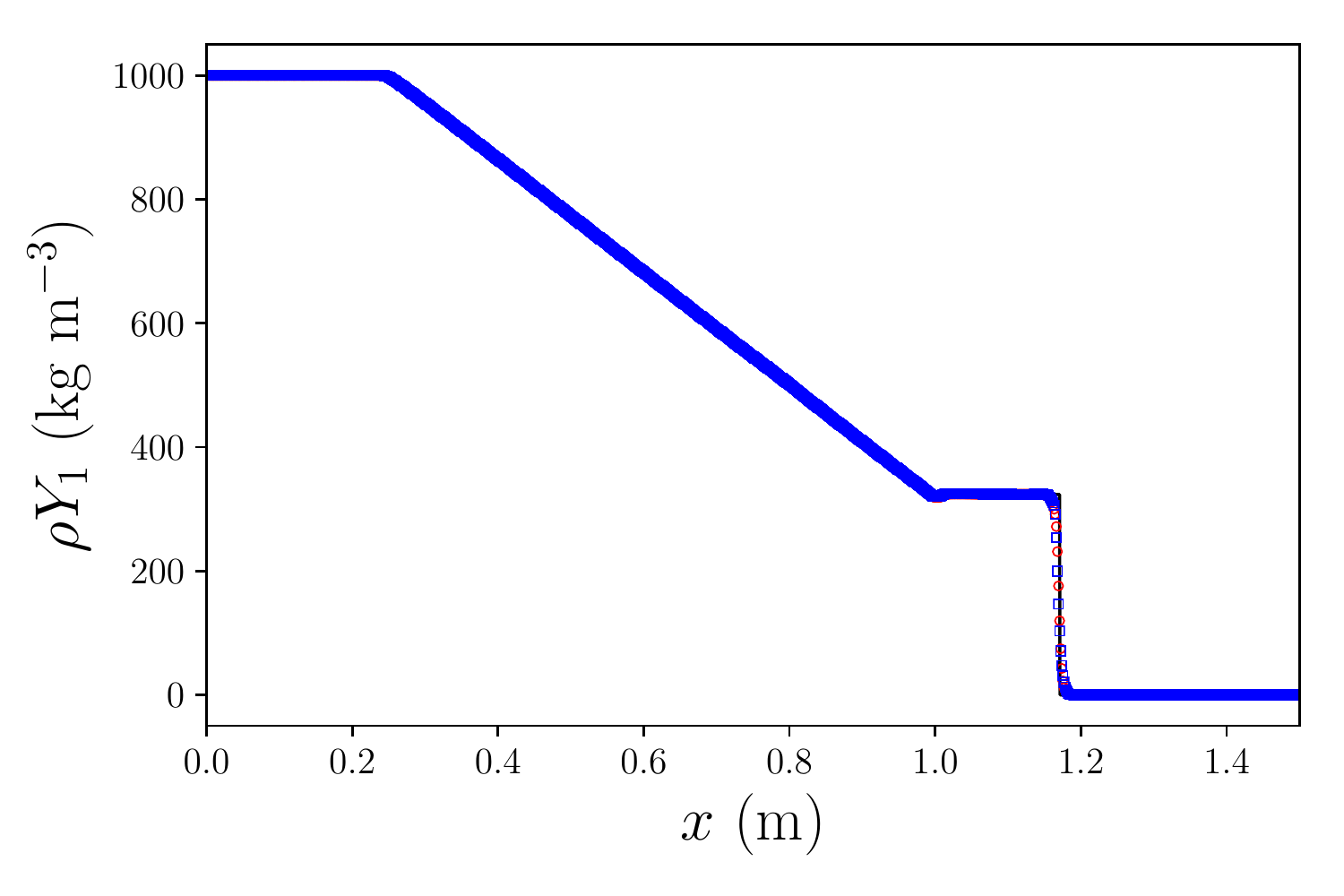}
\label{fig:compare_WCNS_extreme_gas_liquid_rhoY1_global}}
\subfigure[Partial density of gas profile]{%
\includegraphics[width=0.45\textwidth]{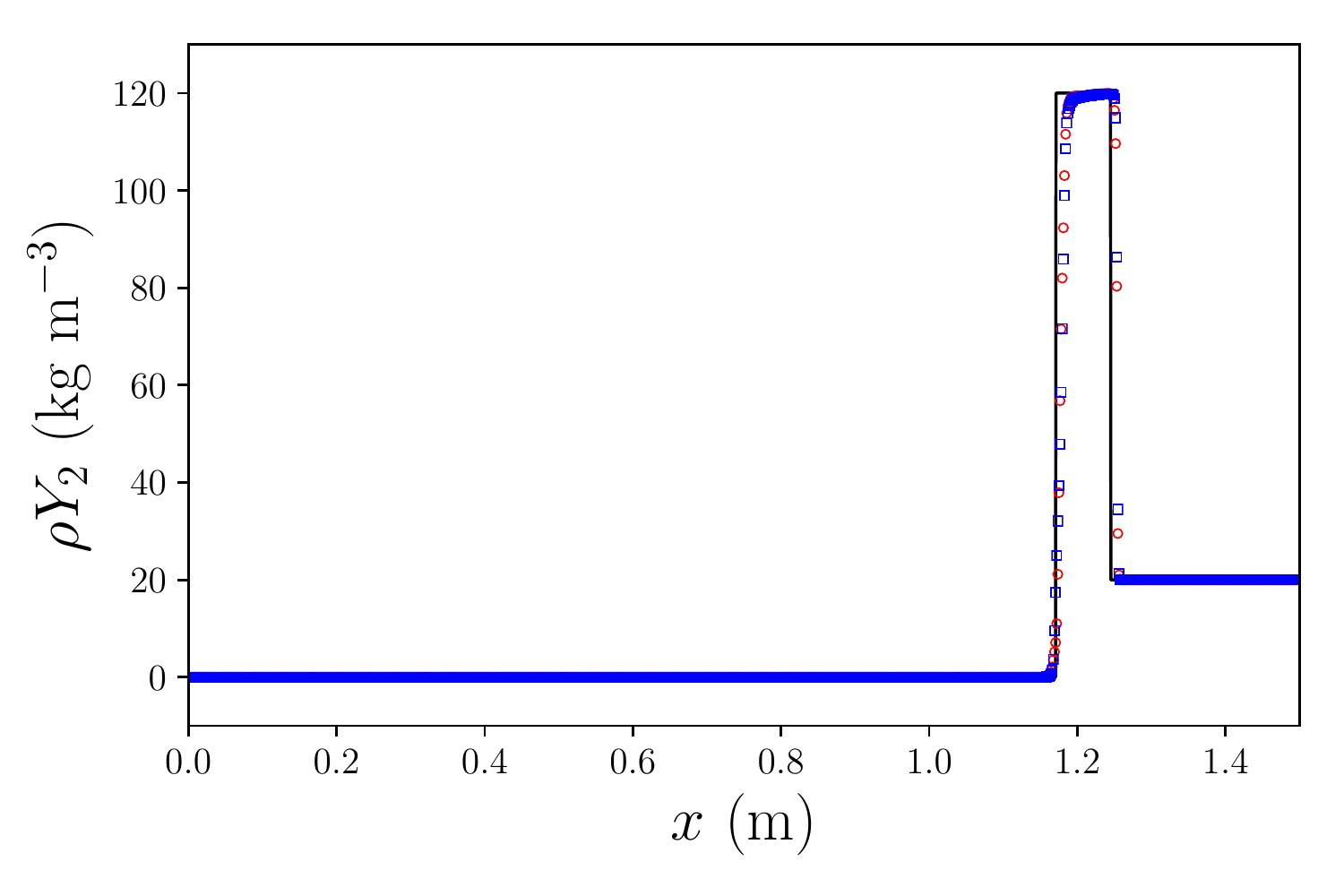}
\label{fig:compare_WCNS_extreme_gas_liquid_rhoY2_global}}
\subfigure[Velocity profile]{%
\includegraphics[width=0.45\textwidth]{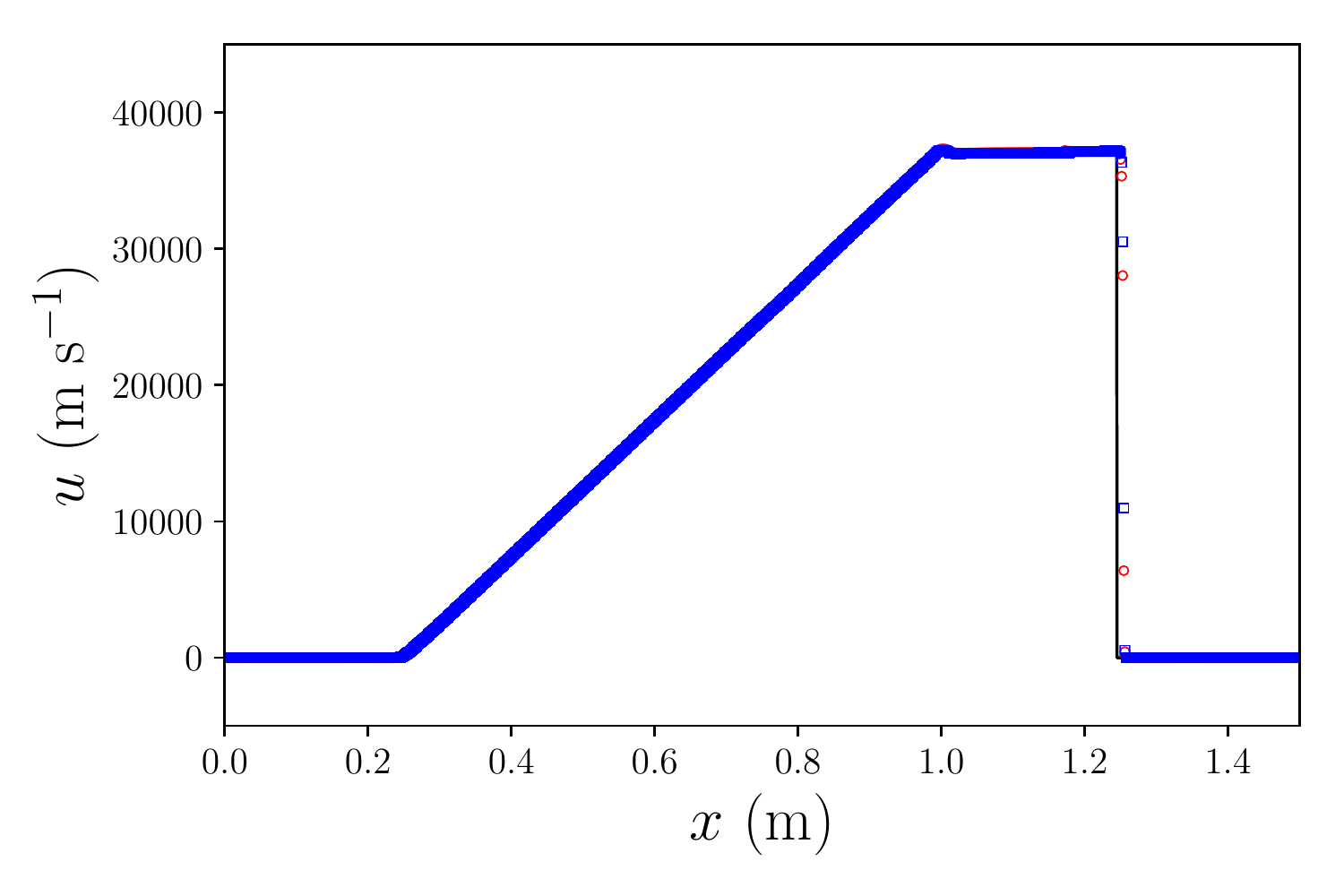}
\label{fig:compare_WCNS_extreme_gas_liquid_u_global}}
\subfigure[Pressure profile]{%
\includegraphics[width=0.45\textwidth]{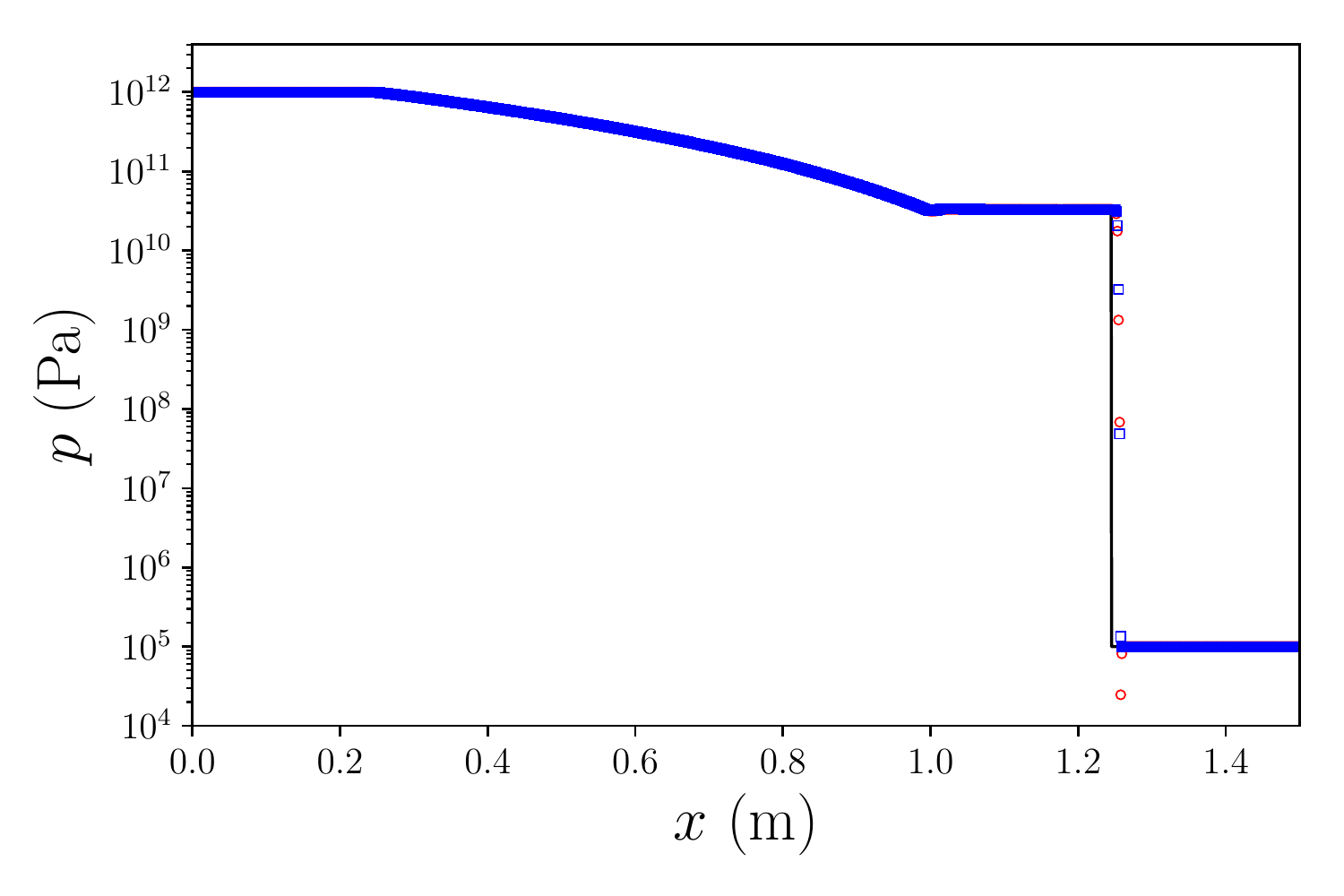}
\label{fig:compare_WCNS_extreme_gas_liquid_p_global}}
\subfigure[Temperature profile]{%
\includegraphics[width=0.45\textwidth]{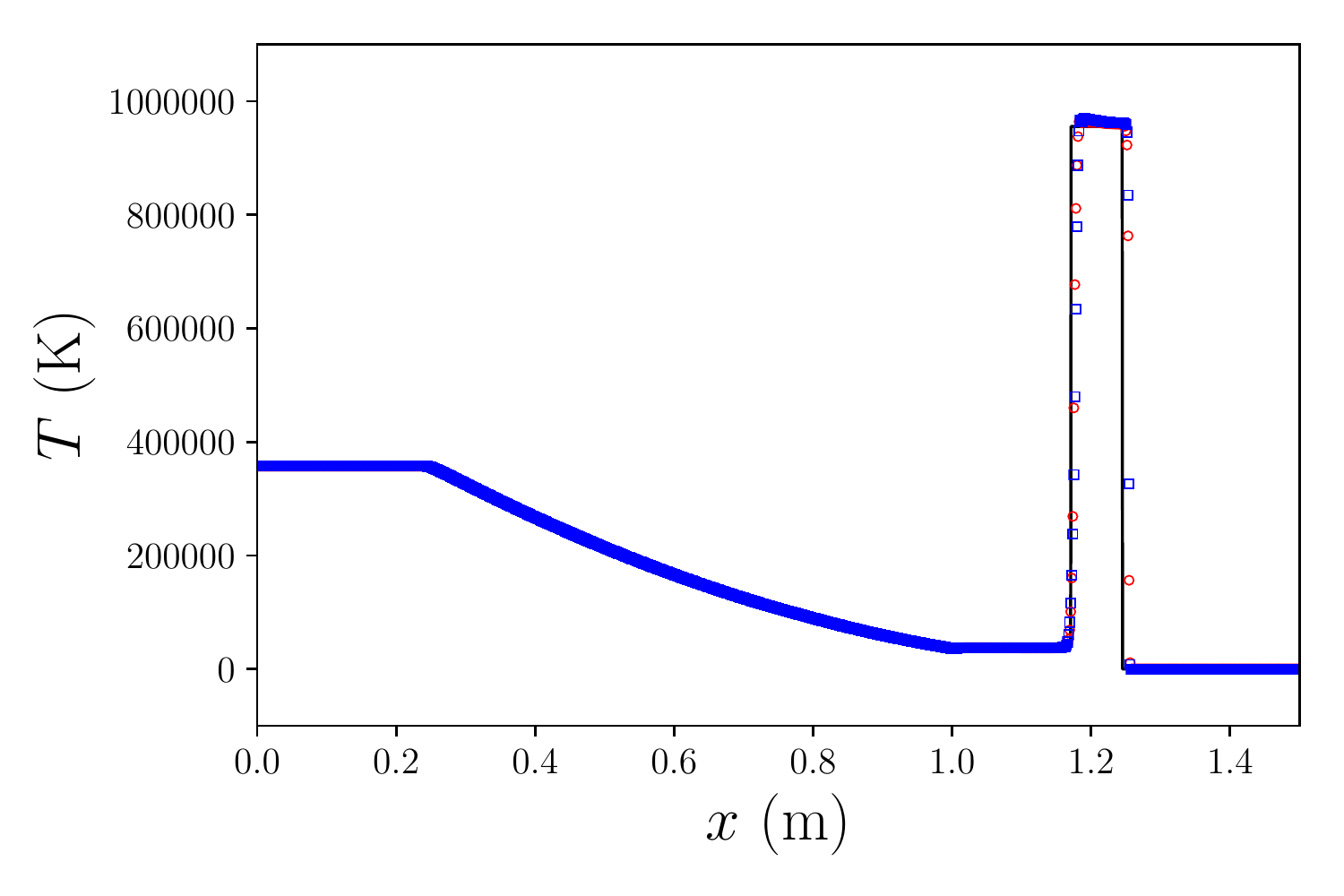}
\label{fig:compare_WCNS_extreme_gas_liquid_T_global}}
\subfigure[Volume fraction of liquid profile]{%
\includegraphics[width=0.45\textwidth]{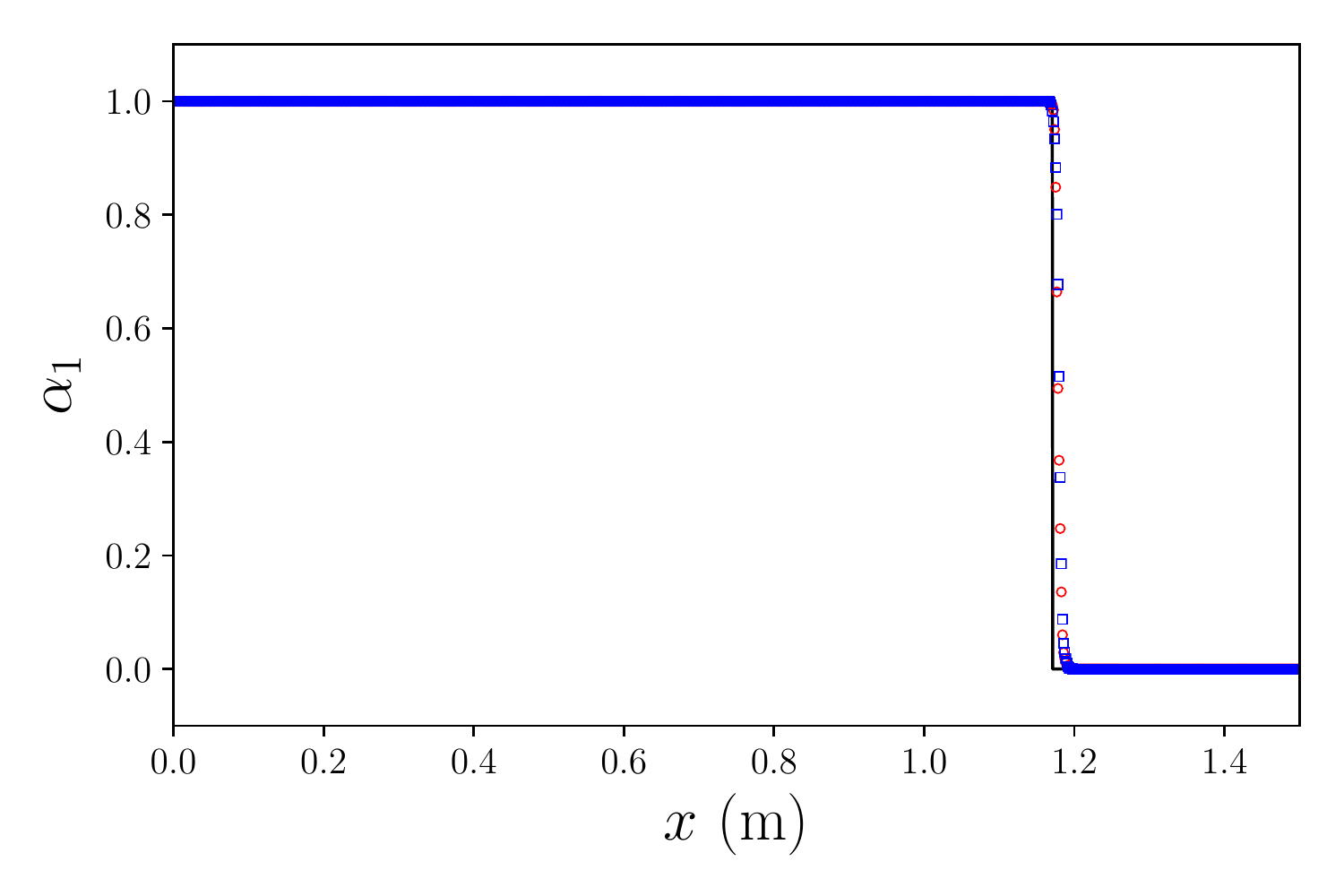}
\label{fig:compare_WCNS_extreme_gas_liquid_vol_frac_global}}
\caption{Extreme gas/liquid shock tube problem at $t = 1\mathrm{e}{-5} \ \mathrm{s}$ using the PP-WCNS-IS without/with flux blending. Black solid line: exact; red circles: without flux blending; blue squares: with flux blending.}
\label{fig:compare_WCNS_extreme_gas_liquid}
\end{figure}

Figure~\ref{fig:compare_WCNS_extreme_gas_liquid} compares the numerical solutions obtained with and without the flux/velocity blending. It can be seen that except the pressure field, all other fields do not have much sensitivities on the blending. Clearly, the flux/velocity blending eliminates a large undershoot in the pressure field that can be seen near the shock in the solutions when blending is completely turned off. 

\section{Incremental-stencil interpolation \label{appendix:weno_IS}}

The incremental-stencil interpolation approximates the midpoint values for the computation of the flux vector and velocity at midpoints with nonlinear weighting of interpolated values from different sub-stencils. For simplicity, only the interpolation of left-biased midpoint values in a 1D domain is presented. The interpolation of right-biased midpoint values is similar due to symmetry. Consider the left-biased interpolated values, $\tilde{u}_{i+1/2}^{k}$, of a variable $u$ with different sub-stencils $S_k$:
\begin{align}
    S_0: \quad \tilde{u}_{i+\half}^{0} =&
      \frac{1}{2}\left(u_i + u_{i+1} \right),
      \label{eq:interpolation_stencil_0} \\
    S_1: \quad \tilde{u}_{i+\half}^{1} =&
      \frac{1}{2}\left(-u_{i-1} + 3u_{i} \right),
      \label{eq:interpolation_stencil_1} \\
    S_2: \quad \tilde{u}_{i+\half}^{2} =&
      \frac{1}{8}\left(3u_{i} + 6u_{i+1} - u_{i+2} \right),
      \label{eq:interpolation_stencil_2} \\
    S_3: \quad \tilde{u}_{i+\half}^{3} =&
      \frac{1}{8}\left(3u_{i-2} - 10u_{i-1} + 15u_{i} \right).
      \label{eq:interpolation_stencil_3}
\end{align}
The interpolated values from stencils $S_0$ and $S_1$ are second order accurate and those from stencils $S_2$ and $S_3$ are third order accurate. 
As explained in a previous section, the primitive variables projected to the characteristic fields are chosen in the interpolation process for better robustness.

The left-biased fifth order 5-point linear interpolation with full stencil $S_5$ can be obtained from the linear combination of the interpolations with different sub-stencils:
\begin{equation}
    S_5: \quad \tilde{u}_{i+\half}^{5} =
      \sum_{k=0}^{3} d_k \tilde{u}_{i+\half}^{k},
      \label{eq:WCNS_stencil_5}
\end{equation}
where the linear weights are given by:
\begin{equation}
    d_0 = \frac{15}{32}, \quad
    d_1 = \frac{5}{32}, \quad
    d_2 = \frac{5}{16}, \quad
    d_3 = \frac{1}{16}.
\end{equation}

In order to reduce the formal order of accuracy for robustness near discontinuities, a nonlinear WENO interpolation is employed through the replacement of the linear weights with the incremental-stencil nonlinear weights~\cite{wong2021positivity}:
\begin{equation}
    \tilde{u}_{i+\half} =
    \sum_{k=0}^{3} \omega_k \tilde{u}_{i+\half}^{k} ,
    \label{eq:WENO_nonlinear_interpolation}
\end{equation}
where the nonlinear weights have the following forms~\cite{wang2018incremental,wong2021positivity}:
\begin{align}
    \omega_k &= \frac{\eta_k}{\sum_{s=0}^{3} \eta_s}, \\
    \eta_k &=
      \left\{ \begin{array}{ll} 
        d_k \left( 1 + \frac{\tau_5}{\beta_k + \epsilon}
          \cdot \frac{\tau_5}{\beta_{01} + \epsilon} \right),
          & \text{if }  k<2,  \\ 
        d_k \left( 1 + \frac{\tau_5}{\beta_k + \epsilon} \right),
          & \text{otherwise} .
      \end{array}\right.
\end{align}
$\epsilon = 1.0\mathrm{e}{-15}$ is used to avoid division by zero.
The smoothness indicators $\beta_k$ and $\beta_{01}$ are given by:
\begin{align}
    \beta_0 &= \left( u_{i} - u_{i+1} \right)^2, \label{eq:beta_0_IS} \\
    \beta_1 &= \left( u_{i-1} - u_{i} \right)^2, \label{eq:beta_1_IS} \\
    \beta_{01} &= \frac{13}{12} \left( u_{i-1} - 2u_{i} + u_{i+1} \right)^2 +
        \frac{1}{4} \left( u_{i-1} - u_{i+1} \right)^2, \label{eq:beta_01_IS} \\
    \beta_2 &= \frac{13}{12} \left( u_{i} - 2u_{i+1} + u_{i+2} \right)^2 +
        \frac{1}{4} \left( 3u_{i} - 4u_{i+1} + u_{i+2} \right)^2, \label{eq:beta_2_IS} \\
    \beta_3 &= \frac{13}{12} \left( u_{i-2} - 2u_{i-1} + u_{i} \right)^2 +
        \frac{1}{4} \left( u_{i-2} - 4u_{i-1} + 3u_{i} \right)^2 . \label{eq:beta_3_IS} 
\end{align}
The reference smoothness indicator $\tau_5$ is defined as:
\begin{equation}
    \tau_5 = \frac{13}{12} \left( u_{i+2} - 4u_{i+1} + 6u_{i} - 4u_{i-1} + u_{i-2} \right)^2
      + \frac{1}{4} \left( u_{i+2} - 2u_{i+1} + 2u_{i-1} - u_{i-2} \right)^2 . \label{eq:tau_5_IS}
\end{equation}


\section{Characteristic decomposition \label{appendix:char_decomp}}

The variables used for WENO interpolation and reconstruction are very critical to avoid spurious oscillations near discontinuities, especially around the material interfaces. Primitive variables that include the pressure and velocity instead of conservative variables were suggested~\cite{johnsen2006implementation,coralic2014finite,wong2017high} for any interpolation or reconstruction in order to better maintain the pressure and velocity equilibria across material interfaces. Moreover, WENO reconstruction and interpolation of characteristic variables projected from primitive variables can avoid the interaction of discontinuities in different characteristic fields and improve the robustness.
To show how the primitive variables can be projected to the characteristic fields,
the 2D governing equations in the quasi-linear primitive form for the hyperbolic step are first considered by following previous works~\cite{coralic2014finite,wong2017high,wong2021positivity}:
\begin{equation} \label{eq:quasi-conservative_eqn}
    \frac{\partial{\Vv}}{\partial{t}} + \Av(\Vv)\frac{\partial{\Vv}}{\partial{x}} + \Bv(\Vv) \frac{\partial{\Vv}}{\partial{y}} = 0,
\end{equation}
where $\Vv$ is the vector of primitive variables and is defined as:
\begin{equation}
    \Vv = \left(   \alpha_1 \rho_1 \
                    \alpha_2 \rho_2 \
                    \cdots \
                    \alpha_N \rho_N \
                    u \
                    v \
                    p \
                    \alpha_1 \
                    \alpha_2 \
                    \cdots \
                    \alpha_{N-1} \right) ^{T}
\end{equation}
The matrices $\Av$ and $\Bv$ are given by:
\begin{equation}
    \Av = \begin{pmatrix} 
                    u      & 0      & \cdots & 0      & \alpha_1 \rho_1 & 0      & 0              & 0      & 0      & \cdots & 0  \\
                    0      & u      & \cdots & 0      & \alpha_2 \rho_2 & 0      & 0              & 0      & 0      & \cdots & 0  \\
                    \vdots & \vdots & \ddots & \vdots & \vdots          & \vdots & \vdots         & \vdots & \vdots & \ddots & \vdots  \\
                    0      & 0      & \cdots & u      & \alpha_N \rho_N & 0      & 0              & 0      & 0      & \cdots & 0  \\
                    0      & 0      & \cdots & 0      & u               & 0      & \frac{1}{\rho} & 0      & 0      & \cdots & 0  \\
                    0      & 0      & \cdots & 0      & 0               & u      & 0              & 0      & 0      & \cdots & 0  \\
                    0      & 0      & \cdots & 0      & \rho c^2        & 0      & u              & 0      & 0      & \cdots & 0  \\
                    0      & 0      & \cdots & 0      & 0               & 0      & 0              & u      & 0      & \cdots & 0  \\
                    0      & 0      & \cdots & 0      & 0               & 0      & 0              & 0      & u      & \cdots & 0  \\
                    \vdots & \vdots & \ddots & \vdots & \vdots          & \vdots & \vdots         & \vdots & \vdots & \ddots & \vdots  \\
                    0      & 0      & \cdots & 0      & 0               & 0      & 0              & 0      & 0      & \cdots & u
                    \end{pmatrix}, \quad
    \Bv = \begin{pmatrix} 
                    v      & 0      & \cdots & 0      & 0      & \alpha_1 \rho_1 & 0              & 0      & 0      & \cdots & 0 \\
                    0      & v      & \cdots & 0      & 0      & \alpha_2 \rho_2 & 0              & 0      & 0      & \cdots & 0 \\
                    \vdots & \vdots & \ddots & \vdots & \vdots & \vdots          & \vdots         & \vdots & \vdots & \ddots & \vdots \\
                    0      & 0      & \cdots & v      & 0      & \alpha_N \rho_N & 0              & 0      & 0      & \cdots & 0 \\
                    0      & 0      & \cdots & 0      & v      & 0               & 0              & 0      & 0      & \cdots & 0 \\
                    0      & 0      & \cdots & 0      & 0      & v               & \frac{1}{\rho} & 0      & 0      & \cdots & 0 \\
                    0      & 0      & \cdots & 0      & 0      & \rho c^2        & v              & 0      & 0      & \cdots & 0 \\
                    0      & 0      & \cdots & 0      & 0      & 0               & 0              & v      & 0      & \cdots & 0 \\
                    0      & 0      & \cdots & 0      & 0      & 0               & 0              & 0      & v      & \cdots & 0 \\
                    \vdots & \vdots & \ddots & \vdots & \vdots & \vdots          & \vdots         & \vdots & \cdots & \ddots & \vdots \\
                    0      & 0      & \cdots & 0      & 0      & 0               & 0              & 0      & 0      & \cdots & v
                    \end{pmatrix} .
\end{equation}
The eigenvectors of the matrices $\Av$ and $\Bv$ have to be determined first in order to transform primitive variables to characteristic variables. The eigenvalue decompositions of the matrices are given by:
\begin{equation}
    \Av = \Rv_{\Av} \mathbf{\Lambda}_\Av \Rv_{\Av}^{-1}, \quad
    \Bv = \Rv_{\Bv} \mathbf{\Lambda}_\Bv \Rv_{\Bv}^{-1},
\end{equation}
where $\Rv_{\Av}$, $\Rv_{\Av}^{-1}$ and $\mathbf{\Lambda}_\Av$ are given by:
\begin{equation}
\begin{split}
    \Rv_{\Av} &= \begin{pmatrix}
        -\frac{\alpha_1 \rho_1}{2c} & 1      & 0      & \cdots & 0      & 0      & 0      & 0      & \cdots & 0      & \frac{\alpha_1 \rho_1}{2c} \\
        -\frac{\alpha_2 \rho_2}{2c} & 0      & 1      & \cdots & 0      & 0      & 0      & 0      & \cdots & 0      & \frac{\alpha_2 \rho_2}{2c} \\
        \vdots                      & \vdots & \vdots & \ddots & \vdots & \vdots & \vdots & \vdots & \ddots & \vdots & \vdots \\
        -\frac{\alpha_N \rho_N}{2c} & 0      & 0      & \cdots & 1      & 0      & 0      & 0      & \cdots & 0      & \frac{\alpha_N \rho_N}{2c} \\
        \frac{1}{2}                 & 0      & 0      & \cdots & 0      & 0      & 0      & 0      & \cdots & 0      & \frac{1}{2} \\
        0                           & 0      & 0      & \cdots & 0      & 1      & 0      & 0      & \cdots & 0      & 0 \\
        -\frac{\rho c}{2}           & 0      & 0      & \cdots & 0      & 0      & 0      & 0      & \cdots & 0      & \frac{\rho c}{2} \\
        0                           & 0      & 0      & \cdots & 0      & 0      & 1      & 0      & \cdots & 0      & 0 \\
        0                           & 0      & 0      & \cdots & 0      & 0      & 0      & 1      & \cdots & 0      & 0 \\
        \vdots                      & \vdots & \vdots & \vdots & \vdots & \vdots & \vdots & \vdots & \ddots & \vdots & \vdots \\
        0                           & 0      & 0      & \cdots & 0      & 0      & 0      & 0      & \cdots & 1      & 0
    \end{pmatrix} , \\
    \Rv_{\Av}^{-1} &= \begin{pmatrix}
        0      & 0      & \cdots & 0      & 1      & 0      & -\frac{1}{\rho c}                 & 0      & 0      & \cdots & 0 \\
        1      & 0      & \cdots & 0      & 0      & 0      & -\frac{\alpha_1 \rho_1}{\rho c^2} & 0      & 0      & \cdots & 0 \\
        0      & 1      & \cdots & 0      & 0      & 0      & -\frac{\alpha_2 \rho_2}{\rho c^2} & 0      & 0      & \cdots & 0 \\
        \vdots & \vdots & \ddots & \vdots & \vdots & \vdots & \vdots                            & \vdots & \vdots & \ddots & \vdots \\
        0      & 0      & \cdots & 1      & 0      & 0      & -\frac{\alpha_N \rho_N}{\rho c^2} & 0      & 0      & \cdots & 0 \\
        0      & 0      & \cdots & 0      & 0      & 1      & 0                                 & 0      & 0      & \cdots & 0 \\
        0      & 0      & \cdots & 0      & 0      & 0      & 0                                 & 1      & 0      & \cdots & 0 \\
        0      & 0      & \cdots & 0      & 0      & 0      & 0                                 & 0      & 1      & \cdots & 0 \\
        \vdots & \vdots & \ddots & \vdots & \vdots & \vdots & \vdots                            & \vdots & \vdots & \ddots & \vdots \\
        0      & 0      & \cdots & 0      & 0      & 0      & 0                                 & 0      & 0      & \cdots & 1 \\
        0      & 0      & \cdots & 0      & 1      & 0      & \frac{1}{\rho c}                  & 0      & 0      & \cdots & 0
    \end{pmatrix} , \quad
    \mathbf{\Lambda}_{\Av} = \begin{pmatrix}
        u - c  & 0      & 0      & \cdots & 0      & 0 \\
        0      & u      & 0      & \cdots & 0      & 0 \\
        0      & 0      & u      & \cdots & 0      & 0 \\
        \vdots & \vdots & \vdots & \ddots & \vdots & \vdots \\
        0      & 0      & 0      & \cdots & u      & 0 \\
        0      & 0      & 0      & \cdots & 0      & u + c
      \end{pmatrix} .
\end{split}
\end{equation}
$\Rv_{\Bv}$, $\Rv_{\Bv}^{-1}$ and $\mathbf{\Lambda}_\Bv$ have similar corresponding forms.

Consider the interpolation in the $x$ direction in a 1D domain, the projection matrices, $\Rv_{\Av,i+1/2}$ and $\Rv^{-1}_{\Av,i+1/2}$, at a midpoint $x_{i+1/2}$ are first approximated with the arithmetic averages of partial densities, mixture density and speed of sound at $x_{i}$ and $x_{i+1}$. All $\Vv_{i}$ in the stencil of interpolation are then transformed to characteristic variable vectors $\Uv_{i}$ with the projection matrix $\Rv^{-1}_{\Av,i+1/2}$:
\begin{equation}
    \Uv_{i} = \Rv^{-1}_{\Av,i+\half} \Vv_{i} .
\end{equation}
After the left-biased and right-biased interpolated values, $\tilde{\Uv}_L$ and $\tilde{\Uv}_R$, are obtained, $\tilde{\Uv}_L$ and $\tilde{\Uv}_R$ can be transformed to $\tilde{\Vv}_L$ and $\tilde{\Vv}_R$ using the projection matrix $\Rv_{\Av,i+1/2}$:
\begin{align}
    \tilde{\Vv}_L &= \Rv_{\Av,i+\half} \tilde{\Uv}_L , \\
    \tilde{\Vv}_R &= \Rv_{\Av,i+\half} \tilde{\Uv}_R .
\end{align}
The flux $\Gxtv_{i+1/2}$ and velocity $\tilde{u}_{i+1/2}$ at the midpoint
can then be obtained with $\tilde{\Vv}_L$ and $\tilde{\Vv}_R$ using the approximate Riemann solver.


%
%

\bibliographystyle{spmpscinat}
\bibliography{references.bib}



\end{document}